\documentclass[10pt,logo,copyright]{nvidiatechreport}
\usepackage[utf8]{inputenc} % allow utf-8 input
\usepackage[T1]{fontenc}    % use 8-bit T1 fonts

\makeatletter
\g@addto@macro{\UrlBreaks}{\UrlOrds}
\makeatother

\usepackage{parskip}        % no paragraph indents
\usepackage{url}            % simple URL typesetting
\usepackage{booktabs}       % professional-quality tables
\usepackage{amsfonts}       % blackboard math symbols
\usepackage{nicefrac}       % compact symbols for 1/2, etc.
\usepackage{microtype}      % microtypography
\usepackage{xcolor}         % colors
\usepackage{listings}       % code listings with syntax highlighting
\usepackage{graphicx}
\usepackage{animate}        % for 360 video in teaser
\usepackage{subcaption}
\usepackage{tabularx}
\usepackage{makecell}
\usepackage{adjustbox}
\usepackage{setspace}
\newcolumntype{M}[1]{>{\centering\arraybackslash}m{#1}}
\usepackage{float}
\usepackage{tikz}
\usepackage{amsmath,amsfonts,bm, bbm,leftindex}
\usepackage{multirow}
\usetikzlibrary{arrows.meta, positioning, fit}
\usepackage{wrapfig}
\usepackage[stable,bottom]{footmisc}

\usepackage[most]{tcolorbox}
\definecolor{nvidiagreen}{RGB}{118, 185, 0}
\newtcolorbox{notebox}{
  enhanced,
  breakable,
  colback=nvidiagreen!8,
  colframe=nvidiagreen,
  boxrule=0pt,
  leftrule=3pt,
  arc=0pt,
  outer arc=0pt,
  left=10pt,
  right=10pt,
  top=8pt,
  bottom=8pt,
  before skip=\baselineskip,
  after skip=\baselineskip,
  fonttitle=\bfseries,
  coltitle=nvidiagreen!80!black,
  attach title to upper={\par\smallskip},
  title={Note}
}

\newtcolorbox{asciibox}[1][]{
  enhanced,
  breakable,
  colback=gray!5,
  colframe=gray!40,
  boxrule=0.5pt,
  arc=2pt,
  left=6pt,
  right=6pt,
  top=6pt,
  bottom=6pt,
  before skip=0.8\baselineskip,
  after skip=0.8\baselineskip,
  fontupper=\small\ttfamily,
  #1
}

\lstdefinestyle{asciiStyle}{
  basicstyle=\small\fontfamily{pcr}\selectfont,
  breaklines=false,
  columns=fullflexible,
  keepspaces=true,
  frame=single,
  framerule=0.5pt,
  rulecolor=\color{gray!40},
  backgroundcolor=\color{gray!5},
  xleftmargin=8pt,
  xrightmargin=8pt,
  framexleftmargin=6pt,
  framexrightmargin=6pt,
  framesep=6pt,
  aboveskip=0.8\baselineskip,
  belowskip=0.8\baselineskip,
}

\lstnewenvironment{asciiart}{\lstset{style=asciiStyle}}{}

\def\figref#1{figure~\ref{#1}}
\def\Figref#1{Figure~\ref{#1}}

\def\secref#1{section~\ref{#1}}
\def\eqref#1{equation~\ref{#1}}
\def\1{\bm{1}}

\DeclareMathAlphabet{\mathsfit}{\encodingdefault}{\sfdefault}{m}{sl}
\SetMathAlphabet{\mathsfit}{bold}{\encodingdefault}{\sfdefault}{bx}{n}
\usepackage{graphicx} % Required for inserting images
\usepackage{amsmath} % For math symbols if needed
\usepackage{amssymb}
\usepackage{comment}
\usepackage{hyperref} % For hyperlinks if needed
\usepackage{booktabs}
\usepackage{array, booktabs, tabularx}
\usepackage{threeparttable}
\usepackage{multirow}
\usepackage{xcolor}

\titleformat{\section}{\Large\bfseries}{\thesection.}{0.5em}{#1}
\titlespacing*{\section}{0pt}{3.5ex plus1ex minus.2ex}{2.3ex plus.2ex}

\titleformat{\subsection}{\large\bfseries}{\thesubsection.}{0.5em}{#1}
\titlespacing*{\subsection}{0pt}{3ex plus1ex minus.2ex}{1.5ex plus.2ex}

\titleformat{\subsubsection}{\fontsize{11}{13}\selectfont\bfseries}{\thesubsubsection.}{0.5em}{#1}
\titlespacing*{\subsubsection}{0pt}{2.5ex plus1ex minus.2ex}{1ex plus.2ex}

\titleformat{\paragraph}[runin]{\normalsize\bfseries\itshape}{\theparagraph}{1em}{#1}
\titlespacing*{\paragraph}{0pt}{2ex plus.5ex minus.2ex}{1em}

\title{Scalable Training of Mixture-of-Experts Models with Megatron Core\texorpdfstring{\\[0pt]{\large Technical Report}}{}}
\author{{\vspace{-10pt}\bfseries\large NVIDIA}\footnote{For the complete list of authors, please refer to the Contributions and Acknowledgments section. Corresponding authors: \texttt{\{zijiey, juney, jiajiey\}@nvidia.com}.}}

\date{\today} % Or specify a date like December 2024

\renewcommand{\absfont}{\linespread{1.15}\fontsize{10}{12}\selectfont}

\renewcommand{\copyrightext}{}

\renewcommand{\footrule}{}

\begin{document}
% Header shows only main title (without "Technical Report" subtitle)
\fancyhead[C]{\footerfont Scalable Training of Mixture-of-Experts Models with Megatron Core}
\maketitle
% Add copyright as unnumbered footnote in the same footnote area
\vspace{-15pt}
\begingroup\renewcommand\thefootnote{}\footnotetext{\par\vspace*{2pt}\textcopyright\, \the\year{} NVIDIA. All rights reserved.}\endgroup
% \vspace{-5pt}
\begin{abstract}
Scaling Mixture-of-Experts (MoE) training introduces systems challenges absent in dense models. Because each token activates only a subset of experts, this sparsity allows total parameters to grow much faster than per-token computation, creating coupled constraints across memory, communication, and computation. Optimizing one dimension often shifts pressure to another, demanding co-design across the full system stack.

\vspace{0.5em}
We address these challenges for MoE training through integrated optimizations spanning memory (fine-grained recomputation, offloading, etc.), communication (optimized dispatchers, overlapping, etc.), and computation (Grouped GEMM, fusions, CUDA Graphs, etc.). The framework also provides Parallel Folding for flexible multi-dimensional parallelism, low-precision training support for FP8 and NVFP4, and efficient long-context training. On NVIDIA GB300 and GB200, it achieves 1,233/1,048 TFLOPS/GPU for DeepSeek-V3-685B and 974/919 TFLOPS/GPU for Qwen3-235B. As a performant, scalable, and production-ready open-source solution, it has been used across academia and industry for training MoE models ranging from billions to trillions of parameters on clusters scaling up to thousands of GPUs.

\vspace{0.5em}
This report explains how these techniques work, their trade-offs, and their interactions at the systems level, providing practical guidance for scaling MoE models with Megatron Core.
\end{abstract}
\abscontent%
% Graphical Abstract - Visual overview of Megatron-Core MoE
\vspace{1ex}
\begin{figure}[H]
    \centering
    \vspace{-0ex}
    \includegraphics[width=0.93\textwidth,trim=0 0pt 0 2pt,clip]{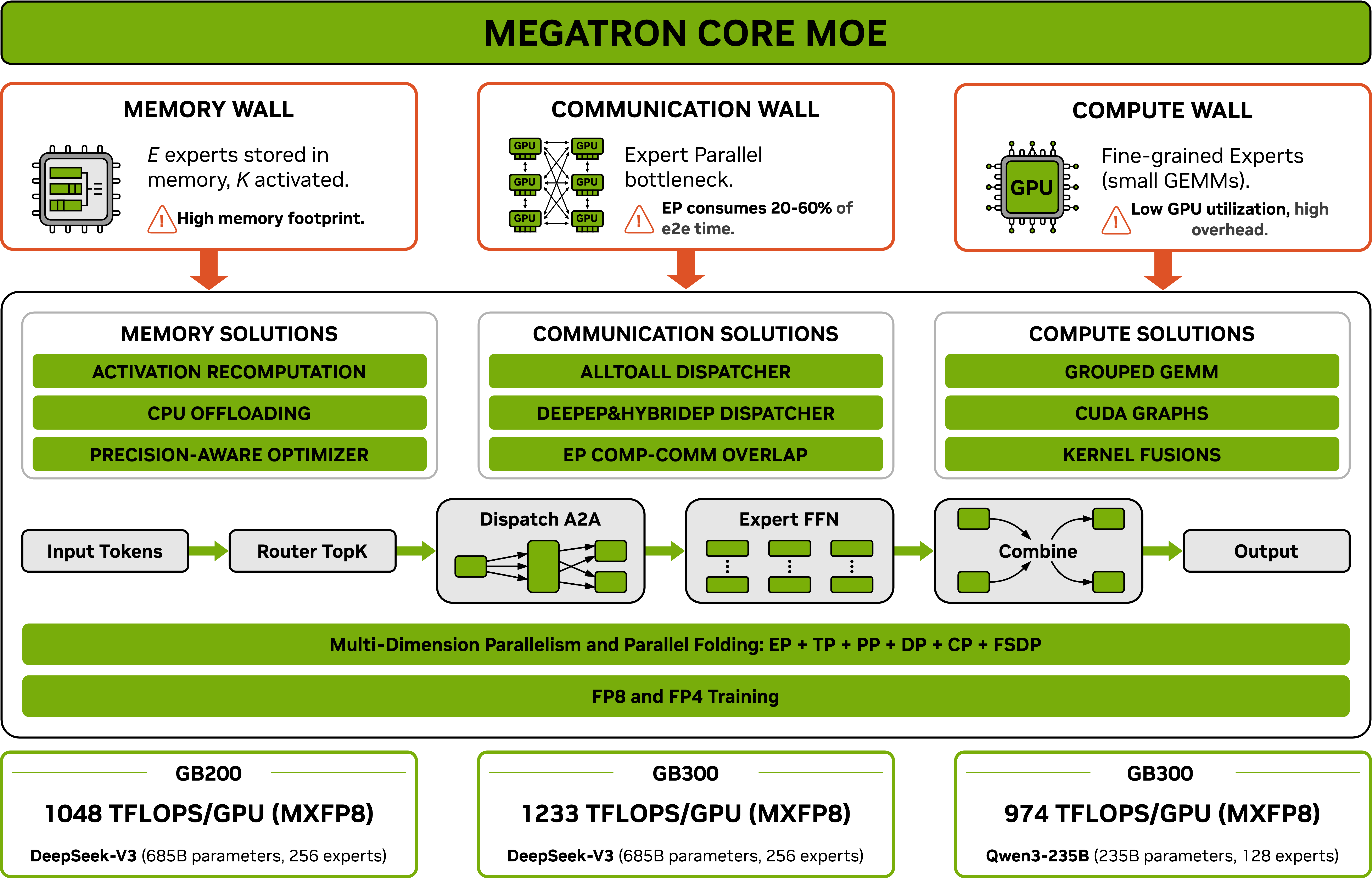}
    \label{fig:graphical-abstract}
\end{figure}
\vspace{2ex}
\tableofcontents % Add a table of contents
\newpage

\section{Introduction}

Training dense transformers at scale requires computation that grows linearly with model size \cite{kaplan2020scaling,hoffmann2022chinchilla}. Mixture of Experts (MoE) models follow a different scaling pattern: by routing each token to a selected subset of expert networks rather than activating all parameters, per-token computation grows sub-linearly with model size \cite{shazeer2017,fedus2022switchtransformersscalingtrillion}. Recent MoE models have demonstrated order-of-magnitude reductions in training cost relative to quality-matched dense models \cite{jiang2024mixtralexperts,deepseekai2025deepseekv3technicalreport,Cai_2025}.

However, training MoE at scale introduces systems challenges that dense-model frameworks were not designed for. This report presents Megatron-Core MoE, the MoE training stack within Megatron-Core~\cite{shoeybi2020megatronlmtrainingmultibillionparameter}, covering the architecture, parallelism strategies, and system optimizations required to train trillion-parameter-class MoE models at high throughput.

\subsection{Mixture of Experts}

A Mixture of Experts (MoE) model augments a standard neural network with a collection of specialized sub-networks called \emph{experts}, together with a lightweight \emph{router} (or gating network) that dynamically selects which experts process each input \cite{shazeer2017,jacobs1991adaptive,eigen2013learning}. In the context of Transformer-based language models \cite{vaswani2017attention}, MoE layers typically replace the dense Feed-Forward Network (FFN) blocks: instead of a single FFN applied to all tokens, an MoE layer contains multiple FFN experts, and each token is routed to a small subset (e.g., top-$k$) of these experts based on learned routing weights.

Formally, given an input token representation $\mathbf{x}$, the router computes a probability distribution over $E$ experts:
\[
\mathbf{p}(\mathbf{x}) = \text{Softmax}(\mathbf{W}_r \mathbf{x})
\]
The output of the MoE layer is then computed as a weighted combination of the selected experts' outputs:
\[
\text{MoE}(\mathbf{x}) = \sum_{i \in \text{TopK}(\mathbf{p}(\mathbf{x}))} p_i(\mathbf{x}) \cdot E_i(\mathbf{x})
\]
where $E_i$ denotes the $i$-th expert network. This architecture offers three key advantages: model capacity can grow independently of computational cost by adding more experts (\emph{scalable capacity}); only a fraction of parameters activate per token, reducing FLOPs relative to a dense model of equivalent size (\emph{computational efficiency}); and different experts can specialize on different input types (\emph{specialization}). Appendix~\ref{sec:notation} provides a notation reference for all symbols and abbreviations used in this report.

While the concept dates back to the early 1990s \cite{jacobs1991adaptive,eigen2013learning}, the integration of MoE with modern Transformers has driven renewed interest. GShard pioneered distributed MoE training at scale, introducing Expert Parallelism and load-balancing auxiliary losses \cite{lepikhin2020gshard}. Switch Transformer demonstrated that MoE could scale to trillion parameters while maintaining training stability \cite{fedus2022switchtransformersscalingtrillion}. GLaM showed that MoE could match dense model quality at a fraction of the training cost \cite{du2022glamefficientscalinglanguage}. Frameworks including Tutel \cite{hwang2023tutel} and DeepSpeed-MoE \cite{rajbhandari2022deepspeedmoe} further advanced MoE training systems.

MoE adoption has accelerated across research and industry. Mixtral-8x7B showed that open-weight MoE models can match proprietary dense models while reducing inference cost \cite{jiang2024mixtralexperts}. DeepSeek-V2 and DeepSeek-V3 extended this with fine-grained expert architectures, using hundreds of small experts to maximize the capacity-to-compute ratio \cite{dai2024deepseekmoe,deepseekai2024deepseekv2strongeconomicalefficient,deepseekai2025deepseekv3technicalreport}. NVIDIA's Nemotron-3 family~\cite{nvidia2025nemotron3} adopts a hybrid Mamba-Transformer MoE architecture with LatentMoE~\cite{elango2026latentmoe}, trained at scale using Megatron-Core. Scaling law studies confirm that fine-grained MoE achieves favorable compute-optimal trade-offs \cite{krajewski2024scaling,ludziejewski2025jointmoescalinglaws}, accelerating this trend---while amplifying the systems challenges it creates.

\subsection{Challenges in Training Large-Scale MoE Models}\label{sec:challenges}

As models push toward hundreds of experts with smaller individual capacity, the systems challenges of MoE training grow in proportion. These challenges stem from a single root cause: MoE's \textbf{sparsity}---which manifests as a \emph{Parameter-Compute Mismatch} (total parameters far exceeding active computation, this section) and a \emph{Dense-Sparse Mismatch} (attention and MoE layers requiring conflicting parallelism configurations, Section~\ref{parallelism-strategies}).

\textbf{The core asymmetry comes from sparsity.} In a dense transformer, every parameter participates in every training step. A model with $N_{\text{total}}$ parameters requires roughly $6 N_{\text{total}}$ FLOPs per token (forward and backward combined), so parameters and per-token computation scale in lockstep. Splitting the model across more GPUs usually also splits computation in the same proportion, keeping each GPU busy enough that communication overhead stays small.

For MoE, sparsity causes this coupling to break: only $K$ of $E$ total experts activate per token, so per-token computation is roughly $6 N_{\text{active}}$ rather than $6 N_{\text{total}}$, where $N_{\text{active}}$ scales with $K$ while $N_{\text{total}}$ scales with $E$, and $K \ll E$. This creates a fundamental parameter-compute mismatch: compared with a dense model matched on active parameters per token, an MoE model has far more total parameters, often by an order of magnitude. DeepSeek-V3 illustrates this concretely: 685B total parameters but only 37B active per token, an 18$\times$ gap.

With so little computation per token, model partitioning requires more care than for dense models---naively sharding expert matrices (as Tensor Parallelism would) fragments already-small computations, making them even less efficient. Because MoE experts are independent networks, the natural strategy is \emph{Expert Parallelism} (EP): placing different experts on different GPUs, preserving full-size expert GEMMs. EP introduces all-to-all communication to route tokens to their assigned GPUs (Section~\ref{parallelism-strategies} details why EP is preferred and how Parallel Folding addresses the resulting challenges). This parameter-compute mismatch and EP's communication demands together create three tightly coupled challenges---the \emph{Three Walls}---that constrain every MoE training step:

\textbf{The Memory Wall.} All $E$ experts' parameters, gradients, and optimizer states must reside in memory during training, even though only $K$ activate per token. This creates memory pressure far exceeding that of a dense model with comparable per-token compute~\cite{fedus2022switchtransformersscalingtrillion,Cai_2025}. Reducing this pressure requires spending elsewhere: distributing parameters across more devices costs \emph{communication bandwidth}; recomputing activations instead of storing them costs \emph{extra computation}; offloading to host memory costs \emph{PCIe bandwidth}. Dynamic routing further complicates matters: uneven token distributions cause unpredictable memory spikes when some experts receive disproportionate load~\cite{lepikhin2020gshard,wang2024auxiliarylossfreeloadbalancingstrategy}.

\textbf{The Communication Wall.} EP requires all-to-all collectives to dispatch tokens to their assigned experts and collect results~\cite{lepikhin2020gshard}. The per-GPU send volume in each all-to-all is approximately $T \cdot K \cdot h \cdot \frac{EP-1}{EP}$, where $T$ is the local token count, $K$ the top-$k$, and $h$ the hidden dimension; a full dispatch-and-combine cycle doubles this. As EP grows, this volume saturates but the communication increasingly moves from high-bandwidth intra-node links (e.g., NVLink) to narrower inter-node interconnects, where available bandwidth drops by an order of magnitude~\cite{deepep2025}. The sparse activation pattern, meanwhile, provides limited computation to overlap with this communication. In architectures like DeepSeek-V3, where experts span multiple nodes, unoptimized all-to-all can consume up to 60\% of total training time.

\textbf{The Compute Efficiency Wall.} MoE introduces computational inefficiencies absent in dense models:
\begin{itemize}[nosep,leftmargin=*]
    \item \textbf{Small GEMMs.} Fine-grained experts produce many small matrix multiplications that underutilize GPU compute units~\cite{megablocks}. In our measurements, GEMMs account for $\sim$70\% of execution time in Llama-3 405B (dense) but under 50\% in DeepSeek-V3 (MoE). The remainder is consumed by operations that scale with tensor count rather than FLOP count.
    \item \textbf{Routing and permutation overhead.} Token routing and permutation, absent in dense models, add $\sim$9\% to layer execution time even after optimization.
    \item \textbf{Load imbalance.} Dynamic routing assigns uneven token counts to experts, leaving some overloaded while others sit idle, wasting compute capacity~\cite{wang2024auxiliarylossfreeloadbalancingstrategy}.
    \item \textbf{Host overhead.} MoE launches more kernels for the same amount of FLOPs because of sparsity and routing, and each launch carries fixed host-side cost, these add up and leave the GPU idle between kernels. In dropless MoE, dynamic tensor shapes further require costly host-device synchronization.
\end{itemize}

These three walls are tightly coupled: optimizing one often shifts pressure to another. Increasing batch size improves GEMM utilization but amplifies memory pressure and communication volume. CUDA Graphs eliminate host overhead but require static tensor shapes, conflicting with dropless routing. Grouping tokens across experts improves compute efficiency but complicates load balancing. Section~\ref{parallelism-strategies} provides the detailed parallelism analysis underlying EP and Parallel Folding; Section~\ref{key-features-optimizations} presents Megatron-Core's integrated approach to addressing all three walls while managing their interactions.

\subsection{Megatron-Core MoE}
Built within Megatron-Core, a PyTorch-based library for large-scale transformer training~\cite{shoeybi2020megatronlmtrainingmultibillionparameter,paszke2019pytorchimperativestylehighperformance}, this MoE training stack tackles all three walls simultaneously:

\textbf{Multi-Dimensional Parallelism.} Expert Parallelism (EP) integrates with tensor, pipeline, sequence, and data parallelism. MoE Parallel Folding \cite{liu2025moeparallelfoldingheterogeneous} decouples attention and MoE layer configurations, breaking the traditional $\text{EP} \leq \text{DP}$ constraint and enabling configurations tailored to specific model architectures and hardware topologies.

\textbf{Memory Optimizations.} Fine-grained activation recomputation, memory-efficient permutation, precision-aware optimizers, and activation offloading reduce memory footprint without sacrificing throughput \cite{korthikanti2022reducingactivationrecomputationlarge,a2023learningwithdistributedoptimization}. Comprehensive reduced-precision training (FP8 and FP4) support in expert GEMMs, activations, and communication further reduces activation storage while maintaining convergence through selective precision strategies.

\textbf{Communication Optimizations.} High-performance token dispatchers (DeepEP, HybridEP) maximize bandwidth utilization. Communication-computation overlap hides all-to-all latency behind expert computation.

\textbf{Compute Optimizations.} Grouped GEMM kernels, kernel fusion, CUDA Graphs, and sync-free execution address the computational fragmentation inherent in fine-grained MoE architectures.

\textbf{Production Features.} Load balancing strategies, token dropping with capacity control, distributed optimizer and FSDP support, distributed checkpointing with flexible resharding, and upcycling from dense checkpoints enable deployment at scale.
The MoE stack's modular design enables rapid experimentation, while its production-grade optimizations support training from research prototypes to trillion-parameter models~\cite{nvidiamcoremoeuserguide}. This report explains what the stack provides, why key design decisions were made, and how they address MoE training challenges, with practical guidance for configuration and tuning.

\subsection{Structure of This Paper}

The remainder of this report follows a progression from architecture to optimization to evaluation:

\begin{itemize}

\item \textbf{Section~\ref{meg-core-architecture}: Megatron-Core MoE Architecture.}
Introduces Megatron-Core MoE's design in two parts: the internal design of the MoE layer itself (router, token dispatcher, experts) and the four-stage forward pass (route, dispatch, compute, combine), followed by the external design covering integration with the transformer model, parallel process group organization, and optimizer handling for expert parameters.

\item \textbf{Section~\ref{parallelism-strategies}: Scaling MoE with Parallel Folding and Multi-Dimensional Parallelism.}
Examines how MoE's sparsity breaks the parallelism assumptions of dense training, why Expert Parallelism (EP) is needed alongside traditional strategies, and how the resulting dense-sparse mismatch between attention and MoE layers is resolved by MoE Parallel Folding, which decouples their parallelism configurations for flexible, efficient mapping at trillion-parameter scale.

\item \textbf{Section~\ref{key-features-optimizations}: Breaking the Memory, Communication, and Compute Efficiency Walls.}
Presents Megatron-Core MoE's solutions to the three fundamental barriers: the Memory Wall (activation management, recomputation strategies, offloading, distributed parameter storage), the Communication Wall (optimized dispatchers including DeepEP and HybridEP, communication-computation overlap), and the Compute Efficiency Wall (Grouped GEMM, kernel fusion, CUDA Graphs, sync-free execution for dropless MoE).

\item \textbf{Section~\ref{sec:reduced-precision-training}: Reduced-Precision Training in FP8/FP4 for MoE.}
Covers reduced-precision training as a cross-cutting optimization that simultaneously impacts all three walls by reducing activation memory, halving communication volume, and accelerating Tensor Core GEMMs, while presenting strategies for selective precision to maintain training stability.

\item \textbf{Section~\ref{sec:long-context-training}: Long-Context MoE Training.}
Examines how long-context scenarios (16K to 64K+ tokens) fundamentally shift the optimization landscape as attention computation dominates, and presents techniques for managing activation memory growth through Context Parallelism and Tensor Parallelism scaling.

\item \textbf{Section~\ref{sec:production-features}: Production Features.}
Describes operational features for production training: load balancing and token dropping for stable training, distributed checkpointing for parallelism-agnostic resharding, upcycling from dense checkpoints, and integration with multi-token prediction.

\item \textbf{Section~\ref{performance-evaluations}: Performance Evaluation.}
Validates the framework's effectiveness through empirical benchmarks on DeepSeek-V3 and Qwen3-235B across GB200 and H100 platforms, demonstrating the impact of the full optimization stack.

\item \textbf{Section~\ref{sec:best-practices}: Performance Best Practices with DeepSeek-V3 Case Study.}
Provides a systematic workflow for identifying optimal parallelism configurations, validated through a detailed DeepSeek-V3 case study that demonstrates how the optimizations work together to achieve state-of-the-art performance.

\item \textbf{Section~\ref{sec:rl}: Megatron-Core MoE in Reinforcement Learning.}
Addresses the emerging RL post-training paradigm, covering challenges unique to RL workloads (variable sequence lengths, memory offloading, online weight export), Megatron-Bridge integration with popular RL frameworks, and RL-specific optimizations including packed sequence support, dynamic context parallelism, and router replay.

\end{itemize}

\section{Megatron-Core MoE Architecture}\label{meg-core-architecture}

This section presents Megatron-Core's MoE implementation architecture in two parts. We first describe the \textbf{internal design} of the MoE layer itself: its modular components (router, token dispatcher, experts) and the four-stage forward pass that transforms input tokens into output representations. We then examine the \textbf{external design}: how parallel process groups are organized to support distributed execution and how the optimizer handles expert parameters differently from dense layers.

\subsection{MoE Layer Architecture and Forward Pass}\label{sec:moe-layer-architecture}

An MoE layer replaces the dense feed-forward network (FFN) in a transformer block with a collection of expert FFNs, only a subset of which process each token. Megatron-Core implements this through three modular components (a \emph{router} for token-to-expert assignment, a \emph{token dispatcher} for cross-GPU communication, and \emph{experts} for computation) connected through the four-stage forward pass shown in \figref{fig:moe-forward-pass}. This separation of concerns enables independent optimization: the router can be fused into CUDA Graphs without affecting dispatcher logic, dispatchers can be swapped between all-to-all and DeepEP without modifying expert computation, and expert implementations can use different GEMM backends transparently.

\subsubsection{Forward Pass: Route, Dispatch, Compute, Combine}

The MoE layer processes input tokens through four sequential stages:

% Fix the typo in this figure
\begin{figure}[ht]
\centering
\includegraphics[width=0.9\textwidth]{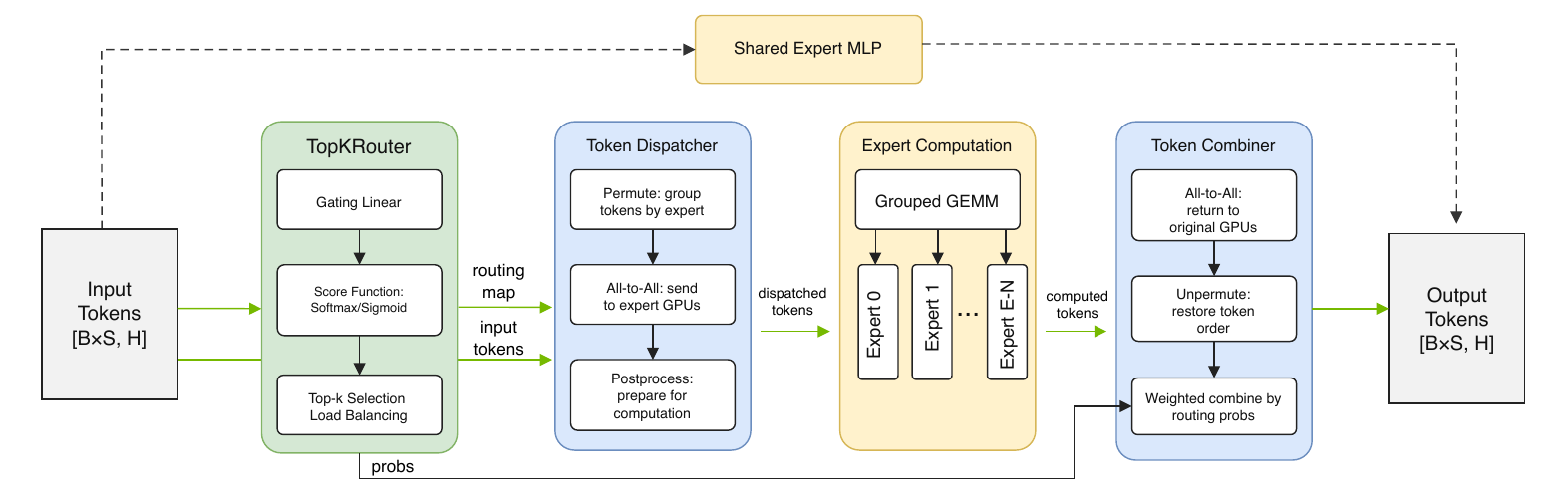} 
\caption{Data flow through an MoE layer: Route, Dispatch, Compute, and Combine stages.}
\label{fig:moe-forward-pass}
\end{figure}

\textbf{Stage 1: Route.} The \texttt{TopKRouter} determines which experts process each token. A learned linear projection maps each token's hidden state to $E$ logits (one per expert), a score function (softmax or sigmoid) converts logits to probabilities, and top-$k$ selection identifies the highest-scoring experts for each token. The router outputs two tensors: \texttt{probs} containing the routing weights, and \texttt{routing\_map}, a boolean mask indicating token-expert assignments. For numerical stability with many experts, the router can operate in FP32 via \texttt{-{}-moe-router-dtype fp32}.

\textbf{Stage 2: Dispatch.} The token dispatcher prepares tokens for cross-GPU communication. It first permutes tokens so that all tokens destined for the same expert are contiguous; this permutation is essential for efficient dense GEMM on the expert side. The dispatcher then moves tokens to the GPUs hosting their assigned experts using one of three backends: AllGather (simple but memory-intensive), all-to-all (standard NCCL-based), or Flex (supporting optimized backends like DeepEP and HybridEP).

\textbf{Stage 3: Expert Computation.} Each GPU executes its local experts on the received tokens. All local experts run in a single Grouped GEMM call via \texttt{TEGroupedMLP}, maximizing GPU utilization even when individual expert workloads are small.

\textbf{Stage 4: Combine.} The inverse communication returns processed tokens to their original GPUs, followed by unpermutation to restore the original sequence order. If a shared expert is configured, its output (optionally computed in parallel with routed experts) is added at this stage.

\subsubsection{Router: Token-to-Expert Assignment}

The router transforms a global token batch into expert-specific workloads through two operations \cite{shazeer2017}:

\begin{enumerate}
    \item \textbf{Gating:} A linear projection $\mathbf{W}_r \in \mathbb{R}^{h \times E}$ maps each token's hidden state $\mathbf{x} \in \mathbb{R}^h$ to logits $\mathbf{l} = \mathbf{W}_r^\top \mathbf{x} \in \mathbb{R}^E$.

    \item \textbf{Top-$k$ Selection:} A score function converts logits to probabilities: softmax ($p_i = e^{l_i} / \sum_j e^{l_j}$) or sigmoid ($p_i = \sigma(l_i) / \sum_j \sigma(l_j)$, used by DeepSeek-V3 \cite{deepseekai2025deepseekv3technicalreport}). The top-$k$ experts with highest probabilities are selected per token.
\end{enumerate}

\begin{figure}[ht]
    \centering
    \includegraphics[width=0.85\textwidth]{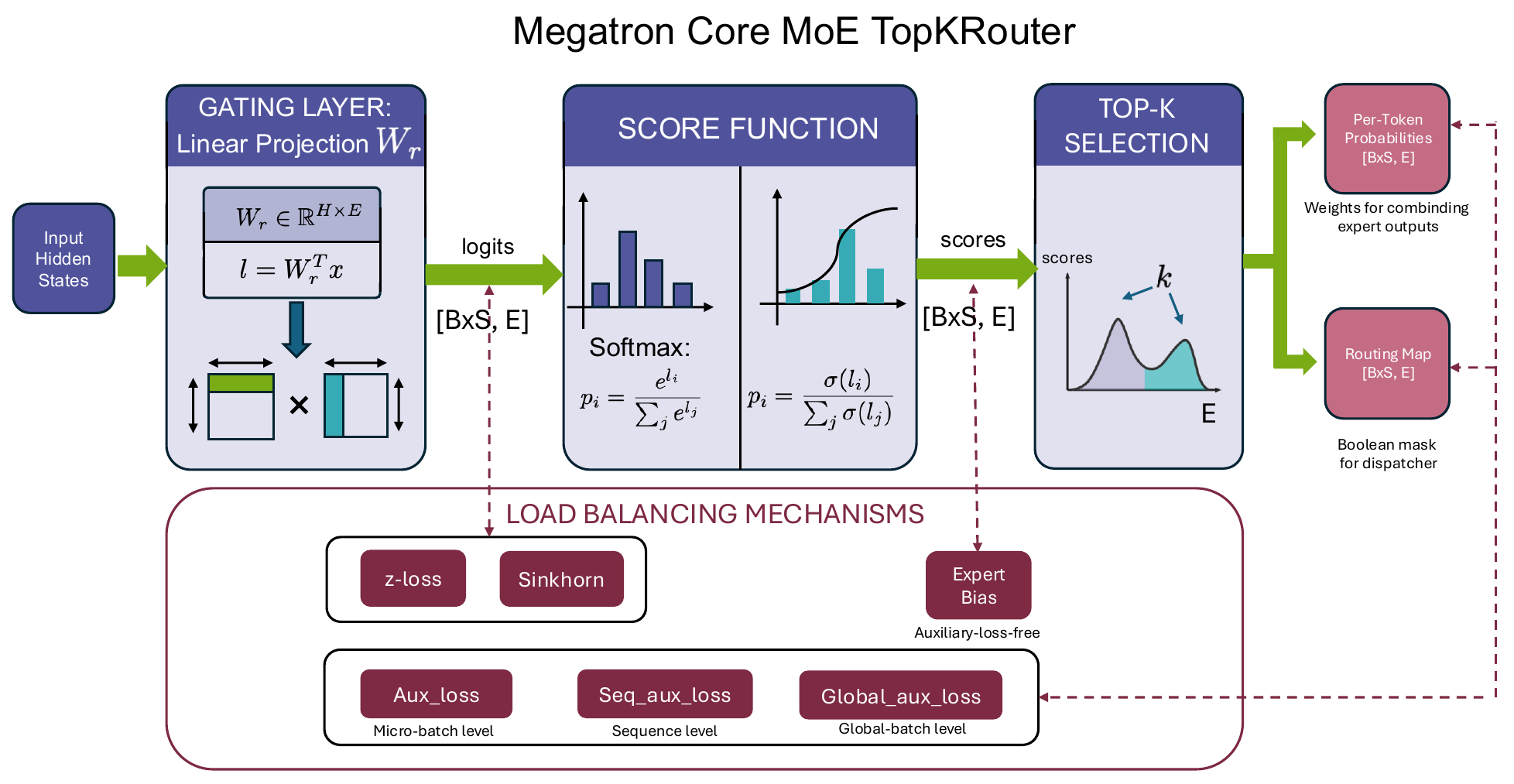}
    \caption{Router architecture: linear projection, score function, top-$k$ selection, and load balancing.}
    \label{fig:router-architecture}
\end{figure}

Uneven expert utilization degrades both training efficiency and model quality. As shown in \figref{fig:router-architecture}, Megatron-Core supports multiple load balancing strategies; these are discussed in Section~\ref{sec:production-features}.

\subsubsection{Token Dispatcher: Communication Abstraction}

The dispatcher manages token movement between GPUs through a six-phase pipeline: \texttt{dispatch\_preprocess} $\rightarrow$ \texttt{token\_dispatch} $\rightarrow$ \texttt{dispatch\_postprocess} (forward), and \texttt{combine\_preprocess} $\rightarrow$ \texttt{token\_combine} $\rightarrow$ \texttt{combine\_postprocess} (backward).

\textbf{Communication Backends.} Three dispatcher types are available:
\begin{itemize}
    \item \textbf{AllGather} (\texttt{allgather}): Each GPU gathers all tokens and filters for local experts. Simple but memory-intensive; suitable for small EP sizes.
    \item \textbf{All-To-All} (\texttt{all-to-all}): Standard NCCL-based point-to-point communication \cite{lepikhin2020gshard}. Each GPU sends only the tokens needed by each destination. Scales well but incurs synchronization overhead.
    \item \textbf{Flex} (\texttt{flex}): Unified design supporting DeepEP (only high-throughput kernels are integrated, hiding NVLink communication latency through overlap) \cite{deepep2025} and HybridEP (high-bandwidth MoE communication kernels that deliver better performance on NVLink-rich topologies like NVL72).
\end{itemize}

\subsubsection{Experts: Computation Module}

Each expert is a two-layer MLP with optional gating (for SwiGLU/GeGLU activations \cite{shazeer2020glu}). Grouped GEMM enables efficient batching of multiple expert computations \cite{megablocks,hwang2023tutel}. Two implementations are available:

\begin{itemize}
    \item \textbf{TEGroupedMLP}: Transformer Engine \cite{nvidia_transformer_engine} optimized implementation supporting FP8 and FP4 quantization.
    \item \textbf{SequentialMLP}: Executes experts one at a time in a loop. Useful for debugging but significantly slower.
\end{itemize}

\textbf{Shared Experts.} Some architectures (DeepSeek-V2/V3 \cite{deepseekai2024deepseekv2strongeconomicalefficient,deepseekai2025deepseekv3technicalreport}, Qwen \cite{qwen2025qwen25technicalreport}) include a shared expert that processes all tokens regardless of routing. Shared expert computation can run in parallel with the dispatch-compute-combine pipeline, hiding its latency. See Section~\ref{sec:production-features} for details.

\subsection{System Integration: Parallelism and Optimizer}\label{sec:moe-system-integration}

Having described how an MoE layer works internally, we now examine how it integrates into the distributed training system: parallel process groups and optimizer handling.

\subsubsection{Parallel Group Management}

MoE layers require distinct process groups from dense layers because different components have different communication patterns. Megatron-Core organizes these groups through \texttt{ProcessGroupCollection}:

\begin{lstlisting}
ProcessGroupCollection
|-- Attention Layer Groups: tp, cp, dp, pp
|-- Expert Layer Groups:    ep, expt_tp, expt_dp, pp
\end{lstlisting}

Each MoE component uses specific groups based on its communication requirements:

\begin{table}[ht]
\centering
\begin{tabular}{lll}
\toprule
\textbf{Component} & \textbf{Groups Used} & \textbf{Reason} \\
\midrule
Router & \texttt{tp}, \texttt{cp}, \texttt{tp\_cp} & Weights duplicated across EP ranks \\
Token Dispatcher & \texttt{ep}, \texttt{tp\_ep} & all-to-all across expert ranks \\
Experts & \texttt{ep}, \texttt{expt\_tp}, \texttt{expt\_dp} & Sharded across EP; gradients reduced in EDP \\
Shared Experts & \texttt{tp} & Same as dense MLP \\
\bottomrule
\end{tabular}
\caption{MoE component to process group mapping.}
\label{tab:moe-process-groups}
\end{table}

This separation enables \textbf{Parallel Folding} (Section~\ref{sec:parallel-folding}): attention and MoE layers can use different TP/DP configurations. For example, attention layers might use TP=4 while MoE layers use ETP=1 with higher EP, optimizing each layer type independently.

\subsubsection{Optimizer and Gradient Handling}

Expert parameters require distinct handling in distributed optimization. Megatron-Core uses a \texttt{Chained\-Optimizer} that wraps separate optimizers for dense and expert parameters. Three key design decisions support correct MoE optimization:

\begin{enumerate}
    \item \textbf{Parameter Identification.} Expert parameters are marked with \texttt{allreduce=False}, distinguishing them from dense parameters that use standard data-parallel gradient reduction.

    \item \textbf{Separate Reduction Groups.} Dense layers reduce gradients across \texttt{dp\_cp\_group} (full data parallelism), while experts reduce across \texttt{expt\_dp\_group} (expert data parallelism). This ensures gradients are averaged over the correct number of replicas.

    \item \textbf{Gradient Scaling.} Expert gradients are scaled by \texttt{edp\_size / dp\_size} to account for the different effective batch sizes seen by experts (which process routing-dependent token subsets) versus dense layers.
\end{enumerate}

This design allows ZeRO-style optimizer state sharding \cite{rajbhandari2020zero} to work seamlessly with MoE: optimizer states for expert parameters are sharded across the EP group, while states for dense parameters follow standard DP sharding.

With the architecture established, Section~\ref{parallelism-strategies} addresses how to distribute it across devices, and Section~\ref{key-features-optimizations} tackles the memory, communication, and compute efficiency challenges that arise at scale.

\section{Scaling MoE: Parallel Folding and Multi-Dimensional Parallelism}\label{parallelism-strategies}

Section~\ref{sec:challenges} established that MoE's sparsity decouples model size from per-token computation, creating a parameter-compute mismatch that makes Expert Parallelism (EP) necessary but introduces all-to-all communication. This section examines how that mismatch concretely affects parallelism design. We first review the parallelism baseline for dense models and the trade-offs it entails, then show how MoE's sparsity breaks the assumptions underlying these strategies. Finally, we present \textbf{MoE Parallel Folding}, Megatron-Core's solution to the resulting \emph{dense-sparse mismatch}: attention and MoE layers have conflicting optimal parallelism configurations, and Parallel Folding decouples their mappings so each can use its optimal topology.

%==============================================================================
% 3.1 Opening: Why Parallelism, and Why MoE is Different
%==============================================================================
\subsection{Why Parallelism, and Why MoE is Different}\label{sec:why-parallelism}

Before examining MoE-specific challenges, we establish the fundamental reasons parallelism is necessary for large model training and the trade-offs it introduces.

\subsubsection{Why Large Model Training Needs Parallelism}

\textbf{Memory is the fundamental constraint.} A single GPU has limited memory, but training a large model requires storing model parameters, optimizer states, gradients, and activations simultaneously \cite{rasley2020deepspeed}. Consider Llama-405B \cite{grattafiori2024llama3herdmodels,touvron2023llama,touvron2023llama2} trained with BF16 precision and Adam optimizer:

\begin{center}
\begin{tabular}{lr}
\toprule
\textbf{Component} & \textbf{Memory (Llama-405B, BF16)} \\
\midrule
Model parameters & $\sim$810 GB \\
Optimizer states (Adam) & $\sim$4860 GB \\
Gradients & $\sim$1620 GB \\
Activations (8K sequence) & $\sim$5575 GB \\
\midrule
\textbf{Total} & \textbf{$\sim$12865 GB} \\
\bottomrule
\end{tabular}
\end{center}

This far exceeds any single GPU's capacity. Multiple GPUs are not optional; they are \emph{required} to hold the model.

\textbf{Compute throughput is the efficiency reason.} Beyond memory, a single GPU has limited compute capacity. Large model training requires astronomical FLOPs; aggregating multiple GPUs increases throughput and reduces wall-clock training time from years to weeks.

\subsubsection{The Trade-off of Parallelism}

Parallelism is not free. Every parallelism strategy introduces overhead:

\begin{itemize}
    \item \textbf{Communication overhead}: Data exchange between GPUs consumes time and bandwidth.
    \item \textbf{Synchronization}: Fast GPUs must wait for slow GPUs at synchronization points.
    \item \textbf{Pipeline bubbles}: Pipeline parallelism introduces idle time at pipeline boundaries.
    \item \textbf{Reduced compute intensity}: Sharding reduces per-GPU matrix sizes, lowering GEMM efficiency.
\end{itemize}

The result: Model FLOP Utilization (MFU) is heavily influenced by parallelism strategy, and effective parallelism design can minimize the gap between ideal and actual MFU.

\textbf{Key insight for dense models}: The ``cost'' and ``benefit'' of parallelism scale proportionally. More parameters require more GPUs, but more parameters also mean more computation per forward-backward pass. Because computation grows with model size, communication takes a smaller share of each step, keeping MFU relatively stable as models scale.

MoE's sparsity disrupts this balance. Because only $K$ of $E$ experts activate per token, total parameters scale with $E$ while per-token compute scales only with $K$. More GPUs are needed for memory, but per-token computation does not grow to match, leaving communication overhead exposed. The following section examines how this asymmetry breaks the assumptions underlying traditional parallelism strategies and why a new parallelism dimension is needed.

%==============================================================================
% 3.2 The Challenge of MoE Parallelism
%==============================================================================
\subsection{The Challenge of MoE Parallelism}\label{sec:moe-parallelism-challenge}

Traditional parallelism strategies were designed for dense Transformers. MoE models have fundamentally different computation patterns that require a new parallelism dimension, Expert Parallelism (EP), and create unique challenges when combining EP with existing strategies.

\subsubsection{The Parallelism Paradox of MoE}\label{sec:parallelism-paradox}

\begin{figure}[t]
    \centering
    \includegraphics[width=0.95\textwidth]{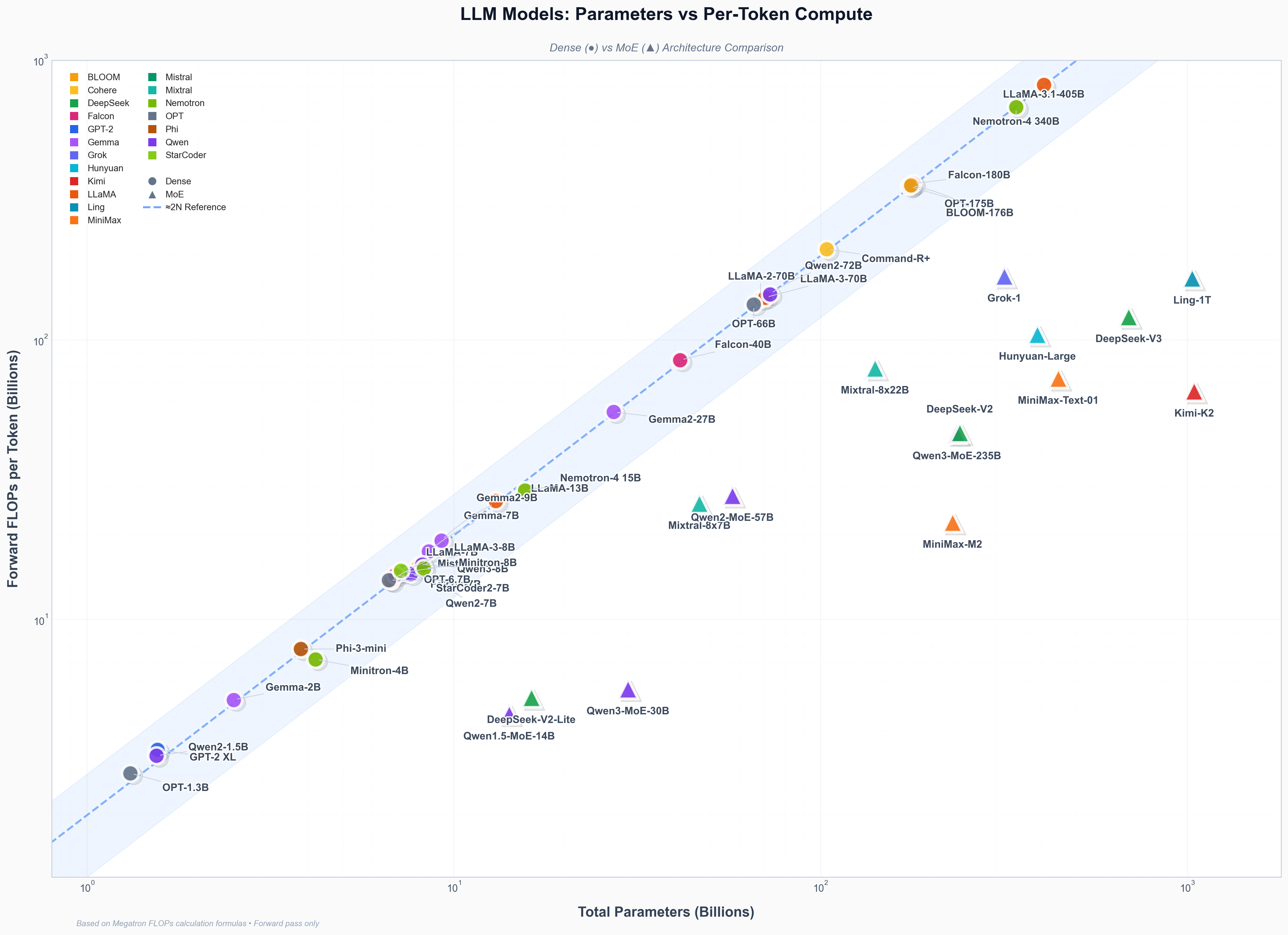}
    \caption{Dense Model vs MoE Model parameter/compute scaling.}
    \label{fig:parallelism-paradox}
\end{figure}

\Figref{fig:parallelism-paradox} illustrates this paradox by plotting total parameters against \emph{forward} FLOPs per token for major LLMs. Dense models (circles) follow the $\approx$2N reference line, exhibiting a \emph{virtuous cycle} as they scale: increasing parameters requires more GPUs for memory, but more parameters also means more computation per token. Because computation grows with model size, communication takes a smaller share of each step, keeping MFU stable.

MoE models (triangles) break this cycle. They fall significantly below the 2N line, achieving equivalent capability with far fewer FLOPs per token. Consider the fundamental asymmetry:

\begin{center}
\begin{tabular}{lccc}
\toprule
\textbf{Model} & \textbf{Total Params} & \textbf{Active Params} & \textbf{Ratio} \\
\midrule
Llama-70B (Dense) & 70B & 70B & 1:1 \\
DeepSeek-V3 (MoE) & 685B & 37B & \textbf{18:1} \\
\bottomrule
\end{tabular}
\end{center}

{\footnotesize \textbf{Note:} DeepSeek-V3 comprises 671B Main Model weights and 14B Multi-Token Prediction (MTP) Module weights (685B total). The table shows the total model parameters including MTP.}

DeepSeek-V3 has 18$\times$ more parameters than its active computation suggests, visible in \Figref{fig:parallelism-paradox} as the gap between DeepSeek-V3's position and where a dense model of equivalent parameters would lie. This creates a \emph{compounding effect}:

\begin{enumerate}
    \item \textbf{Memory grows fast}, forcing distribution across many GPUs: All $E$ experts' parameters, gradients, and optimizer states must reside in memory, and larger top-$k$ further increases activation memory through token replication.
    \item \textbf{More communication}: Distributing experts via EP introduces all-to-all traffic whose volume scales with $K$ (each token is dispatched to $K$ experts across EP ranks).
    \item \textbf{But compute stays low}: Per-token FLOPs scale only with $N_{\text{active}}$ ($\propto K$), not $N_{\text{total}}$ ($\propto E$), leaving insufficient computation to overlap with the growing communication.
\end{enumerate}

The consequence: \textbf{MoE training is fundamentally communication-bound} unless parallelism is designed specifically for this asymmetry. This is not a matter of degree; it is a \emph{qualitative difference} from dense models.

\subsubsection{Traditional Parallelism Strategies}\label{sec:traditional-parallelism}

Dense Transformer training typically combines four parallelism strategies:

\textbf{Tensor Parallelism (TP)} shards weight matrices across GPUs along the hidden dimension \cite{shoeybi2020megatronlmtrainingmultibillionparameter}. Each GPU computes a partial result, then AllGather or ReduceScatter collectives combine results. TP works well when matrices are large enough to offset communication overhead.

\textbf{Pipeline Parallelism (PP)} splits the model by layers across GPUs \cite{huang2019gpipe,narayanan2019pipedream,shoeybi2020megatronlmtrainingmultibillionparameter,qi2023zerobubblepipelineparallelism,qi2024pipelineparallelismcontrollablememory,zhang2022chimera,guan2025pipeoptimensuringeffective1f1b}. Micro-batches flow through the pipeline, with point-to-point communication between stages. PP introduces pipeline bubbles (idle time at boundaries) and P2P communication overhead, but scales better across nodes than TP.

\textbf{Data Parallelism (DP)} replicates the model across GPUs, with each GPU processing different data batches \cite{shallue2019measuringeffectsdataparallelism,bai2022moderndistributeddataparallellargescale}. Gradients are synchronized via AllReduce. DP is simple but requires each GPU to hold the full model.

\textbf{Context Parallelism (CP)} \cite{korthikanti2023sequence,liu2023ringattention,jacobs2023deepspeedulysses} partitions the input sequence across GPUs along the sequence dimension. Each GPU processes a contiguous chunk of the sequence, with communication required only for attention computation where tokens must attend across chunk boundaries. CP is essential for long-context training where activation memory scales quadratically with sequence length.

\textbf{Why traditional parallelism fails for MoE:}

\begin{itemize}
    \item \textbf{TP on MoE experts}: Expert hidden dimensions are small; applying TP creates even smaller shards, reducing compute efficiency while increasing communication's share of total time. High TP can distribute attention efficiently but \emph{hurts} MoE.
    \item \textbf{PP with many experts}: MoE models have massive parameter counts, requiring many pipeline stages if using PP alone. This creates excessive pipeline bubbles and reduces throughput.
    \item \textbf{DP alone}: DP replicates the full model on each GPU. For a trillion parameter model, this is impossible because DP cannot partition parameters, only data.
\end{itemize}

\subsubsection{Expert Parallelism: The Fifth Dimension}\label{sec:expert-parallelism}

Traditional parallelism partitions \emph{layers} (PP), \emph{weight matrices} (TP), \emph{sequence} (CP), or \emph{data} (DP). But MoE has a unique structure: \textbf{experts are independent sub-networks}. This enables a fifth parallelism dimension that partitions \emph{experts themselves} across GPUs.

\textbf{Expert Parallelism (EP)} distributes experts across GPUs \cite{lepikhin2020gshard}. With EP degree equal to $E$, each GPU holds $E/\text{EP}$ experts. The forward pass proceeds as:

\begin{enumerate}
    \item \textbf{Route}: The router selects top-$k$ experts for each token.
    \item \textbf{Dispatch}: all-to-all communication sends tokens to the GPUs holding their assigned experts.
    \item \textbf{Compute}: Each GPU processes tokens using only its local experts.
    \item \textbf{Combine}: all-to-all communication returns results to original GPUs.
\end{enumerate}

\textbf{EP's unique trade-off}:
\begin{itemize}
    \item \textbf{Communication}: all-to-all collectives. Volume scales with token count, not expert count.
    \item \textbf{Compute}: Each GPU runs fewer experts, but each expert processes its \emph{full} hidden dimension.
    \item \textbf{Memory}: Higher EP = fewer experts per GPU = lower memory pressure.
\end{itemize}

\begin{figure}[ht]
    \centering
    \includegraphics[width=0.9\textwidth]{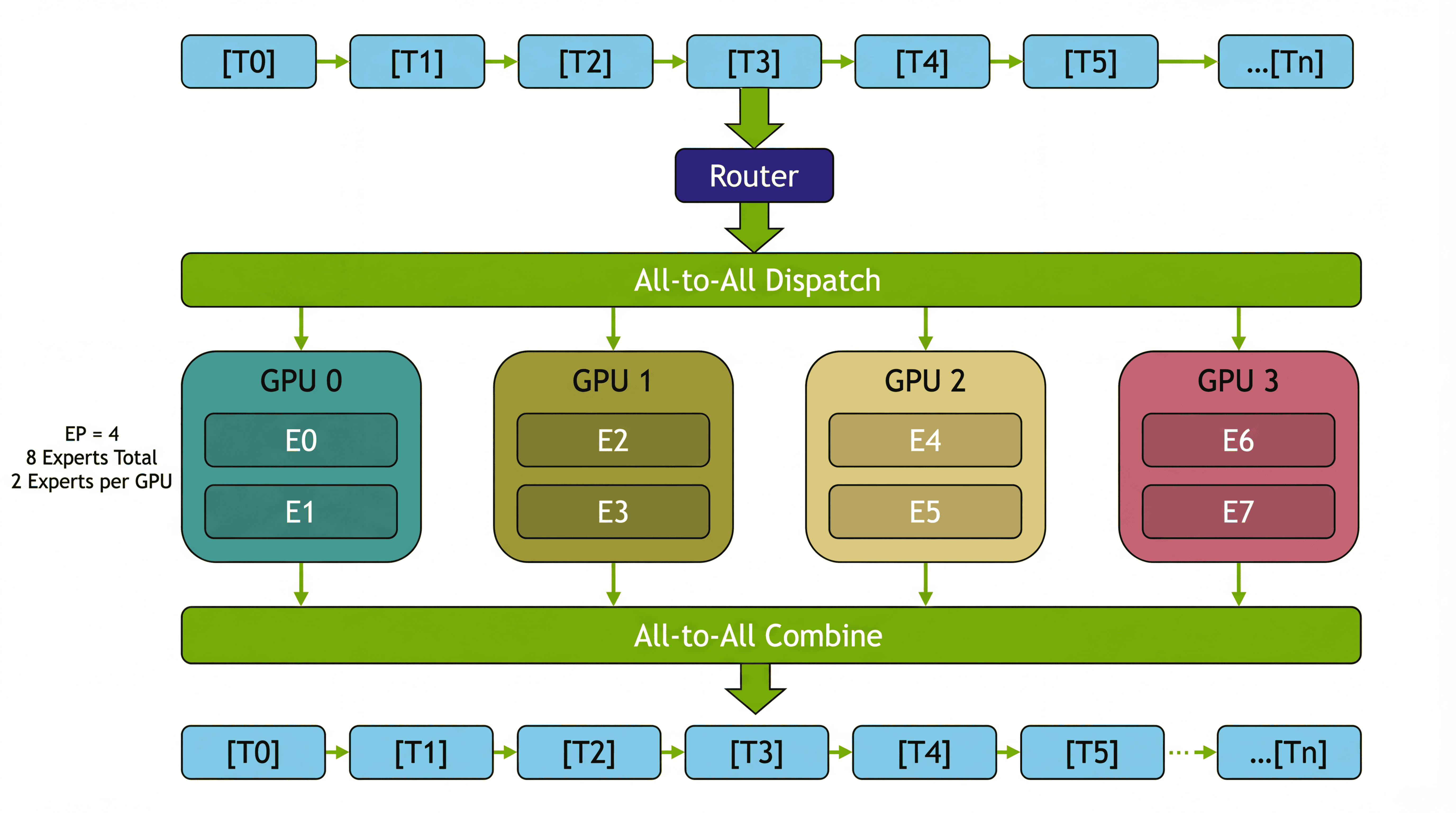}
    \caption{Expert Parallelism (EP) distributes experts across GPUs. The all-to-all communication dispatches tokens to their assigned experts and combines results.}
    \label{fig:expert-parallelism-overview}
\end{figure}

EP provides two key benefits (\figref{fig:expert-parallelism-overview}). First, grouping tokens from different GPUs to a single expert increases computation intensity, improving GEMM efficiency. Second, all-to-all communication volume remains constant as expert count increases; only the number of GPUs changes.

\textbf{But EP alone is not enough.} EP applies only to MoE layers; attention layers have no experts and still require TP, CP, or other strategies. Moreover, both layer types need Pipeline Parallelism to further partition model parameters for large-scale MoE training. This heterogeneity creates the dense-sparse mismatch discussed next.

\subsubsection{The Challenges of Combining EP with Traditional Parallelism}\label{sec:combining-parallelism}

A single Transformer block contains \textbf{two fundamentally different computation patterns} \cite{Singh_2023}, summarized in Table~\ref{tab:dense-sparse-comparison}:

\begin{table}[ht]
\centering
\caption{Contrasting parallelism requirements of attention and MoE layers within a single Transformer block.}
\label{tab:dense-sparse-comparison}
\begin{tabular}{lp{0.4\textwidth}p{0.4\textwidth}}
\toprule
\textbf{Aspect} & \textbf{Attention (Dense)} & \textbf{MoE (Sparse)} \\
\midrule
Computation & Every token attends to all others & Each token routes to $K$ of $E$ experts \\
TP & Large QKV matrices benefit from high TP & Small per-expert dimensions make high TP counterproductive \\
CP & Long sequences benefit from high CP & No sequence dependency; CP is irrelevant \\
EP & Not applicable (no experts) & Essential for distributing many experts \\
\bottomrule
\end{tabular}
\end{table}

\textbf{The Dense-Sparse Mismatch.} Traditional frameworks force \emph{one} parallelism configuration for \emph{both} layer types, but their optimal configurations conflict directly: high TP benefits attention but fragments expert shards; high CP helps long-context attention but provides no benefit to MoE layers; and high EP is essential for MoE but irrelevant to attention. Like the parameter-compute mismatch (Section~\ref{sec:challenges}), this stems from MoE's sparsity, but manifests at the parallelism configuration level: it is a \textbf{structural mismatch} between dense and sparse layers' needs, not a tuning problem.

Prior MoE frameworks treat EP as a sub-dimension of DP \cite{lepikhin2020gshard,fedus2022switchtransformersscalingtrillion}:
\[
\text{World Size} = \text{TP} \times \text{CP} \times \text{PP} \times \text{DP}, \quad \text{where } \text{EP} \subseteq \text{DP}
\]

This constraint exists because frameworks assumed uniform parallelism: attention layers use (TP, CP, PP, DP), and MoE layers simply carve EP out of the DP group. This design creates three critical challenges:

\textbf{Challenge 1: Multiplicative GPU Requirements.} Traditional frameworks require $\text{TP} \times \text{CP} \times \text{PP} \times \text{DP}$ GPUs at minimum. Since EP $\subseteq$ DP, requesting EP=8 forces DP $\geq$ 8. Combined with CP=8 for long sequences, the minimum becomes $1 \times 8 \times 1 \times 8 = 64$ GPUs---even if attention and MoE could theoretically share the same 8 GPUs. This inflates the entry barrier for MoE training.

\textbf{Challenge 2: Forced Suboptimal Parallelism.} Since attention and MoE share the same TP configuration, practitioners must choose between two suboptimal configurations: use high TP (e.g., TP=8) to efficiently shard large attention matrices, which fragments small experts into inefficient shards; or use low TP (e.g., TP=1) to preserve expert computation efficiency, which leaves attention layers underparallelized. Neither option achieves optimal performance for both layer types.

\textbf{Challenge 3: Cross-Node Communication.} With EP constrained within DP, high EP often forces all-to-all communication to cross node boundaries, where bandwidth is 5--10$\times$ lower than NVLink. Meanwhile, CP communication for attention may also span nodes. Without the ability to independently map EP and CP to high-bandwidth domains, communication overhead dominates training time.

These challenges are not independent tuning problems. They stem from the fundamental assumption that all layers must share one parallelism configuration. \textbf{The question becomes}: How do we break these constraints while still allowing each layer type to use its optimal parallelism?

%==============================================================================
% 3.3 Megatron-Core's Solution: Parallel Folding
%==============================================================================
\subsection{Megatron-Core's Solution: Parallel Folding and Multi-Dimensional Framework}\label{sec:parallel-folding}

\textbf{Parallel Folding} is Megatron-Core's answer to the dense-sparse mismatch. Rather than forcing attention and MoE layers to share the same parallelism configuration, Parallel Folding \textbf{decouples their parallelism mappings}, allowing each layer type to use its optimal topology \cite{liu2025moeparallelfoldingheterogeneous}. This section covers the key concepts and benefits; implementation details such as parallelism transition handling and token dispatch design are described in the companion paper \cite{liu2025moeparallelfoldingheterogeneous}.

\subsubsection{Decoupled Parallelism Mappings}

The core idea is simple: \emph{do not force attention and MoE to share parallelism}. Let each use its optimal configuration independently.

\begin{figure}[ht]
    \centering
    \includegraphics[width=0.9\textwidth]{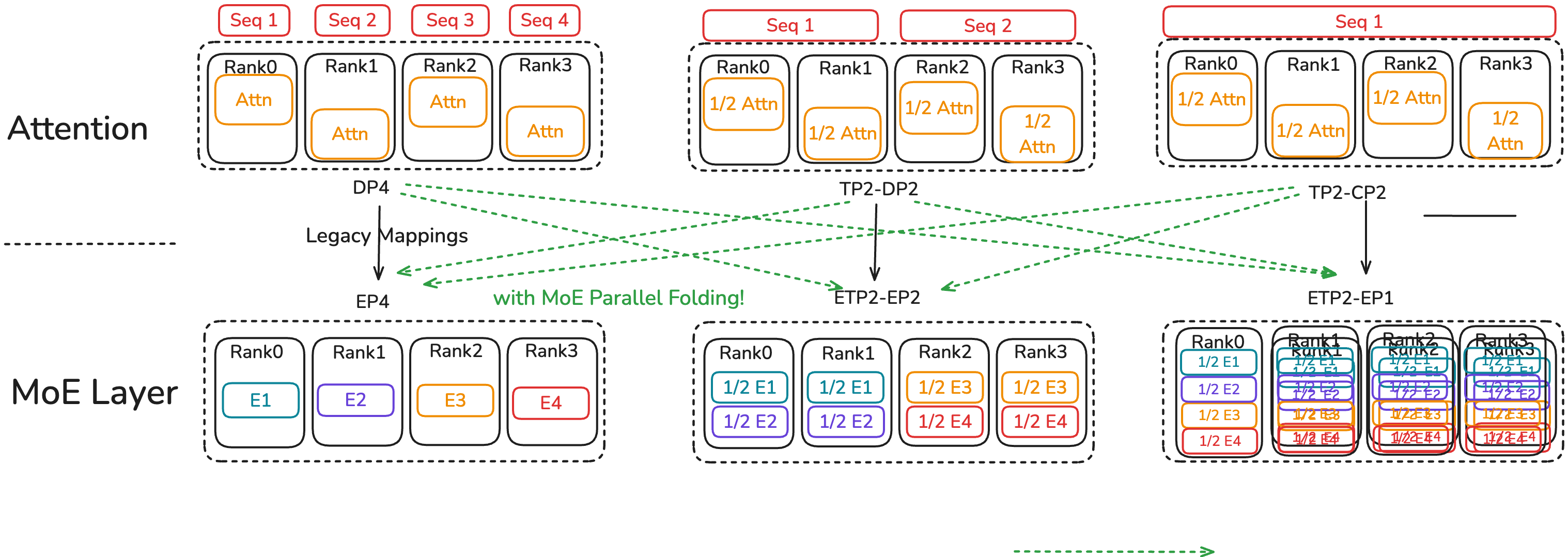}
    \caption{Parallelism mappings: traditional constraints vs. MoE Parallel Folding decoupling.}
    \label{fig:moe-parallel-folding}
\end{figure}

Parallel Folding introduces separate parallelism groups for attention and MoE layers:

\begin{itemize}
    \item \textbf{Attention layers} form groups over $\text{TP} \times \text{CP} \times \text{DP} \times \text{PP}$, optimized for sequence-level dense computation.
    \item \textbf{MoE layers} form groups over $\text{ETP} \times \text{EP} \times \text{EDP} \times \text{PP}$, where ETP (Expert Tensor Parallelism) and EDP (Expert Data Parallelism) are MoE-specific dimensions.
\end{itemize}

The sole constraint: \textbf{Pipeline Parallelism (PP) must remain consistent} across both layouts to ensure correct gradient flow through the model.
\subsubsection{Complete Multi-Dimensional Parallelism Stack}\label{sec:multi-dim-parallelism}

With Parallel Folding as the foundation, Megatron-Core orchestrates five parallelism dimensions:

\begin{center}
\begin{tabular}{lll}
\toprule
\textbf{Dimension} & \textbf{Applies To} & \textbf{Purpose} \\
\midrule
TP (Tensor) & Attention & Shard large QKV/projection matrices \\
CP (Context) & Attention & Distribute long sequences \\
DP (Data) & Attention & Process different batches \\
PP (Pipeline) & Both & Split model by layers (must be consistent) \\
\midrule
EP (Expert) & MoE & Distribute experts across GPUs \\
ETP (Expert Tensor) & MoE & Shard within experts (rarely used) \\
EDP (Expert Data) & MoE & Replicate experts for throughput \\
\bottomrule
\end{tabular}
\end{center}

\textbf{Key configuration principles}:
\begin{itemize}
    \item \textbf{Attention layers}: Optimize for large matrices (high TP) and long sequences (high CP).
    \item \textbf{MoE layers}: Optimize for many small experts (high EP, typically ETP=1).
    \item \textbf{PP}: Must remain consistent across both to ensure correct data flow.
\end{itemize}

\begin{figure}[ht]
    \centering
    \includegraphics[width=0.9\textwidth]{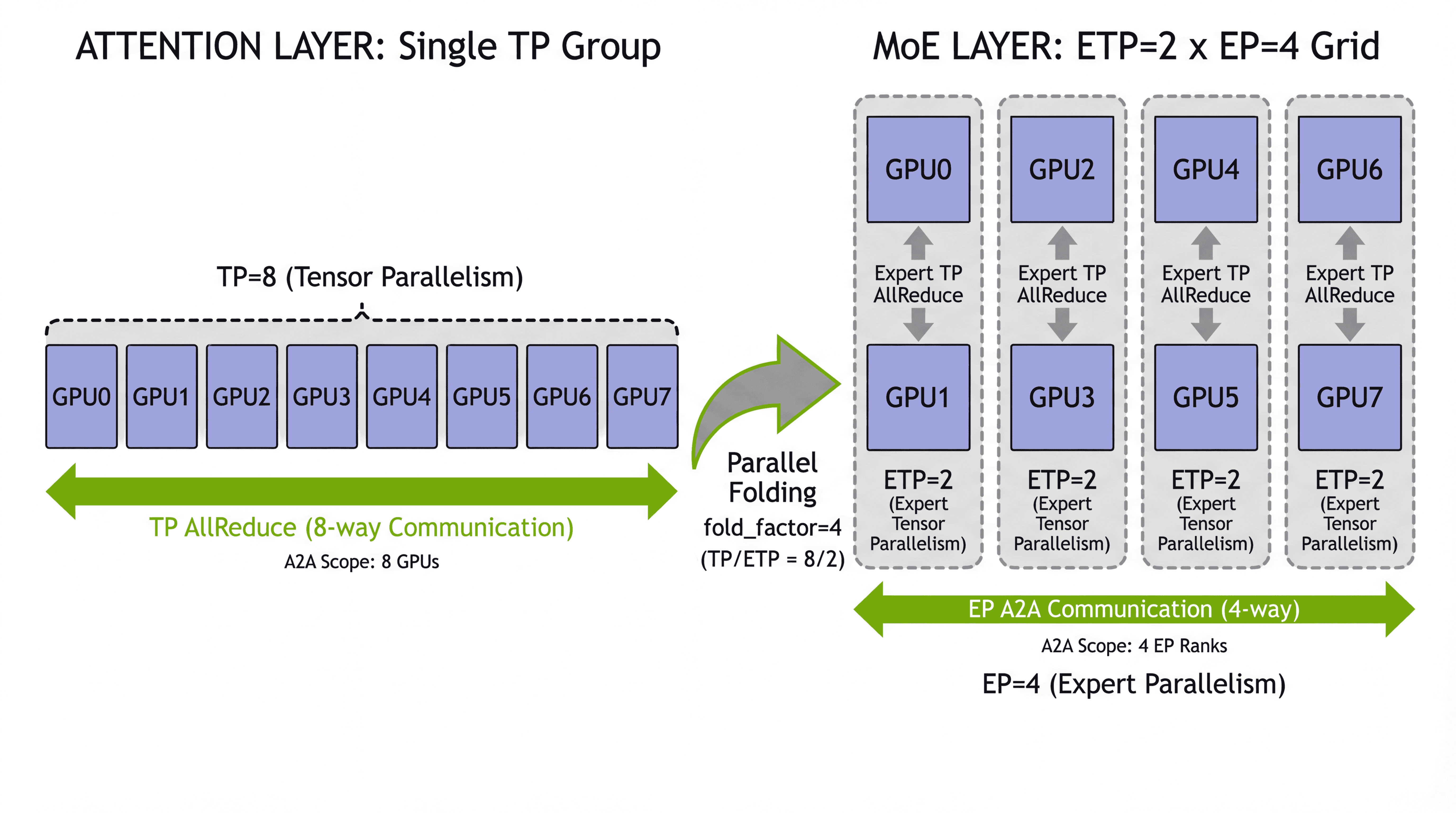}
    \caption{Parallel Folding: decoupled attention and MoE parallelism mappings.}
    \label{fig:parallel-folding-detailed}
\end{figure}

Beyond the parallelism strategies described above (\figref{fig:parallel-folding-detailed}), Megatron-Core's Parallel Folding framework also integrates Distributed Optimizer and FSDP with EP support to further reduce memory footprint:

\textbf{Distributed Optimizer + EP.} Only weights and gradients for local experts reside on each rank; optimizer states are sharded among replicas of the same expert (via EDP). This removes redundant optimizer memory for non-local experts and confines gradient synchronization to minimal groups.

\textbf{FSDP + EP.} For even greater memory efficiency, Megatron-Core's custom FSDP (Megatron-FSDP) fully shards parameters, gradients, and optimizer states across data/expert groups via a dual DeviceMesh architecture, reducing memory footprint while overlapping AllGather and ReduceScatter with computation. It is compatible with TP/EP/CP and mixed precision (BF16, FP8, FP4). See \secref{sec:fsdp-ref} for the full design.

\subsubsection{Benefits of Parallel Folding}

As illustrated in \figref{fig:moe-parallel-folding}, MoE Parallel Folding eliminates the EP $\leq$ DP limitation by allowing EP to ``fold'' across arbitrary sub-groups of the attention parallelism configuration. This provides four key advantages:

\begin{enumerate}
    \item \textbf{Breaks the EP $\leq$ DP constraint}: EP can now exceed DP by folding across TP$\times$CP groups. Consider attention configured with TP=4, CP=2, DP=8, PP=4 (total 256 GPUs):
    \begin{itemize}
        \item Traditional: EP $\leq$ DP = 8, so maximum EP is 8.
        \item With Parallel Folding: MoE uses ETP=1, EP=64, EDP=1 (same PP=4). EP ``folds'' across the TP$\times$CP$\times$DP groups, enabling 8$\times$ higher expert parallelism while attention layers maintain their optimal TP=4, CP=2 configuration.
    \end{itemize}
    
    \item \textbf{Reduces minimum GPU requirements}: Traditional configurations with CP=8, EP=8 require at least 64 GPUs. With Folding, CP and EP share the same GPU group, requiring only 8 GPUs.
    
    \item \textbf{Enables independent optimization}: Attention can use high TP for large matrices while MoE uses ETP=1 for full expert width and better GEMM efficiency.
    
    \item \textbf{Keeps high-bandwidth communication in NVLink domain}: Both CP (for attention) and EP (for MoE) all-to-all communication can remain within the NVLink-connected GPU group, avoiding slower cross-node transfers.
\end{enumerate}

% fix typo

\subsubsection{Summary}

This section addressed the \emph{distribution challenge} for MoE training. The core problem is the \emph{dense-sparse mismatch}, rooted in MoE's sparsity: attention and MoE layers have conflicting optimal parallelism configurations, yet prior frameworks forced them to share one configuration. \textbf{MoE Parallel Folding} solves this by decoupling parallelism mappings, allowing attention to use high TP/CP while MoE uses high EP independently. Together with Distributed Optimizer and FSDP, Megatron-Core orchestrates all parallelism dimensions into a unified framework that scales MoE training to thousands of GPUs.

However, scalable distribution alone does not guarantee training efficiency. The three fundamental barriers identified in Section~\ref{sec:challenges}, the Memory, Communication, and Compute Efficiency Walls, must still be addressed. The following section presents Megatron-Core's solutions to break through these walls.

\section{Scaling MoE: Breaking the Memory, Communication, and Compute Efficiency Walls}\label{key-features-optimizations}

Section~\ref{parallelism-strategies} established \emph{how} to distribute MoE training across thousands of GPUs. This section addresses \emph{how to make it efficient}. The challenges are significant: Section~\ref{sec:challenges} identified three fundamental barriers (the Memory Wall, Communication Wall, and Compute Efficiency Wall) that emerge from MoE's sparsity-driven parameter-compute mismatch. These walls are not merely inconveniences; without addressing them, training large-scale MoE models is either infeasible (memory) or prohibitively slow (communication, compute).

\textbf{Why These Three Walls?} These barriers emerge from the fundamental resources of GPU-accelerated training: memory stores parameters and activations, interconnects transfer data between devices, and compute units execute operations. Every training system must acquire sufficient memory, move data where it is needed, and perform computations efficiently. Other potential bottlenecks, such as storage I/O for checkpointing or host-side data loading, exist but operate at timescales that allow them to be overlapped with training iterations. The three walls represent the hard constraints on every forward-backward pass.

\textbf{The Walls Interact.} These challenges are not independent; they form a tightly coupled system where fixing one wall can expose or worsen another. Consider a concrete optimization trajectory: A team training a 1000B MoE model first encounters the memory wall: activations alone exceed GPU capacity. They enable activation recomputation (Section~\ref{sec:memory-wall}), trading memory for computation by regenerating activations during the backward pass. Memory constraints resolved, profiling reveals that all-to-all communication now dominates; the compute wall was hiding the communication wall. They implement communication-computation overlap (Section~\ref{sec:comm-wall}), pipelining all-to-all transfers with expert execution. But fine-grained experts complete faster than communication, limiting overlap effectiveness. They enable reduced-precision training (Section~\ref{sec:reduced-precision-training}) to reduce memory footprint, allowing larger batch sizes that provide more computation to hide communication. This works, but introduces quantization kernels that fragment the execution stream and increase kernel launch overhead. The system becomes host-bound: GPUs idle between kernel launches. They deploy CUDA Graphs (Section~\ref{sec:compute-wall}) to reduce launch overhead, but graphs require static tensor shapes, conflicting with dropless routing's dynamic expert assignments. Each solution ripples through the system, demanding awareness of all three walls simultaneously. Addressing walls in isolation leads to suboptimal solutions; effective optimization requires treating memory, communication, and compute as a unified system.

Megatron-Core addresses these challenges through integrated system design. The remainder of this section presents solutions organized by the wall they primarily address:
\begin{itemize}
    \item \textbf{Section~\ref{sec:memory-wall}: Breaking the Memory Wall.} How do we fit training within GPU memory constraints? This section covers activation management, recomputation strategies, offloading, and distributed parameter storage.
    \item \textbf{Section~\ref{sec:comm-wall}: Breaking the Communication Wall.} How do we minimize time lost to inter-device communication? This section covers optimized dispatchers, communication-computation overlap, and pipeline integration.
    \item \textbf{Section~\ref{sec:compute-wall}: Breaking the Compute Efficiency Wall.} How do we keep GPUs saturated with work? This section covers kernel fusion, CUDA Graphs, and sync-free execution for dropless MoE.
\end{itemize}

Reduced-precision training in FP8/FP4, which provides benefits across all three walls simultaneously, is covered separately in Section~\ref{sec:reduced-precision-training}. Long-context MoE training, which changes the optimization balance when attention dominates computation, is addressed in Section~\ref{sec:long-context-training}.

\subsection{Breaking the Memory Wall}\label{sec:memory-wall}

Memory is the first hard constraint in MoE training: if the combined footprint of parameters, optimizer states, and activations exceeds GPU capacity, training cannot proceed. As model and expert counts grow, memory requirements grow quickly, making memory optimization essential for practical training. Understanding \emph{where} memory is consumed is essential for effective optimization.

\subsubsection{Memory Anatomy: Sources of Memory Consumption}

MoE models typically consume substantially more memory than equivalent dense models, owing to several distinct sources of overhead. Consider DeepSeek-V3 trained with BF16 precision using a $\text{PP}4 \times \text{VPP}4 \times \text{EP}64$ configuration across 256 GPUs. Table~\ref{tab:memory-breakdown} presents the per-GPU memory breakdown, revealing a total requirement of 199.5 GB, well beyond the 80 GB capacity of an H100 GPU without optimization.

\begin{table}[ht]
\centering
\caption{Memory breakdown per GPU for DeepSeek-V3 with BF16 training ($\text{PP}4 \times \text{VPP}4 \times \text{EP}64$, 256 GPUs).}
\label{tab:memory-breakdown}
\begin{tabular}{lrl}
\toprule
\textbf{Component} & \textbf{Memory per GPU} & \textbf{Optimization Techniques} \\
\midrule
Weights \& Gradients & 36.4 GB & PP, EP, or TP sharding \\
Main Weights \& Optimizer States & 32.1 GB & Distributed optimizer, BF16 moments \\
Activations & 131.0 GB & Low Precision, Recomputation, Offloading \\
\midrule
\textbf{Total} & \textbf{199.5 GB} & \\
\bottomrule
\end{tabular}
\end{table}

Three components contribute to this footprint:

\textbf{Weights and Gradients (36.4 GB).} All $E$ expert parameters must stay in memory even though only $K$ activate per token. DeepSeek-V3's 256 experts with top-8 routing illustrate this: 685B parameters but only 37B activated per token, an 18$\times$ gap.

\textbf{Optimizer States (32.1 GB).} Adam stores momentum and variance per parameter, tripling parameter memory in FP32. Mixed-precision training with BF16 moments reduces this overhead but does not eliminate it.

\textbf{Activations (131.0 GB).} The largest memory consumer, exceeding weights and optimizer states combined. MoE activations scale with layer depth, hidden dimension, and top-$k$, as well as batch size and sequence length. This makes activations the primary optimization target.

\begin{notebox}
Activations dominate memory consumption in large-scale MoE training, often exceeding the combined memory of weights, gradients, and optimizer states. This makes activation memory optimization the highest priority for enabling larger batch sizes or more flexible parallelism configurations.
\end{notebox}

Additional MoE-specific factors add to memory pressure: dynamic routing can create load imbalance where certain experts temporarily receive too many tokens, and tokens must often be padded to fit efficient computation kernels, consuming memory beyond the actual data. These are activation memory costs that can be reduced through the recomputation techniques discussed below. Load imbalance is further addressed by ECHO (Section~\ref{sec:echo}), which dynamically clones popular experts to balance token distribution across ranks.

Model architecture cannot be changed to reduce memory, so optimization must target how data is stored and managed. Four complementary strategies address memory constraints:

\begin{enumerate}
    \item \textbf{Reduce storage precision.} Lower-precision formats (FP8/FP4 instead of BF16) reduce activation memory with minimal impact on accuracy. Memory-Efficient Permutation eliminates redundant intermediate tensors entirely.
    \item \textbf{Trade computation for storage.} Activation recomputation \cite{chen2016training,korthikanti2022reducingactivationrecomputationlarge,beaumont2021optimal} discards intermediate results during the forward pass and regenerates them during backward, exchanging compute cycles for memory capacity.
    \item \textbf{Offload to host memory.} When GPU memory is exhausted, activations can be transferred to CPU memory during forward pass and retrieved during backward, trading PCIe bandwidth for GPU memory \cite{ren2021zerooffload}.
    \item \textbf{Distribute across devices.} Fully Sharded Data Parallel (FSDP) \cite{zhao2023pytorchfsdp,rajbhandari2020zero,rajbhandari2021zeroinf,rajbhandari2023zeropp} partitions parameters, gradients, and optimizer states across data-parallel ranks, enabling training of models that exceed single-device capacity. 
\end{enumerate}

The following subsections present these techniques in two groups. First, activation memory optimizations ordered by overhead: zero-overhead (Memory-Efficient Permutation), precision trade-offs (FP8/FP4 vs. BF16), computational trade-offs (recomputation), and bandwidth trade-offs (offloading). Then, optimizations targeting weights and optimizer states: precision-aware optimizer and distribution strategies (FSDP).

\subsubsection{Memory-Efficient Permutation: Zero-Overhead Activation Reduction}\label{sec:mem-eff-permute}

The most desirable memory optimizations are those with no computational overhead. Memory-Efficient Permutation achieves exactly this by eliminating redundant intermediate tensors through a simple algebraic rearrangement. The technique is zero-overhead because it merely changes \emph{when} router probabilities are applied, not \emph{whether} they are applied; the pre-activation buffers required for the backward pass would be stored anyway in conventional implementations.

As described in Section~\ref{sec:moe-layer-architecture}, the router assigns each token to its top-$k$ experts with a learned routing weight, and the weighted expert outputs are combined to produce the final result.

Consider a token $\mathbf{x}$ routed to its top-$k$ experts. Let $\mathcal{T}(\mathbf{x}) \subset \{1,\dots,E\}$ denote the selected expert set with routing weights $\{p_i\}_{i \in \mathcal{T}(\mathbf{x})}$. Each expert $E_i$ is a two-layer MLP with weight matrices $\mathbf{W}_1^{(i)}, \mathbf{W}_2^{(i)}$ and nonlinear activation $\phi$ (e.g., SwiGLU):
\[
E_i(\mathbf{x}) = \mathbf{W}_2^{(i)}\,\phi\!\bigl(\mathbf{W}_1^{(i)}\mathbf{x}\bigr).
\]
In the \textbf{standard formulation}, routing weights are applied \emph{after} expert computation:
\begin{equation}\label{eq:moe-standard-combine}
\mathbf{y} = \sum_{i \in \mathcal{T}(\mathbf{x})} p_i \cdot \mathbf{W}_2^{(i)}\,\phi\!\bigl(\mathbf{W}_1^{(i)}\mathbf{x}\bigr).
\end{equation}
\textbf{Memory-Efficient Permutation} absorbs $p_i$ into the activation, applying it \emph{before} the second linear layer:
\begin{equation}\label{eq:moe-mem-eff-combine}
\mathbf{y} = \sum_{i \in \mathcal{T}(\mathbf{x})} \mathbf{W}_2^{(i)}\,\bigl(p_i \cdot \phi\!\bigl(\mathbf{W}_1^{(i)}\mathbf{x}\bigr)\bigr).
\end{equation}
When the experts have no bias terms, $\mathbf{W}_2^{(i)}$ is a pure linear map and scalar multiplication commutes: $p_i \cdot \mathbf{W}_2^{(i)}\mathbf{h} = \mathbf{W}_2^{(i)}(p_i \cdot \mathbf{h})$ for any vector $\mathbf{h}$, so \eqref{eq:moe-standard-combine} and \eqref{eq:moe-mem-eff-combine} are mathematically equivalent.

\begin{figure}[ht]
    \centering
    \includegraphics[width=0.6\linewidth]{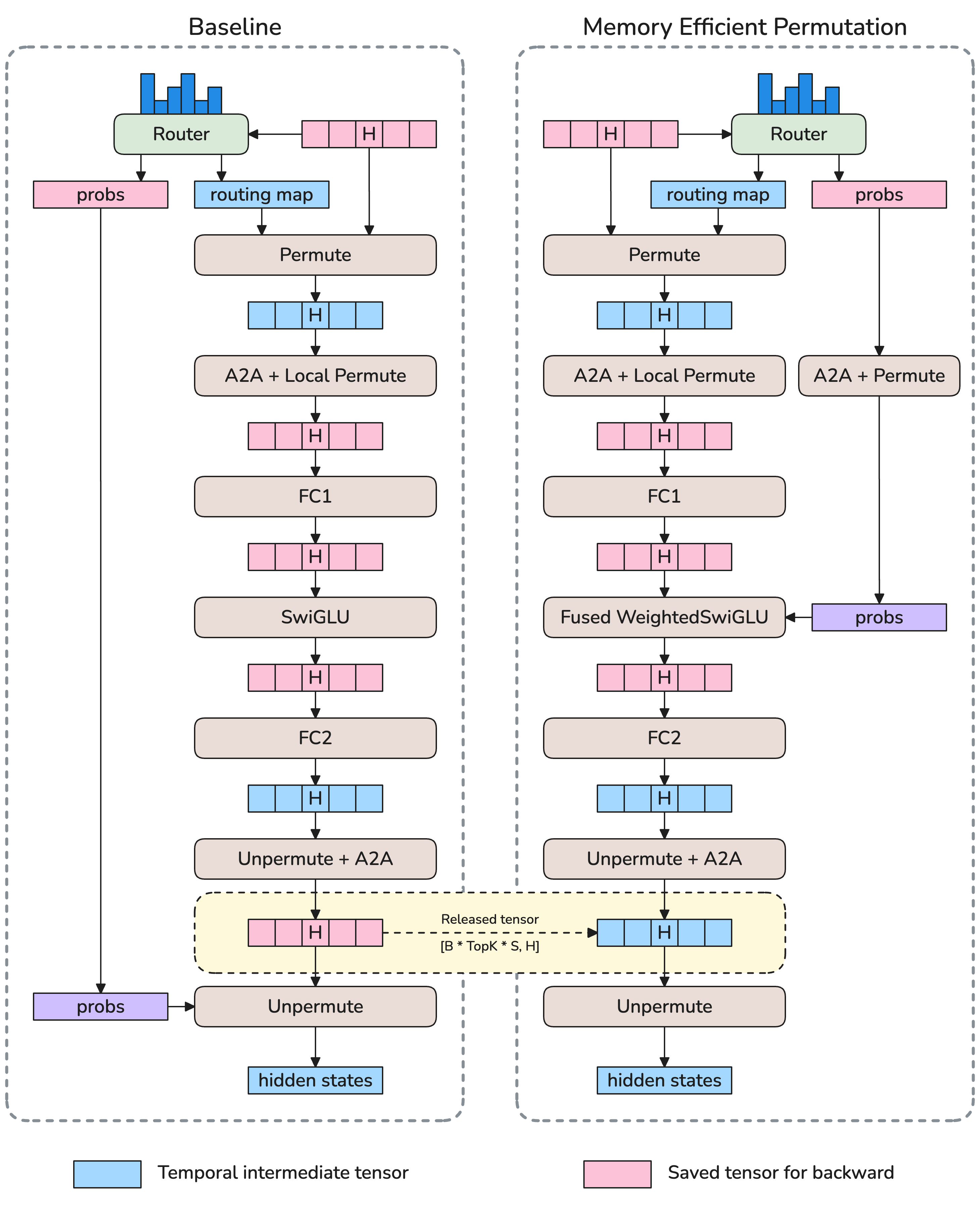}
        \caption{Memory-Efficient Permutation.}
    \label{fig:mem-eff-permute}
\end{figure}

This rearrangement reduces peak memory by eliminating saved tensors needed for the router's backward pass. Let $\mathbf{z}_i = \mathbf{W}_1^{(i)}\mathbf{x}$ denote the pre-activation input. In the standard formulation, computing $\partial \mathcal{L}/\partial p_i$ requires retaining each expert output $E_i(\mathbf{x})$ throughout the backward pass. In the memory-efficient formulation, $p_i$ multiplies $\phi(\mathbf{z}_i)$ directly, so $\partial \mathcal{L}/\partial p_i$ only depends on $\phi(\mathbf{z}_i)$, which a fused backward kernel recomputes from $\mathbf{z}_i$ on the fly. Since $\mathbf{z}_i$ must already be saved for the SwiGLU activation's own backward pass regardless of whether Memory-Efficient Permutation is used, no additional buffers are introduced, yielding a net reduction in peak memory with zero computational overhead. \Figref{fig:mem-eff-permute} illustrates this transformation.

For DeepSeek-V3 (Table~\ref{tab:memory-breakdown}), Memory-Efficient Permutation saves approximately 26.3 GB of activation memory per GPU, a significant reduction with zero computational cost.

\subsubsection{Reduced-Precision Training: FP8/FP4 for Activation Memory Reduction}\label{sec:fp8-activation}

Storing activations in FP8/FP4 instead of BF16 reduces their memory footprint with minimal impact on model quality.

During the forward pass, the input to each linear layer must be retained to compute weight gradients in the backward pass. In Transformer models, these linear layer inputs make up most of the activation memory: attention projections (Q, K, V, output) and MLP layers (including expert MLPs in MoE) each save their inputs for gradient computation. By storing these input tensors in FP8/FP4 instead of BF16, each tensor's memory footprint is reduced by 50\%/75\%.

For the DeepSeek-V3 configuration in Table~\ref{tab:memory-breakdown}, enabling FP8 training reduces approximately 16 GB of activation memory, representing roughly 12\% of the 131 GB activation budget. This corresponds to halving approximately 32 GB of linear layer inputs that are eligible for FP8 storage. The remaining activations (attention scores, normalization intermediates, routing tensors) either require higher precision for numerical stability or are not stored across the forward-backward boundary. This reduction is orthogonal to other activation optimizations such as recomputation and offloading, allowing them to be combined for cumulative savings.

\textit{For details on reduced-precision training mechanisms, including recipes, quantization strategies, and MoE-specific optimizations, see Section~\ref{sec:reduced-precision-training}.}

\subsubsection{Recomputation: Trading Compute for Memory}\label{sec:recomputation}

Activation recomputation (or activation checkpointing) is a well-established technique that discards intermediate activations during forward pass and recomputes them during backward pass \cite{chen2016training}. However, trivial full-layer recomputation can add $\sim$33\% computational overhead, and for MoE layers it is even more costly because recomputing expert computation also re-triggers EP all-to-all communication. Megatron-Core introduces \textit{fine-grained} recomputation that targets only the most memory-intensive yet computationally cheap operations, achieving significant memory savings with minimal overhead \cite{korthikanti2022reducingactivationrecomputationlarge}.

Two techniques compose to form this strategy (\figref{fig:selective-recomputation}):

\begin{figure}[ht]
    \centering
    \includegraphics[width=0.5\linewidth]{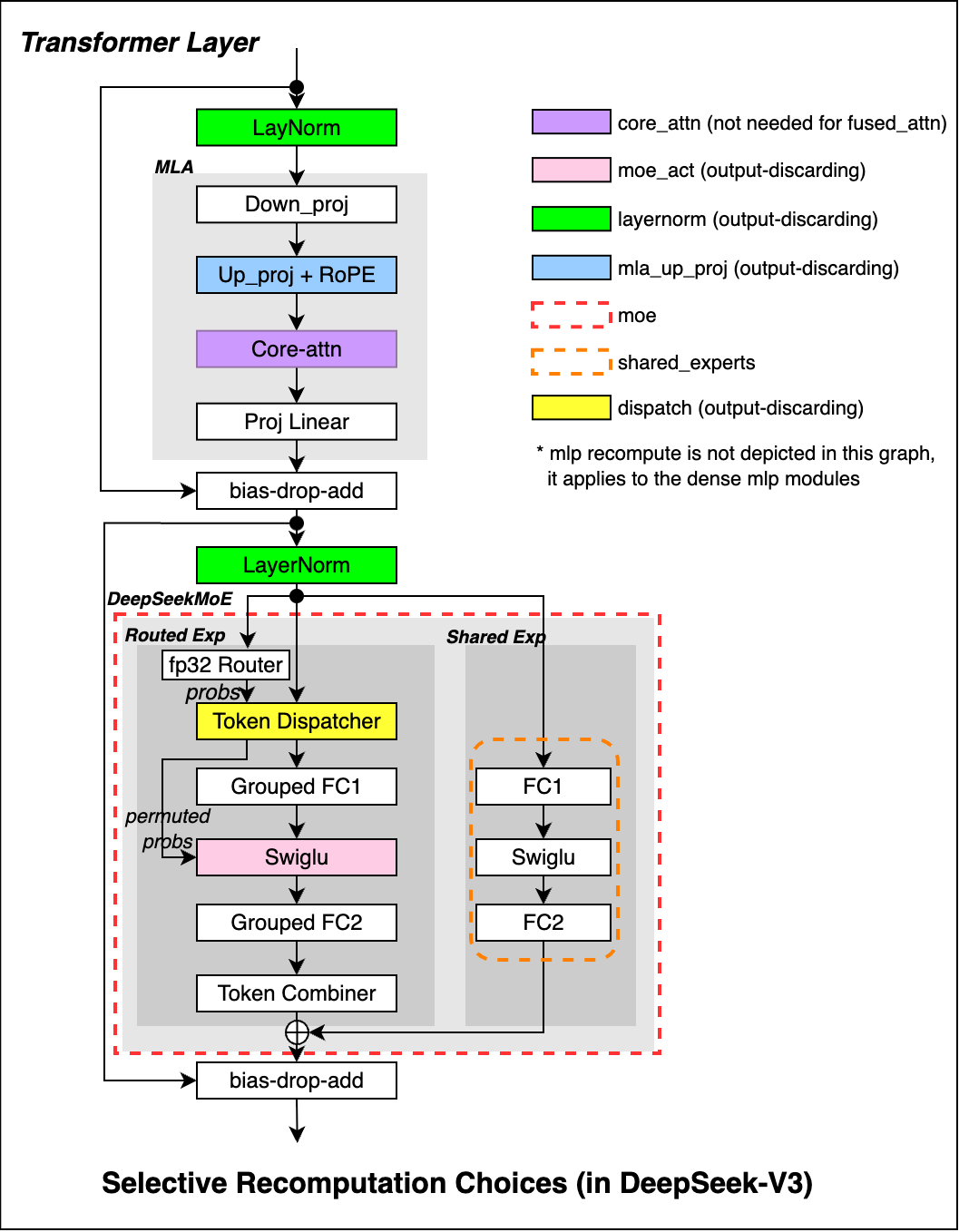}
        \caption{Selective Recomputation.}
    \label{fig:selective-recomputation}
\end{figure}

\textbf{Granular recomputation.} Rather than applying activation checkpointing to large, monolithic sections, users specify exactly which computations to recompute in the backward pass. For instance, one may recompute only the activation functions within expert MLPs, the LayerNorm modules, or the up-projection within Multi-Latent Attention (MLA). By recomputing individual operations or submodules, significant memory savings are achievable with only a modest increase in computational workload, as only the selected portions require recalculation (less than 5\% additional compute overhead).

\textbf{Output-discarding recomputation.} Conventional activation checkpointing workflows pass the outputs of checkpointed modules to downstream layers, storing these outputs for potential use in backpropagation. However, since these outputs will be recomputed during the backward pass, this storage is redundant. To avoid this, Megatron-Core MoE promptly releases the outputs of checkpointed modules after they are consumed by subsequent layers. During the backward pass, the outputs are restored from recomputed results. This strategy ensures that memory is not unnecessarily reserved for activations that can be cheaply restored, reducing memory footprint without compromising gradient correctness or training dynamics.

Table~\ref{tab:recomputation-savings} summarizes the memory reduction achieved by different recomputation targets for the DeepSeek-V3 configuration in Table~\ref{tab:memory-breakdown}.

\begin{table}[ht]
\centering
\caption{Memory reduction per GPU from fine-grained recomputation for DeepSeek-V3 ($\text{PP}4 \times \text{VPP}4 \times \text{EP}64$, 256 GPUs).}
\label{tab:recomputation-savings}
\begin{tabular}{lr}
\toprule
\textbf{Recomputation Target} & \textbf{Memory Saved per GPU} \\
\midrule
MLA Up-Projection & 30.4 GB \\
Activation Function (SwiGLU) & 3.8 GB \\
LayerNorm & 8.2 GB \\
\midrule
\textbf{Total} & \textbf{42.4 GB} \\
\bottomrule
\end{tabular}
\end{table}

\subsubsection{Fine-grained Activation Offloading}\label{sec:offloading}

% \footnote{This is collaborative work between NVIDIA and Rednote.} the footnote has been removed
When GPU memory remains insufficient even after precision and recomputation optimizations, \textit{offloading} activations to CPU memory provides additional capacity. Unlike recomputation which trades compute cycles for memory, offloading trades PCIe bandwidth instead. The challenge is hiding transfer latency behind computation so that offloading appears ``free''.

\paragraph{Motivation}
Fine-grained MoE models have extreme parameter inflation: DeepSeek-V3 activates only 37B per token (18$\times$ ratio, Table~\ref{tab:memory-breakdown}); Kimi-K2 reaches 1T total with 32B active (31$\times$). This parameter-compute mismatch (Section~\ref{sec:challenges}) is especially acute for offloading because activation memory does not decrease with Expert Parallelism (EP) or Pipeline Parallelism (PP); these strategies reduce parameter memory, not activation memory.

Transformer Engine provides layer-level offloading, but this coarse granularity limits effectiveness. Different modules within a layer have significantly different memory-to-compute ratios: LayerNorm activations are small and cheap to recompute, while \texttt{expert\_fc1} inputs are large but computationally expensive. Layer-level offloading cannot distinguish these cases. It either offloads everything (wasting bandwidth) or nothing (wasting memory).

\begin{figure}[ht]
    \centering
    \includegraphics[width=0.9\textwidth]{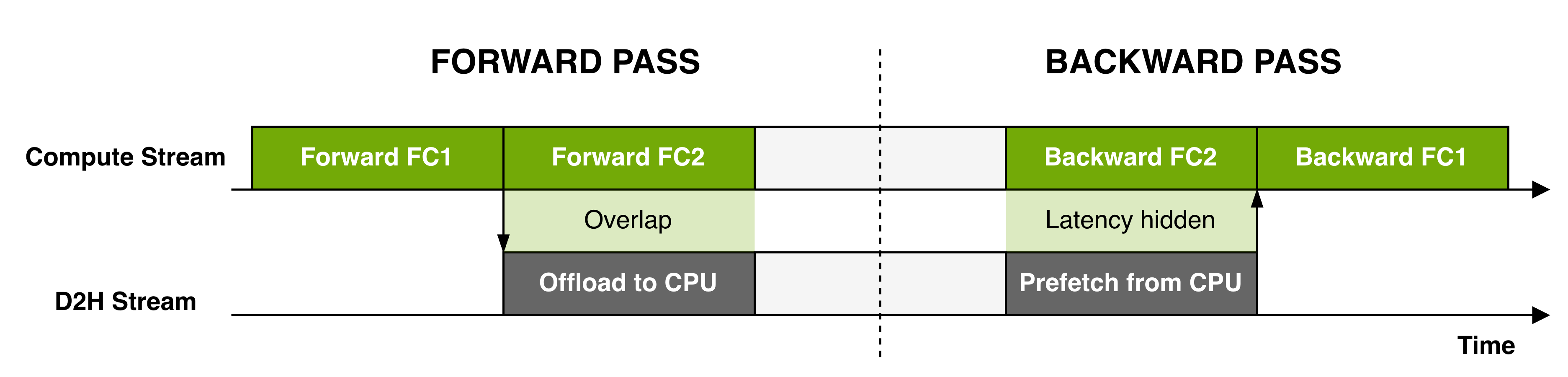}
    \caption{Fine-grained activation offloading: stream overlap for forward and backward passes.}
    \label{fig:activation-offloading-timeline}
\end{figure}

\paragraph{High-Level Idea: Overlap and Prefetch.}
The GPU's Copy Engine and Compute Engine operate independently. When a module's computation time exceeds its activation transfer time, the D2H copy can run concurrently with subsequent computation at zero cost. \Figref{fig:activation-offloading-timeline} illustrates the stream overlap mechanism for both forward and backward passes.

\paragraph{Forward Pass.}
During forward propagation, input activations are offloaded to CPU immediately after module computation, running in parallel with the next module's computation via a dedicated D2H stream. One exception: the last layer's activations are not offloaded because they are needed immediately during backward, so no computation is available to hide the transfer latency.

\paragraph{Backward Pass.}
During backward propagation, activation reloading follows a \textit{Layer-Staggered Reload} pattern: the system reloads the activation of the same module (e.g., \texttt{expert\_fc1}) from the \textit{next} layer while computing gradients for the current layer. The reload occurs \textit{after} backward completes for each module, so only one activation per module type resides in GPU memory at any time, avoiding the need to double activation storage. This is essential when a single module has very large activations where 2$\times$ memory footprint would cause unexpected memory peaks.

\paragraph{PP and VPP Handling.}
For PP/VPP scenarios, a \texttt{Chunk\-Offload\-Handler} manages the offloading/reloading logic for each (microbatch, VPP stage) combination, identical to the PP=1 case. The main challenge is managing data dependencies and execution order between virtual pipeline chunks. Handlers are enqueued into a deque with VPP stages in reverse order (FILO) and microbatches in normal order (FIFO). During backward, the popped handler automatically matches the VPP chunk execution order, correctly pairing offloaded activations with their corresponding backward passes.

\paragraph{Key Technical Features.}
\begin{itemize}[itemsep=2pt,topsep=4pt]
    \item \textbf{Module-level granularity:} Users specify which modules to offload via \verb|--offload-modules|, enabling mixed strategies. Lightweight modules (LayerNorm, activations) use recomputation while expensive ones (attention, experts) use offloading.
    \item \textbf{Asynchronous transfers:} Dedicated D2H/H2D CUDA streams run transfers in parallel with computation; CUDA events coordinate synchronization only when necessary.
    \item \textbf{Recomputation integration:} Combines with fine-grained recomputation (\secref{sec:recomputation}). For \texttt{moe\_act}, both strategies apply: recompute the activation while offloading its input, releasing the entire \texttt{fc1}$\rightarrow$\texttt{act} chain.
    \item \textbf{CUDA Graphs compatibility:} Uses external events rather than stream synchronization, allowing offloading modules to be hidden by computations outside the CUDA graph.
    \item \textbf{Full-scenario compatibility:} Supports PP=1/PP$>$1/VPP$>$1, all precisions (BF16/FP8/MXFP8/NVFP4), 1F1B with all-to-all overlap, and MoE/MLA architectures.
\end{itemize}

\paragraph{Peak Memory Advantage over Full Recomputation.}
Full recomputation stores each layer's \textit{input} on GPU while releasing intermediate activations; backward recomputes intermediates from these stored inputs. For an $L$-layer model, GPU peak memory is $L \times \text{layer\_input} + 1 \times \text{layer\_intermediate}$. In contrast, offloading moves layer inputs to CPU; during backward, each input is reloaded just before use and released immediately after. This reduces GPU peak memory up to $1 \times \text{layer\_input} + 1 \times \text{layer\_intermediate}$, independent of model depth. For deep models (e.g., 60+ layers), this represents a fundamental memory advantage that full recomputation cannot achieve.

\begin{figure}[ht]
    \centering
    \includegraphics[width=0.8\linewidth]{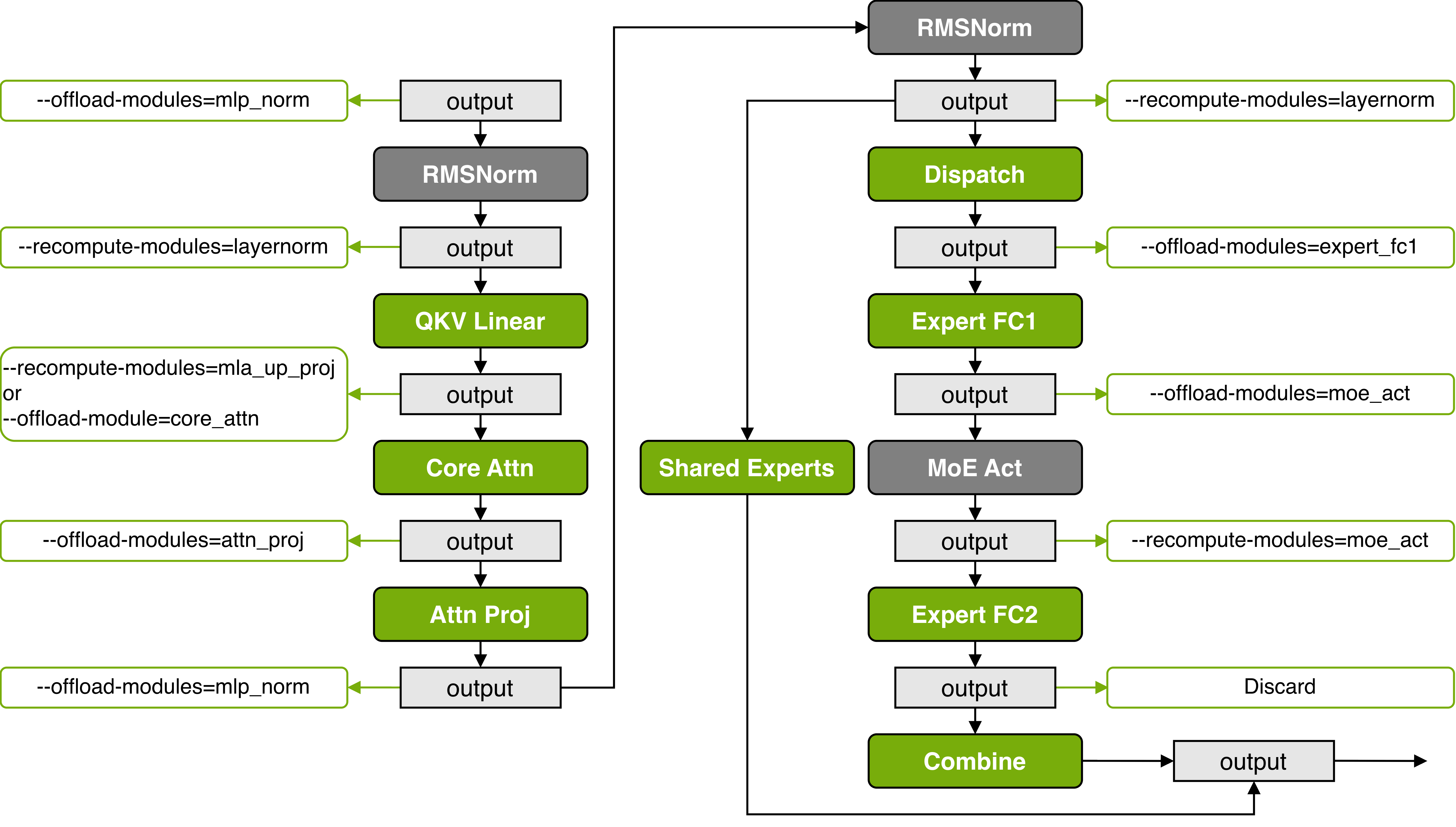}
    \caption{Fine-grained offloading and recomputation: complementary memory optimization strategies.}
    \label{fig:offloading-recomputation}
\end{figure}

\paragraph{Performance.}
Fine-grained offloading and recomputation work together as complementary strategies (\figref{fig:offloading-recomputation}): lightweight operations like LayerNorm use recomputation, while expensive modules like attention and experts use offloading. Asynchronous transfers overlap with computation to hide PCIe latency. Table~\ref{tab:offloading-results} shows results across multiple configurations. Fine-grained offloading reduces memory by 10--18\% with only 1.6--2\% throughput overhead. In the case of training Qwen3-235B, offloading enables reducing Tensor Parallelism degree, which \textit{improves} throughput by 15.0\% with nearly the same memory cost.

\begin{table}[ht]
\centering
\caption{Memory and throughput impact of fine-grained activation offloading.}
\label{tab:offloading-results}
\begin{tabular}{lccccc}
\toprule
\textbf{Model \& Config} & \textbf{Baseline} & \textbf{+Offload} & \textbf{Mem $\Delta$} & \textbf{Throughput $\Delta$} \\
\midrule
DeepSeek-V3 full & 169 GB & 151 GB & $-$10.7\% & $-$1.6\% \\
{\small TP1PP8EP32VPP4, MXFP8} & 945 TF/s & 930 TF/s & & \\
\midrule
Qwen3-235B  & 172 GB & 175 GB & $+$1.7\% & $+$15.0\% \\
{\small TP2$\rightarrow$TP1 + EP16$\rightarrow$EP64} & 800 TF/s & 920 TF/s & & \\
\bottomrule
\end{tabular}
\end{table}

\subsubsection{Weight and Optimizer Optimization: Low-Precision Storage and Offloading}\label{sec:precision-opt}

The preceding optimizations target activation memory, which dominates the memory breakdown. However, Table~\ref{tab:memory-breakdown} shows that main weights and optimizer states consume 32.1 GB per GPU, representing 16\% of the total memory footprint. For models with hundreds of billions of parameters, this component becomes a significant optimization target. Megatron-Core provides two techniques: precision-aware optimization that reduces storage requirements, and CPU offloading that moves inactive state off-GPU.

\paragraph{Precision-Aware Optimizer.}
The Adam optimizer \cite{kingma2014adam} maintains two state tensors per parameter: first moment (\texttt{exp\_avg}) and second moment (\texttt{exp\_avg\_sq}). Traditional implementations store these in FP32, consuming 8 bytes per parameter, which poses a significant memory bottleneck for large-scale training. The key insight is that optimizer states can tolerate lower storage precision without affecting convergence, provided that the actual update computation remains in higher precision.

The precision-aware optimizer decouples storage precision from computation precision \cite{dettmers2022optimizers8bit}. The first and second moments can be stored in BF16 (2 bytes each) or even FP8 (1 byte each), reducing per-parameter storage from 8 bytes to 4 bytes (BF16) or 2 bytes (FP8). During each optimizer step, these lower-precision states are dynamically cast to FP32 within TransformerEngine's FusedAdam kernel; the update is computed with full-precision arithmetic to maintain numerical stability.

The implementation provides four configurable precision levels: main gradients, main parameters, first moment, and second moment precision. In typical configurations, main parameters and gradients remain in FP32 to ensure high-quality gradient updates, while moment estimates are stored in BF16 \cite{liu2024deepseek}. This achieves approximately 50\% reduction in optimizer state memory (roughly 10--12 GB savings from the 32.1 GB budget in Table~\ref{tab:memory-breakdown}) with negligible impact on training dynamics.

When combined with the distributed optimizer, which shards optimizer states across data-parallel ranks of size $d$, the theoretical memory requirement per rank is further reduced. In DeepSeek-V3 training with BF16 moments, the memory consumption per parameter per DP rank decreases from $6+12/d$ bytes to $6+8/d$ bytes.

\paragraph{State Offloading.}
During forward and backward passes, optimizer states occupy GPU memory but remain inactive; offloading reclaims this memory for other operations. Optimizer state offloading keeps the optimizer step on GPU but transfers optimizer states (\texttt{exp\_avg}, \texttt{exp\_avg\_sq}) and master weights to CPU after \texttt{optimizer.step()} and reloads them before the next step. This approach uses GPU compute while reclaiming memory during forward and backward passes.

State offloading is particularly effective on systems with high-bandwidth interconnects. On GB200 with NVLink-C2C, asynchronous transfers overlap with computation, and pinned memory enables maximum bandwidth utilization. For DeepSeek-V3, state offloading saves 15--20 GB of GPU memory (47--62\% of the 32.1 GB optimizer and weight budget) with only 0.1--0.2 seconds per iteration overhead.

These two techniques offer trade-offs that work well together. The precision-aware optimizer incurs no performance overhead since the FP32 cast occurs within the fused Adam kernel; it reduces optimizer state memory by up to 50\%. CPU offloading saves more memory (all the optimizer state and master weights) but introduces modest transfer overhead. Importantly, the techniques compose well: using precision-aware storage with BF16 moments reduces the state size that must be offloaded, thereby decreasing transfer time and making offloading more practical even on systems without the highest-bandwidth interconnects.

\subsubsection{FSDP for MoE}\label{sec:fsdp-ref}

Fully Sharded Data Parallelism (FSDP) \cite{zhao2023pytorchfsdp,rajbhandari2020zero} shards model parameters, gradients, and optimizer states across data-parallel ranks so that each GPU holds only a local shard. Expert parameters often dominate memory in MoE models, making FSDP a natural complement to Expert Parallelism. However, the two must compose correctly.

\paragraph{Why FSDP for MoE?}

Megatron-FSDP composes seamlessly with multiple parallel strategies, including Expert Parallelism (EP), Tensor Parallelism (TP), and Context Parallelism (CP). For large-scale MoE models, combining FSDP with EP turns the general advantages of FSDP into concrete benefits for expert-heavy workloads.

\begin{itemize}[itemsep=2pt]
  \item \textbf{Reduced Memory and Communication}: EP assigns each GPU a subset of experts, and FSDP shards those local experts across the expert data-parallel (EDP) group rather than the full DP group. Per-GPU memory and collective volume both scale with EDP size instead of total DP size, so MoE models can support more experts or larger batches under the same hardware budget.

  \item \textbf{PP-free Flexibility}: FSDP+EP avoids several engineering pain points of pipeline parallelism, including uneven PP/VPP stage balancing, placement of output and MTP layers in DeepSeek-style models, and partitioning vision encoders in multimodal settings. Configuration reduces to choosing EP size and FSDP sharding degree, without complex pipeline-stage design.

  \item \textbf{Broad Model Compatibility}: Megatron-FSDP supports two model implementation paths: models built with Megatron-Core's own modules and PyTorch-native models composed from standard \texttt{torch.nn} modules. Both paths receive the same FSDP+EP sharding, communication optimizations, and checkpointing support. For interoperability with the HuggingFace ecosystem, Megatron-Bridge handles online weight conversion between HuggingFace models and Megatron FSDP, so users can load a pretrained HuggingFace checkpoint, train with FSDP+EP, and export weights back without manual format conversion.
\end{itemize}

\paragraph{FSDP+EP: Dual DeviceMesh Design}

The core design challenge is that dense and expert layers need different sharding scopes. Dense modules (attention, normalization) benefit from FSDP sharding over the full DP group. EP partitions expert modules first, so each GPU holds only its assigned experts. FSDP should therefore shard within each expert's data-parallel (EDP) group, not globally.

Megatron-FSDP resolves this through a \emph{dual DeviceMesh} architecture. A primary DeviceMesh governs DP-Shard, DP-Outer, TP, and CP for dense modules. An auxiliary Expert DeviceMesh manages EP modules with FSDP scoped to the EDP dimension. Each transformer layer routes its sub-modules to the appropriate mesh automatically. Attention and normalization use the primary mesh, while MoE expert FFN layers use the Expert DeviceMesh. As a result, AllGather and ReduceScatter for expert parameters stay within small EDP groups instead of spanning all DP ranks.

% \begin{figure}[ht]
%     \centering
%     \includegraphics[width=0.95\textwidth]{figures/ai_gen/ai_gen_fsdp_ep_dual_devicemesh.png}
%     \caption{Dual DeviceMesh for FSDP+EP: primary mesh for dense layers, expert mesh for MoE layers.}
%     \label{fig:fsdp-ep-mesh}
% \end{figure}

For large-scale deployments, this design extends to \emph{Hybrid Sharded Data Parallelism} (HSDP), which adds an outer DP replication dimension. HSDP fully shards parameters within a subset of ranks and replicates across subsets, bounding AllGather to the intra-group size. The dual DeviceMesh keeps separate outer-DP groups for expert and non-expert parameters, exploiting the bandwidth gap between NVLink (intra-node) and the scale-out interconnect (inter-node).

\paragraph{Zero-Copy Communication}

The dual DeviceMesh determines which ranks participate in each FSDP collective, but collectives themselves still carry overhead from buffer management and data copying. Megatron-FSDP eliminates this overhead through two optimizations.

\begin{figure}[ht]
  \centering
  \begin{subfigure}[b]{0.45\textwidth}
    \centering
    \includegraphics[width=\textwidth]{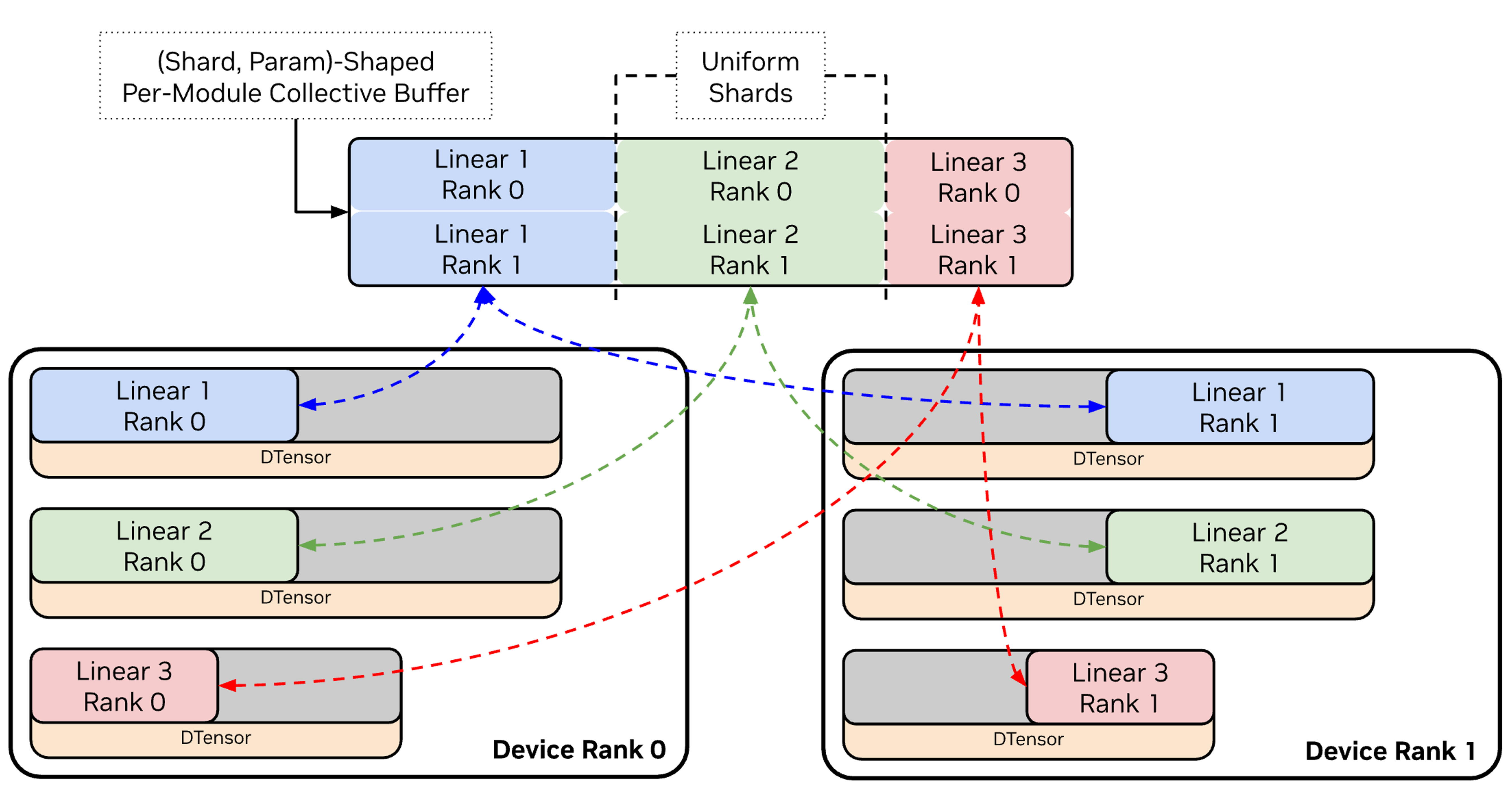}
    \caption{FSDP2 per-parameter uniform sharding.}
    \label{fig:fsdp2-sharding}
  \end{subfigure}
  \hfill
  \begin{subfigure}[b]{0.45\textwidth}
    \centering
    \includegraphics[width=\textwidth]{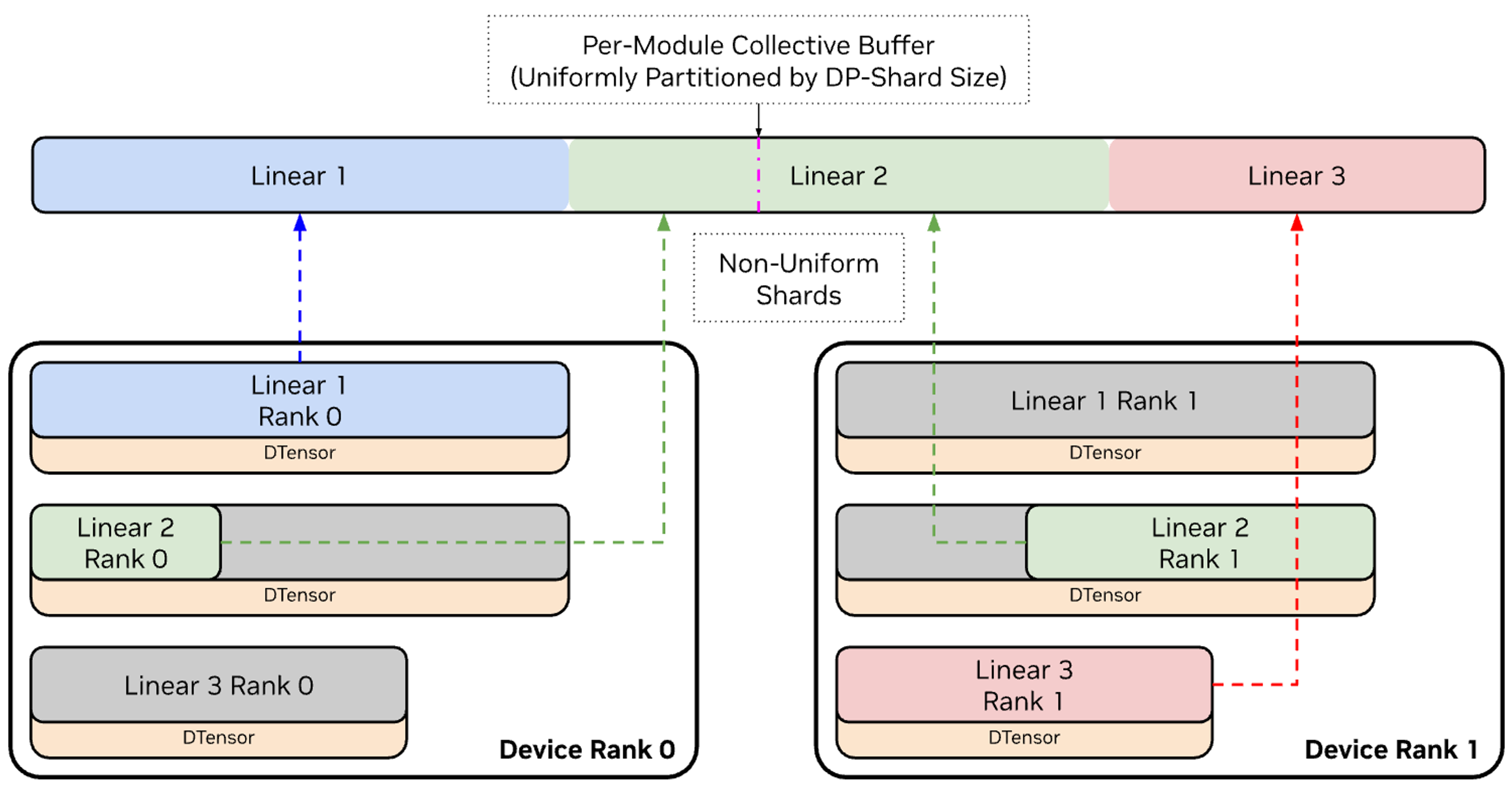}
    \caption{Megatron-FSDP per-module non-uniform sharding.}
    \label{fig:mfsdp-sharding}
  \end{subfigure}
  \caption{Comparison of sharding strategies: (a)~FSDP2 shards each parameter uniformly; (b)~Megatron-FSDP flattens per-module and shards non-uniformly, aligning with communication buffers.}
  \label{fig:sharding-comparison}
\end{figure}

\textbf{1. Non-uniform sharding:} Standard FSDP2 shards each parameter independently along its primary dimension, producing uniform per-parameter shards (Figure~\ref{fig:fsdp2-sharding}). Megatron-FSDP instead flattens and concatenates all parameters within a module, then applies non-uniform sharding across devices (Figure~\ref{fig:mfsdp-sharding}). Shard boundaries then align with communication buffer layouts, and collectives read directly from flattened storage without redundant copying. In Llama3 405B training, this reduces communication overhead by roughly 10\%.

\begin{figure}[ht]
  \centering
  \includegraphics[width=0.85\textwidth]{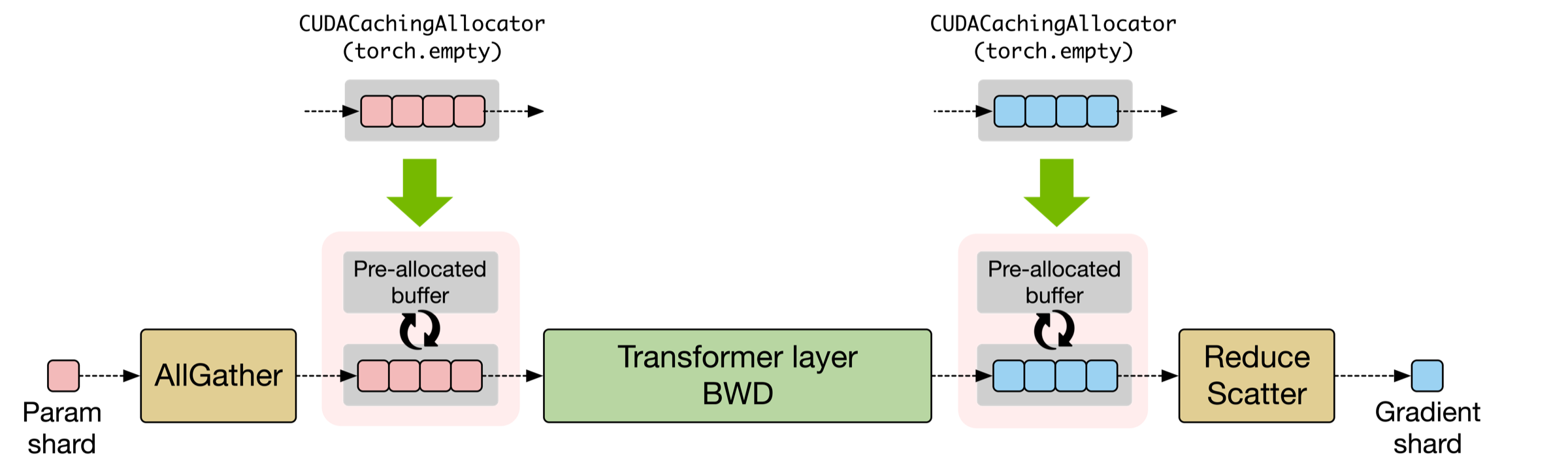}
  \caption{Persistent double-buffer design: two pre-allocated buffers are cycled across FSDP collectives, eliminating allocation overhead and enabling NCCL User Buffer Registration.}
  \label{fig:double-buffer}
\end{figure}

\textbf{2. Persistent double buffers with NCCL User Buffer Registration:} Baseline FSDP frequently allocates and frees communication buffers, and NCCL copies data between user buffers and its internal staging area on every collective. Megatron-FSDP pre-allocates two persistent buffers at training start and cycles between them (Figure~\ref{fig:double-buffer}), eliminating allocation churn. Megatron-FSDP then registers these buffers with NCCL via User Buffer Registration (UBR), so NCCL reads and writes the pre-registered memory directly without intermediate copies. The combined effect is true zero-copy communication. On NVLink systems, the SM footprint of communication kernels drops from 8--32 SMs to 1--4 SMs. On SHARP-enabled InfiniBand, network switches handle reductions and free GPU SMs entirely.

\paragraph{Computation--Communication Overlap}

Even with zero-copy collectives, AllGather and ReduceScatter still occupy the network while GPUs wait for parameters or gradient synchronization. Megatron-FSDP pipelines these collectives with computation using dedicated CUDA streams: AllGather for the next FSDP unit is issued while the current unit's forward or backward pass is still executing, and ReduceScatter for gradients runs concurrently with the backward pass of subsequent layers. Increasing the micro-batch size extends the computation window available for hiding communication, though at the cost of higher memory usage. Users can tune this trade-off through the \texttt{overlap\_param\_gather} and \texttt{overlap\_grad\_reduce} flags.

% \paragraph{DTensor-Based Checkpointing}

% Megatron-FSDP represents all model parameters as PyTorch DTensor objects, each carrying a DeviceMesh and Placement that encode its sharding layout. Dense parameters reference the primary DeviceMesh and expert parameters reference the Expert DeviceMesh, so the dual-mesh separation is preserved at the checkpoint level. Checkpointing uses PyTorch's Distributed Checkpoint (DCP) API end-to-end, with no intermediate format conversions. Because DTensor can also express traditional 3D-parallel (TP+PP+DP) layouts, Megatron-FSDP ships conversion tools that translate existing 3D-parallel checkpoints into DTensor-based formats with no manual resharding.

% Summary of memory optimizations
\subsubsection{Summary}
Table~\ref{tab:memory-opt-summary} summarizes the memory optimization techniques described in this section and their primary targets.

\begin{table}[ht]
\centering
\caption{Summary of memory optimization techniques.}
\label{tab:memory-opt-summary}
\begin{tabular}{lll}
\toprule
\textbf{Technique} & \textbf{Memory Target} & \textbf{Trade-off} \\
\midrule
Reduced-Precision Training & Activations & Numerical Precision and CPU Overhead \\
Memory-Efficient Permutation & Activations & - \\
Fine-grained Recomputation & Activations & Compute Overhead \\
Fine-grained Offloading & Activations & CPU Overhead and Non-Overlapped Copy \\
Precision-aware Optimizer & Optimizer States & Numerical Precision \\
FSDP (with EP) & Params + Optimizer & Communication Overhead \\
\bottomrule
\end{tabular}
\end{table}

These memory optimizations complement each other: Memory-Efficient Permutation eliminates redundant storage with zero overhead; FP8/FP4 activations reduce precision with minimal quality impact; fine-grained recomputation offers favorable compute-memory trade-offs for specific modules; activation offloading provides additional headroom when other techniques are insufficient; low-precision storage and state offloading reduce weight and optimizer memory; and FSDP enables scaling beyond single-device capacity. Together, they reduce memory from a blocking barrier to a manageable constraint.

Memory optimization is not a one-time gate that, once passed, can be forgotten. For large-scale MoE models, memory is a persistently scarce resource throughout the entire optimization lifecycle. Beyond enabling training to proceed at all, memory headroom unlocks other optimizations: larger batch sizes provide more computation to hide communication latency (Section~\ref{sec:comm-overlap}), CUDA Graphs require additional static buffers (Section~\ref{sec:cuda-graphs}), and EP communication overlap must hold activations from multiple microbatches simultaneously. Many optimizations described in the following sections consume memory, and the techniques above are what make that consumption feasible.

\subsection{Breaking the Communication Wall}\label{sec:comm-wall}

Communication overhead directly reduces GPU utilization: every microsecond spent in collective operations represents lost compute capacity. Before optimization, EP all-to-all communication typically consumes 20--60\% of training time, depending on model configuration, EP size, and hardware topology. When EP stays within the NVLink domain (e.g., DeepSeek-V3 with EP64 on GB200 NVL72), the overhead is around 20\%; when EP crosses node boundaries (e.g., DeepSeek-V3 with EP64 on H100 across nodes), it rises to 40--60\%. The techniques in this section target all points on this spectrum.

Expert Parallelism (EP) distributes experts across devices to scale MoE models beyond single-device capacity, but this distribution introduces a unique communication pattern. Unlike AllReduce, whose volume is independent of parallelism degree, all-to-all volume is proportional to token count and hidden dimension, and larger EP pushes this communication cross-node where bandwidth is limited. For fine-grained MoE architectures like DeepSeek-V3 and Kimi-K2, three factors compound this challenge:

\begin{itemize}
    \item \textbf{High frequency}: More experts mean more dispatch/combine operations per layer (DeepSeek-V3 has 58 MoE layers, each requiring two all-to-all operations).
    \item \textbf{Cross-node bottleneck}: With large EP sizes spanning multiple nodes, inter-node all-to-all latency dominates due to lower bandwidth.
    \item \textbf{Low arithmetic intensity}: Small experts complete computation quickly, leaving less time to overlap with communication.
\end{itemize}

\subsubsection{Communication Anatomy: The Expert Parallel Pattern}

\begin{figure}[ht]
    \centering
    \includegraphics[width=0.8\linewidth]{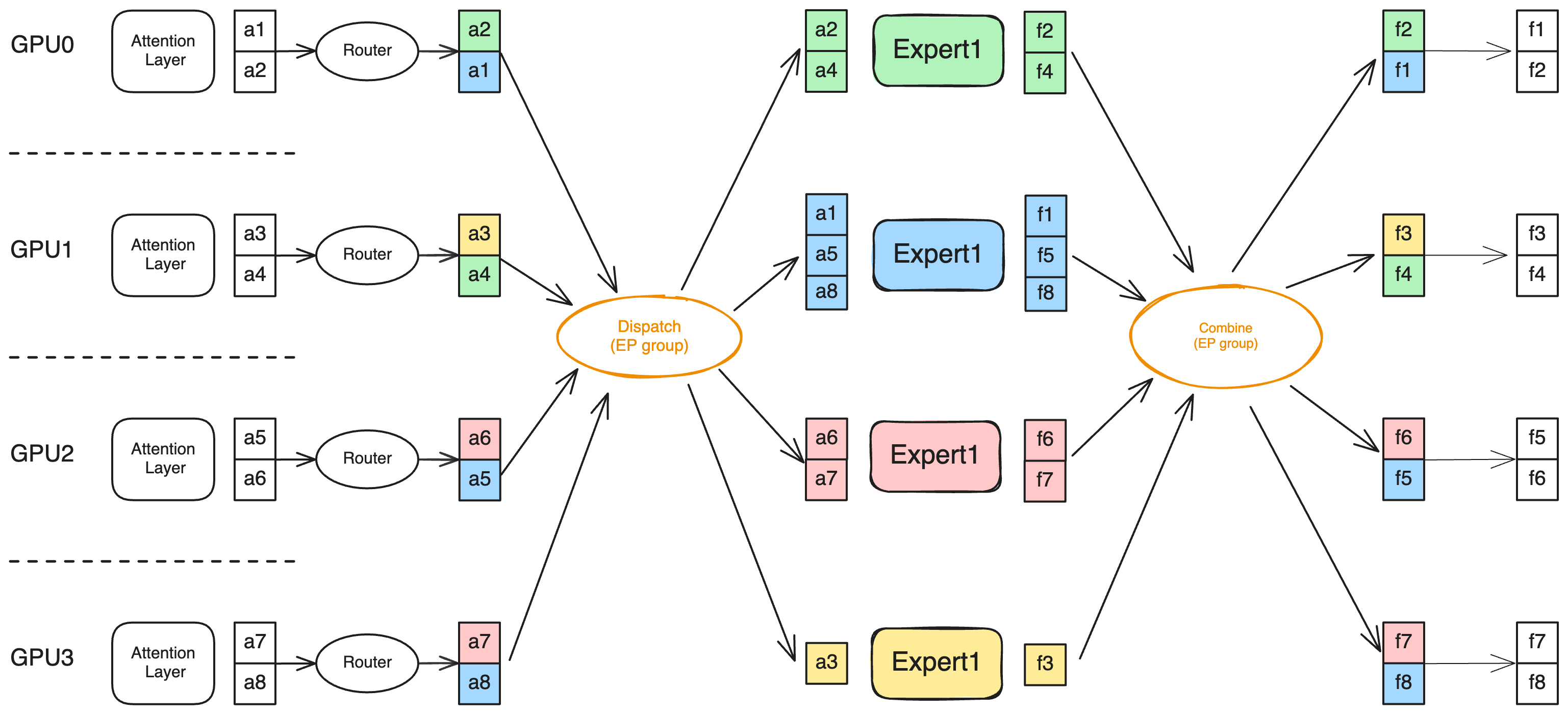}
    \caption{Expert parallelism across 4 GPUs with 4 experts.}
    \label{fig:EP_comm}
\end{figure}

\Figref{fig:EP_comm} illustrates the expert parallelism communication pattern. In standard EP implementations, each MoE layer requires two collective communication operations: \textit{dispatch} sends tokens to their assigned experts, and \textit{combine} returns processed tokens to their original ranks. For a model with $L$ MoE layers, $B$ tokens per batch, hidden dimension $h$, and EP degree $\text{EP}$, each forward pass involves $2L$ dispatch/combine operations, each transferring $O(Bh)$ data across $\text{EP}$ ranks.

At DeepSeek-V3 scale, this translates to 58 MoE layers $\times$ 2 operations/layer = 116 dispatch/combine operations per forward pass. The backward pass doubles this count. At 50 GB/s inter-node bandwidth (e.g., InfiniBand NDR), a single dispatch with 200 MB payload takes several milliseconds, accumulating to hundreds or thousands of milliseconds per iteration.

Communication volume is determined by the EP configuration and cannot be reduced without changing the parallelism strategy. Optimization must therefore target how communication is executed and scheduled. Two strategies work together to address communication overhead:

\begin{enumerate}
    \item \textbf{Maximize bandwidth utilization.} Standard NCCL all-to-all implementations do not fully exploit available bandwidth, particularly for fine-grained MoE workloads. Optimized dispatchers (DeepEP, HybridEP) use specialized kernels that fuse operations and exploit hardware primitives to approach peak bandwidth.
    \item \textbf{Hide latency behind computation.} Even with optimal bandwidth, all-to-all operations take time. By overlapping communication with computation from adjacent microbatches, this latency can be hidden rather than exposed on the critical path.
\end{enumerate}

The following subsections present techniques organized by these strategies: bandwidth optimization (DeepEP, HybridEP), and latency hiding (EP communication overlapping, pipeline integration).

\subsubsection{DeepEP and HybridEP: Maximizing EP Bandwidth}\label{sec:deepep}

In conventional EP implementations, token exchange relies on all-to-all communication. Even with optimized NCCL collectives, this path has inherent limitations: a permutation stage is needed before dispatch, which replicates each token top-$k$ times and creates redundant traffic. In some settings, this preprocessing also surfaces as host overhead.

To address these issues, Megatron-Core provides two token-based dispatch backends, HybridEP and DeepEP \cite{deepep2025}. Token-based dispatch eliminates the permutation step and avoids sending redundant tokens, reducing overall communication volume and improving effective bandwidth.

HybridEP was developed by NVIDIA. It follows the same token-based principle as DeepEP, exploits hardware primitives such as TMA and IBGDA, and targets comparable or higher bandwidth with lower SM usage, including Multi-Node NVLink (MNNVL) deployments.

\begin{figure}[ht]
    \centering
    \includegraphics[width=1\linewidth]{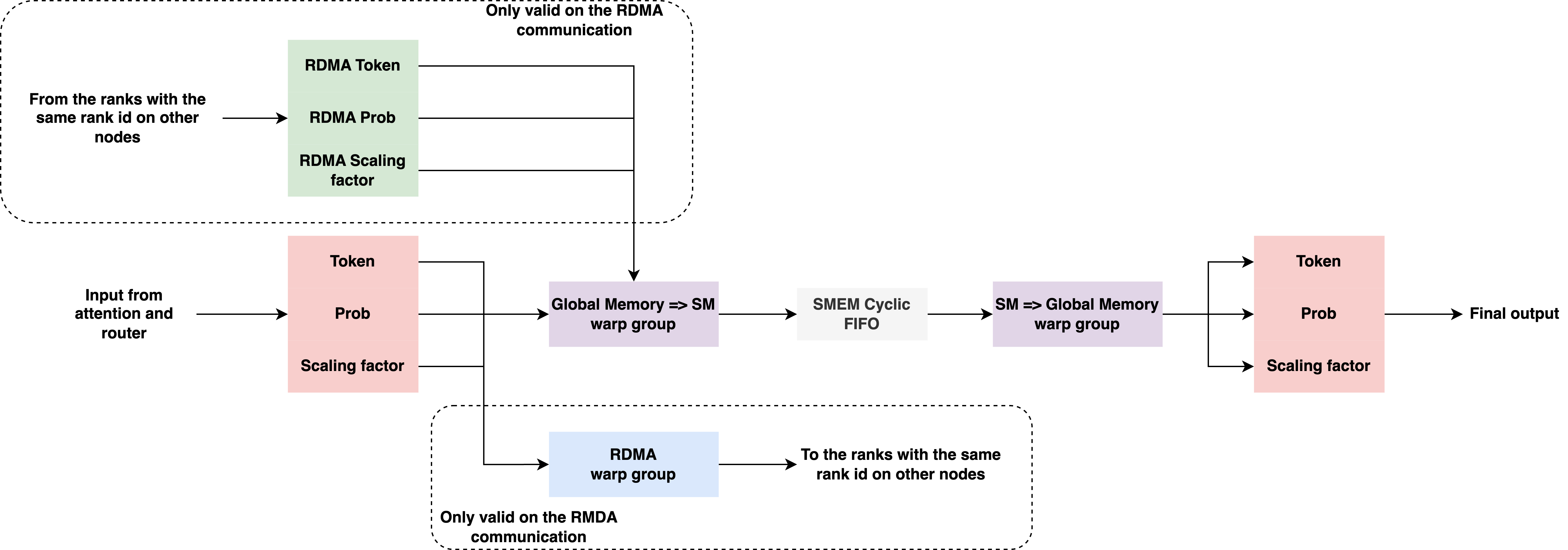}
    \caption{The dispatch kernel design of HybridEP.}
    \label{fig:hybrid_ep_dispatch}
\end{figure}

For dispatch, as illustrated in \figref{fig:hybrid_ep_dispatch}, HybridEP reads data from global memory into shared memory based on routing information, then writes tokens to destinations through a FIFO queue. In the inter-node case, instead of sending duplicated payloads directly through the network interface, HybridEP first uses an RDMA warp group to exchange data between GPUs with the same local index across nodes, then forwards within each node. This reduces cross-node traffic and allows inter-node and intra-node transfers to overlap.

\begin{figure}[ht]
    \centering
    \includegraphics[width=1\linewidth]{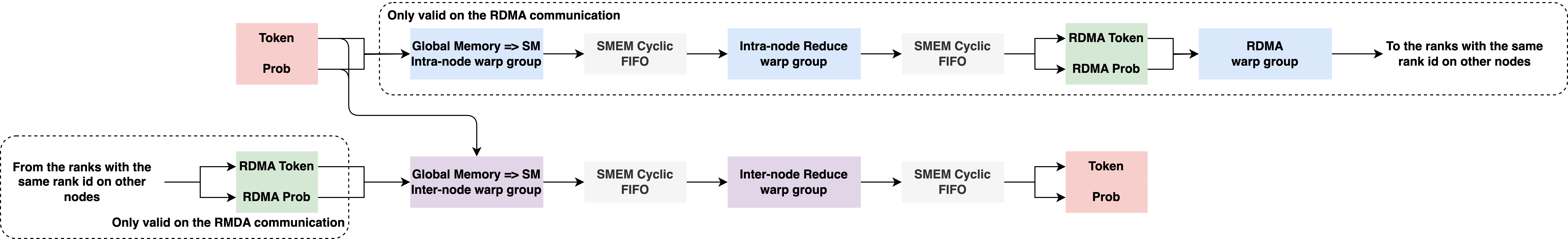}
    \caption{The combine kernel design of HybridEP.}
    \label{fig:hybrid_ep_combine}
\end{figure}

For combine, standard all-to-all dispatch performs communication only, so a separate unpermutation stage is required afterward. HybridEP instead fuses reduction into the communication kernel. As shown in \figref{fig:hybrid_ep_combine}, HybridEP reads data through a FIFO queue, performs reduction, and writes results directly to target locations. In inter-node settings, HybridEP first performs reduction across nodes for data that must be communicated cross-node, then completes a second reduction within each node.

We evaluate all-to-all dispatch and HybridEP on GB200 and H100 with hidden size 7168, sequence length 4096, and 256 experts under multiple EP sizes. Results are summarized in Table~\ref{tab:hybrid-ep}. Across all tested settings, HybridEP consistently outperforms all-to-all, with larger gains in inter-node scenarios. The table reports communication latency only; in end-to-end training, the gap is typically larger once permutation and host overhead are included.

\begin{table}[ht]
\centering
\caption{EP Scaling Performance for HybridEP and all-to-all (in µs).}
\label{tab:hybrid-ep}
\begin{tabular}{llrrrr}
\toprule
 &  & \multicolumn{2}{c}{GB200 (µs)} & \multicolumn{2}{c}{H100 (µs)} \\
 & EP size & HybridEP & all-to-all & HybridEP & all-to-all \\
\midrule
\multirow{4}{*}{dispatch}
 & 8  & 391 & 735 & 661  & 1265 \\
 & 16 & 578 & 743 & 1485 & 5774 \\
 & 32 & 612 & 769 & 3064 & 8059 \\
 & 64 & 675 & 930 & 4626 & 9164 \\
\midrule
\multirow{4}{*}{combine}
 & 8  & 353 & 741 & 624  & 1277 \\
 & 16 & 527 & 765 & 1688 & 5628 \\
 & 32 & 646 & 758 & 3088 & 7815 \\
 & 64 & 744 & 827 & 4398 & 8727 \\
\bottomrule
\end{tabular}
\end{table}

\subsubsection{EP Communication Overlapping: Hiding EP Communication Latency}\label{sec:comm-overlap}

Optimized dispatchers improve all-to-all \textit{bandwidth utilization}, but the fundamental issue remains: all-to-all still lies on the end-to-end critical path. For DeepSeek-V3 with EP64, EP all-to-all communication can still account for 30--40\% of iteration time, directly limiting throughput.

The key observation is that all-to-all latency can be \textit{hidden} behind computation when enough independent work exists in the overlap window \cite{wang2022coconet,wang2024flux}. For fine-grained MoE models such as DeepSeek-V3 and Kimi-K2, this is challenging because cross-node EP communication can consume about half of layer time, while adjacent operators (for example, GEMMs) are often too short to hide it.

To address this imbalance, Megatron-Core uses a dedicated 1F1B all-to-all overlap scheme \cite{1f1boverlap} that merges forward and backward passes from neighboring micro-batches and interleaves compute and all-to-all kernels across CUDA streams \cite{narayanan2019pipedream,huang2019gpipe}. Concretely, backward all-to-all for one micro-batch overlaps with forward attention/MLP for another. Conceptually, this is a DualPipe-like \cite{hfdualpipe} bidirectional schedule built on top of standard 1F1B while preserving Megatron-Core compatibility.

To enable all-to-all overlap in 1F1B, we merge adjacent micro-batch forward and backward passes and evaluate two patterns:

\begin{enumerate}
    \item \textbf{Merged FWD-FWD / BWD-BWD}: This strategy merges passes of the same type from two micro-batches (\figref{fig:fwd_fwd_merge}). In \textit{FWD-FWD}, forward passes for micro-batches 0 and 1 run in parallel and overlap all-to-all with computation. \textit{BWD-BWD} applies the same principle to backward passes. This approach has clear trade-offs:
    \begin{itemize}[leftmargin=*]
        \item \textbf{Memory Overhead}: 2x peak activation memory.
        \item \textbf{all-to-all Overlap}: Less opportunity to hide all-to-all because forward computation is roughly half of backward computation.
    \end{itemize}
    
    \begin{figure}[ht]
        \centering
        \includegraphics[width=0.8\linewidth]{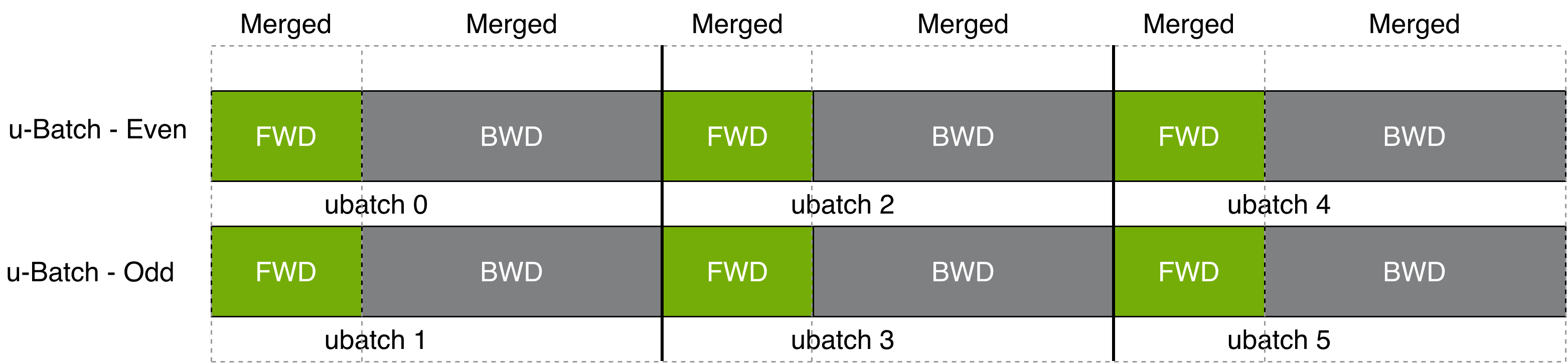}
        \caption{Merged FWD-FWD Timeline with all-to-all Overlapping.}
        \label{fig:fwd_fwd_merge}
    \end{figure}

    \item \textbf{Merged FWD-BWD (DualPipe Equivalent)}: This is the preferred strategy (\figref{fig:fwd_bwd_merge}). It merges the forward pass of one micro-batch with the backward pass of another (for example, FWD of micro-batch 1 with BWD of micro-batch 0). Compared with \textit{FWD-FWD}, it offers:
    \begin{itemize}[leftmargin=*]
        \item \textbf{Memory Overhead}: No additional memory overhead (activations from the forward pass are reused for the backward pass).
        \item \textbf{Compatibility}: Matches DualPipe's design but avoids complex scheduling.
        \item \textbf{Limitation}: The first FWD and last BWD remain on the end-to-end critical path, so their all-to-all cannot be hidden.
    \end{itemize}

    \begin{figure}[ht]
        \centering
        \includegraphics[width=0.8\linewidth]{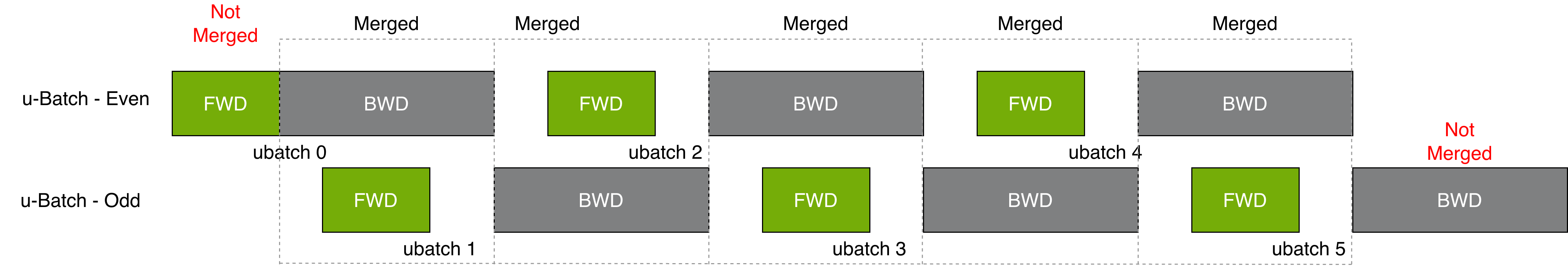}
        \caption{Merged FWD-BWD Timeline with all-to-all Overlapping.}
        \label{fig:fwd_bwd_merge}
    \end{figure}
\end{enumerate}

\begin{figure}[ht]
    \centering
    \includegraphics[width=0.9\textwidth]{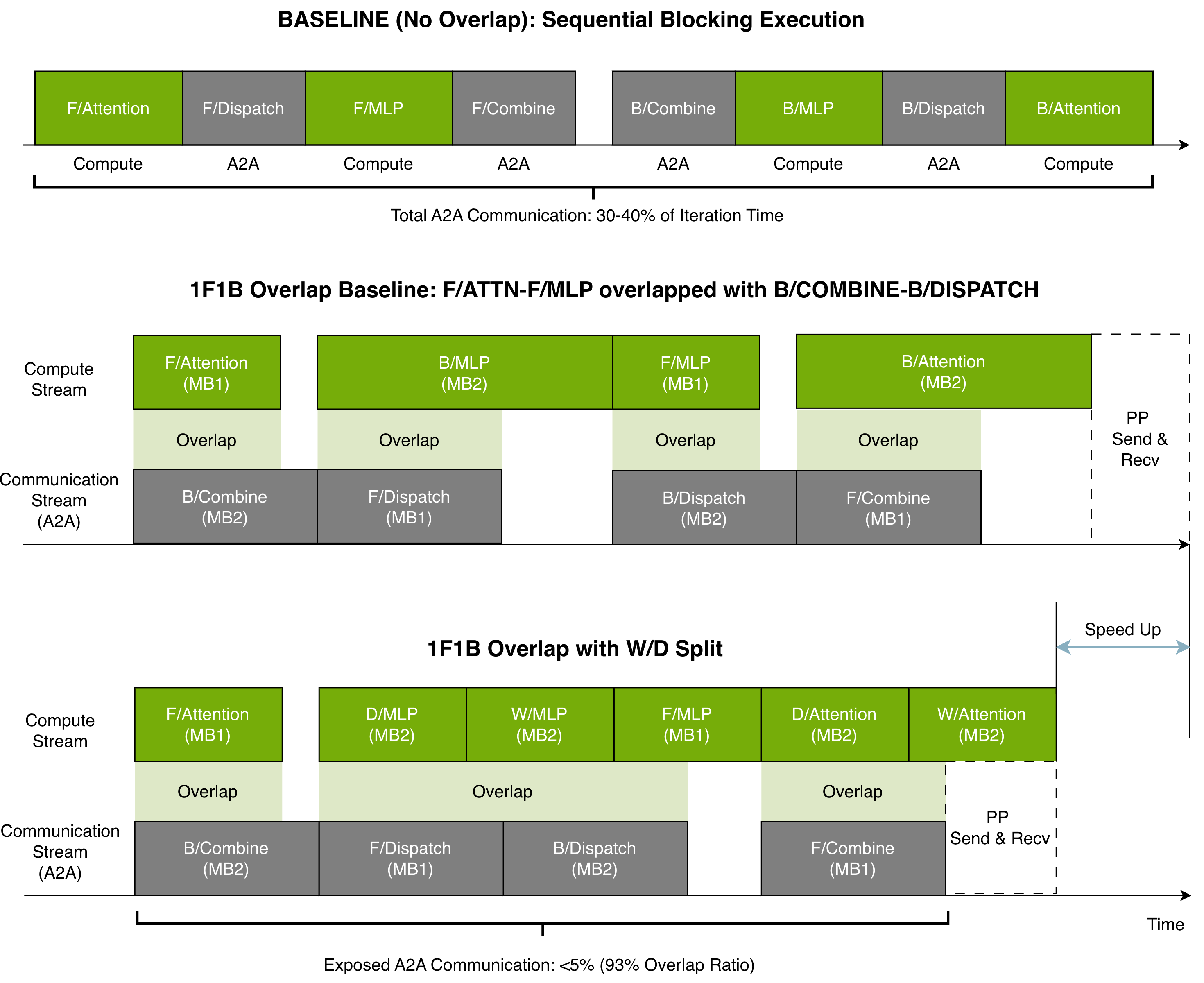}
    \caption{EP all-to-all communication overlap strategies: baseline vs. 1F1B with W/D split.}
    \label{fig:ep-all-to-all-overlap}
\end{figure}

To maximize all-to-all hiding in merged FWD-BWD, we use two key optimizations:

\begin{enumerate}
    \item \textbf{Stream Separation}: We split workloads into two CUDA streams:
    \begin{itemize}[leftmargin=*]
        \item \textit{Compute Stream}: Runs forward/backward computations (e.g., attention, expert MLP).
        \item \textit{Comm Stream}: Runs all-to-all communication (e.g., token dispatch/combine for EP).
    \end{itemize}
    By alternating tasks across streams, all-to-all runs in parallel with computation—minimizing idle cycles.

    \item \textbf{W/D Split (Weight-Gradient / Data-Gradient Split)} \cite{deepseekai2025deepseekv3technicalreport}: A key dependency blocks overlap: backward dispatch (B/dispatch) requires outputs from backward MLP (B/mlp). To break this dependency, we split backward MLP into:
    \begin{itemize}[leftmargin=*]
        \item \textit{W/mlp}: Weight gradient calculation (independent of B/dispatch).
        \item \textit{D/mlp}: Data gradient calculation (feeds into B/dispatch).
    \end{itemize}
    This split reduces compute-stream idle time: W/mlp can overlap with F/mlp to hide B/dispatch when F/mlp alone is too short.
\end{enumerate}
\Figref{fig:ep-all-to-all-overlap} summarizes these optimizations. The baseline (top) executes sequentially, where all-to-all blocks computation. The 1F1B FWD-BWD scheme (bottom) interleaves adjacent micro-batches so communication can be hidden behind compute. The W/D split further increases overlap opportunities. Together, these methods reduce EP communication overhead from 30--40\% (after DeepEP) to under 5\% of iteration time in DeepSeek-V3 training on H100.

% \begin{figure}[ht]
%     \centering
%     \includegraphics[width=0.8\linewidth]{figures/1f1b_split_wgrad.pdf}
%     \caption{Compute/Comm Stream Schedules}
%     \label{fig:compute_comm_streams}
% \end{figure}

\begin{figure}[ht]
    \centering
    \includegraphics[width=0.9\linewidth]{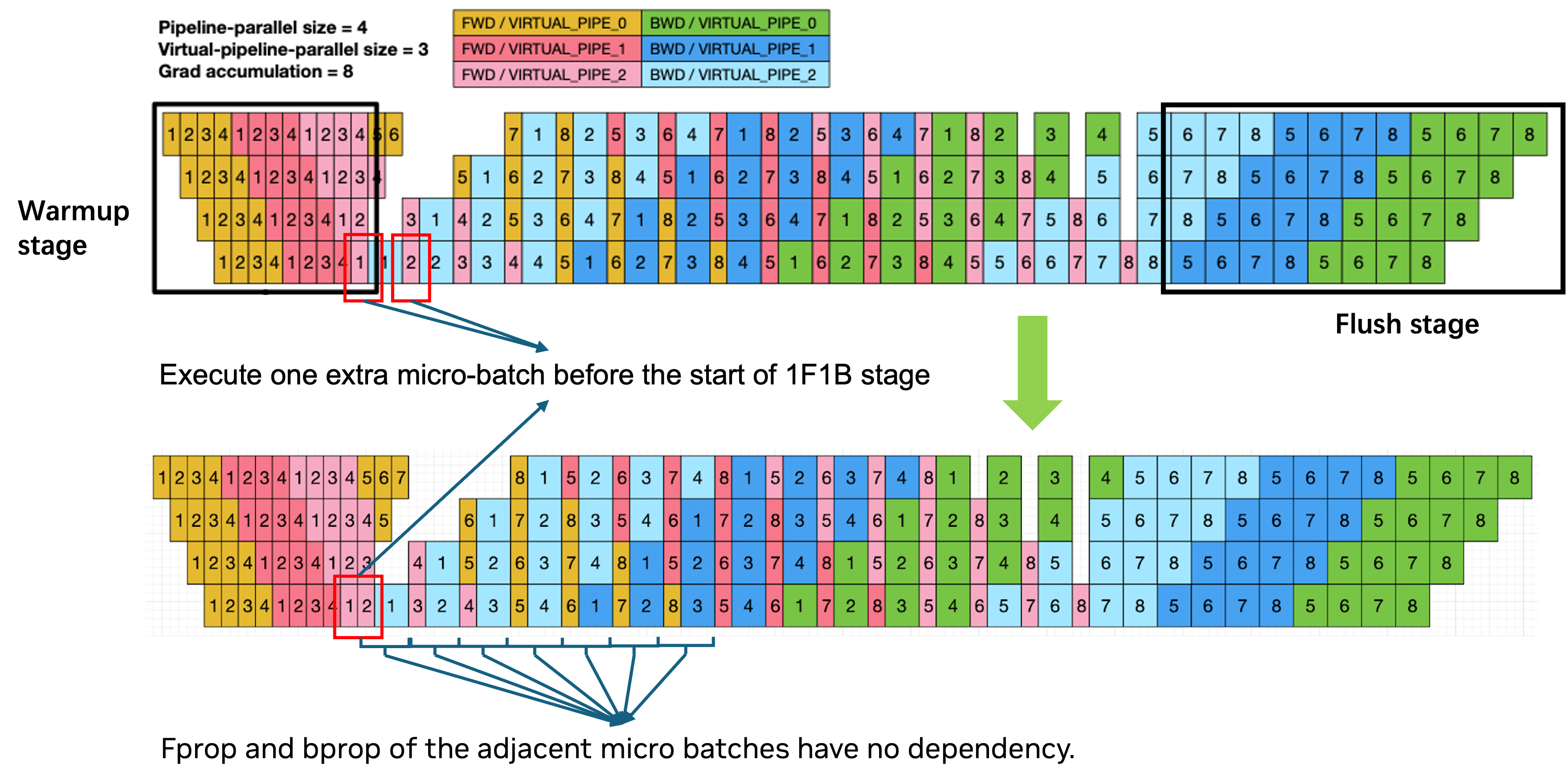}
    \caption{Interleaved PP Timeline with all-to-all Overlapping.}
    \label{fig:interleaved_pp}
\end{figure}

To scale overlap to large models, merged FWD-BWD can be combined with \textit{Interleaved Pipeline Parallelism} (Interleaved PP), which partitions the model into virtual pipeline stages (VPP) and expands overlap opportunities across pipeline ranks. \Figref{fig:interleaved_pp} shows an interleaved 1F1B schedule with three phases: warmup, 1F1B, and flush. Adjacent FWD-BWD pairs in the 1F1B phase can apply the EP overlap pattern above. However, adjacent pairs may still have data dependencies when they belong to the same micro-batch. To avoid this, we run \textit{one extra micro-batch} at the end of warmup before entering 1F1B, ensuring adjacent FWD/BWD pairs are dependency-free and enabling full all-to-all overlap across virtual stages.

A latency comparison between merged FWD-BWD (1F1B) and DualPipeV (a refined DualPipe variant) shows:
\begin{itemize}
    \item \textit{1F1B is faster} with large VPP sizes (more virtual stages enable more overlapping opportunities).
    \item The latency gap shrinks at large PP sizes and micro-batch counts (both strategies reach near-optimal overlap).
    \item For hybrid models (e.g., DeepSeek-V3 with mixed dense/MoE layers), workload balance across PP stages is critical. Megatron-Core's \textit{Flexible Asymmetric VPP} supports custom per-stage layer placement to maximize overlap.
\end{itemize}

Several factors limit the achievable speedup from all-to-all overlap:
\begin{itemize}
    \item \textbf{Proportion of overlapped batches}: More micro-batches increase the proportion of overlapped batches.
    \item \textbf{all-to-all Proportion}: EP overlap gains are larger when all-to-all dominates, especially for cross-node EP and fine-grained MoE models with low compute-to-communication ratios.
    \item \textbf{SM Carve-out Overhead}: Reserving SMs for all-to-all can reduce GEMM efficiency. In our DeepSeek-V3 benchmark, DeepEP uses 20 SMs per GPU and introduces roughly 20\% GEMM-efficiency overhead.
\end{itemize}

On fine-grained MoE models such as DeepSeek-V3, combining all-to-all overlap with optimized dispatchers (Section~\ref{sec:deepep}) achieves a 93\% overlap ratio for expert communication latency, reducing expert communication share from 30--40\% of iteration time to under 5\%.

% \subsubsection{FP8 EP Communications: Reducing Communication Volume}\label{sec:fp8-comm}

% The optimizations above hide all-to-all latency behind computation. An orthogonal approach is to reduce the all-to-all \textit{volume} itself. By performing dispatch communication in FP8 instead of BF16, we halve the data transferred per all-to-all operation.

% However, FP8 dispatch requires careful handling depending on the FP8 recipe:
% \begin{itemize}
%     \item \textbf{Blockwise/Subchannel FP8}: Tokens must be quantized before dispatch. In backward pass, we dequantize to reconstruct high-precision input, then re-quantize because forward and backward use different quantization dimensions. Transposition may be required for Hopper Tensor Core layouts.
%     \item \textbf{MXFP8}: Similar to blockwise, though transposition is unnecessary as Blackwell Tensor Cores accept any layout, backward computation requires the input tokens to be quantized along a different dimension. As a result, dequantization and re-quantization are still needed.
% \end{itemize}

% Note that \textit{combine} communication typically remains in higher precision to avoid gradient accumulation errors. The dispatch-only FP8 strategy provides 50\% all-to-all volume reduction with minimal accuracy impact. See Section~\ref{sec:fp8-dispatch-comm} for more details.

\subsubsection{Summary}
The communication wall optimizations work together: DeepEP/HybridEP maximize bandwidth utilization; FWD-BWD overlap hides latency behind computation; Interleaved PP extends overlap opportunities across pipeline stages; Flexible VPP enables optimal load balance. Combined, they reduce all-to-all's contribution to training time under 10\%.

With communication latency hidden behind computation, high GPU utilization might be expected. However, profiling reveals the Compute Efficiency Wall, which has two distinct aspects: \textit{kernel efficiency} (fine-grained experts produce small GEMMs that underutilize GPU resources) and \textit{host overhead} (numerous small operations create gaps between kernel executions where the GPU awaits work from the host).

\subsection{Breaking the Compute Efficiency Wall}\label{sec:compute-wall}

Compute efficiency is the final constraint: even with sufficient memory and hidden communication, GPU resources can remain underutilized if kernels are inefficient or the host cannot dispatch work fast enough. As model and expert counts grow, compute inefficiencies compound, making optimization essential for achieving peak hardware utilization.

\subsubsection{Compute Anatomy: Sources of Inefficiency}

Compute inefficiency in MoE training stems from two distinct sources:

\textbf{Kernel Efficiency.} Fine-grained MoE architectures produce workloads that underutilize GPU resources. DeepSeek-V3's 256 small experts yield GEMMs with M dimensions of $\sim$128 tokens per expert, far from the thousands needed for peak Tensor Core utilization. Beyond expert computation, MoE introduces complex routing and dispatching operations composed of many small kernels that individually cannot saturate GPU compute capacity.

\textbf{Host Overhead.} Numerous small operations create kernel launch overhead, and the CPU cannot dispatch work fast enough to keep the GPU saturated. Three factors make this worse:
\begin{itemize}
    \item \textbf{Fine-grained experts}: Many individual GEMMs rather than few large ones, each requiring a separate kernel launch.
    \item \textbf{Reduced-precision training}: Additional quantization kernels add to the launch overhead.
    \item \textbf{Dropless routing}: Dynamic token counts require host-device synchronization to determine tensor shapes, placing the CPU on the critical path.
\end{itemize}

The result is \textit{host-boundedness}: GPU kernels separated by microseconds of host-side latency, visible as gaps in profiling traces where compute resources sit idle.

Kernel shapes are determined by model architecture and cannot be changed without modifying the model. Optimization must therefore target how computation is organized and scheduled. Five strategies work together to address compute inefficiency:

\begin{enumerate}
    \item \textbf{Fuse related operations.} Kernel fusion consolidates routing, permutation, and auxiliary loss computations into fewer, larger kernels.
    \item \textbf{Accelerate with low precisions.} Reduced-precision training in FP8/FP4 uses faster Tensor Core operations, increasing throughput for the same kernel shapes.
    \item \textbf{Eliminate per-iteration CPU logic.} CUDA Graphs capture kernel sequences and replay them with minimal host involvement.
    \item \textbf{Remove host-device synchronization.} Sync-free execution enables GPU kernels to proceed without waiting for shape information from the host.
    \item \textbf{Balance expert load.} ECHO dynamically clones hot experts to underutilized ranks, reducing load imbalance that causes some ranks to wait while others compute.
\end{enumerate}

The following subsections present techniques organized by these strategies: batching and fusion for kernel efficiency (Section~\ref{sec:kernel-fusion}), low-precision acceleration (Section~\ref{sec:fp8-training}), CUDA Graphs for host overhead elimination (Section~\ref{sec:cuda-graphs}), sync-free execution for dropless MoE (Section~\ref{sec:sync-free-moe}), and load balancing via ECHO (Section~\ref{sec:echo}).

\subsubsection{Grouped GEMM and Kernel Fusion: Improving Kernel Efficiency}\label{sec:kernel-fusion}

The most direct approach to improving kernel efficiency is combining small operations into larger ones. Grouped GEMM batches expert computations for better hardware utilization \cite{megablocks}, while kernel fusion consolidates routing and permutation operations into fewer kernels \cite{chen2018tvm}. Megatron-Core implements these optimizations at three levels: expert computation (Grouped GEMM), token routing (Permutation Fusion), and router logic (Router and Aux-Loss Fusion).

\paragraph{Grouped GEMM}\label{sec:grouped-gemm}

The computation of experts is essentially a series of independent GEMMs. Grouped GEMM improves performance over separate sequential GEMMs by overlapping the wave tail effect of kernels.

Megatron-Core provides two Grouped GEMM implementations:

\begin{itemize}
\item[i.] Multi-stream launch of cuBLASLt GEMMs.
By launching individual cuBLASLt GEMMs into multiple CUDA streams, they can overlap with each other. This method supports various precision and scaling modes, including BF16, per-tensor FP8, blockwise FP8, MXFP8, and NVFP4, with performance comparable to highly optimized Grouped GEMM implementations.

\item[ii.]\label{gmm:2} CUTLASS Grouped GEMM.
By fusing individual GEMMs into a single kernel, CUTLASS Grouped GEMM achieves better performance when the number of GEMMs is large. However, different precisions, scaling modes, and hardware platforms require individual development effort, and additional tuning is needed for different problem sizes. The current implementation in TE only supports BF16 on the Hopper platform.
\end{itemize}

Two additional implementations are under development:

\begin{itemize}
\item[iii.] cuBLASLt Grouped GEMM via \texttt{CUBLASLT\_BATCH\_MODE\_GROUPED}.
The cuBLASLt Grouped GEMM assumes the shape information on device and conducts the computation in one single kernel, which unblocks applying CUDA Graphs to the expert part. It covers all precisions and scaling modes with built-in heuristics, and is intended to supersede (i) and (ii) where supported.

\item[iv.] cuteDSL Grouped GEMM with fusions.
Grouped GEMM fused with activation functions, quantization (for the next layer), and scaling factor swizzling via cuteDSL. These kernels are optimized for MXFP8 and NVFP4~\cite{abecassis2025nvfp4} on the Blackwell platform, targeting the \texttt{fprop} GEMM of FC1 and \texttt{dgrad} GEMM of FC2. By consolidating expert computation, activation, and quantization into a single kernel, this approach significantly reduces kernel count and is compatible with CUDA Graphs.
\end{itemize}

\subsubsection{Permutation Fusion}

Currently, Grouped GEMM requires tokens assigned to the same expert to be stored contiguously (i.e., permuted tokens). If all-to-all collective communication is enabled, further permutation is necessary to ensure tokens assigned to each rank are grouped together. Implementing efficient permutation using native PyTorch is challenging, as it tends to launch many small kernels and incurs additional CPU overhead.

The permute fusion pipeline in the training workload (\figref{fig:permute_fusion}) consists of three stages:

\begin{figure}[ht]
    \centering
    \includegraphics[width=0.6\linewidth]{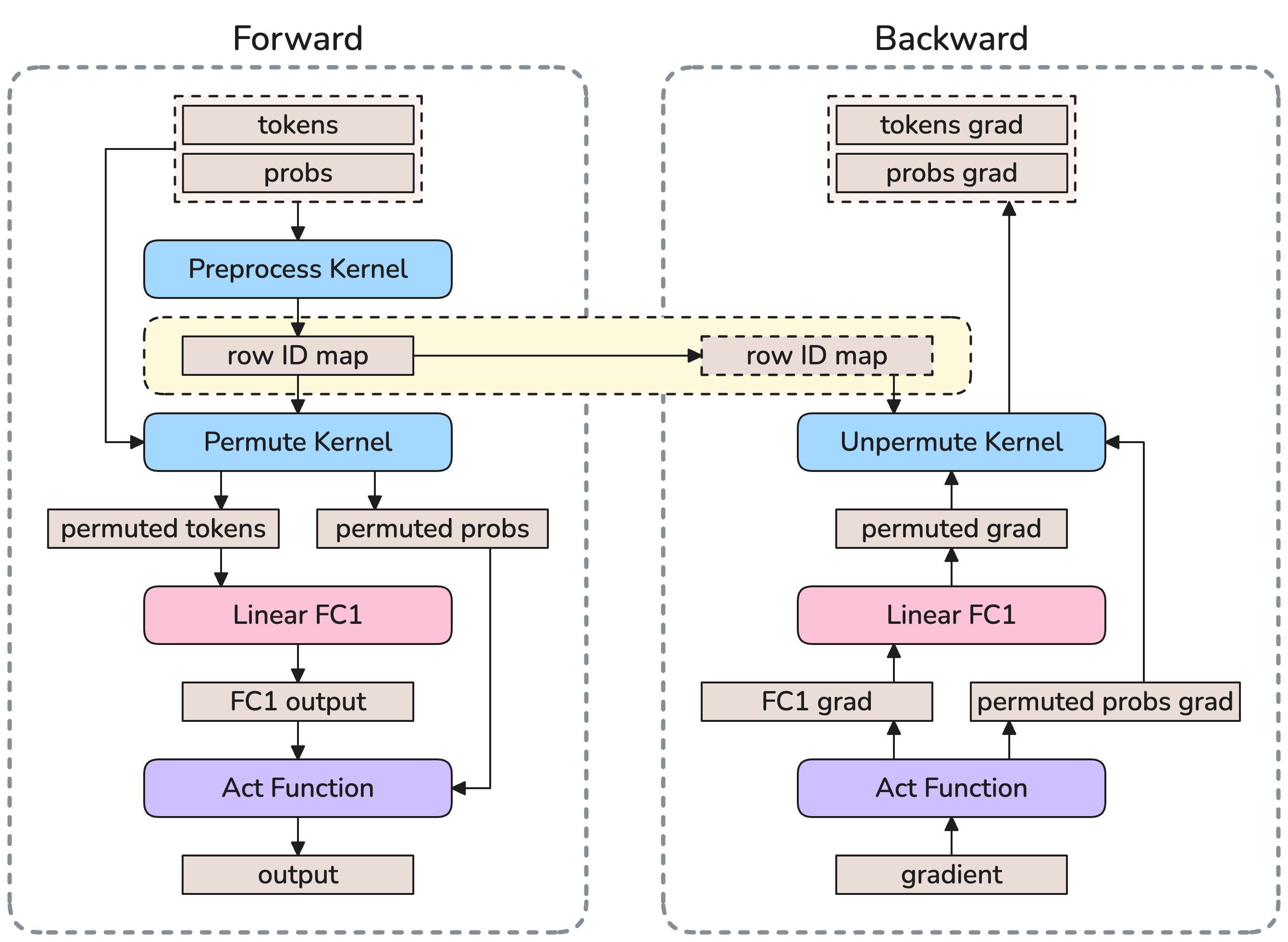}
    \caption{The pipeline for permute fusion in the training process.}
    \label{fig:permute_fusion}
\end{figure}

\begin{itemize}
    \item Preprocessing: Permutation is fundamentally a data transfer process that requires tokens to be stored consecutively in the buffer corresponding to each expert. The purpose of the preprocessing step is to generate an offset map (Row ID map in \figref{fig:permute_fusion}), which indicates the offset of each token in the input and output buffers. This ensures that permute and unpermute operations can be executed efficiently. The preprocessing kernel is called only once before permute during the forward pass, while other components reuse the results.
    \item Permute: The functionality of the permute kernel is straightforward: it moves tokens from the input buffer to the output buffer according to the offset map, and then passes them to the subsequent expert MLP. In memory-efficient permute, probabilities also need to be permuted, with the results directly entering the activation part of the expert.
    \item Unpermute: The unpermute step is the inverse of the permute operation. Since a token will be copied multiple times and sent to different destinations during permutation, these tokens must be summed together during the combine phase. In memory-efficient permute, the combine step simply adds them up; otherwise, the probabilities are used as weights. All accumulation is performed in FP32 precision.
\end{itemize}

\subsubsection{Router and Aux-Loss Fusion}

The router and auxiliary loss \cite{fedus2022switchtransformersscalingtrillion,zoph2022stmoe,hwang2023tutel,rajbhandari2022deepspeedmoe} computations are other areas that generate many small kernels and introduce considerable CPU overhead. As device computing power grows, fusion becomes increasingly important.

\begin{figure}[ht]
    \centering
    \includegraphics[width=1\linewidth]{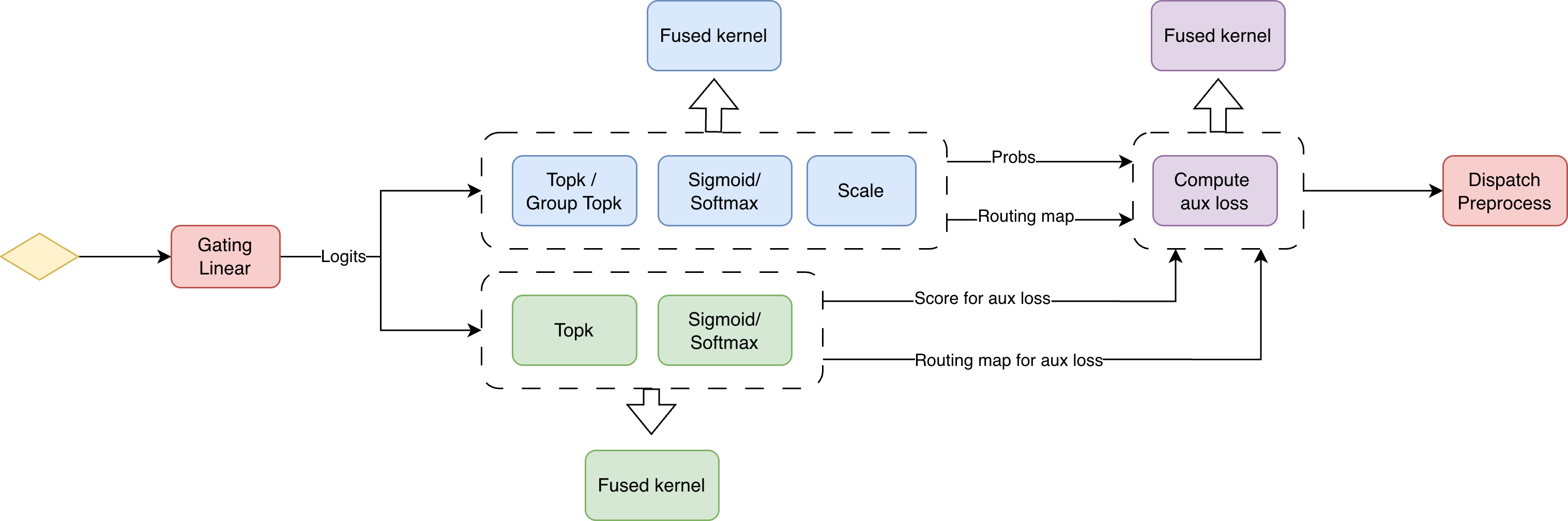}
    \caption{The workflow of the router fusion.}
    \label{fig:router_fusion}
\end{figure}

The challenge is that the router contains components difficult to fuse, such as GEMM and communication. We decompose the remaining operations into three sections, each fused into a single kernel (\figref{fig:router_fusion}).
\begin{itemize}
    \item Score computation, including top-$k$ and softmax/sigmoid: This is the most complex function, with various branches for different models. It includes different top-$k$ algorithms (naive top-$k$, group top-$k$ \cite{deepseekai2024deepseekv2strongeconomicalefficient,deepseekai2025deepseekv3technicalreport}), score functions (sigmoid, softmax), their combinations, and possible scaling operations.
    \item Score computation for auxiliary loss: This is a subset of the first section. After the score function and naive top-$k$ computation, the input for auxiliary loss calculation is generated.
    \item Computation of MoE auxiliary loss: Building on step 2, the auxiliary loss computation is fused into a single kernel.

\end{itemize}
After enabling router fusion, the workflow within the router module is shown in \figref{fig:router_fusion}. Future work may incorporate communication into the fusion scope.

\subsubsection{Reduced-Precision Training: Low-Precision Acceleration}\label{sec:fp8-training}

Fine-grained MoE workloads with small expert GEMMs are often memory-bandwidth bound, limiting Tensor Core utilization. Reduced-precision training addresses this issue by reducing data movement and using faster FP8/FP4 Tensor Core operations on Hopper and Blackwell GPUs.

From a computation efficiency perspective, reduced-precision training provides a key benefit: \textbf{FP8/FP4 GEMMs} maximize Tensor Core utilization by performing Linear Layer computation in FP8/FP4. Together with the memory benefits in Sections~\ref{sec:fp8-activation}, FP8 training achieves approximately \textbf{10\% to 25\% end-to-end performance improvement} for large-scale MoE training on different hardware platforms, while FP4 gives it even more speedup.

For complete coverage of reduced-precision training recipes, MoE-specific challenges, and quantization strategies, see Section~\ref{sec:reduced-precision-training}.

\subsubsection{CUDA Graphs: Eliminating Host Overhead}\label{sec:cuda-graphs}

Kernel fusion reduces the \emph{number} of kernel launches; CUDA Graphs eliminate the \emph{per-iteration cost} of those launches. This subsection presents how CUDA Graphs address CPU overhead, the different strategies for drop-and-pad versus dropless MoE, and the memory optimizations required to make CUDA Graphs practical.

In MoE training, CPU overhead often becomes the dominant performance bottleneck. This overhead stems from three sources:
\begin{enumerate}
    \item \textbf{Python execution}: Interpreter overhead, C foreign function interface (CFFI), and garbage collection introduce latency.
    \item \textbf{Framework overhead}: Even simple operations such as \texttt{torch.empty} add microseconds of host-side cost; library layers (TransformerEngine, cuBLAS, cuDNN) add more.
    \item \textbf{Kernel launch overhead}: Each kernel launch incurs several microseconds of CPU-side cost.
\end{enumerate}

Three trends amplify this in modern training:
\begin{itemize}
    \item \textbf{Faster GPUs}: As kernel execution speeds increase, less time remains to overlap CPU work, making CPU overhead increasingly visible.
    \item \textbf{MoE complexity}: MoE models add substantial complexity to FFN layers---routers, dispatch, Grouped GEMMs, and combine operations---yielding many small kernels beyond the GEMMs themselves and thus substantial CPU overhead.
    \item \textbf{Reduced-precision training}: Extra quantization kernels, especially in fine-grained MoE where operation count scales with the number of experts.
\end{itemize}

\begin{figure}[ht]
    \centering
    \includegraphics[width=0.8\textwidth]{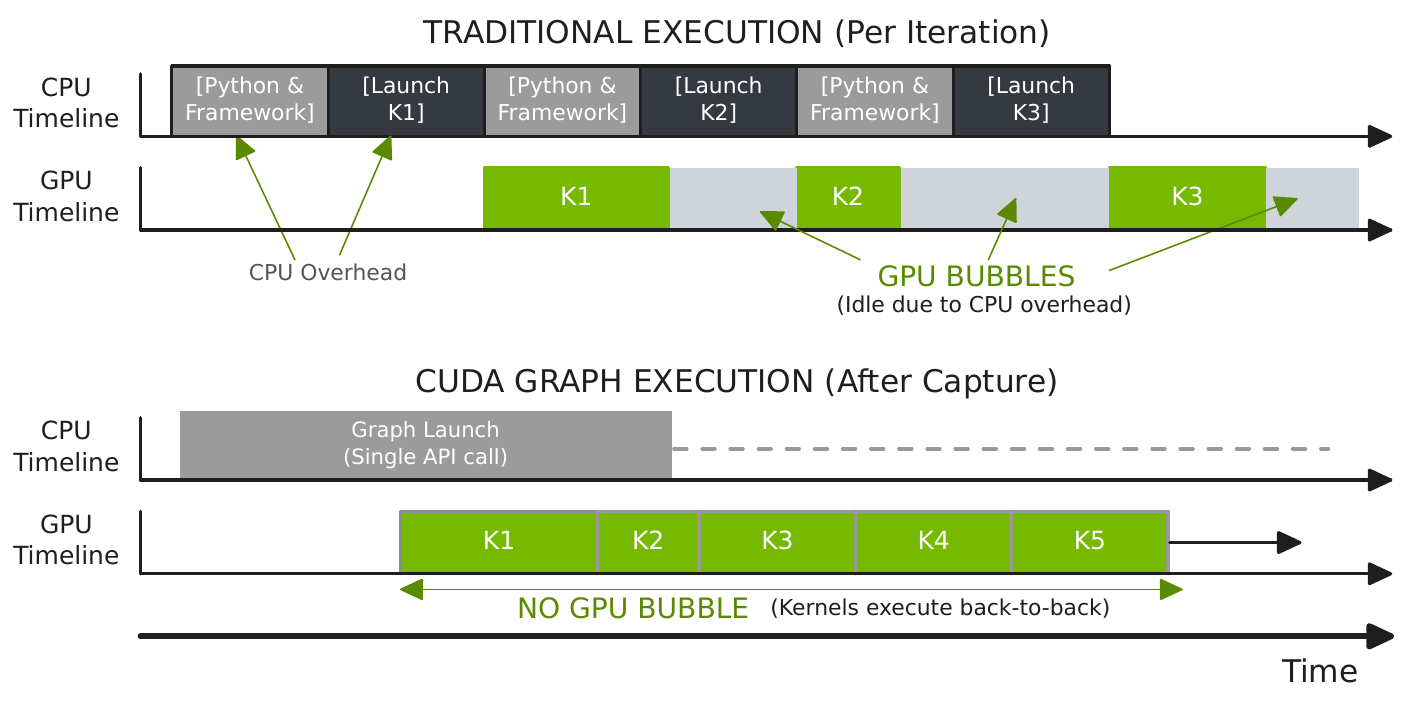}
    \caption{Traditional execution (top) versus CUDA Graph execution (bottom).}
    \label{fig:cuda-graph-execution-comparison}
\end{figure}

% Nsight Systems profiling of non-CUDA Graph MoE training (the upper half of \figref{fig:partial_cudagraph_nsys}) 
\Figref{fig:cuda-graph-execution-comparison} shows the large CPU overhead and the resulting GPU bubbles in a traditional execution pipeline, revealing large gaps between GPU kernels. This clearly indicates that the CPU cannot launch kernels fast enough to keep the GPU fully utilized, a phenomenon known as CPU overhead or host-boundedness.
CUDA Graphs address this by capturing a sequence of GPU kernels into a replayable graph during an initial iteration \cite{nvidia_cudagraphs,ansel2024pytorch2}. Subsequent iterations issue a single \textit{Graph Launch} call, largely bypassing per-op Python/framework overhead and per-kernel launch overhead, thereby reducing CPU-side latency.

However, CUDA Graphs require all captured operations to have \textbf{static, predetermined shapes}. Dynamic shapes or control flow cause graph capture to fail. This creates a fundamental tension with MoE, where token counts per expert vary dynamically. The next paragraph describes Megatron-Core's two CUDA Graph modes---full CUDA Graphs and layer-wise CUDA Graphs---and how they address this dynamic-shape constraint in MoE.

\begin{figure}[ht]
    \centering
    \includegraphics[width=0.95\linewidth]{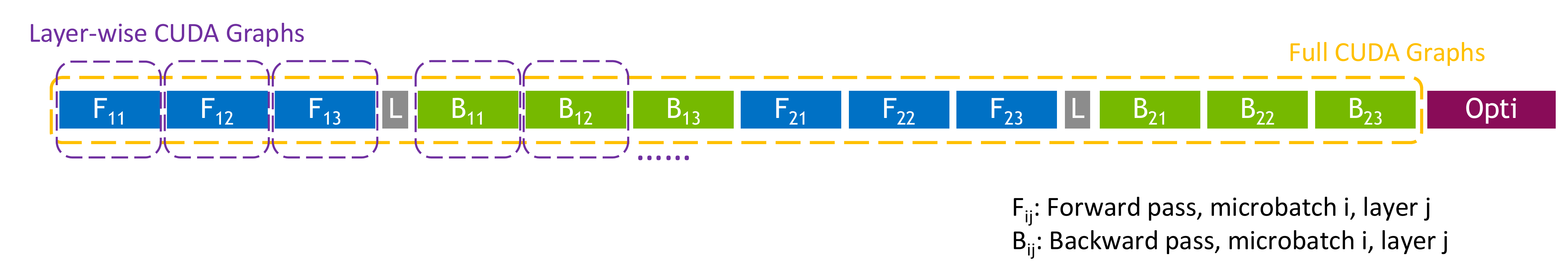}
    \caption{Full versus layer-wise CUDA Graphs in one training iteration (three layers, two microbatches).}
    \label{fig:full_iter_layerwise}
\end{figure}

\paragraph{Full CUDA Graphs vs Layer-wise CUDA Graphs}

As shown in \figref{fig:full_iter_layerwise}, two types of CUDA Graphs are supported in Megatron-Core: full CUDA Graphs and layer-wise CUDA Graphs. \textit{Full} CUDA Graphs, also noted as full-iteration CUDA Graphs, capture the entire forward-backward pass into a single CUDA Graph. This is only supported for dense or MoE models using \textbf{token dropping with padding}, which means each expert has a fixed receiving capacity; tokens exceeding capacity are dropped, and inputs are padded to ensure constant tensor shapes. With all shapes static, the whole iteration can be graphed, maximizing CPU overhead reduction. As for \textbf{dropless} MoE models such as DeepSeek-V3, where no routed tokens can be dropped, full CUDA Graphs are not feasible because the number of tokens assigned to each expert varies dynamically based on routing decisions. Expert GEMM shapes change from iteration to iteration, violating the static-shape requirement. Additionally, \textit{device-host synchronization} is required to retrieve per-expert token counts for dispatching. Section~\ref{sec:sync-free-moe} describes our effort to overcome the dynamic dispatching and synchronization issues so that the full CUDA Graphs can also be applied to dropless MoE. Alternatively, here we introduce a simpler method based on layer-wise CUDA Graphs, called \textit{partial} CUDA Graphs.

\begin{comment}
Megatron-Core supports full-iteration CUDA Graph for drop-and-pad MoE models via the \texttt{local} implementation:

\begin{lstlisting}
--moe-expert-capacity-factor 1.0 \
--moe-pad-expert-input-to-capacity \
--cuda-graph-impl local \
--cuda-graph-scope full_iteration
\end{lstlisting}
\end{comment}

% Instead of capturing the entire forward-backward pass into a single CUDA Graph, layer-wise CUDA Graphs capture each transformer layer separately. Note that both full CUDA Graphs and layer-wise CUDA Graphs are natively \textbf{not feasible for dropless MoE models} such as DeepSeek-V3, where no routed tokens can be dropped, so the number of tokens assigned to each expert varies dynamically based on routing decisions. Expert GEMM shapes change from iteration to iteration, violating the static-shape requirement. Additionally, \textit{device-host synchronization} is required to retrieve per-expert token counts for dispatching. Section~\ref{sec:sync-free-moe} describes our effort to overcome the dynamic dispatching and synchronization issues so that the full CUDA Graphs (either the full-iteration CUDA Graphs or full-layer-wise CUDA Graphs) can also be applied to dropless MoE. Alternatively, here we introduce a simpler method based on layer-wise CUDA Graph, called partial CUDA Graphs, which selectively captures only the \textbf{static components} of each MoE layer while leaving dynamic parts outside the graph, achieving significant performance gains for dropless MoE despite not capturing the entire model.

\begin{figure}[ht]
    \centering
    \includegraphics[width=0.6\linewidth]{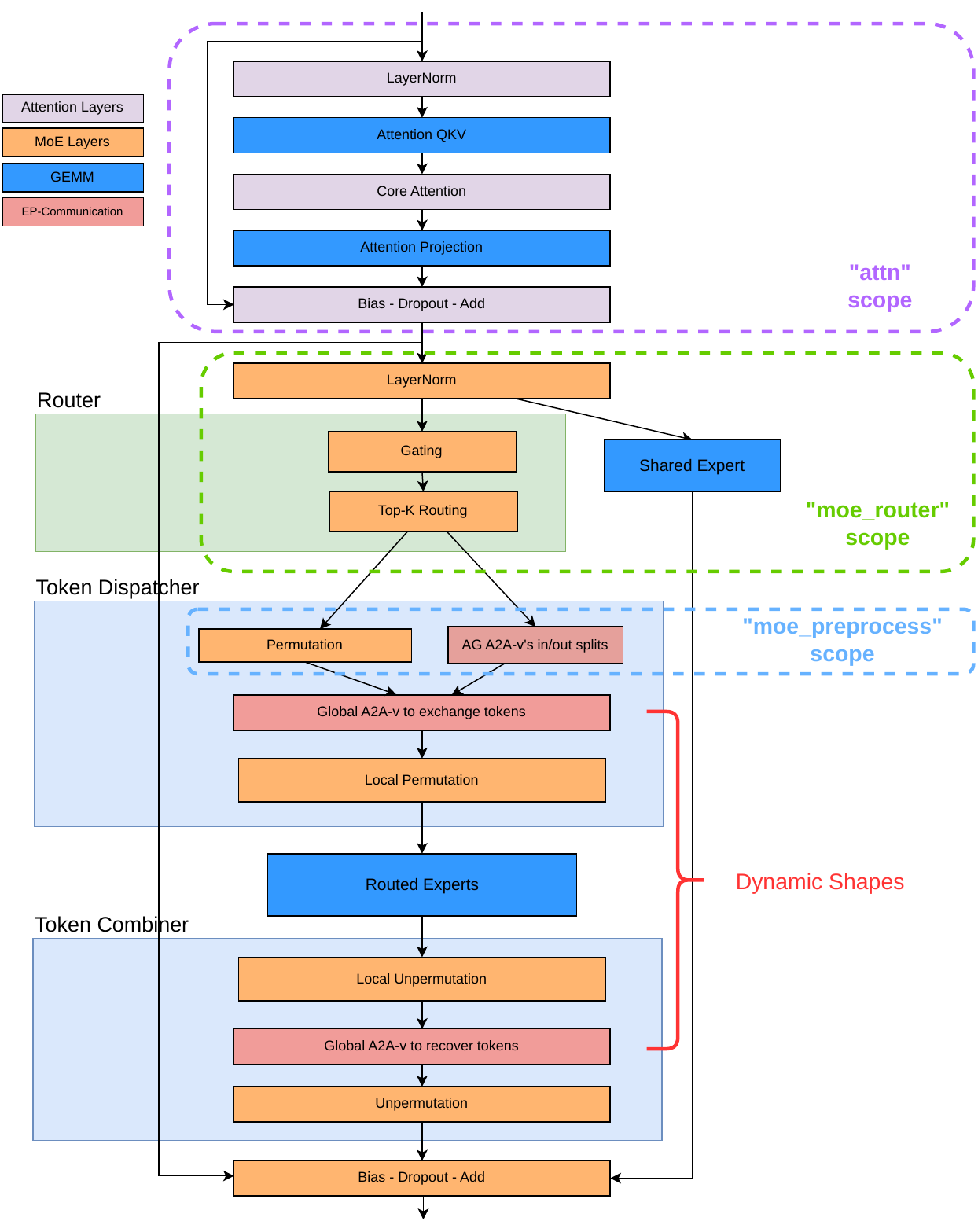}
    \caption{Partial CUDA Graphs capture static components (attention, shared experts, router, preprocessing) while leaving dynamic expert computation outside the graph.}
    \label{fig:partial_cudagraph}
\end{figure}

Layer-wise CUDA Graphs capture each transformer layer separately. As with full CUDA Graphs, capturing an entire transformer layer still runs into dynamic shapes in the MoE part. Partial CUDA Graphs avoid this by capturing only the \textbf{static components} of each layer while leaving dynamic parts outside the graph, yielding large performance gains for dropless MoE even without capturing the full model. In a dropless MoE layer:

\begin{itemize}
    \item \textbf{Static components} (can be graphed):
    \begin{itemize}
        \item Attention layers (process a fixed number of tokens)
        \item Router computation (fixed input/output shapes)
        \item Expert Parallelism (EP) preprocessing (static permutation metadata)
        \item Shared experts (if present, process all tokens with fixed shapes)
        \item MLP layers in dense transformer blocks
    \end{itemize}

    \item \textbf{Dynamic components} (cannot be graphed):
    \begin{itemize}
        \item Token dispatch (from fixed shape to dynamic shape)
        \item Expert GEMM operations (dynamic M dimension based on token assignment)
        \item Token combine (from dynamic shape to fixed shape)
    \end{itemize}
\end{itemize}

Partial CUDA Graphs capture the static portions of each layer independently, leaving routed expert computation to execute normally.
As shown in \figref{fig:partial_cudagraph}, the forward path of each transformer layer is split into several ``scopes'', and the user can choose which scopes to graph. With the "\textit{attn+moe\_router+moe\_preprocess}" scopes enabled, one graph captures all static components before the dispatching operation for that layer: attention, router (projection, top-$k$ selection, and auxiliary loss computation), and EP preprocessing (token metadata calculation and permutation). A shared expert, if present, is also captured in this graph. This approach eliminates CPU overhead while preserving correctness by allowing dynamic expert computation to execute with varying shapes. Despite not capturing the entire model, this achieves significant performance gains.

\begin{figure}[ht]
    \centering
    \includegraphics[width=0.95\linewidth]{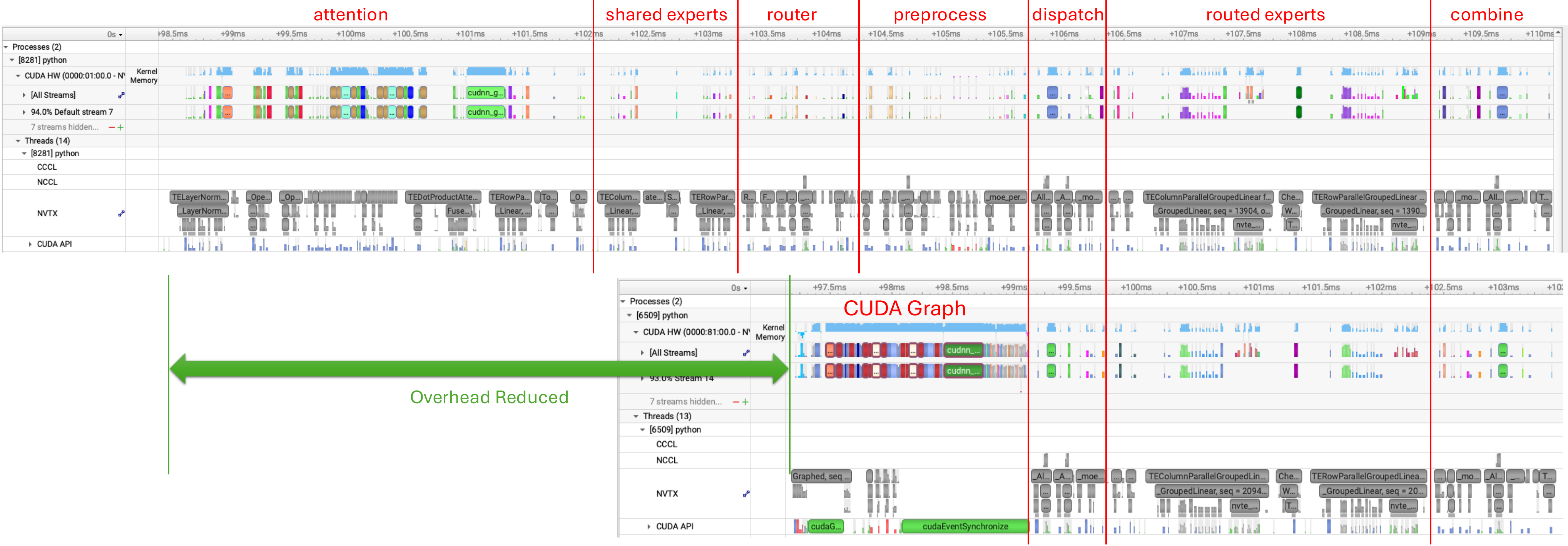}
    \caption{Transformer layer forward pass: without (upper) and with (lower) partial CUDA Graphs. CPU overhead is largely eliminated for static components.}
    \label{fig:partial_cudagraph_nsys}
\end{figure}

Megatron-Core has been extensively adapted so that CUDA Graphs work correctly with key features like Multi-Token Prediction (MTP), EP communication overlapping, fine-grained offloading and checkpointing, and flexible PP layouts. Nsight Systems profiling with partial CUDA Graphs enabled shows greatly reduced CPU overhead (\figref{fig:partial_cudagraph_nsys}). The remaining CPU overhead is primarily confined to the region between token dispatch and combine operations, which cannot be graphed due to their dynamic nature. See Section~\ref{sec:sync-free-moe} for more details on how to address this remaining bottleneck by making the whole layer graphable.

\paragraph{Memory Optimizations for CUDA Graphs}

CUDA Graphs add memory overhead from three sources:

\begin{enumerate}
    \item \textbf{Graph structure}: Memory for storing the graph topology (typically negligible).
    \item \textbf{Separate memory pools}: Graphed and non-graphed operations need separate PyTorch memory pools and cannot share memory. With partial CUDA Graphs, a single transformer layer needs two pools: one for graphed work and one for non-graphed work.
    \item \textbf{Static buffers}: Each graph needs dedicated input/output static buffers that, once allocated, are taken out of PyTorch's memory allocator.
\end{enumerate}

\begin{figure}[ht]
    \centering
    \includegraphics[width=0.9\textwidth]{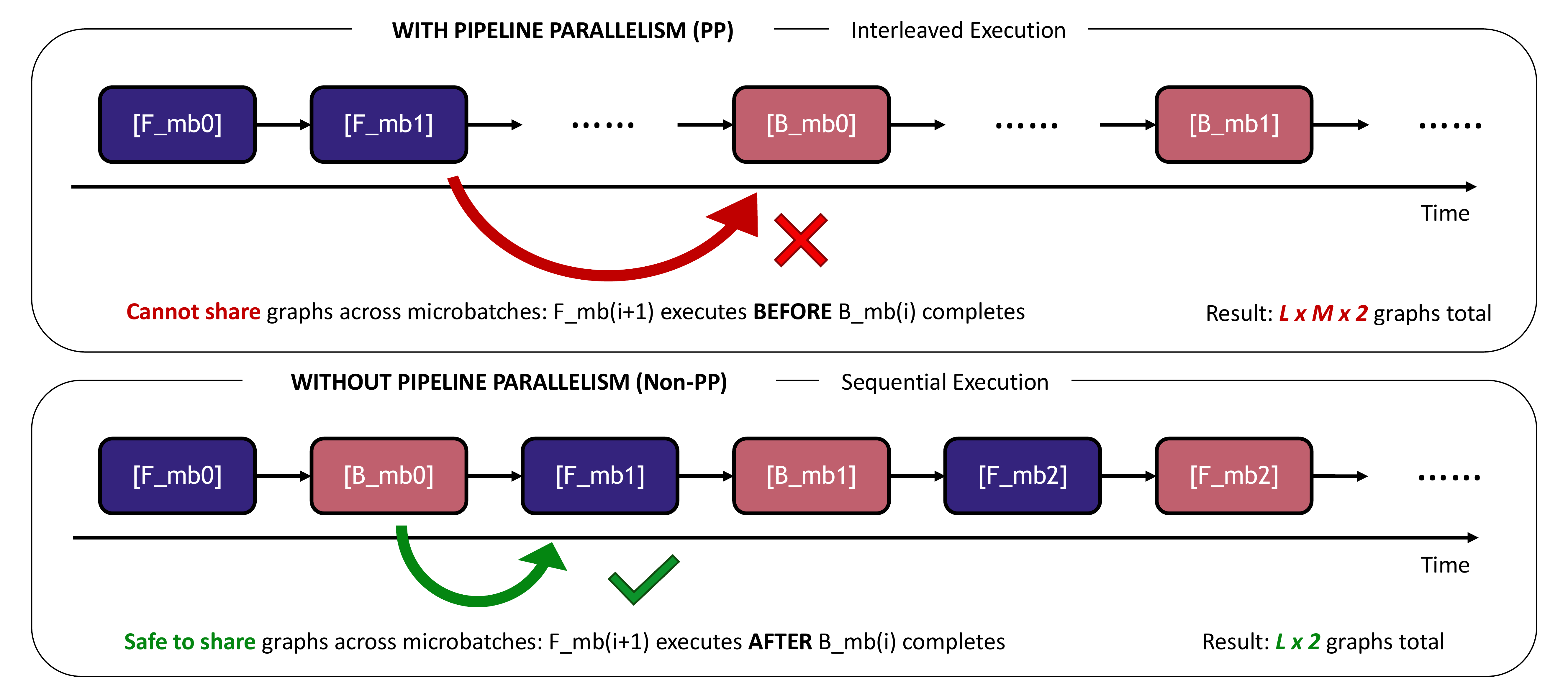}
    \caption{Why Pipeline Parallelism prevents CUDA Graphs from being shared across microbatches. \textbf{With PP} (top): Execution is interleaved---multiple forward passes run before any backward pass. If microbatches share a graph, F\_mb1 overwrites saved context of F\_mb0 before B\_mb0 uses it, causing memory corruption. Each microbatch needs its own graph ($L \times M \times 2$ graphs total). \textbf{Without PP} (bottom): Execution is sequential---each microbatch completes (forward+backward) before the next starts. Context is consumed before being overwritten, so microbatches can safely share graphs (only $L \times 2$ graphs total).}
    \label{fig:pp-microbatch-graph-sharing}
\end{figure}

If not carefully optimized, partial CUDA Graphs would incur much higher memory overhead from the above three sources. Megatron-Core and Transformer Engine implement several optimizations to minimize this overhead:

\textit{Reducing the number of graphs.}
With pipeline parallelism (PP), \textbf{each microbatch} requires a dedicated CUDA Graph. The reason is that if microbatches share a graph, microbatch $i+1$'s forward pass will overwrite the saved context before microbatch $i$'s backward pass executes, causing memory corruption (\figref{fig:pp-microbatch-graph-sharing}). This results in $L \times M \times 2$ graphs in total (where $L$ = layers per GPU, $M$ = microbatches, $\times 2$ for forward/backward).

On the contrary, if PP is not used, microbatches can share the same graph, reducing the count to $L \times 2$. To enable graph sharing, an \texttt{is\_first\_microbatch} GPU flag is introduced to control microbatch-specific behaviors within the shared graph, e.g., quantization that runs only on the first microbatch. Setting this flag to 0 or 1 before replay causes the captured quantization kernel to be skipped or executed as needed.

\textit{Pool sharing.}
Although graphed and non-graphed operations require separate memory pools, all graphs can share one pool when captured in execution order. Transformer Engine's \texttt{make\_graphed\_callables()} API accepts an \texttt{\_order} argument that specifies the inter-microbatch scheduling, allowing sequential capture with correct pool sharing.

\textit{Buffer reuse.}
Static input/output buffers can be reused across graphs according to the PP execution order. Once a microbatch finishes its backward pass of a layer, its forward buffers and backward input buffer become available for the next forward and backward graphs to reuse. Only the backward output buffer must be kept until the data is sent to the next PP stage.

With these memory optimizations, we achieve 10\% end-to-end speedup with about 7~GB extra memory on DeepSeek-V3 GB200 training.

\paragraph{The Remaining Problem: Dynamic Expert Computation}

Full CUDA Graphs do not support dropless MoE. Partial CUDA Graphs eliminate CPU overhead for static components, but \textbf{the dynamic expert computation region remains outside the graph}. Is there any way we could make CUDA Graphs \textit{cover} the dynamic part? There are at least two challenges we must solve to achieve this:
\begin{enumerate}
    \item \textbf{Dynamic shapes require host--device synchronization}: The host must query per-expert token counts from the GPU to launch token dispatching and Grouped GEMMs.
    \item \textbf{Memory allocation requires known buffer sizes}: Worst-case buffer allocation wastes memory; actual-size allocation requires synchronization.
\end{enumerate}

The next subsection presents how Megatron-Core achieves full CUDA Graphs coverage for dropless MoE through three complementary techniques: sync-free kernels, ECHO, and Paged Stashing.

\subsubsection{Full CUDA Graphs Coverage for Dropless MoE via Sync-Free Kernels, ECHO, and Paged Stashing}\label{sec:sync-free-moe}

While kernel fusion and CUDA Graphs help reduce host overhead, dropless MoE introduces a new challenge: dynamic tensor shapes require host-device synchronization, preventing CUDA Graphs from covering expert computation.

In dropless MoE, token counts received by each EP rank vary dynamically based on routing decisions. The router generates this shape information on the GPU, but the host needs it to launch subsequent operations and allocate memory. Obtaining this information requires a device-to-host copy and synchronization, placing the CPU on the critical path and preventing CUDA Graphs capturing of the dynamic portions. To enable sync-free MoE for dropless training with dynamic routing, there are two fundamental challenges:

\textbf{Challenge 1: Kernel launch without knowing the actual problem size.}
Conventionally, many GPU operators assume that (dynamic) shape information is available on the host. The host-side code determines launch configurations, such as grid size and tile size, as well as the amount of work the kernel must perform based on the shape information obtained at run time. In dropless MoE, this creates a host-device synchronization point: the host must query per-expert token counts from the device before launching Grouped GEMM or communication kernels. To tackle this issue, the GPU kernels handling dynamic shapes need to be redesigned. We developed \textbf{device-initiated GPU kernels}, including device-initiated Grouped GEMM and sync-free dispatch with HybridEP.

\textbf{Challenge 2: Memory allocation without knowing the actual size.}
Sync-free execution means the size of buffers needs to be pre-determined on the host. To avoid token overflow, buffers often need to be over-sized, e.g., based on the worst-case size to accommodate the case where all tokens are routed to the same expert. This over-provisioning, however, leads to severe memory fragmentation: if the actual token distribution is balanced, the vast majority of the preallocated memory remains unused, yet it cannot be reclaimed by other operations within the graph. The worst-case buffer potentially requires $O(\text{EP\_size})$ times more memory than the actual working set.

This memory fragmentation can be mitigated through two complementary strategies: \emph{reducing load imbalance} so that worst-case provisioning is closer to actual usage, and \emph{dynamic memory management} within the CUDA Graph to reclaim unused buffer space. \textbf{ECHO} employs the first strategy by dynamically cloning popular experts on underutilized ranks, reducing the variance in per-rank token counts and thus the gap between worst-case and average memory requirements. \textbf{Paged Stashing} adopts the second strategy by enabling fine-grained memory management within CUDA Graph, allowing unused portions of preallocated buffers to be repurposed for other operations. ECHO and Paged Stashing are detailed below.

\paragraph{Device-Initiated GPU Kernels}
\label{sec:device-init}

To eliminate the mandatory host-device synchronization required by conventional host-initiated GPU operators, kernels must be \emph{device-initiated}. This requires three things:
\begin{itemize}
\item[i.] The GPU kernel needs to read shape information from GPU memory and use it to guide the computation.
\item[ii.] The GPU kernel needs to decouple the actual amount of work it performs, which is only known at runtime, from its static launch configuration.
\item[iii.] The GPU kernel needs to skip unnecessary computation, such as operations on padded data.
\end{itemize}
To satisfy these requirements, existing kernels must be redesigned. The Grouped GEMM kernel and HybridEP kernel serve as two concrete examples below.

\textbf{Device-Initiated Grouped GEMM.}
As discussed in Section~\ref{sec:grouped-gemm}, TE offers two Grouped GEMM implementations: the multi-stream cuBLASLt GEMMs and the CUTLASS Grouped GEMM. Both are host-initiated. They require a CPU-side list of per-expert token counts to determine each GEMM's shape and launch configuration, creating a host-device synchronization barrier. To eliminate this, TE introduces device-initiated Grouped GEMM, which reads shape information directly from device memory. Two implementations are provided:

\begin{itemize}
\item[i.] \textbf{cuBLASLt Grouped GEMM.}
Since CUDA 13.1, cuBLASLt Grouped GEMM supports passing matrix shapes as device arrays. This implementation includes built-in heuristics for selecting optimal kernel configurations, now covering various precisions and scaling modes across recent CUDA releases.

\item[ii.] \textbf{cuteDSL Grouped GEMM with Fused Activation and Quantization.}
In the cuteDSL-based implementation, the SwiGLU activation can be fused into the epilogue stage, and for FP8 training, quantization can also be fused. While current support is limited to specific precisions and fusion patterns, ongoing development continues to expand coverage across new configurations.
\end{itemize}

\textbf{Sync-Free Dispatch with HybridEP.}
HybridEP, introduced in Section~\ref{sec:deepep}, provides an efficient implementation of MoE communication and also plays an important role in sync-free MoE. After each dispatch and permutation, the number of tokens received by a given rank is dynamic, normally requiring synchronization to obtain buffer sizes. By estimating an upper bound and passing it to HybridEP, the dispatcher can pre-allocate output buffers according to this bound, eliminating all synchronizations at the cost of additional GPU memory.

Even with device-initiated kernels, worst-case buffer allocation wastes memory. Two complementary strategies address this: \textbf{ECHO} reduces load imbalance so worst-case provisioning approaches actual usage, while \textbf{Paged Stashing} enables fine-grained memory management to reclaim unused buffer space.

\paragraph{ECHO: Elastic Cloning for Hot Experts.}\label{sec:echo}\footnote{Implementation details: \url{https://github.com/NVIDIA/Megatron-LM/pull/2368}}

Load imbalance is inherent in MoE: popular experts receive far more tokens than others, creating two problems. First, EP ranks hosting hot experts become compute bottlenecks, causing other ranks to wait at synchronization points. Second, high load variance means worst-case buffer provisioning wastes significant memory compared to actual usage. ECHO addresses both by dynamically cloning hot experts to spare slots on underutilized ranks.

\begin{figure}[ht]
    \centering
    \includegraphics[width=0.85\linewidth]{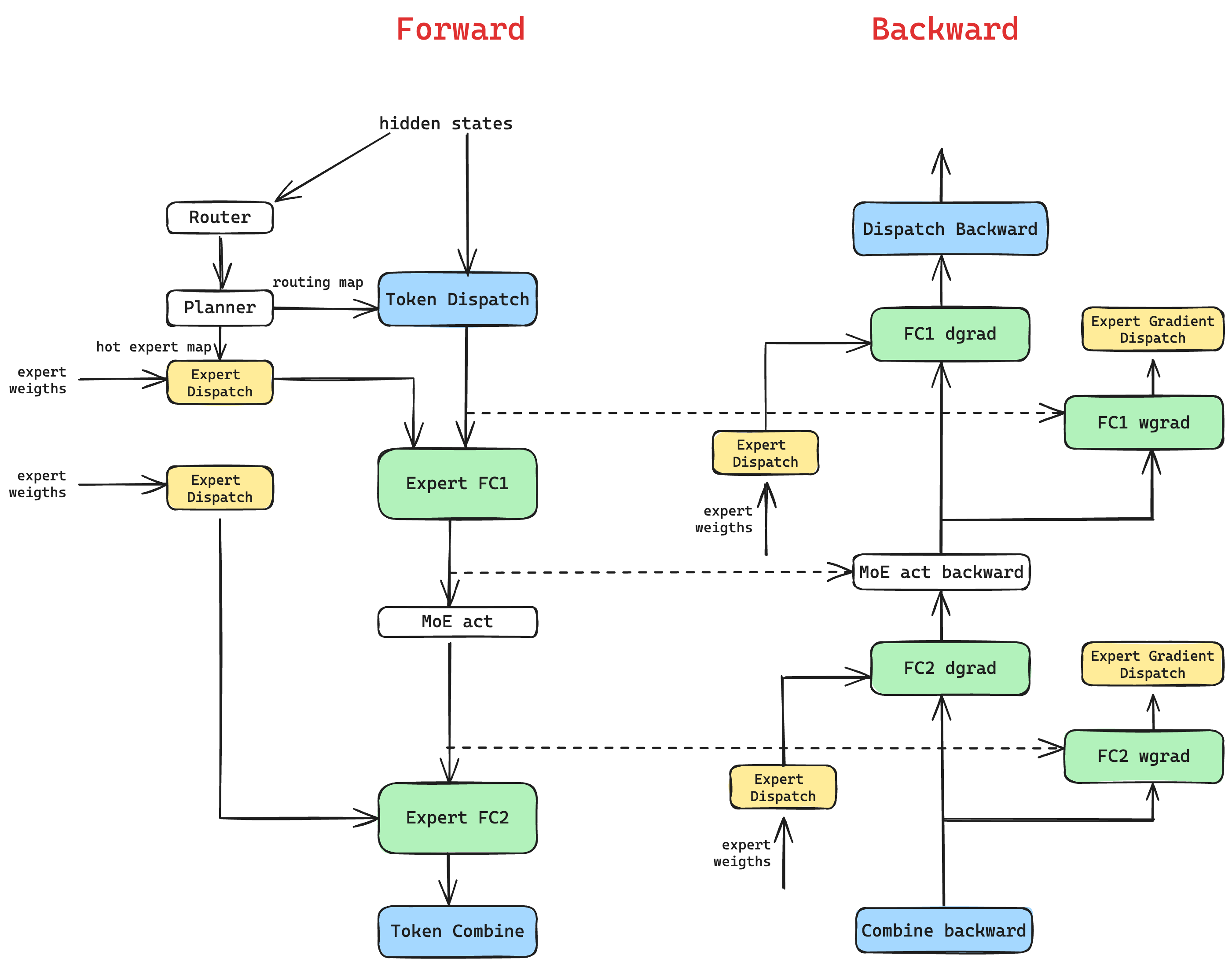}
    \caption{ECHO workflow for forward and backward passes. The planner generates routing and hot expert maps. Expert Dispatch clones hot expert weights to spare slots; Expert Gradient Dispatch reduces gradients back to home experts.}
    \label{fig:echo}
\end{figure}

\figref{fig:echo} illustrates the ECHO workflow. In the forward pass, the ECHO planner identifies hot experts and generates two outputs: a \emph{hot expert map} indicating which experts to clone to which spare slots, and an updated \emph{routing map} redirecting overflow tokens to clones. Expert Dispatch copies hot expert weights to spare slots via HybridEP-based sync-free communication; Token Dispatch routes tokens to both home and cloned experts; expert computation proceeds on all experts. In the backward pass, Expert Gradient Dispatch collects gradients from clones and reduces them to home experts, ensuring consistency. Cloned experts are discarded after computation to save memory.

Cloning experts for training is much more expensive than for inference: each cloned expert requires weight communication during forward and gradient reduction during backward to ensure consistency. Cloning all hot experts to all spare slots would maximize load balance but introduce excessive communication overhead. The goal of the ECHO planner is to minimize the number of expert clones while achieving sufficient load balance. It computes the \emph{spillover} for each expert (tokens exceeding the average load per EP rank) and the \emph{spare capacity} on each rank (available capacity below average). Using a bin-packing algorithm, the planner efficiently matches spillover tokens to spare capacity, determining which specific experts to clone and to which ranks, producing the hot expert map and routing map with minimal cloning overhead.

ECHO provides two key benefits: (1) \emph{reduced memory fragmentation for CUDA Graphs}, since reducing load variance across EP ranks makes worst-case buffer sizing closer to actual usage, enabling smaller static buffers; and (2) \emph{improved compute efficiency}, since balanced workloads reduce the straggler problem, improving overall GPU utilization.

\paragraph{Paged Stashing.}\footnote{Implementation details: \url{https://github.com/NVIDIA/Megatron-LM/pull/2690}}

While ECHO reduces load imbalance to narrow the gap between worst-case and typical memory requirements, Paged Stashing addresses the complementary problem: enabling fine-grained memory management within CUDA Graph to reduce internal memory fragmentation.

\begin{figure}[ht]
    \centering
    \includegraphics[width=0.95\linewidth]{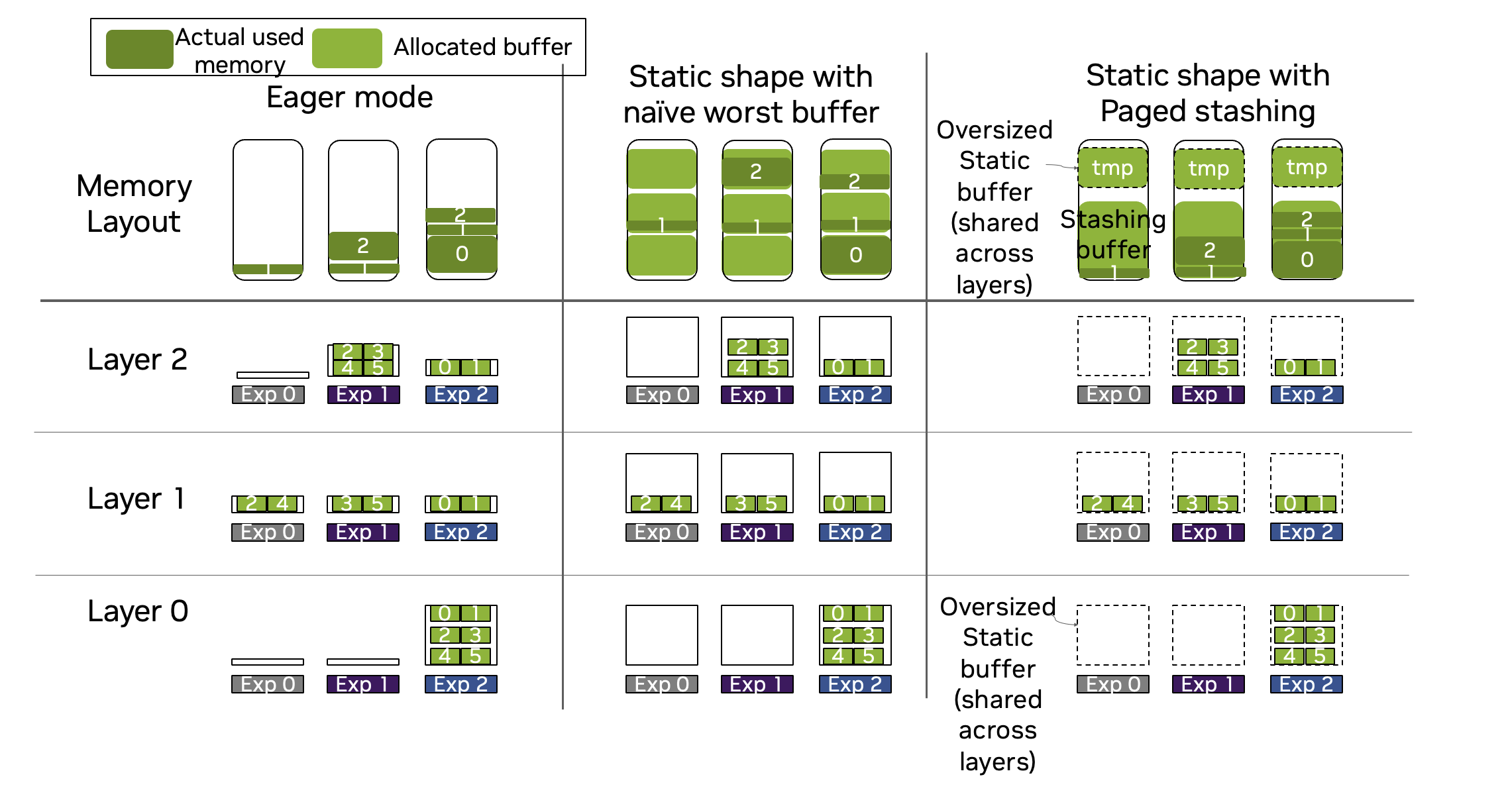}
    \caption{Memory layout comparison across three execution modes. \textbf{Left}: Eager mode allocates memory dynamically based on actual usage. \textbf{Middle}: Baseline static shape requires worst-case sized buffers for each layer independently, causing severe fragmentation when actual usage is lower. \textbf{Right}: Paged Stashing uses a single worst-case tmp buffer shared across layers for computation, while a paged stashing buffer stores only the actual tokens, significantly reducing total memory footprint.}
    \label{fig:paged_stash}
\end{figure}

The fragmentation problem arises because sizes of buffers involved in CUDA Graphs need to be pre-determined. To avoid undesired overflow, the buffers usually need to be over-sized. In a straightforward approach, buffers in all layers are allocated according to the worst-case size (\figref{fig:paged_stash}, middle). For dropless MoE, this means each layer reserves a buffer based on maximum possible tokens each rank may receive, resulting in $O(\text{layers} \times \text{worst\_case})$ memory overhead even when actual token counts are much lower.

Paged Stashing is based on a key observation: there is a vast gap (often more than an order of magnitude) between the actual memory needed to store activations for backward computation and the worst-case memory allocated to avoid overflow at each layer. Paged Stashing addresses this by decoupling these two memory buffers. This is illustrated in \figref{fig:paged_stash} (right). Instead of allocating worst-case-sized buffers for every layer, the paged-stashing manager maintains: (1) a single \emph{tmp buffer} sized for the worst-case token counts, which is shared across all layers for computation, and (2) a \emph{stashing buffer} organized in the form of pages that stores only the actual tokens used by each layer. When a layer completes its forward computation, its activations are copied from the tmp buffer to the stashing buffer, storing only the actual token count rather than the worst-case allocation. The tmp buffer can then be readily reused for the computation in following layers. This significantly reduces memory consumption from $O(\text{layers} \times \text{worst\_case})$ to $O(\text{worst\_case} + \text{actual\_total})$.

The stashing buffer uses the classic idea of paging to manage memory in a flexible yet efficient way. The \texttt{PagedStashBuffer} is organized as pages of fixed size (default 64 tokens per page). The page stashing manager tracks available pages through a free list implemented by a circular buffer. Paging operations are implemented as device-initiated \emph{stash kernels}. The stash kernel copies activations to free pages allocated from the head of the free list and records the target pages. The \emph{reload kernel} retrieves activations from the recorded pages and returns the unused pages to the tail of the free list.

\begin{figure}[ht]
    \centering
    \includegraphics[width=0.9\textwidth]{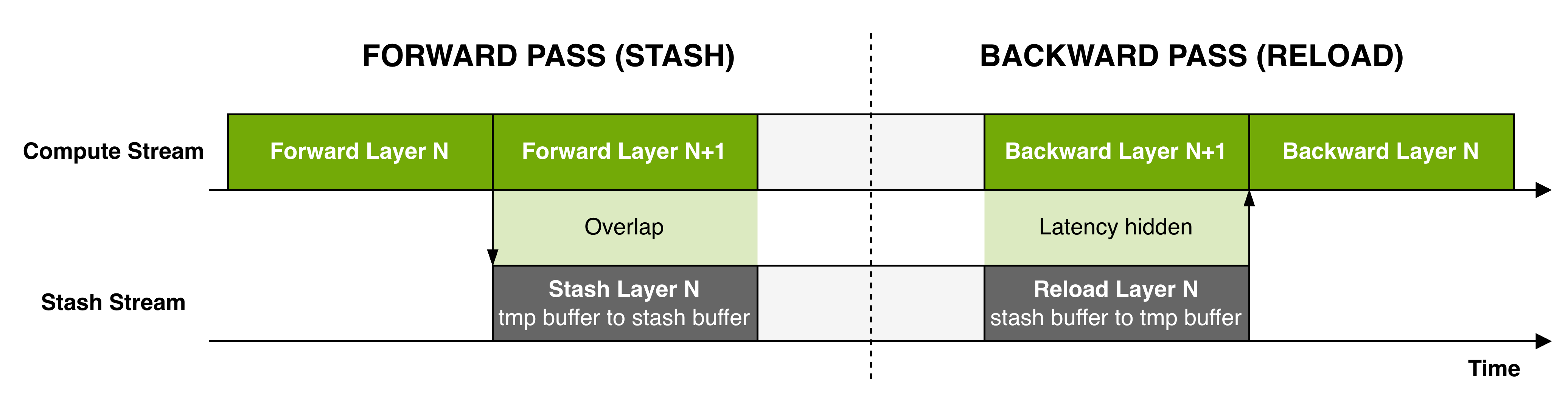}
    \caption{Paged Stashing stream overlap. \textbf{Forward pass}: After Layer N computes, its activations are stashed (copied from tmp buffer to paged stashing buffer) on a dedicated Pack stream while Layer N+1 computes on the main Compute stream---the stash is completely overlapped. \textbf{Backward pass}: Activations for Layer N are pre-fetched (reloaded from stashing buffer to tmp buffer) on the Unpack stream before Layer N backward starts, hiding reload latency.}
    \label{fig:paged-stashing-overlap}
\end{figure}

To minimize overhead, Paged Stashing overlaps memory transfers with computation using dedicated CUDA streams (\figref{fig:paged-stashing-overlap}). The system pre-fetches activations for the next backward layer while current computation executes, hiding reload latency. This overlap requires slight memory overhead due to double buffering.

\paragraph{Putting It Together: Full CUDA Graphs Coverage}

Together, these three techniques enable full CUDA Graphs coverage for dropless MoE, eliminating CPU overhead while preserving the flexibility of dynamic routing:
\begin{itemize}
    \item \textbf{Device-initiated kernels} remove host-device synchronization entirely by reading dynamic shapes directly from GPU memory, making all MoE operations capturable by CUDA Graphs.
    \item \textbf{ECHO} balances expert workloads across EP ranks, narrowing the gap between worst-case and actual buffer requirements. This enables smaller static allocations and mitigates the straggler problem that otherwise limits GPU utilization.
    \item \textbf{Paged Stashing} reduces memory from $O(\text{layers} \times \text{worst\_case})$ to $O(\text{worst\_case} + \text{actual\_total})$ by sharing a single worst-case tmp buffer across layers and storing only actual activations in a paged stashing buffer.
\end{itemize}

\noindent\textbf{Overhead.}
These benefits come with measured trade-offs.
ECHO introduces extra communication for cloning hot experts in the forward pass and reducing their gradients in the backward pass; in practice, only a small fraction of experts are cloned (determined by the spillover above average load), so the additional communication remains modest relative to the standard all-to-all dispatch.
Paged Stashing adds memory copy operations (stash in the forward pass, reload in the backward pass) and a persistent worst-case-sized tmp buffer with double-buffering overhead; these copies run on dedicated CUDA streams overlapped with layer computation, so the latency is largely hidden and the primary cost is additional memory bandwidth.
Both overheads are relatively small compared to the gains from eliminating per-iteration host-device synchronization and enabling static memory planning with CUDA Graphs.

\subsubsection{Summary}

The compute efficiency optimizations address both sources of inefficiency. For kernel efficiency: Grouped GEMM batches expert computations, kernel fusion consolidates routing and permutation operations, and reduced-precision training accelerates Tensor Core throughput. For host overhead: CUDA Graphs eliminate per-iteration CPU logic, and Sync-Free MoE removes host-device synchronization barriers. ECHO addresses load imbalance that compounds both issues. Together, these optimizations transform compute inefficiency from a dominant bottleneck into a manageable constraint.

\subsection{Summary: Breaking the Three Walls}

This section addressed the three fundamental barriers to efficient MoE training, each a hard constraint that, unaddressed, prevents practical training at scale.

\begin{itemize}
    \item \textbf{Memory Wall}: Memory is the first constraint. If parameters, optimizer states, and activations exceed the GPU capacity, training cannot proceed. Memory-Efficient Permutation, FP8/FP4 activations, fine-grained recomputation, offloading, precision-aware optimizers, and FSDP together transform the 199.5 GB per-GPU requirement of DeepSeek-V3 into a feasible training configuration.

    \item \textbf{Communication Wall}: Communication overhead directly reduces GPU utilization. Optimized dispatchers (DeepEP, HybridEP) maximize bandwidth and FWD-BWD overlap hides latency behind computation. Together, these reduce all-to-all overhead from up to 60\% of the training time to less than 10\% in DeepSeek-V3 training.

    \item \textbf{Compute Efficiency Wall}: Compute inefficiency stems from two sources: small kernels that underutilize GPU resources, and host overhead that leaves GPUs idle. Grouped GEMM and kernel fusion improve kernel efficiency; CUDA Graphs and Sync-Free MoE eliminate host overhead. These transform computation inefficiency from the dominant bottleneck into a manageable constraint.
\end{itemize}

These optimizations are not independent. They form a coherent system where solutions to one wall enable or enhance solutions to others. Reduced-precision training shows this clearly: it reduces memory (activations) and improves compute efficiency (faster Tensor Core operations), while requiring careful integration to avoid introducing new bottlenecks (quantization kernel overhead).

The following sections address two cross-cutting concerns: Section~\ref{sec:reduced-precision-training} covers reduced-precision training in FP8/FP4, which simultaneously impacts all three walls, and Section~\ref{sec:long-context-training} addresses long-context MoE training, which changes the optimization balance when attention dominates computation.

%==============================================================================
% SECTION 5: Low PRECISION TRAINING - THE FP8 JOURNEY
%==============================================================================

\section{Reduced-Precision Training in FP8/FP4 for MoE}\label{sec:reduced-precision-training}

The preceding sections addressed each of the three walls with targeted optimizations. Reduced-precision training is unique: it simultaneously impacts memory, communication, and computation efficiency, making it a cross-cutting optimization that deserves dedicated treatment.

Mixed-precision training has been a cornerstone of efficient deep learning, with BF16 as standard practice \cite{micikevicius2018mixed,peng2023fp8lmtrainingfp8large,xi2024coat}. Reduced-precision training in FP8/FP4 represents a more aggressive step, reducing the precision from 16 bits to 8 or even 4, with correspondingly larger benefits and risks \cite{micikevicius2022fp8formatsdeeplearning,fishman2025scalingfp8trainingtrilliontoken}. For DeepSeek-V3, FP8 training reduces activation memory by approximately 16 GB per GPU, improves expert GEMMs through faster Tensor Core operations, and accelerates the parameter AllGather communication. These gains compound across the three walls, making FP8 essential for efficient large-scale MoE training.

However, low-precision training introduces convergence risks that must be systematically addressed. This section presents Megatron-Core's approach to reduced-precision training: understanding where precision matters, selecting appropriate quantization recipes, and implementing MoE-specific optimizations that capture benefits of reduced-precision training while maintaining training stability.

\subsection{Why Reduced-Precision Training Matters for MoE}\label{sec:precision-anatomy}

MoE architectures amplify both the benefits and risks of low-precision training compared to dense models:

\textbf{Amplified Benefits.} With hundreds of experts, activation memory scales proportionally. FP8/FP4 activations provide larger absolute memory savings than in dense models. The communication volume for parameter AllGather is halved\footnote{For the NVFP4 recipe on Blackwell, since we need to gather both the row-wise and column-wise weight tensors, the saved communication volume is also 50\%.}, and expert GEMMs (which dominate MoE computation) benefit from faster Tensor Core throughput.

\textbf{Amplified Risks.} Router decisions depend on precise scores to assign tokens to experts. Quantization noise could destabilize expert selection, leading to training instability, degraded model quality, or expert collapse \cite{nagel2021whitepaper}. The numerically sensitive components must be protected from aggressive quantization.

\textbf{The Strategy: Selective Precision.} The solution is precision where it matters, efficiency everywhere else. Three principles guide the deployment of reduced-precision training in MoE:
\begin{enumerate}
    \item \textbf{Protect routing decisions.} The router remains in FP32 to ensure stable expert selection.
    \item \textbf{Preserve precision for key components.} Embeddings, output layers, main gradients, master weights, and optimizer states remain in their original precision to maintain model quality.
    \item \textbf{Quantize bulk computation.} Expert GEMMs, which constitute the majority of computation, use reduced-precision training with carefully designed quantization schemes (called \textit{recipe}).
\end{enumerate}

\begin{notebox}
The key to successful reduced-precision MoE training is selective precision: keep numerically sensitive components (router, embeddings, output layers, gradients, and optimizer states) in higher precision while aggressively quantizing the bulk of computation (expert GEMMs). This strategy captures most benefits of the reduced-precision training while avoiding convergence issues.
\end{notebox}

\subsection{The Impact from Reduced-Precision Training on All Three Walls}

Reduced-precision training provides benefits across all three performance walls, making it a unifying optimization. Table~\ref{tab:fp8-impact-summary} summarizes these impacts.

\begin{table}[ht]
\centering
\begin{threeparttable}
\caption{Summary of the impact from reduced-precision training on the Three Walls.}
\label{tab:fp8-impact-summary}
\begin{tabular}{lll}
\toprule
\textbf{Wall} & \textbf{Reduced-Precision Training Benefit} & \textbf{Details} \\
\midrule
Memory & 50\% (FP8)/75\% (FP4) activation reduction & Section~\ref{sec:fp8-activation} \\
        & Eliminate BF16 weight copies & Section~\ref{subsec:fp8-primary-weights} \\
        & BF16 optimizer states & Section~\ref{sec:precision-opt} \\
\midrule
Communication & 50\% parameter AllGather \tnote{1} & Section~\ref{subsec:fp8-primary-weights} \\
\midrule
Compute & Faster Tensor Core GEMMs & Section~\ref{sec:fp8-training} \\
        & Quantization kernel overhead & Section~\ref{sec:kernel-fusion} \\
\bottomrule
\end{tabular}
\begin{tablenotes}
    \footnotesize
    \item[1] For NVFP4, we need to gather both row-wise and column-wise FP4 weights, so the reduced traffic of parameter AllGather is also 50\%.
\end{tablenotes}
\end{threeparttable}
\end{table}

\subsubsection{Breaking the Memory Wall with Reduced-Precision Training}

Reduced-precision training reduces memory consumption in three ways:

\textbf{Activation Memory (detailed in Section~\ref{sec:fp8-activation}).} The primary source of activation memory savings comes from input tensors of linear layers saved for backward computation. By storing these tensors in FP8/FP4 instead of BF16, the memory footprint is reduced by 50\%/75\%. For example, FP8 saves approximately 16 GB of activation memory per GPU for DeepSeek-V3.

\textbf{Parameter Memory.} Conventional reduced-precision training maintains three parameter copies: FP32 master weights, BF16 model weights, and FP8/FP4 computation weights. Our \emph{native FP8/FP4} approach (detailed in Section~\ref{subsec:fp8-primary-weights} below) eliminates the BF16 copy by casting weights directly from FP32 to FP8/FP4 and doing parameter AllGather in reduced precisions.

\textbf{Optimizer State Memory (detailed in Section~\ref{sec:precision-opt}).} Adam optimizer states (first and second moments) can be stored in BF16 instead of FP32, reducing optimizer memory by 50\%. This is orthogonal to reduced-precision training and is also applicable to BF16.

\subsubsection{Breaking the Communication Wall with Reduced-Precision Training}
\label{sec:comm_wall_quantized_training}

% \textbf{FP8 EP Dispatch (detailed in Section~\ref{sec:fp8-dispatch-comm}).} For fine-grained MoE models with large EP sizes, all-to-all communication becomes a primary bottleneck. By dispatching tokens in FP8 instead of BF16, the communication volume is halved. The key challenge is managing the dequantization-quantization cycle required because forward and backward passes use different quantization dimensions.

\textbf{Parameter AllGather in FP8/FP4.} When using FP8/FP4 primary weights with distributed optimizer, parameter AllGather communication is reduced by 50\% (1 byte vs 2 bytes per parameter). Note that
\begin{itemize}
    \item For NVFP4, we need to gather both row-wise and column-wise FP4 weights, so the reduced traffic of parameter AllGather is also 50\%.
    \item For MXFP8, even with FP8 primary weights, we first copy the FP32 master weights into a temporary BF16 buffer and communicate in BF16 precision. This is because MXFP8 requires different quantization directions for forward and backward passes. Communicating MXFP8 weights would require both row-wise and column-wise quantized versions, effectively communicating two bytes per parameter and eliminating any advantage over communicating BF16. Hence, for MXFP8, we communicate parameters in BF16 precision. 
\end{itemize}

\subsubsection{Breaking the Compute Efficiency Wall with Reduced-Precision Training}

\textbf{Faster GEMMs.} FP8 (Ada/Hopper and later) and FP4 (Blackwell and later) Tensor Cores provide higher throughput than BF16, accelerating both forward and backward passes.

\textbf{The Trade-off: Quantization Overhead.} Reduced-precision training introduces additional quantization kernels that increase CPU overhead, which is particularly problematic for fine-grained MoE where many small operations already stress the CPU. This overhead is managed through:
\begin{itemize}
    \item \textbf{Kernel fusion} that fuses quantization with other operations, such as normalization and activation functions.
    \item \textbf{Fuse padding/unpadding} with the permutation kernel or routing map to avoid explicit padding/unpadding.
    \item \textbf{Grouped quantization} that fuses the quantization of multiple input tensors into one single kernel.
    \item \textbf{CUDA Graphs} that captures quantization kernels to eliminate launch overhead. See Section~\ref{sec:cuda-graphs} for more details.
\end{itemize}

\subsection{Reduced-Precision Training Recipes: Per-Tensor FP8, Blockwise FP8, MXFP8, and NVFP4}

Having established the impact of reduced-precision training on all three walls, we now examine the technical components that define a reduced-precision training configuration.

\begin{figure}[ht]
    \centering
    \includegraphics[width=0.9\textwidth]{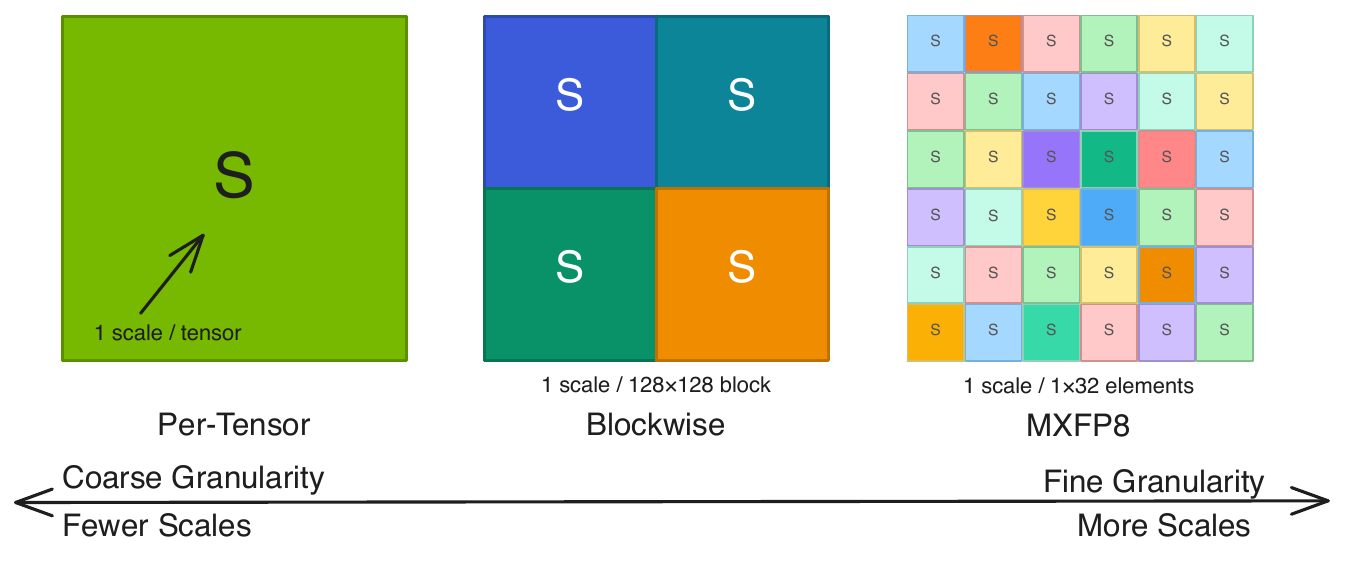}
    \caption{FP8 training recipes: Per-Tensor Scaling, Blockwise FP8, and MXFP8.}
    \label{fig:fp8-recipes-overview}
\end{figure}

A reduced-precision training recipe consists of:
\begin{itemize}
    \item \textbf{Data format.} There are two types of FP8 format: E4M3 and E5M2 \cite{micikevicius2022fp8formatsdeeplearning,kuzmin2022fp8quantization}. Usually there are two combinations used in training:
    \begin{itemize}
        \item \textbf{E4M3}: Inputs, weights, and gradients are all quantized in the E4M3 format.
        \item \textbf{Hybrid}: Inputs and weights are quantized in E4M3, while gradients are quantized in E5M2.
    \end{itemize}
    While for FP4, E2M1 is the only supported format.
    \item \textbf{Scaling granularity.} Choices include per-tensor and per-block (also called group or tile). For per-block scaling, the block size also varies between recipes.
    \item \textbf{Which parts are quantized.} Usually all linear layers are quantized, while keeping embedding, output layer (LM head), and optimizers in their original high precision. Part of the communication can be in FP8/FP4 (e.g., TP AllGather, DP AllGather, EP all-to-all). Current reduced-precision training recipes usually keep SDPA in BF16.
    \item \textbf{Additional mechanisms}, such as stochastic rounding, Random Hadamard Transforms (RHT), etc. These mechanisms are recipe-specific and require careful verification. 
\end{itemize}

Megatron-Core provides three FP8 recipes, each validated at scale (\figref{fig:fp8-recipes-overview}). The evolution from per-tensor to blockwise to MXFP8 reflects both hardware advances and lessons learned from large-scale deployments. \textbf{Per-Tensor Scaling} uses a single scale factor per tensor, simple but limited precision, suitable for experimentation. \textbf{Blockwise FP8} (128$\times$128 blocks) provides finer granularity with proven stability at scale, recommended for Hopper. \textbf{MXFP8} (1$\times$32 granularity) offers native Blackwell Tensor Core support with hardware-accelerated scaling, recommended for Blackwell. Besides, Megatron-Core also provides an NVFP4 recipe \cite{nvidia2025pretraininglargelanguagemodels} on Blackwell.

\subsubsection{Per-Tensor FP8 Recipe}
The per-tensor FP8 recipe, supported on Hopper and Blackwell, usually adopts the hybrid format (E4M3 for inputs/weights, E5M2 for gradients). It calculates the absolute max value (amax) of all values in the input tensor and uses that to quantize the tensor into FP8.

There are two variants of per-tensor scaling:
\begin{itemize}
    \item \textbf{Delayed scaling}: Use the amax from a history window. This breaks the data dependency between the calculation of amax and scaling, offering the best performance at the risk of losing precision due to the estimated amax.
    \item \textbf{Current (live) scaling}: Calculate the amax in a just-in-time manner, providing better precision.
\end{itemize}

\begin{figure*}[t!]
    \centering
    \begin{subfigure}[t]{0.45\textwidth}
        \centering
        \includegraphics[width=1.0\linewidth]{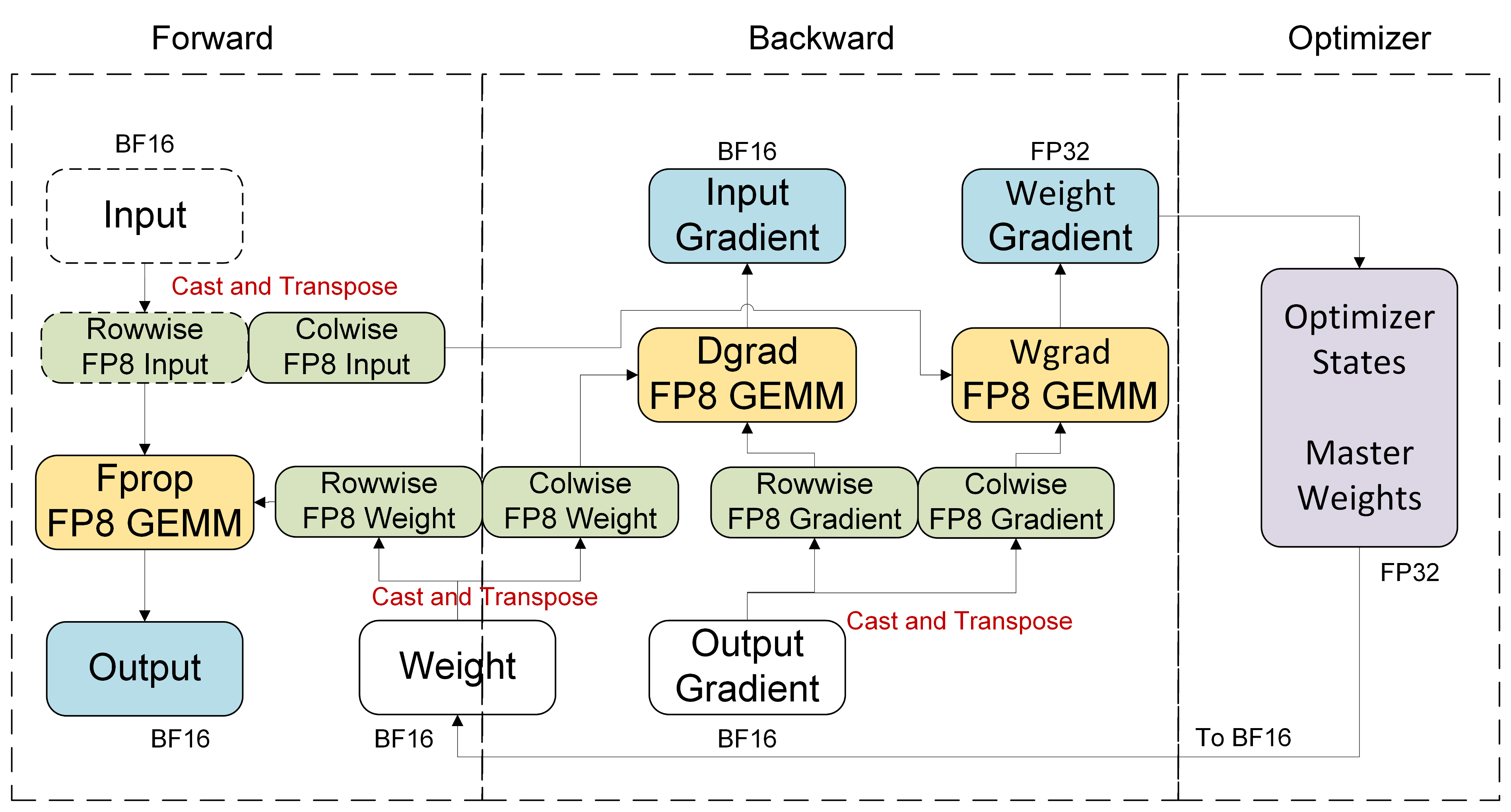}
        \caption{Per-tensor Current scaling on the Hopper platform.}
        \label{fig:fp8-cs-hopper}
    \end{subfigure}
    \begin{subfigure}[t]{0.45\textwidth}
        \centering
        \includegraphics[width=1.0\linewidth]{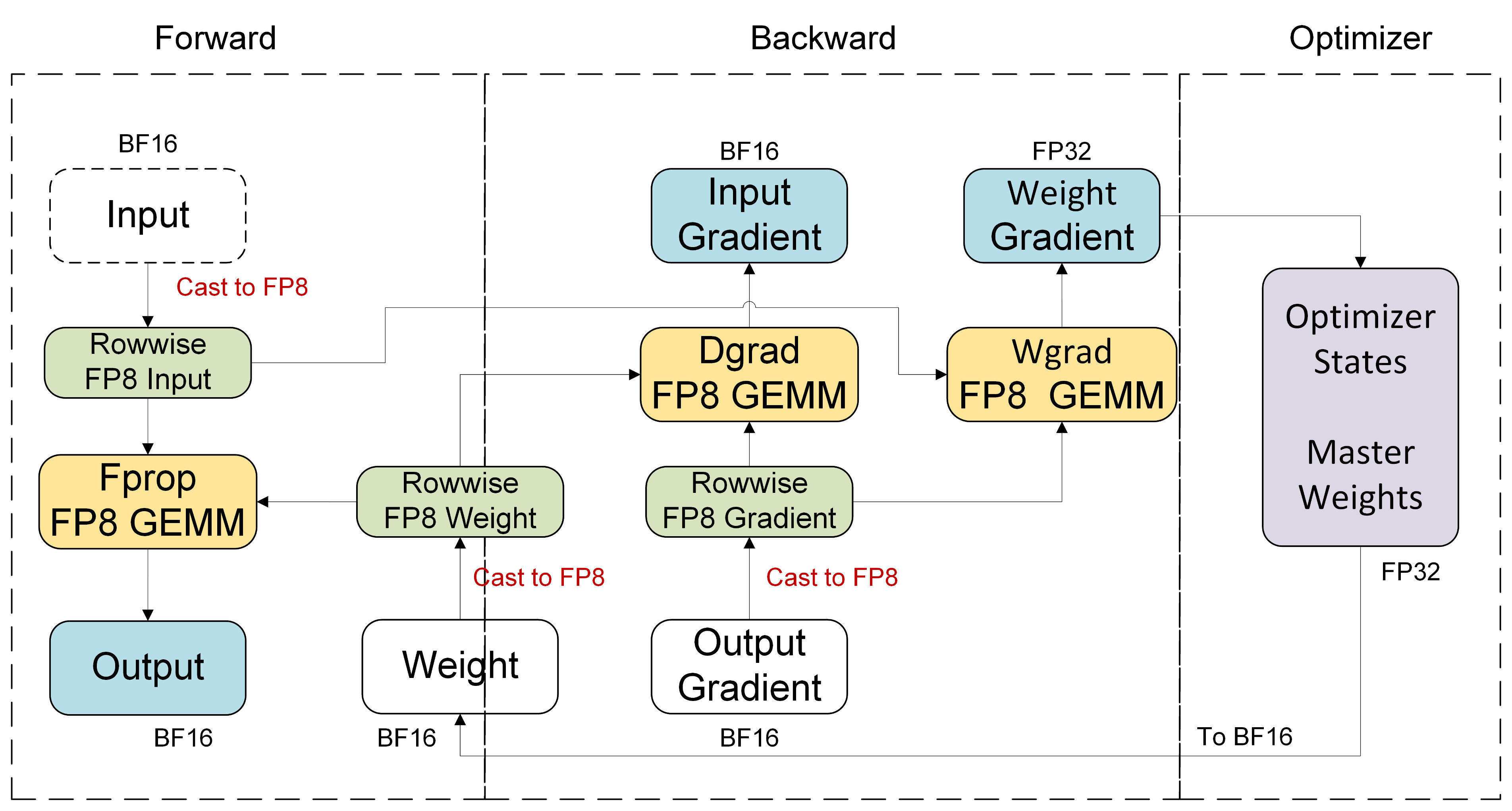}
        \caption{Per-tensor Current scaling on the Blackwell platform.}
        \label{fig:fp8-cs-blackwell}
    \end{subfigure}
    \begin{subfigure}[t]{0.45\textwidth}
        \centering
        \includegraphics[width=1.0\linewidth]{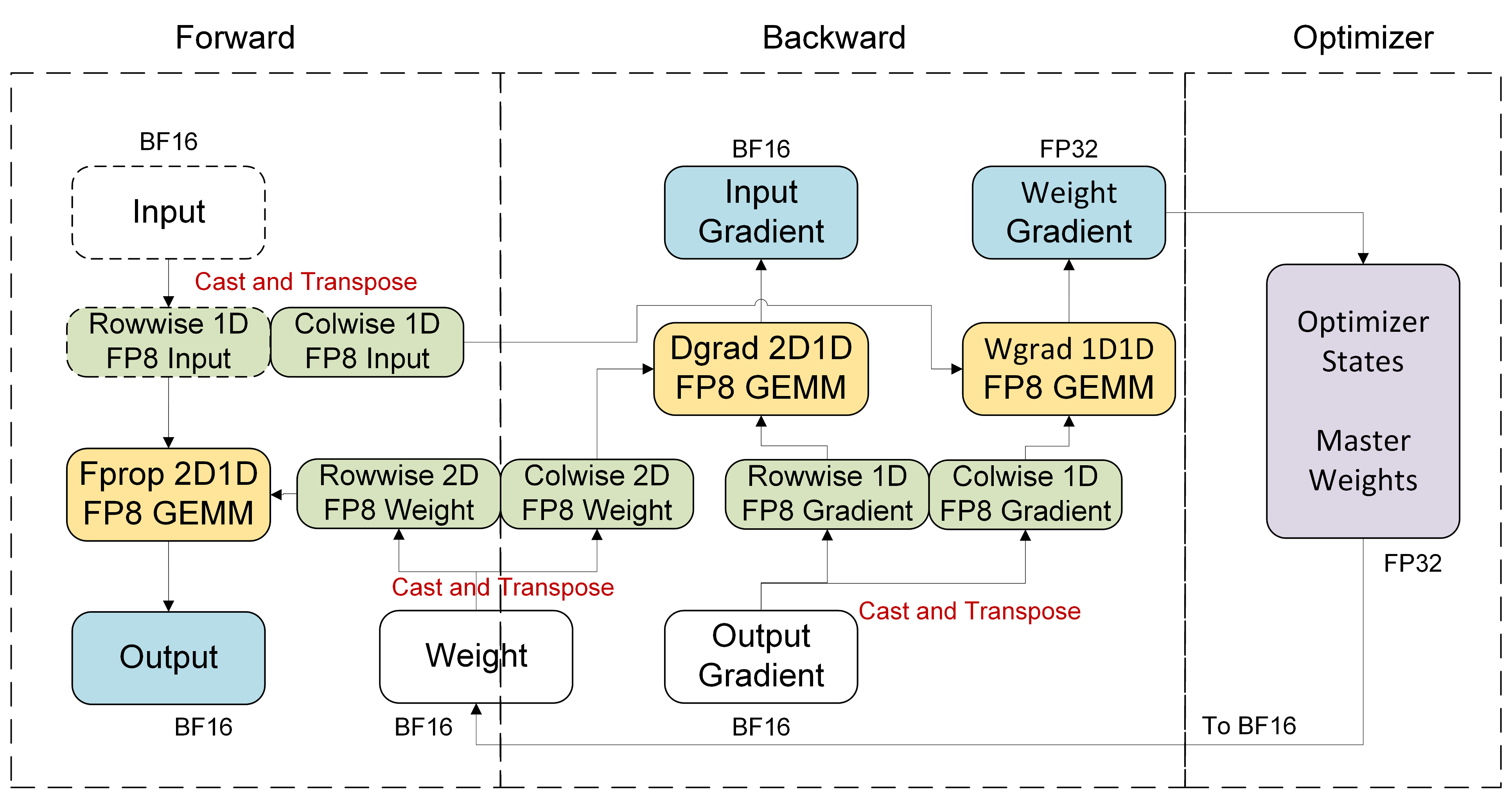}
        \caption{Blockwise FP8 recipe on Hopper.}
        \label{fig:fp8-blockwise-hopper}
    \end{subfigure}
    \begin{subfigure}[t]{0.45\textwidth}
        \centering
        \includegraphics[width=1.0\linewidth]{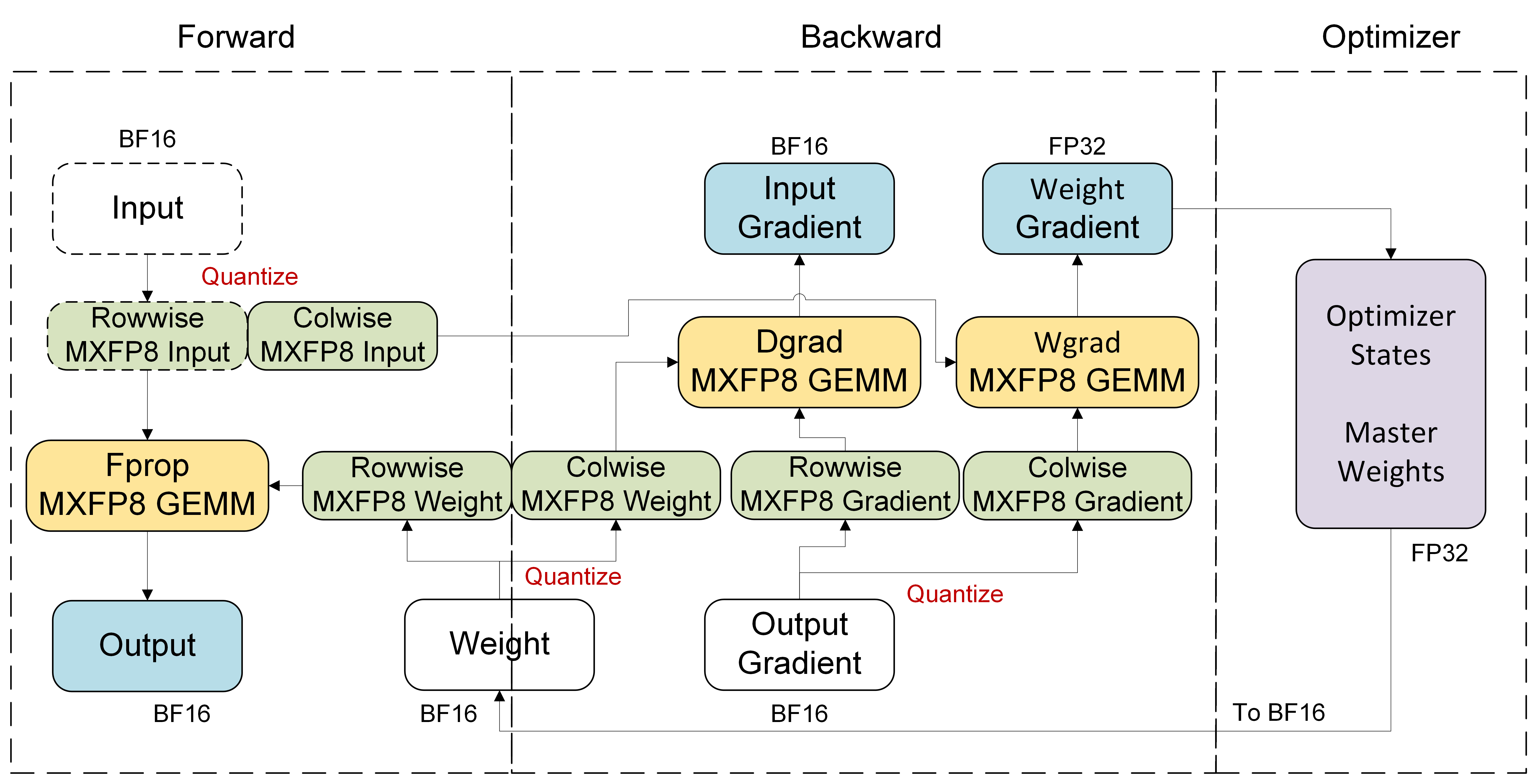}
        \caption{MXFP8 recipe on the Blackwell platform.}
        \label{fig:fp8-mxfp8}
    \end{subfigure}
    \caption{The computation of a linear layer with various FP8 recipes. Note the differences in quantization granularity and tensor layout requirements across platforms.}
    \label{fig:fp8-recipes}
\end{figure*}

We do not recommend delayed scaling due to precision issues; current scaling provides better precision and model convergence.
Figures~\ref{fig:fp8-cs-hopper} and~\ref{fig:fp8-cs-blackwell} illustrate the computation of a linear layer with per-tensor current scaling on the Hopper and Blackwell platforms, respectively. Since only TN layout FP8 GEMM is supported on Hopper, we must store the transposed FP8 activations for backward calculation. In Transformer Engine, the quantization kernels fuse the cast and transpose into a single kernel to reduce global memory access. On the Blackwell platform, FP8 GEMMs in all layouts are supported, so we do not need the transposed version, further reducing the memory footprint.

Per-tensor scaling is a good starting point for migrating to FP8 training due to its simplicity and relatively good precision and performance.

\subsubsection{Blockwise FP8 on Hopper}
The blockwise FP8 recipe adopts the E4M3 format for all tensors, quantizing activations and gradients in $1 \times 128$ tiles and weights in $128 \times 128$ blocks. Blockwise scaling uses fine-grained scaling granularity for better precision and has been \textbf{proven successful in production} at very large-scale MoE models, including DeepSeek-V3 \cite{deepseekai2025deepseekv3technicalreport}, Minimax-M2 \cite{minimax_m2}, Ant Ling-2.0 \cite{li2025every}, etc. As a result, blockwise FP8 is the recommended FP8 recipe on the Hopper platform.

Transformer Engine provides highly optimized quantization kernels and GEMM kernels (via cuBLAS) for the blockwise recipe, as well as layer-level APIs and an autocast context to enable FP8 training with just a few lines of code.

\Figref{fig:fp8-blockwise-hopper} illustrates the computation of a linear layer with the blockwise FP8 recipe on the Hopper platform, which is very similar to the per-tensor current scaling on Hopper, except for tile-based quantization.

\subsubsection{MXFP8 on Blackwell}
On the Blackwell platform, thanks to native fifth-generation Tensor Core support for the MXFP8 format \cite{rouhani2023microscalingdataformatsdeep}, we adopt MXFP8, a more fine-grained quantization scheme for training. Both activations and weights are quantized at $1 \times 32$ granularity, and E8M0 is used for the scaling factor. Theoretically, MXFP8 should be more precise due to the finer-grained scaling granularity, and has better performance due to the native support of MXFP8 in the Tensor Core. Therefore, MXFP8 is the default FP8 recipe on the Blackwell platform.

\Figref{fig:fp8-mxfp8} illustrates the computation of a linear layer with the MXFP8 recipe on the Blackwell platform. Although FP8 GEMMs with all layouts are supported on Blackwell, we need to store additional column-wise quantized FP8 data for activations and weights due to quantization in the other direction.

Ongoing optimizations for the MXFP8 recipe in MoE training include:
\begin{enumerate}
    \item Grouped quantization to reduce the quantization overhead.
    \item Fuse activation and quantization (for the following GEMM) with the GEMM kernel.
    \item Ensure that the whole pipeline of MXFP8 quantization, scaling factor swizzling, and Grouped GEMM is CUDA-graphable.
\end{enumerate}

\subsubsection{NVFP4 on Blackwell}

In \cite{nvidia2025pretraininglargelanguagemodels}, we presented NVFP4 as a 4-bit microscaling format for LLM training that improves numerical fidelity while preserving the efficiency benefits of FP4. 

NVFP4 uses FP4 elements in E2M1 format, so values are quantized with scaling because raw FP4 has a very limited representable range. A central design choice is two-level microscaling. NVFP4 applies a per-tensor FP32 scale and a per-block 8-bit scale in E4M3. The tensor-level FP32 scale first remaps the tensor distribution into a range compatible with block scaling, and the block-level E4M3 scale then maps each block into FP4 range. NVFP4 uses blocks of 16 contiguous elements, and each block’s amax is scaled to the FP4 maximum.

Building on this format design, we also found that stable NVFP4 training depends on several algorithmic choices around quantization. In particular, we add three practical techniques:
\begin{itemize}
\item \textbf{Random Hadamard Transforms (RHT)}: Applied to weight gradient computation to reduce the impact of outliers.
\item \textbf{2D scaling}: Specifically, 16x16 weight block scaling (with the FP32 tensor scale retained) is used to keep weight quantization more consistent between forward and backward passes and reduce forward/backward quantization mismatch.
\item \textbf{Stochastic rounding}: Used on gradients to reduce rounding bias during FP4 conversion, as deterministic rounding introduces bias that hurts convergence.
\end{itemize}
These additions are a key part of the training recipe used in the paper and are important for convergence at larger scales.

\begin{figure}[ht]
    \centering
    \includegraphics[width=1\linewidth]{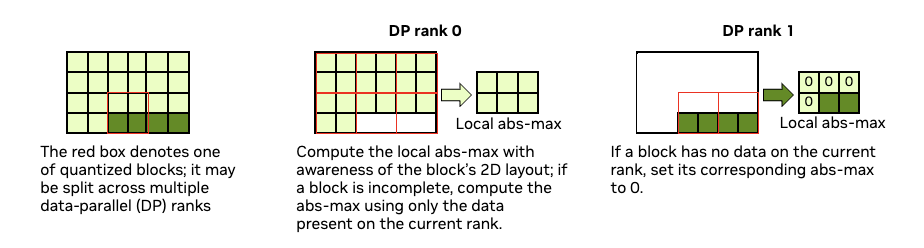}
    \caption{FP8 primary weight quantization scheme for blockwise scaling.}
    \label{fig:native_fp8_blockwise}
\end{figure}

\subsubsection{FP8/FP4 Primary Weights: Eliminating Redundant Storage}\label{subsec:fp8-primary-weights}
Conventional reduced-precision training maintains a three-tier parameter hierarchy: FP32 master weights for optimizer updates, BF16 model weights as an intermediate representation, and FP8/FP4 weights for forward and backward GEMM computation. This introduces redundant memory overhead by maintaining both BF16 and FP8/FP4 copies of parameters.

\textbf{Native FP8/FP4} eliminates this redundancy by establishing a direct casting path from FP32 master weights to FP8/FP4 computation weights, bypassing the BF16 intermediate layer entirely, reducing memory footprint and accelerating parameter AllGather.

The core challenge in implementing native FP8/FP4 lies in managing the quantization metadata required for FP8/FP4 tensors. We provide a unified interface supporting different FP8/FP4 recipes (delayed scaling, current scaling, blockwise scaling, MXFP8 and NVFP4) that handles version-specific differences in TransformerEngine's implementations, ensuring compatibility across versions.

During each optimizer step in the distributed optimizer, the FP32 parameter shards are updated based on reduced gradients. Subsequently, these updated FP32 shards are directly quantized to FP8/FP4 format using the appropriate scaling factors. The quantization process computes the amax from the FP32 values, updates the scaling factors across data-parallel ranks to maintain consistency, and writes the quantized FP8/FP4 values to the model parameter buffers. This direct path eliminates the memory overhead of maintaining BF16 parameters, a reduction particularly impactful for large language models where parameter memory can exceed hundreds of gigabytes.

\begin{figure}[ht]
    \centering
    \includegraphics[width=1\linewidth]{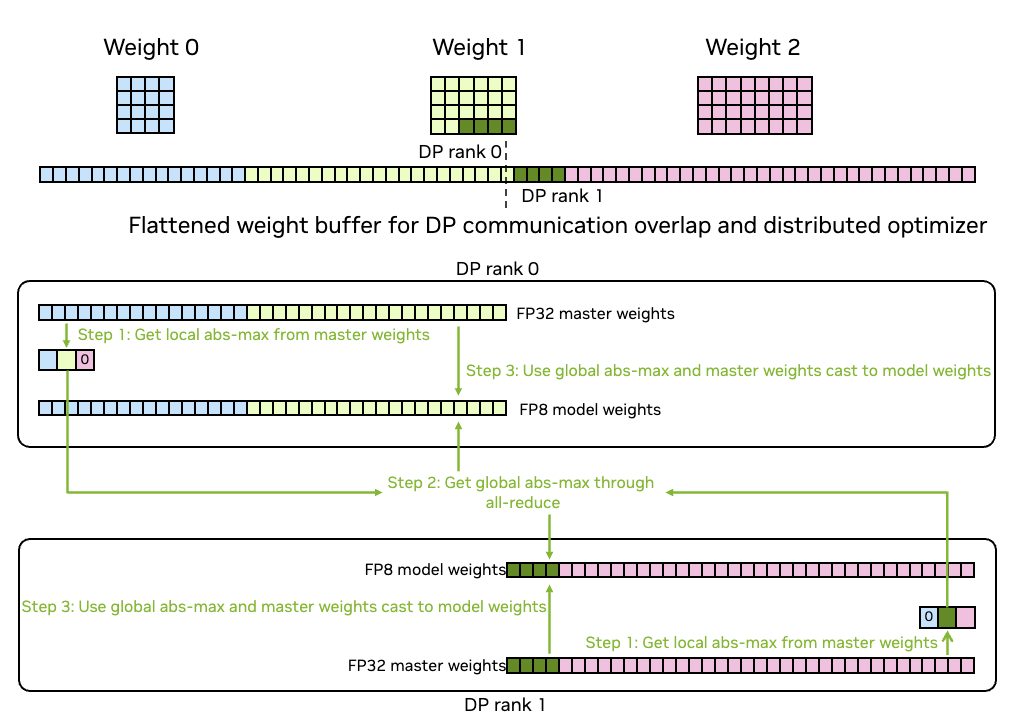}
    \caption{FP8 primary weight quantization scheme for delayed scaling and per-tensor current scaling.}
    \label{fig:native_fp8_current_scaling}
\end{figure}

More specifically, the quantization of FP32 shards for delayed scaling and current scaling proceeds in three steps, see \figref{fig:native_fp8_current_scaling}:
\begin{itemize}
    \item Step 1: Get local abs-max from master weights. If a weight has no sharded part on the current rank, set its corresponding abs-max to 0.
    \item Step 2: Get global abs-max through AllReduce.
    \item Step 3: Use global abs-max and master weights to do partial cast.
\end{itemize}

In the Blockwise recipe, the abs-max of a weight is computed over a 2D block, so we cannot directly use a reduction kernel to obtain the local abs-max. We implemented specialized kernels that are aware of the weight's 2D layout and the correspondence between the master weight and the original weight to compute the abs-max, see \figref{fig:native_fp8_blockwise}.

Another advantage of FP8/FP4 primary weights is that the parameter AllGather communication volume is greatly reduced (see Section \ref{sec:comm_wall_quantized_training}).

\subsection{MoE-Specific Challenges and Solutions of Reduced-Precision Training}

The preceding subsections covered the fundamentals of reduced-precision training applicable to any deep learning model. MoE architectures introduce unique challenges for reduced-precision training that require specialized solutions beyond standard implementations.

\subsubsection{Fusion of Padding and Unpadding: Dynamic Shape Alignment}
FP8 and FP4 GEMMs require tensor dimensions to be aligned to specific multiples: 16 for per-tensor and blockwise recipes, and 32 for MXFP8 and NVFP4. These padding requirements arise from the block scaling granularity of each quantization recipe, as well as the requirement that Tensor Memory Accelerator (TMA) accesses maintain 16-byte alignment along the GEMM dot-product dimension. In the forward pass, the hidden dimension K is the dot-product dimension and is usually already aligned by model design. However, in the weight-gradient GEMM, the dot-product dimension corresponds to the token dimension M, which varies dynamically and often does not satisfy the alignment requirement. As a result, zero-padding along the token dimension is necessary. To further enable grouped quantization kernels with lower CPU overhead and CUDA Graph compatibility, we increase the tokens-per-expert padding to 128. Section \ref{sec:fp4-quant-fusion} explains why this larger padding is required for the NVFP4 grouped quantization kernel; the same principle also applies to MXFP8 grouped quantization. In summary, the final padding applied to the token dimension is determined jointly by the requirements of all kernels involved in the routed expert layer.

Our baseline solution is the fused padding and unpadding kernels for multiple input tensors for different experts. However, explicit padding and unpadding kernels introduce non-negligible overheads. We propose two solutions:
\begin{itemize}
    \item \textbf{Routing map padding}, which pads the routing map instead of the received tokens. This ensures the number of tokens per expert is aligned to the requirement at the cost of sending only a small number of extra tokens, avoiding expensive per-tensor padding operations.
    \item \textbf{Fusing padding into permutation} avoids one-pass of global memory read and write and should provide the best performance. It is the default choice when available.
\end{itemize}

\subsubsection{Grouped Quantization and Grouped GEMM}

\paragraph{Grouped Quantization}

Due to the different padding requirements between the quantized data and scaling factors, and the complex data layouts introduced by both row-wise and column-wise quantization, a naive way to quantize the input tensors for different experts is to apply the quantization kernel to them one by one. Nevertheless, it introduces plenty of tiny kernels, which adds to the CPU overhead and is not efficient from the GPU perspective.

To tackle this problem, we implement grouped quantization kernels to fuse the quantization of multiple tensors into one single kernel, which greatly reduces the CPU overhead, improves the GPU utilization and is CUDA-Graphable.

\paragraph{Grouped GEMM}
Optimized and CUDA-Graphable grouped GEMM in Section \ref{sec:device-init}.

\subsubsection{NVFP4 Quantization Fusion}\label{sec:fp4-quant-fusion}

NVFP4 quantization is particularly more complicated because of the numerical techniques we use to preserve numerical stability, so aggressive kernel fusion is necessary to keep quantization overhead under control. In practice, our NVFP4 quantization kernel is not just a simple “scale + cast” kernel: it needs to carefully absorb the training-recipe logic, including Random Hadamard Transform, 2D scaling and stochastic rounding. 

Among these, RHT fusion is the most latency-sensitive. If implemented as a separate kernel, the Hadamard transform would introduce an additional full-precision (BF16) read/write of the tensor in global memory, significantly increasing bandwidth cost. By fusing RHT with NVFP4 quantization, we perform the Hadamard transform and FP4 quantization in a single kernel, avoiding that extra BF16 traffic.

A second implementation challenge is that Blackwell NVFP4 Tensor Core GEMMs are TN-oriented, while Wgrad uses transposed activations and gradients. Therefore, when RHT is fused into NVFP4 quantization for the Wgrad path, the kernel must also absorb the transpose, rather than relying on a separate BF16 transpose kernel (which would again incur a full BF16 tensor read and write through global memory).

As a result, the fused quantization pipeline must support multiple outputs from the same high-precision input: (1) standard FP4 quantization for forward-pass GEMMs, and (2) transpose + RHT + FP4 quantization for the backward/Wgrad path. In training, we launch this fused kernel during the forward pass to generate two FP4 copies: one consumed immediately by the forward GEMM, and one saved for backward. The original high-precision input is then discarded, so we avoid storing BF16 activations while still preparing the backward path efficiently.

A further complication comes from the per-tensor FP32 scale in NVFP4. Current NVFP4 pretraining recipe, the tensor-wide abs-max is measured after the Hadamard transform, which means we need a dedicated Hadamard+amax kernel that computes only the amax (without materializing a transformed BF16 output buffer). As a result, the Hadamard transform is effectively computed twice—first in the Hadamard-amax kernel and then again in the fused quantization kernel. Although this duplicates transform compute, it is still faster end-to-end than writing out a full transformed BF16 tensor to global memory, because it avoids the extra high-bandwidth BF16 read/write traffic.

In contrast, 2D quantization is less visible in end-to-end throughput because it applies only to weights: we can quantize weights once (for example, on the first microbatch) and cache the quantized weights and scales for reuse across later microbatches, so its cost is largely amortized. 

Stochastic rounding, however, adds a kernel-level requirement on the critical path: the quantization kernel must also support an optional stochastic rounding path with random number generated in kernel (enabled for gradient tensors such as dY, disabled otherwise) so that FP4 quantization remain fused in a single pass. To keep this efficient, we use NVIDIA cuRANDDx\cite{curanddx} for device-side random number generation inside the fused CUDA quantization kernel.

\paragraph{Grouped NVFP4 Quantization for MoE}

Supporting NVFP4 quantization for MoE layers is especially challenging due to the algorithmic complexity of the NVFP4 recipe itself. Under a full-iteration CUDA Graph requirement, the MoE quantization path must also be CUDA Graph safe: the host cannot depend on dynamic expert-token counts on the CPU, and can only rely on a device-side tokens-per-expert tensor. This means that input activation can no longer be splitted and quantized individually. We must implement grouped quantization kernels for the full NVFP4 pipeline. The memory allocation of NVFP4 output must be allocated as a whole flat buffer without knowing the shapes in between. 

A key observation is that much of the dense NVFP4 activation quantization kernel can be reused for grouped quantization, because tokens for different experts are continuously packed in memory after routing/permutation. The main differences are implementation constraints needed to preserve performance and graph safety. 

\begin{enumerate}

\item A quantization thread block should not span tokens from multiple experts, since that introduces significant control-flow and indexing overhead; therefore, we zero-pad each expert’s token count to an integer multiple of the thread block shape in the token dimension (typically 128). 

\item The NVFP4 transpose must be performed per expert (transpose each expert’s packed activation independently, then concatenate), which is not equivalent to transposing the entire grouped buffer as one tensor. 

\item NVFP4 GEMM requires scale-factor swizzling: scaling factors must be padded to a 128×4-aligned shape and swizzled into the 32×16 layout before GEMM as described in the cuBLAS document\cite{cublas_sf_layout}. This scaling factor shape constraint is applied to per-expert GEMM, which means that determining the required padding size would implicitly require CPU visibility into tokens-per-expert, which is not CUDA Graph safe. To avoid that, we enforce 128-token alignment per expert by construction.
\end{enumerate}

Both point 1) and point 3) imposed 128 multiple tokens-per-expert, which requires a specific zero padding kernel to satisfy. To eliminate such zero padding overhead, we fuse the per-expert zero-padding capability into the token-permute kernel, so the routed expert activations are produced in an already aligned and zero padded layout, and downstream grouped NVFP4 quantization/GEMM kernels can assume the required alignment without additional preprocessing.

For the per-tensor FP32 second-level scale in the NVFP4 recipe, the per-tensor second-level scale in MoE means per-expert second-level scale instead of making every expert share the amax to preserve numerical stability during training. These amax values cannot be known ahead of time: they must be computed online from the routed tokens and generated as distinct per-expert amax values on every iteration. To address this efficiently, we leverage the 128-aligned tokens-per-expert guarantee and adapt the dense-input implementation into a CUDA-Graph-safe grouped Hadamard-amax kernel. 

%==============================================================================
% SECTION 6: LONG-CONTEXT MOE TRAINING
%==============================================================================

\section{Long-Context MoE Training}\label{sec:long-context-training}

The preceding sections addressed the three walls (memory, communication, and compute efficiency) under the assumption that MoE layers dominate computational cost. This assumption holds for typical training tasks with sequence lengths of 4K--8K tokens, where expert computation and all-to-all communication are the primary bottlenecks. However, the rise of reasoning models (OpenAI o1, DeepSeek-R1 \cite{deepseekai2025deepseekr1}, etc.) and RL-based training have created a new frontier: sequences of 16K, 64K, or even longer. At these lengths, a fundamental shift occurs in the performance characteristics of MoE training: attention dominates the computation, and the relative importance of the three walls changes accordingly. This section examines how long-context training reshapes the three walls for MoE models, and how Megatron-Core addresses the new challenges.

\begin{figure}[ht]
    \centering
    \includegraphics[width=1\linewidth]{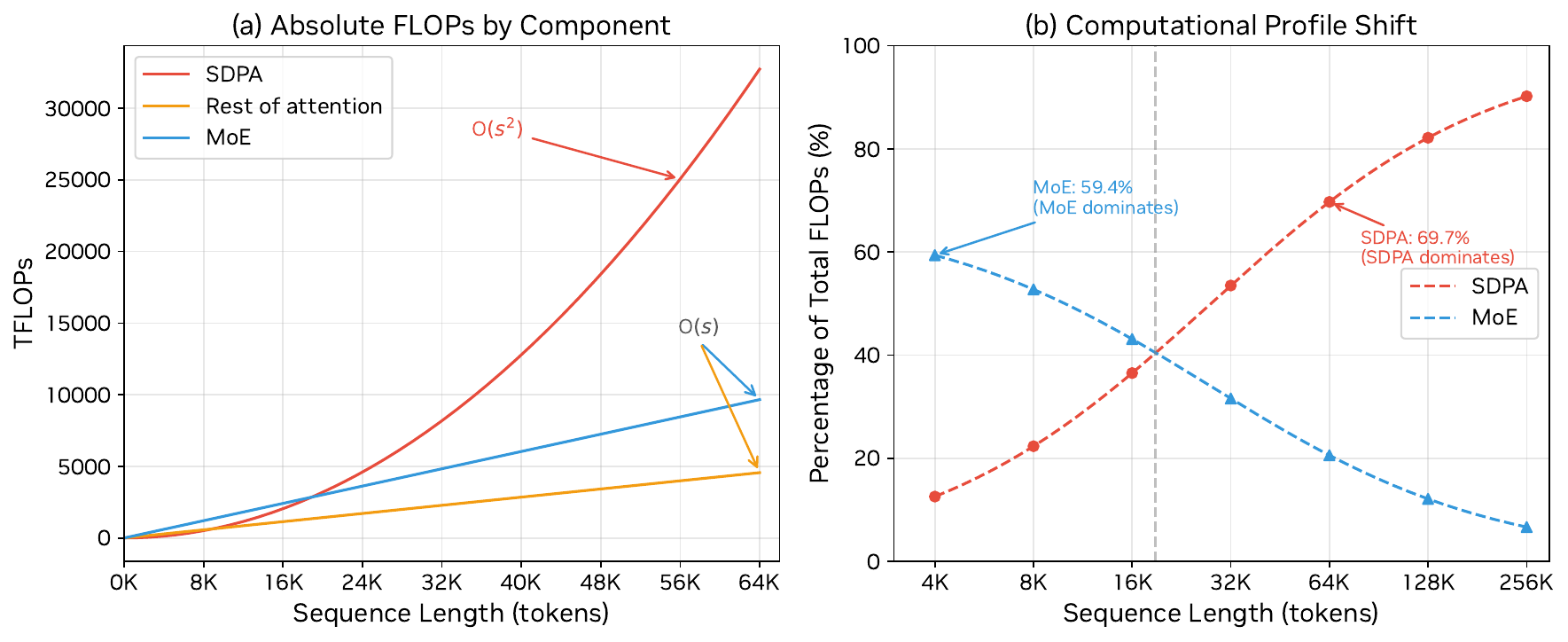}
    \caption{SDPA exhibits $O(s^2)$ complexity, while MoE and the remaining attention operations exhibit $O(s)$ complexity. Therefore, SDPA dominates the computation at longer sequence lengths.}
    \label{fig:long_context_flops_ratio}
\end{figure}

\subsection{When Attention Dominates: The Computational Shift}

In Section~\ref{key-features-optimizations}, we established that MoE training faces three fundamental walls: memory, communication, and compute efficiency. The optimizations presented there (activation management, dispatcher optimization, Grouped GEMM, and CUDA Graphs) target scenarios where MoE layers dominate the computational profile. Long-context training fundamentally alters this assumption. The key insight is the differing computational complexity between MoE and attention layers, as shown in \figref{fig:long_context_flops_ratio}:
\begin{itemize}
    \item \textbf{MLP Components}: Computational cost scales linearly with sequence length, exhibiting $O(s)$ complexity.
    \item \textbf{Attention Components}: Scaled dot-product attention (SDPA) dominates the computational cost, scaling quadratically with sequence length, exhibiting $O(s^2)$ complexity. This quadratic scaling has motivated extensive research into efficient attention mechanisms~\cite{beltagy2020longformer,zaheer2020bigbird}.
\end{itemize}

For example, at 64K tokens, SDPA consumes 69\% of FLOPs, compared to only 10--15\% in short-sequence scenarios. \textbf{Attention becomes dominant in computation.} Fortunately, SDPA is highly optimized in most GPU-accelerated libraries such as FlashAttention \cite{dao2022flashattention,dao2023flashattention2,shah2024flashattention3} and cuDNN (Table~\ref{tab:sdpa-cudnn-perf}), so SDPA does not become a performance bottleneck. The optimization focus therefore shifts to memory and communication: as long as we address these two challenges without introducing excessive overhead, training performance is preserved.

\begin{table}[ht]
\centering
\caption{SDPA performance in cuDNN for DeepSeek-V3.}
\label{tab:sdpa-cudnn-perf}
\begin{tabular}{cccc}
\toprule
\textbf{Platform} & \textbf{Sequence Length} & \textbf{Forward TFLOPS} & \textbf{Backward TFLOPS} \\
\midrule
\multirow{2}{*}{Hopper}
 & 4096  & 553  & 422 \\
 & 16384 & 638  & 523 \\
\midrule
\multirow{2}{*}{Blackwell}
 & 4096  & 1324 & 1083 \\
 & 16384 & 1698 & 1298 \\
\bottomrule
\end{tabular}
\end{table}

\subsection{Managing Activation Memory Growth}

Activation memory growth with sequence length is the primary challenge in long-context training. To address this intensified memory wall, we apply a set of techniques that work together, informed by large-scale MoE workloads, including DeepSeek-V3 \cite{deepseekai2025deepseekv3technicalreport} and Qwen3 \cite{yang2025qwen3technicalreport}. These techniques extend the memory-optimization principles introduced in Section~\ref{sec:memory-wall}.

\textbf{Context Parallelism and Tensor Parallelism.}
Combining Context Parallelism (CP) and Tensor Parallelism (TP) distributes activation memory across devices, enabling sequence lengths that would otherwise exceed GPU capacity. The key principle is to keep sub-sequence length (sequence length per CP/TP shard) approximately constant, typically 4096 or 8192. Scaling $\text{CP} \times \text{TP}$ with sequence length keeps per-device memory near baseline levels. This makes the workload resemble short-context training, except for SDPA and TP/CP-specific communication, so most short-context optimizations remain applicable.

\textbf{Optimizer CPU Offloading.}
Optimizer states often consume tens of gigabytes per GPU in large models. CPU offloading can reclaim this memory almost entirely, at the cost of transfer and host-side optimizer overhead. The choice is therefore a trade-off between memory headroom and throughput. For DeepSeek-V3 on 256 H100 GPUs, at around 50\% MFU and sequence lengths of at least 16K, the worst-case overhead is about 2\%. Although the exact trade-off depends on model size, sequence length, and cluster scale, optimizer CPU offloading is usually worthwhile for long-context training.

\textbf{Selective Recomputation.}
Recomputation trades compute for memory by discarding intermediate activations in forward and regenerating them in backward. The key is module-level selectivity based on memory-to-compute trade-offs. In long-context settings, SDPA dominates total computation, so recomputing SDPA (core attention recomputation in Megatron-Core) is usually too expensive. In contrast, recomputing lower-cost components such as MLP-related modules often provides better net benefit. For DeepSeek-V3 at 64K sequence length, SDPA contributes up to 72\% of total compute; recomputing it adds about 18\% compute overhead, causes about 16\% performance loss, and saves only 9 GB memory. Recomputing non-SDPA components instead saves 89.8 GB globally with comparable or lower performance impact. We therefore recommend disabling core attention recomputation and prioritizing recomputation of other modules.

In practice, CP and TP are the primary mechanisms for long-context memory efficiency. Optimizer CPU offloading and selective recomputation then provide additional headroom, especially on memory-constrained platforms. These techniques work well together and can be combined as needed.

\subsection{Context Parallelism vs. Tensor Parallelism}

Because CP and TP are the two most effective strategies in long-context training, the practical question is how to combine them when attention dominates computation.

\begin{figure}[ht]
    \centering
    \includegraphics[width=1\linewidth]{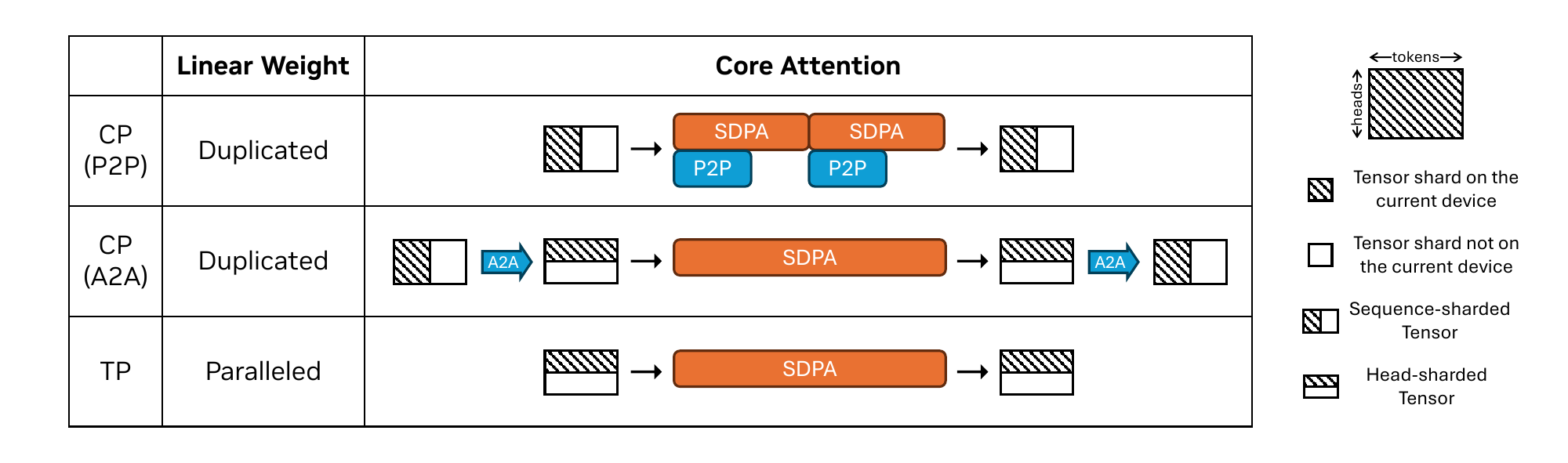}
    \caption{Communication and computation patterns of TP and two types of CP.}
    \label{fig:cp-comm-type}
\end{figure}

Both CP and TP reduce activation memory in long-context training, but their communication patterns and memory effects differ (Figure~\ref{fig:cp-comm-type}). CP partitions activations along the sequence dimension and reduces activation memory by a factor of $\text{CP}$. For SDPA, CP has two communication modes: point-to-point (P2P) and all-to-all. P2P CP exchanges the KV cache within the CP group in a ring-style pattern that overlaps with SDPA computation~\cite{liu2023ringattention,brandon2023striped,jacobs2023deepspeedulysses}. All-to-all CP transforms tensors from sequence-sharded to head-sharded form before SDPA, then restores sequence-sharded form afterward. TP additionally shards linear weights, reducing parameter memory but introducing extra collectives in linear layers. In both all-to-all CP and TP, SDPA runs on head-sharded tensors, which increases per-shard sub-sequence length and can improve SDPA kernel efficiency.

In practice, P2P CP is the more common CP mode. Relative to TP, its main advantage is that communication overlaps naturally with computation, while TP communication is often harder to hide. TP, however, reduces dense parameter memory and can improve SDPA efficiency. As a result, TP is usually preferred within a node, where communication is fast and memory benefits are strong. Across nodes, P2P CP often becomes preferable because TP communication overhead grows while P2P overlap remains effective. All-to-all CP sits between the two: it avoids multi-step ring exchange used by P2P CP and avoids linear-weight sharding overhead introduced by TP. In practice, all-to-all CP is often combined with TP for intra-node execution when ring-style exchange is undesirable and TP communication cost is high.

Megatron-Core supports combining both CP backends as \textbf{hierarchical CP}, enabling topology-aware pairing with TP. A practical starting point is to use all-to-all CP with TP inside nodes to improve SDPA kernel performance and reduce parameter memory, while using P2P CP across nodes to preserve communication-computation overlap. These are guidelines rather than hard rules, but they work well as an initial configuration for TP/CP tuning.

\subsection{Packed Sequences for Variable-Length Training}\label{subsec:packed-sequence}

The preceding subsections examined how the Three Walls (memory, communication, and compute efficiency) shift when sequence length increases. These analyses assumed fixed-length sequences within each batch. However, practical training scenarios, particularly reinforcement learning and supervised fine-tuning, often involve variable-length sequences that introduce additional efficiency challenges.

Traditional batching requires padding all sequences to the maximum length within each batch, leading to substantial waste when sequence lengths vary widely~\cite{krell2021efficient}. For instance, if a batch contains sequences of lengths 4K, 8K, and 32K tokens, all sequences must be padded to 32K, wasting $\sim$60\% of computation and memory on padding tokens. This inefficiency compounds the memory and compute challenges discussed above.

To address this, Megatron-LM introduces \textbf{packed sequences} to process variable-length sequences within a single batch without padding waste. Furthermore, to address the DP imbalance and CP inefficiency caused by variable-length sequences, we introduce \textbf{Dynamic Context Parallelism (Dynamic-CP)}, which adaptively selects the effective CP size on a per-microbatch basis, jointly with the sequence packing plan.

\subsubsection{Packed Sequence Support}\label{subsec:packed_sequence_support}

Megatron-LM supports packed sequences, enabling multiple variable-length sequences to be concatenated and processed within a single batch without inter-sequence padding. This is achieved through the THD (Total tokens $\times$ Heads $\times$ Dimension) tensor format, as opposed to the conventional SBHD (Sequence $\times$ Batch $\times$ Heads $\times$ Dimension) format. THD represents attention tensors as \texttt{[total\_tokens, num\_heads, head\_dim]}, where sequences from different samples are concatenated along the token dimension.

The core mechanism relies on cumulative sequence length tracking, which marks the start and end positions of each individual sequence within the packed tensor. These parameters are propagated through the attention mechanism to Transformer Engine's fused kernels, enabling efficient SDPA and RoPE operations that respect sequence boundaries. The attention computation uses cumulative sequence lengths to ensure queries and keys from different sequences do not attend to each other, maintaining correctness while eliminating padding overhead.

The implementation in Megatron-LM supports packed sequences across multiple training scenarios:
\begin{itemize}
    \item \textbf{Reinforcement Learning}: Bin-packing algorithms group variable-length trajectories into fixed-size bins, maximizing GPU utilization.
    \item \textbf{Multimodal Training}: Vision-language models with variable image token counts use packed sequences to handle heterogeneous sequence lengths from different numbers of visual patches combined with text tokens.
    \item \textbf{Context Parallelism Integration}: When Context Parallelism is enabled with padding requirements, the system automatically switches to THD format and provides both padded and unpadded cumulative lengths to handle communication alignment while preserving computational efficiency.
\end{itemize}

The benefits of packed sequence support are particularly pronounced in long-context scenarios. For RL training with thinking models, where sequence lengths can vary from hundreds to tens of thousands of tokens, packed sequences can reduce memory usage by 40--60\% and improve training throughput by 1.5--2$\times$ compared to traditional padding-based approaches. This efficiency gain becomes increasingly critical as context lengths extend beyond 32K tokens, where padding waste would otherwise dominate memory consumption and significantly reduce effective batch sizes.

\subsubsection{Dynamic Context Parallelism for Packed Sequences}
\label{subsec:dynamic-cp}

\begin{figure}[ht]
      \centering
      \begin{subfigure}[b]{0.48\textwidth}
        \centering
        \includegraphics[width=0.8\textwidth]{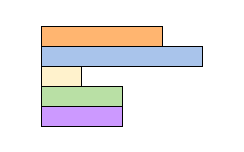}
        \caption{Unpacked sequences.}
        \label{fig:unpacked-sequence}
      \end{subfigure}
      \hfill
      \begin{subfigure}[b]{0.48\textwidth}
        \centering
        \raisebox{0.8cm}{\includegraphics[width=0.8\textwidth]{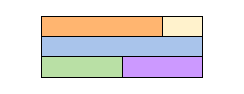}}
        \caption{Packed sequences.}
        \label{fig:packed-sequence}
      \end{subfigure}
      \caption{Unpacked vs. Packed sequences.}
      \label{fig:packing-sequences}
    \end{figure}

\begin{figure}[ht]
      \centering

        \includegraphics[width=0.8\textwidth]{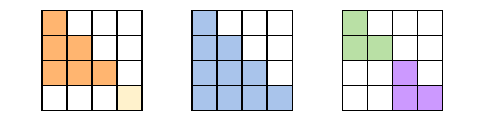}
        \caption{Compute imbalance in causal attention over packed sequences.}
      \label{fig:packing-sequence-imbalance}
    \end{figure}

While packed sequences in THD format remove padding waste, they do not eliminate computational imbalance across data-parallel (DP) ranks. Even when packed samples have identical total token counts, their attention workload can still vary substantially due to the quadratic complexity of dot-product attention with respect to sub-sequence lengths. For example, \Figref{fig:packing-sequences} shows sequences packed to equal total lengths, yet as \Figref{fig:packing-sequence-imbalance} reveals, their compute workloads differ significantly.

This DP imbalance leads to GPU idling at gradient synchronization points and can further amplify pipeline-parallel bubbles. Moreover, when Context Parallelism (CP) is enabled, a static CP size is typically chosen to satisfy the memory requirement of the longest packed sample in the batch. Consequently, shorter packed samples that would fit on fewer devices are still forced to use the same CP size, incurring unnecessary CP communication. In practice, CP communication is expected to be hidden by attention computation; however, under packed sequences that consist of many short sub-sequences, the compute per CP shard may become insufficient to hide CP collectives, especially when CP spans inter-node links. We refer to this effect as CP inefficiency.

To address both DP imbalance and CP inefficiency, we introduce Dynamic Context Parallelism (Dynamic-CP) \cite{nvidia_dynamiccp_blog}, which adaptively selects the effective CP size on a per-microbatch basis, jointly with the packing plan. The key observation is that resizing CP is comparatively lightweight: it only changes how token slices are partitioned and which CP communication group is used by attention operators, without requiring any parameter redistribution or optimizer-state migration. Therefore, Dynamic-CP provides a practical form of dynamic parallelism for variable-length training with minimal framework overhead. Related work, including ByteScale~\cite{ge2025bytescale} and WLB-LLM~\cite{wang2025wlb}, addresses similar load-balancing challenges.

\begin{figure}[ht]
      \centering
      \begin{subfigure}[b]{0.19\textwidth}
        \centering
        \raisebox{0.8cm}{\includegraphics[width=0.8\textwidth]{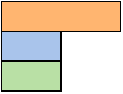}}
        \caption{Input sequences.}
        \label{fig:input-sequences}
      \end{subfigure}
      \hfill
      \begin{subfigure}[b]{0.3\textwidth}
        \centering
        \makebox[\linewidth][c]{\raisebox{0.2cm}{\hspace*{-0.8cm}\includegraphics[width=1.4\textwidth]{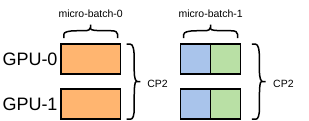}}}
        \caption{Standard CP2.}
        \label{fig:standard-cp}
      \end{subfigure}
        \hfill
      \begin{subfigure}[b]{0.3\textwidth}
        \centering
        \makebox[\linewidth][c]{\raisebox{0.2cm}{\hspace*{-0.8cm}\includegraphics[width=1.4\textwidth]{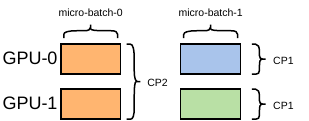}}}
        \caption{Dynamic-CP.}
        \label{fig:dynamic-cp}
      \end{subfigure}

      \caption{Dynamic Context Parallelism for Packed Sequences.}
      \label{fig:dynamic-cp-example}
    \end{figure}

For example, consider the case in \Figref{fig:input-sequences} with three sequences. We assume only two GPUs are available. If we adopt the standard CP2 configuration (i.e., DP=1, CP=2), the workload is executed in two micro-batches as illustrated in \Figref{fig:standard-cp}. In microbatch-0, the orange sequence is sufficiently long that it must be partitioned across the two GPUs. In microbatch-1, the packed blue and green sequences are also partitioned across the two GPUs. However, this introduces an unnecessary split for the blue and green sequences, since both sequences can fit on a single GPU without partitioning.

As shown in \Figref{fig:dynamic-cp}, with Dynamic-CP, the orange sequence in microbatch-0 still requires CP=2. In microbatch-1, we no longer need to split the blue and green sequences; instead, we let each of them run with CP=1 independently.

To make per-microbatch CP resizing practical in Megatron-Core, Dynamic-CP avoids any expensive redistribution of model states. Instead, it treats CP resizing as a lightweight runtime choice: changing only how token slices are partitioned and which CP communication group is used by attention operators. Concretely, rather than binding each rank to a single statically constructed \texttt{cp\_group}, the framework pre-constructs multiple CP groups per rank during initialization, with candidate \texttt{cp\_size} ranging from 1 up to \texttt{dp}$\times$\texttt{cp} (restricted to powers of two). At runtime, the scheduler selects the effective \texttt{cp\_size} for each microbatch and the corresponding pre-built \texttt{cp\_group}, enabling Dynamic-CP without incurring the overhead of dynamically creating communication groups.

Dynamic-CP is designed for the THD layout, where variable-length sequences are packed under a length constraint and the original batch/sequence dimensions collapse into a token dimension. A direct consequence is that the number of microbatches is no longer a fixed quantity derived from \texttt{global\_batch\_size} and \texttt{micro\_batch\_size}; it can vary across iterations because the number of original sequences packed into each microbatch is not constant. To minimize invasive changes to existing Megatron-Core pipeline schedulers, Dynamic-CP introduces a lightweight \texttt{data\_iterator\_wrapper} around the original \texttt{data\_iterator}. For each global batch, the wrapper (i) reschedules and packs sequences to create balanced workloads across DP ranks, (ii) chooses an effective \texttt{cp\_size} for each microbatch to avoid over-sharding short packed samples, and (iii) returns the effective \texttt{num\_microbatches} for the current iteration. Because only a subset of ranks typically drives scheduling decisions under PP/VPP, the framework broadcasts the dynamic scheduling metadata (including \texttt{num\_microbatches}, \texttt{max\_seqlen}, and \texttt{cu\_seqlens}) to ensure consistent execution across pipeline stages. Furthermore, \texttt{PackedSeqParams} is extended to carry both the selected \texttt{cp\_size} and \texttt{cp\_group}; all CP-dependent components (e.g., position embedding and attention) retrieve CP configuration from \texttt{PackedSeqParams} rather than relying on globally static CP variables.

Given variable-length sequences in THD, different samples contribute different numbers of valid tokens. Therefore, loss is computed on a per-token basis to avoid bias from padding:
\begin{equation}
\mathcal{L} = \frac{\sum_{t\in \mathcal{V}} \ell_t}{|\mathcal{V}|},
\end{equation}
where $\mathcal{V}$ denotes the set of valid (non-padding) tokens in the packed representation. On the planning side, the Dynamic-CP solver jointly determines packing and per-microbatch CP sizes under GPU memory constraints. Since attention compute scales roughly as $\mathcal{O}(S^2)$ with respect to sub-sequence lengths while activation memory scales closer to $\mathcal{O}(S)$, it is difficult to simultaneously balance compute and memory. Dynamic-CP therefore alternates between workload-oriented and memory-oriented decisions: microbatches whose estimated workload exceeds a target quota are assigned larger \texttt{cp\_size} to reduce per-rank compute, after which memory becomes the dominant constraint and remaining capacity is filled by selecting less compute-heavy samples while preserving feasibility.

Finally, Dynamic-CP is engineered to keep runtime overhead negligible. Constructing a plan requires an additional pass over the global batch to probe lightweight shape and sequence-length metadata; this I/O pressure is mitigated by distributing the probing across the cluster and gathering only compact metadata. The solver itself runs asynchronously (e.g., within the \texttt{data\_sampler}) to overlap with training iterations. To keep the search space manageable, exhaustive search is replaced by a one-dimensional grid search over the microbatch count, constrained so that all DP ranks use the same \texttt{num\_microbatches}. This count is swept from \texttt{PP}$\times$1 up to a small multiple of \texttt{PP}, capturing the trade-off between per-microbatch workload and pipeline bubbles; in practice, the search can be further narrowed by selecting the ``knee'' point on the workload-versus-microbatch curve and exploring only its neighborhood.

Overall, Dynamic-CP complements packed sequence training by (i) reducing synchronization stalls caused by DP imbalance and (ii) preventing unnecessary CP communication for microbatches that do not benefit from large CP sizes, thereby improving throughput in long-context variable-length training. According to the benchmark, Dynamic-CP yields a 35--60\% end-to-end performance improvement in real-world scenarios with highly imbalanced sequence length distributions, such as multi-modal training. \footnote{Explore \href{https://github.com/NVIDIA/Megatron-LM/pull/2000}{PR\#2000} and \href{https://github.com/NVIDIA/Megatron-LM/pull/2959}{PR\#2959} to start training models with variable-length sequences using Megatron-Core optimizations.}

\subsection{Summary}

Long-context MoE training represents a distinct optimization regime where the fundamental assumptions of Section~\ref{key-features-optimizations} must be revisited. While the Three Walls (memory, communication, and compute efficiency) remain the fundamental barriers, their relative importance shifts substantially as sequence length increases:

\begin{itemize}
    \item \textbf{Computational Shift}: Attention's $O(s^2)$ complexity displaces MoE layers as the dominant computational bottleneck, consuming $>$ 69\% of FLOPs at $>$ 64K tokens. The MoE-focused optimizations of Section~\ref{sec:compute-wall} remain valuable but no longer address the primary bottleneck.
    
    \item \textbf{Memory Wall Intensified}: Activation memory scales with sequence length, requiring the parallelism strategies from Section~\ref{parallelism-strategies} (specifically CP and TP) combined with the memory techniques from Section~\ref{sec:memory-wall} (recomputation, offloading) to maintain feasible memory footprints.
    
    \item \textbf{Communication Trade-offs}: The choice between CP and TP involves nuanced trade-offs between communication overhead, memory savings, and SDPA computational efficiency, a balance distinct from the MoE-centric all-to-all optimization focus of Section~\ref{sec:comm-wall}.
\end{itemize}

By combining context parallelism, tensor parallelism, selective recomputation, and optimizer offloading, Megatron-Core enables efficient training across diverse sequence lengths. Our experiments on 256 Hopper GPUs at 256K sequence length demonstrate this: DeepSeek-V3 achieves 88\% of its short-context MFU using TP, optimizer CPU offloading, and selective recomputation, while Qwen3-235B-A22B reaches 129\% of its short-context MFU with TP, CP, and selective recomputation (the latter exceeding 100\% because SDPA kernels are highly efficient when dominating the computation at longer sequences). The packed sequence and Dynamic-CP features further extend these capabilities to variable-length workloads in RL and SFT training.

%==============================================================================
% SECTION 7: PRODUCTION Features
%==============================================================================

\section{Production Features}\label{sec:production-features}

The previous sections focused on performance optimizations that address the Three Walls. However, production MoE training also requires robustness, flexibility, and ease of use. This section describes features that address these operational requirements: load balancing to ensure stable training, shared and latent expert architectures, distributed checkpointing for flexible deployment, upcycling to use existing dense models, and integration with advanced training paradigms like multi-token prediction.

\subsection{Load Balancing and Token Dropping}

Dynamic routing in MoE models can lead to significant workload imbalance, where certain experts receive disproportionately more tokens than others. This causes computational inefficiency, memory bottlenecks, and degraded hardware utilization. We address these challenges with two coordinated mechanisms: load balancing and token dropping.

\begin{figure}[ht]
    \centering
    \includegraphics[width=\textwidth]{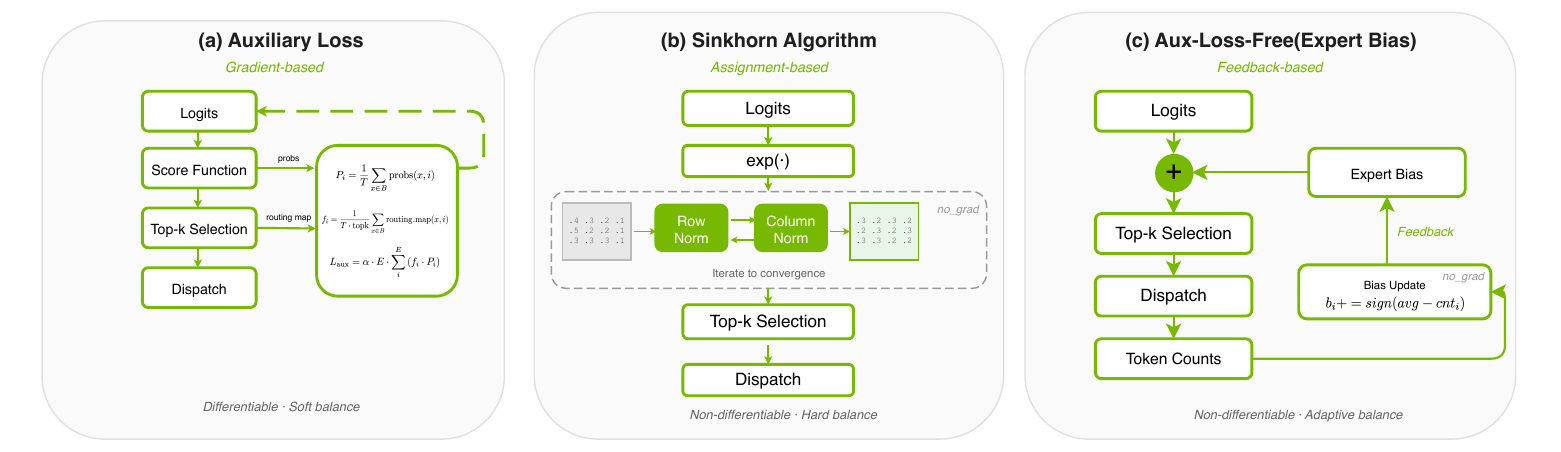}
    \caption{Load balancing strategies in Megatron-Core MoE.}
    \label{fig:load-balancing-strategies}
\end{figure}

\textbf{Load Balancing.}
\Figref{fig:load-balancing-strategies} illustrates the load balancing strategies available in Megatron-Core MoE. 
We provide multiple strategies to encourage uniform token distribution. The primary approach employs an \textbf{auxiliary loss}~\cite{lepikhin2020gshard,fedus2022switchtransformersscalingtrillion}, a differentiable penalty term that discourages routing all tokens to a small subset of experts. We also support \textbf{expert choice routing}~\cite{zhou2022mixtureofexpertsexpertchoicerouting}, which formulates balanced routing as an optimal transport problem, and \textbf{auxiliary-loss-free balancing} via learnable expert bias terms~\cite{wang2024auxiliarylossfreeloadbalancingstrategy} that dynamically adjust routing decisions based on historical load.

\textbf{Token Dropping.}
We support two dispatch strategies with different capacity management policies. In \textit{dropless mode} (the default), all routed tokens are processed without capacity constraints, maximizing model expressiveness at the cost of variable per-expert workload. In \textit{droppable mode}, explicit expert capacity limits are enforced~\cite{lepikhin2020gshard,fedus2022switchtransformersscalingtrillion}: when tokens assigned to an expert exceed its capacity, excess tokens are dropped and bypassed through residual connections. This provides predictable memory bounds, useful during early training when the router is poorly initialized.

\textbf{Pad-to-Max for Static Shapes.}
Droppable mode also enables \textit{pad-to-max} functionality, which pads all expert inputs to the same capacity limit. This converts MoE's inherently dynamic per-expert token counts into static shapes, enabling optimizations like CUDA Graphs (Section~\ref{sec:cuda-graphs}) that require fixed tensor dimensions across iterations.

\subsection{Shared Experts}

\begin{figure}[ht]
    \centering
    \includegraphics[width=0.7\textwidth]{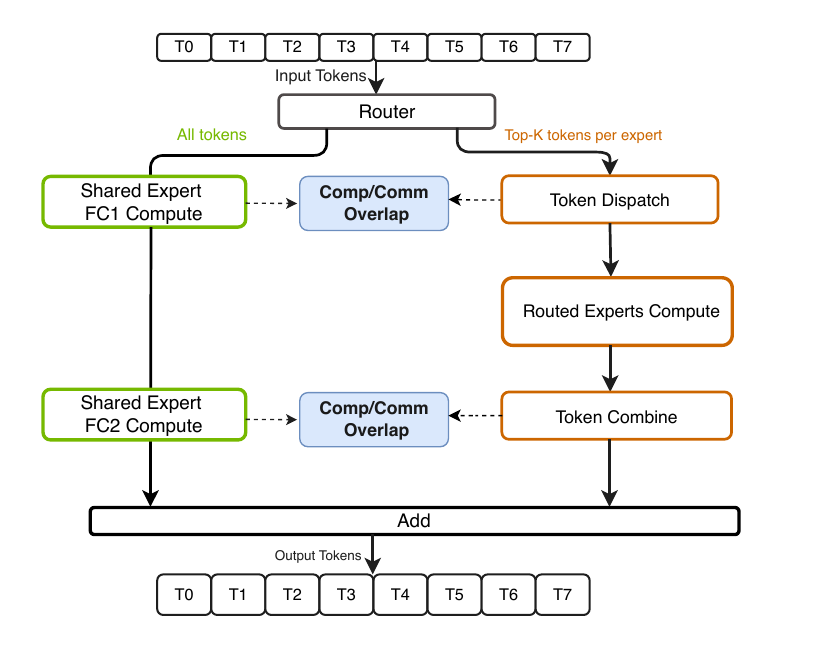}
    \caption{Shared expert architecture in Megatron-Core MoE. The shared expert processes \emph{all} tokens while routed experts process only their assigned tokens. When overlap is enabled, shared expert computation runs in parallel with the token dispatch/combine communication, hiding its latency.}
    \label{fig:shared-expert-architecture}
\end{figure}

Some MoE architectures (DeepSeek-V2/V3 \cite{deepseekai2024deepseekv2strongeconomicalefficient,deepseekai2025deepseekv3technicalreport}, Qwen \cite{qwen2025qwen25technicalreport}) include a shared expert that processes all tokens regardless of routing, providing consistent baseline capacity across all tokens (\figref{fig:shared-expert-architecture}). When overlap is enabled (\texttt{-{}-moe-shared-expert-overlap}), shared expert computation runs in parallel with the all-to-all communication and routed expert computation, hiding its latency behind the dispatch-compute-combine pipeline.

\subsection{Latent MoE}\label{sec:latent-moe}

In standard MoE, all2all communication dispatches tokens at the full hidden dimension \(d\), and each expert is parameterized by weight matrices in \(\mathbb{R}^{m \times d}\) and \(\mathbb{R}^{d \times m}\). LatentMoE~\cite{elango2026latentmoe} reduces both costs by inserting a shared down-projection \(W_\downarrow \in \mathbb{R}^{\ell \times d}\) before dispatch and an up-projection \(W_\uparrow \in \mathbb{R}^{d \times \ell}\) after combine, where \(\ell < d\) is the latent dimension. Routing still operates on the full hidden dimension to preserve routing quality; each routed expert operates entirely in the compressed latent space; shared experts remain at dimension \(d\). The layer output becomes:
\[
\text{output}(\mathbf{x}) = W_\uparrow \cdot \Bigl(\sum_{i \in \mathcal{T}_{K,E}} p_i \, E_i(W_\downarrow \cdot \mathbf{x};\, \ell)\Bigr) + \sum_{j} E_j^{\text{shared}}(\mathbf{x};\, d).
\]

The compression ratio \(\alpha = d / \ell\) reduces all2all communication volume by a factor of \(\alpha\) (tokens are dispatched at dimension \(\ell\) instead of \(d\)) and per-expert weight size by the same factor (expert matrices shrink from \(\mathbb{R}^{m \times d}\) to \(\mathbb{R}^{m \times \ell}\)). Elango et al.~\cite{elango2026latentmoe} define two ways to exploit these savings. In \(\ell\text{-MoE}_{\text{eff}}\), the total expert count \(E\) is scaled by \(\alpha\) while top-\(K\) is unchanged, preserving baseline accuracy at reduced inference cost. In \(\ell\text{-MoE}_{\text{acc}}\) (recommended), both \(E\) and top-\(K\) are scaled by \(\alpha\), which restores inference cost to the standard-MoE level but exponentially expands the combinatorial space of expert selections (\(\binom{\alpha E}{\alpha K} \geq \binom{E}{K}^{\alpha}\)), yielding higher accuracy at iso-cost. At scales up to 95B parameters, \(\ell\text{-MoE}_{\text{acc}}\) consistently outperforms standard MoE in accuracy per FLOP and per parameter. The architecture has been adopted by NVIDIA's Nemotron-3 Super and Ultra models.

\textbf{Implementation.}
LatentMoE is enabled via \texttt{-{}-moe-latent-size}, which sets \(\ell\). The \texttt{MoELayer} instantiates two \texttt{TELinear} projections: \texttt{fc1\_latent\_proj} (down-projection in the \texttt{preprocess} step, after routing but before dispatch) and \texttt{fc2\_latent\_proj} (up-projection in the \texttt{postprocess} step, after combine). Expert backends (\texttt{TEGroupedMLP}, \texttt{SequentialMLP}) automatically adapt their input and output dimensions to \(\ell\).

\subsection{Distributed Checkpoint}

Traditional checkpointing tightly couples saved model state with the specific parallelism configuration, requiring complex offline conversion when changing TP, EP, PP, or other parallelism settings. Our distributed checkpoint library solves this through \textbf{parallelism-agnostic checkpointing} with automatic resharding.

The core abstraction is the \texttt{ShardedTensor} descriptor, which encodes each local tensor's global shape, offset, and sharding pattern. During saving, each rank independently writes its local shard (Fully Parallel Saving), eliminating coordinator bottlenecks. During loading, each rank determines which portions of global tensors it needs based on the \textit{new} sharding specification and reads only those slices.

This enables any-to-any parallelism reconfiguration: a checkpoint saved with $\text{TP}=2$, $\text{EP}=4$ can be loaded with $\text{TP}=4$, $\text{EP}=8$ without offline conversion. The library supports Zarr (default) and PyTorch Distributed~\cite{zhao2023pytorchfsdp} storage backends.

\subsection{Flexible Asymmetric Virtual Pipeline Parallelism}\label{sec:flexible-vpp}

Interleaved pipeline parallelism~\cite{narayanan2021efficient} divides each physical pipeline stage into multiple virtual stages with interleaved scheduling, mitigating pipeline bubbles. Traditional VPP requires uniform layer distribution (e.g., a 24-layer model with $\text{PP}=4$, $\text{VPP}=2$ distributes layers as $[6, 6, 6, 6]$). However, MoE models exhibit substantial workload heterogeneity: MoE layers, dense layers, embedding, loss, and specialized layers like MTP have significantly different computational costs.

We introduce \textbf{Flexible Asymmetric VPP}, which allows different numbers and types of layers per virtual stage. Consider DeepSeek-V3 with 61 decoder layers and 1 MTP layer at $\text{PP}=16$, $\text{VPP}=2$ (Table~\ref{tab:flexible-vpp-example}). The initial rank combines the lightweight embedding with 3 dense decoders (matching the cost of 2 MoE layers). Most ranks hold 2 MoE decoders per stage. The final ranks strategically place the heavy MTP layer and lightweight loss layer to balance workload.

This approach enables VPP for models with arbitrary layer counts, achieves true load balancing by accounting for per-layer computational costs, and provides flexible placement for specialized layers.

\begin{table}[ht]
\centering
\caption{Layer distribution for DeepSeek-V3 with flexible asymmetric VPP ($\text{PP}=16$, $\text{VPP}=2$).}
\label{tab:flexible-vpp-example}
\begin{tabular}{lll}
\toprule
\textbf{PP rank} & \textbf{VPP rank 0} & \textbf{VPP rank 1} \\
\midrule
0       & embedding + 3$\times$ decoder & 2$\times$ decoder \\
1--13   & 2$\times$ decoder             & 2$\times$ decoder \\
14      & 2$\times$ decoder             & MTP \\
15      & 2$\times$ decoder             & loss \\
\bottomrule
\end{tabular}
\end{table}

\begin{figure}[ht]
    \centering
    \includegraphics[width=0.95\textwidth]{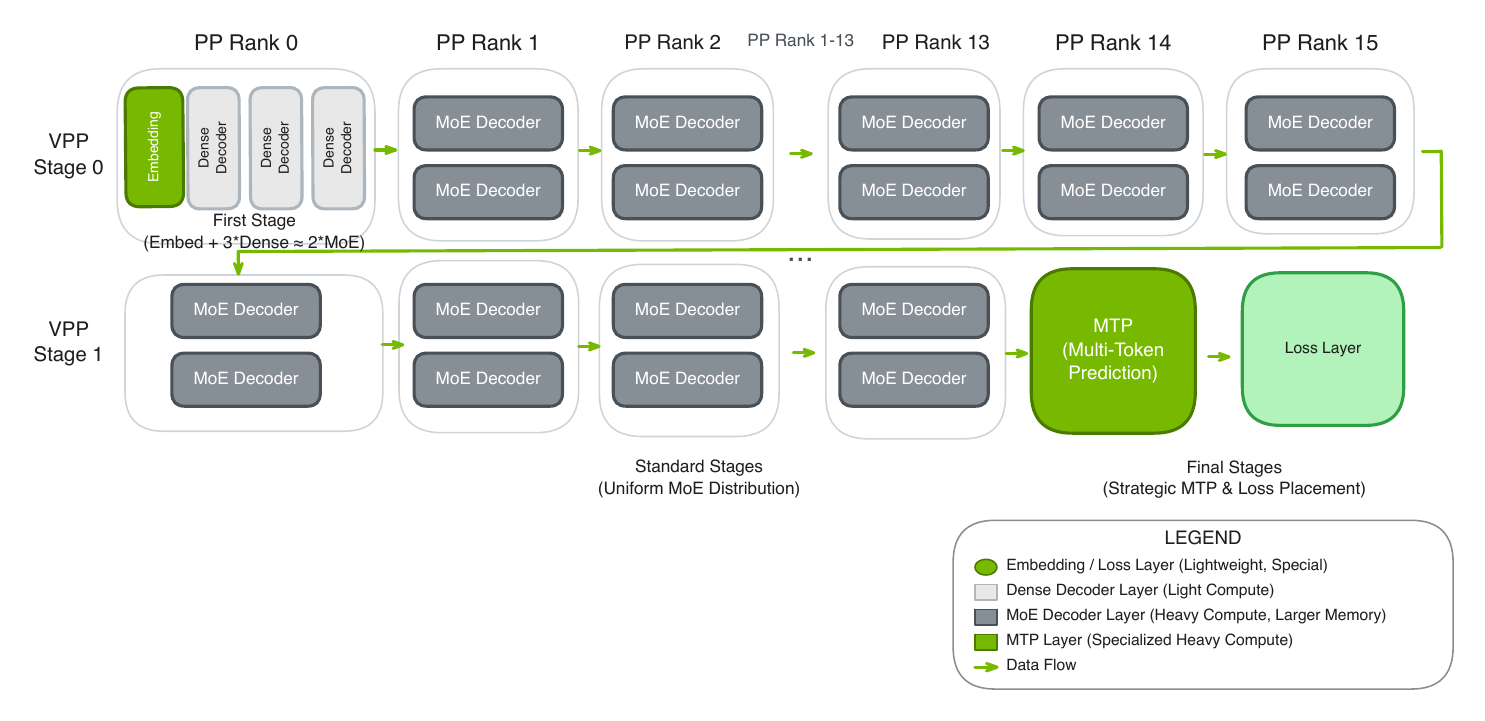}
    \caption{\textbf{Flexible Pipeline Parallel Placement.}}
    \label{fig:flexible-vpp-layout}
\end{figure}

This flexible asymmetric approach (\figref{fig:flexible-vpp-layout}) delivers several key advantages: (1) it enables VPP for models with arbitrary layer counts and compositions, removing artificial constraints on model architecture design; (2) it achieves true computational load balancing by allowing fine-grained control over layer distribution, accounting for the vast differences in computational cost between layer types; and (3) it provides flexible placement strategies for specialized layers (MTP, encoder-decoder structures, etc.), enabling diverse model architectures to fully exploit the latency-hiding benefits of pipeline parallelism. Practitioners can design layouts that distribute computationally expensive layers across ranks to avoid memory pressure and minimize pipeline bubbles.

\subsection{Upcycling}
\begin{figure}[!ht]
    \centering
    \includegraphics[width=1\textwidth]{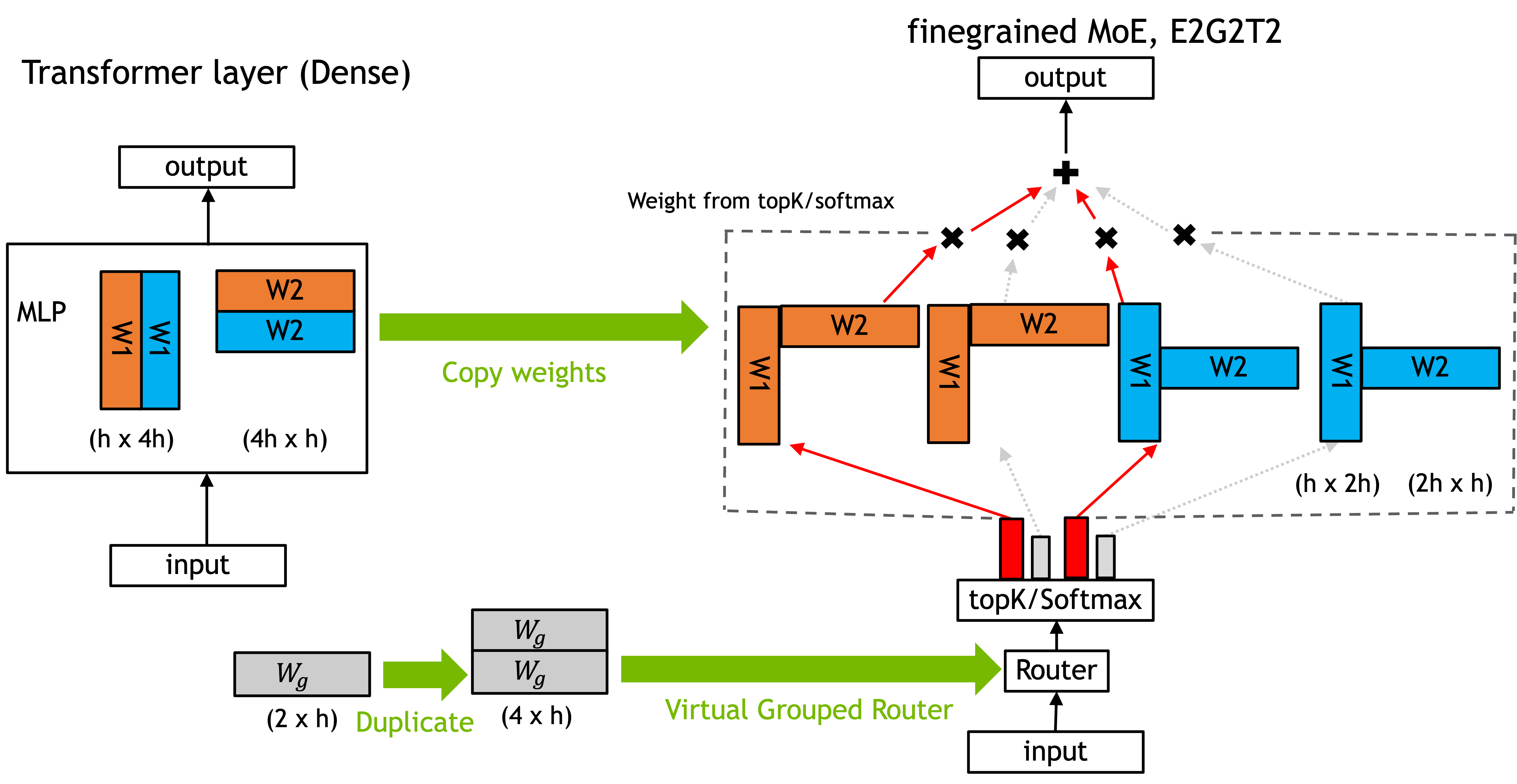}
    \caption{An example of granular upcycling a dense layer into E2G2T2 fine-grained MoE. E2G2T2 denotes 4 experts, top 2, with half intermediate size. (1) We shard MLP weights in the intermediate dimension ($4h \rightarrow 2h$) then duplicate the shards. (2) We initialize half the router weights then duplicate them. This ensures Top2 always selects one of each MLP shard so MoE output is the same as the dense model at the start of training.}
    \label{fig:granular_upcycling_diagram}
\end{figure}

Upcycling converts a pre-trained dense model into a sparse MoE architecture, expanding model capacity without retraining from scratch~\cite{komatsuzaki2022sparse,he2024upcyclinglargelanguagemodels,vavre2024llama}. We support virtual group initialization and expert weight scaling to enable seamless adaptation of dense checkpoints into fine-grained MoE models. The approach employs softmax-then-topK routing, which routes tokens through a selective subset of experts for increased expressivity at constant or reduced inference cost, as illustrated in \figref{fig:granular_upcycling_diagram}.

\subsection{Multi-Token Prediction}
Multi-token prediction (MTP) optimizes a model to predict multiple consecutive future tokens at each position, densifying the supervision signal~\cite{gloeckle2024better,deepseekai2025deepseekv3technicalreport}. Unlike parallel independent predictions, MTP maintains causal dependencies between predictions through hidden state transitions, accelerating convergence and improving generation quality. During inference, the model reverts to standard single-token prediction for deployment compatibility.

We integrate MTP with flexible pipeline parallelism, allowing MTP layers to be placed strategically within the VPP layout for balanced workload distribution.

\subsection{Muon Optimizer}
Unlike traditional optimizers such as AdamW that perform element-wise updates, the Muon optimizer introduces a matrix-aware approach by orthogonalizing entire weight matrices~\cite{liu2025muonscalablellmtraining}. This improves the conditioning of optimization trajectories and significantly reduces training steps compared to AdamW.

Our integration provides production-ready support for large-scale distributed training with several key advantages: (1) full support for split query-key-value (QKV) weight layouts, enabling efficient orthogonalization even when attention projection matrices are stored as separate tensors; (2) seamless integration with the distributed optimizer, allowing optimizer states to be sharded across data-parallel ranks while maintaining correct orthogonalization semantics; and (3) CPU offloading for Muon's orthogonalization buffers when GPU memory is constrained.

\textbf{MuonClip.}
Training trillion-parameter models introduces stability challenges where query-key dot products can grow unbounded, causing attention explosions \cite{kimik2}. MuonClip addresses this with hardware-accelerated implementations in cuDNN, cudnn-frontend, and Transformer Engine.

\section{Performance Evaluation}\label{performance-evaluations}
The preceding sections described Megatron-Core MoE's architecture, parallelism strategies, and optimizations. This section validates these contributions through empirical evaluation, showing the performance of the Megatron-Core MoE stack across diverse model configurations and hardware platforms.

\begin{notebox}
\textbf{Living Document.} The performance results in this section represent a point-in-time snapshot based on Megatron-Core \textbf{v0.16}, not the upper bound of achievable performance. Active optimization efforts are ongoing, and these numbers will improve in future releases. We also plan to expand coverage to additional model architectures as they emerge in the community. For the latest performance data, please refer to the Megatron-LM and Megatron-Bridge repository.
\end{notebox}

\subsection{Experimental Setup}

\textbf{Benchmark Models.}
We evaluate Megatron-Core MoE on two state-of-the-art fine-grained MoE architectures that stress-test the framework's capabilities: DeepSeek-V3-685B~\cite{deepseekai2025deepseekv3technicalreport} and Qwen3-235B~\cite{qwen2025qwen25technicalreport,yang2025qwen3technicalreport}. Both models feature the fine-grained expert design that amplifies the Three Walls challenges: high expert counts strain memory capacity, top-k routing increases all-to-all communication volume, and smaller per-expert computations reduce GEMM efficiency. These characteristics make them demanding benchmarks for validating our optimization strategies.

\textbf{Hardware and Software Stack.}
Benchmarks are conducted on NVIDIA GB200 and H100 GPUs to assess performance across hardware generations. We use the \texttt{dev} branch of Megatron-LM\footnote{\url{https://github.com/NVIDIA/Megatron-LM/tree/dev}} with the latest TransformerEngine\footnote{\url{https://github.com/NVIDIA/TransformerEngine/tree/main}}. The full optimization stack described in Section~\ref{key-features-optimizations} is enabled, including but not limited to:
\begin{itemize}[noitemsep,topsep=0pt]
    \item \textbf{FP8 training:} MXFP8 on GB200/GB300 (Blackwell) with native Tensor Core support and blockwise FP8 on H100 (Hopper)
    \item \textbf{Memory optimizations:} Selective recomputation and activation \& optimizer state offloading to reduce peak memory footprint
    \item \textbf{Optimized token dispatchers:} HybridEP\footnote{\url{https://github.com/deepseek-ai/DeepEP/tree/hybrid-ep}} on GB200/GB300 NVL systems exploiting Multi-Node NVLink, and DeepEP\footnote{\url{https://github.com/deepseek-ai/DeepEP/tree/main}} on H100 systems
    \item \textbf{Communication overlap:} 1F1B all-to-all overlap with expert computation
    \item \textbf{Kernel optimizations:} Grouped GEMM, router fusion, permutation fusion, and CUDA Graphs
\end{itemize}

\textbf{Metrics.}
We report per-GPU throughput in TFLOPS and tokens processed per second per GPU. These metrics provide two views of computational efficiency and practical training throughput.

\subsection{Key Performance Results}

\textbf{Overview.}
Table~\ref{tab:moe_throughput_unified} summarizes per-GPU throughput for representative fine-grained MoE training workloads on NVIDIA GB300, GB200, and H100 systems under the fully enabled optimization stack described in Section~\ref{key-features-optimizations}. All configurations use force-balanced routing to ensure even token distribution across experts.

% \textbf{Mixtral-8$\times$22B (long-context stress test).}
% We additionally include Mixtral-8$\times$22B with a sequence length of 65{,}536 to highlight regimes where long-context training amplifies memory pressure and communication overhead (Table~\ref{tab:moe_throughput_unified}). At this long sequence length, FP8 training improves throughput by roughly 14\% relative to BF16 on both GB200 and H100, and the newer GB200 platform delivers substantially higher per-GPU throughput than H100.

\textbf{DeepSeek-V3 and Qwen3-235B (large-scale training).}
Table~\ref{tab:moe_throughput_unified} reports results for two contemporary, fine-grained MoE training workloads at a standard sequence length of 4{,}096, including a 1{,}024-GPU DeepSeek-V3 run on H100. Across these settings, the measurements demonstrate strong per-GPU efficiency and high sustained throughput across all three platforms.

\def\unifiedthroughputtablescale{1.1}
\begin{table}[t]
    \centering
    \caption{Unified throughput benchmarks (per-GPU figures) for two mixture-of-experts models on NVIDIA GB300, GB200, and H100. All configurations use force-balanced routing. The Dtype column specifies the FP8 recipe: FP8-BLK denotes blockwise FP8 on Hopper, and MXFP8 denotes microscaling FP8 on Blackwell.}
    \scalebox{\unifiedthroughputtablescale}{%
    \begin{tabular}{lcccccc}
    \toprule
    Model & System & \#GPUs & Seqlen & Dtype & Per-GPU TF & Tokens/s/GPU \\
    \midrule
    DeepSeek-V3   & GB300 & 256  &   4,096 & MXFP8    & 1233 & 4,730 \\
    DeepSeek-V3   & GB200 & 256  &   4,096 & MXFP8    & 1048 & 4,020 \\
    DeepSeek-V3   & GB200 & 256  &   4,096 & BF16     & 857  & 3,298 \\
    DeepSeek-V3   & H100  & 1024 &   4,096 & FP8-BLK  & 368  & 1,412 \\
    \midrule
    Qwen3-235B    & GB300 & 256  &   4,096 & MXFP8    & 974  & 6,583 \\
    Qwen3-235B    & GB200 & 256  &   4,096 & MXFP8    & 919  & 6,212 \\
    Qwen3-235B    & GB200 & 256  &   4,096 & BF16     & 750  & 5,100 \\
    Qwen3-235B    & H100  & 256  &   4,096 & BF16     & 320  & 2,132 \\
    \midrule
    Qwen3-235B    & GB300 & 128  & 131,072 & MXFP8    & 1,150  & 1,556 \\
    \bottomrule
    \end{tabular}%
    }
    \label{tab:moe_throughput_unified}
\end{table}

\textbf{FP8 Training Effectiveness.}
As shown in Table~\ref{tab:moe_throughput_unified}, FP8 training achieves 368 TFLOPS per GPU on H100 for DeepSeek-V3, demonstrating that the blockwise FP8 recipe enables efficient training of large-scale MoE models on Hopper architecture. The optimizations discussed in Section~\ref{sec:fp8-training} successfully address the challenges of applying FP8 to MoE models with dynamic expert routing and varying computation patterns, maintaining numerical stability while providing the performance benefits of reduced precision.

\textbf{GB200 and GB300 Performance.}
The GB200 and GB300 platforms deliver approximately 3$\times$ higher token throughput compared to H100 for both MoE models at comparable or smaller GPU counts. This efficiency gain stems from superior memory bandwidth, higher computational capacity, and native MXFP8 Tensor Core support. Combined with our software optimizations (Sections~\ref{sec:memory-wall} and~\ref{sec:compute-wall}), GB200 achieves over 1,048 TFLOPS per GPU for DeepSeek-V3 training, and GB300 further pushes this to 1,233 TFLOPS per GPU.

\textbf{Scalability.}
These results demonstrate that Megatron-Core MoE scales efficiently to production workloads with thousands of GPUs and hundreds of billions of parameters. The successful deployment of DeepSeek-V3 at 1,024-GPU scale validates the effectiveness of parallel folding (Section~\ref{sec:parallel-folding}), optimized dispatchers (Section~\ref{sec:comm-wall}), and kernel optimizations (Section~\ref{sec:compute-wall}).

\textbf{Long-context Stress Test).}
To capture behavior beyond the standard 4{,}096-token regime, Table~\ref{tab:moe_throughput_unified} includes a long-context Qwen3-235B run at sequence length 131{,}072 on GB300. Even in this memory- and communication-intensive setting, the system sustains 1{,}150 TFLOPS per GPU, indicating strong long-context efficiency under the full optimization stack.

\noindent
\textbf{Configuration Details.} The parallelism configurations (TP, PP, CP, EP, DP, VPP, etc.), batch settings, and other optimization details used for each benchmark entry in Table~\ref{tab:moe_throughput_unified} are provided in Appendix~\ref{sec:showcase_settings}. Note that these configurations represent our best-found settings through empirical tuning; they may not be globally optimal, as exhaustively searching the full parallelism and optimization configuration space is infeasible for models of this scale.

Section~\ref{performance-evaluations} demonstrated that Megatron-Core MoE achieves state-of-the-art performance across diverse MoE architectures on the latest GPU platforms. We now examine the systematic methodology behind these results and show how the optimizations from Sections~\ref{sec:memory-wall}--\ref{sec:compute-wall} translate into deployment decisions through detailed case studies.

\section{Performance Best Practices}\label{sec:best-practices}

Sections~\ref{parallelism-strategies}--\ref{sec:production-features} presented a comprehensive optimization arsenal for MoE training: parallel folding, optimized dispatchers, Grouped GEMM, reduced-precision training, CUDA Graphs, and more. However, \textbf{having many optimization techniques does not automatically translate into high performance}. Each technique addresses specific bottlenecks but introduces its own trade-offs. Reduced-precision training reduces memory but requires careful precision management. Communication overlap improves throughput but adds memory overhead. CUDA Graphs eliminate host latency but conflict with dynamic tensor shapes. The challenge is not the availability of optimizations; it is knowing \emph{which} optimizations to apply, \emph{when} to apply them, and \emph{how} they interact with each other.

This complexity means that achieving peak performance requires a systematic methodology, not trial-and-error tuning. Through optimization of diverse MoE models (from Mixtral to DeepSeek-V3) on GB200 and H100, we developed a repeatable workflow for identifying bottlenecks and applying targeted solutions. This section distills those lessons into actionable guidance. Section~\ref{sec:best-practices-methodology} presents the methodology under the Three Walls framework with concrete examples for each phase. Section~\ref{sec:deepseek-case-study} then shows how these principles combine for DeepSeek-V3 on GB200 and H100, including why the same model requires different strategies on different hardware.

\subsection{A Systematic Optimization Workflow}\label{sec:best-practices-methodology}

We present a three-phase workflow for MoE performance optimization. This methodology emerged from tuning Mixtral, DeepSeek-V3, and Qwen3 across GB200 and H100 platforms. The process is inherently \textbf{iterative}: solving one bottleneck often exposes the next, requiring continuous profiling and refinement.

\subsubsection{Phase 1: Establish Memory-Feasible Parallelism}

Memory feasibility is the first constraint. Before optimizing for throughput, the configuration must fit in GPU memory. Table~\ref{tab:parallel-memory-impact} summarizes how each parallelism strategy affects per-GPU memory and communication overhead.

\begin{table}[ht]
\centering
\caption{Impact of parallelism strategies on memory and communication. $d$ = parallelism degree. $^\dagger$Requires distributed optimizer (\texttt{-{}-use-distributed-optimizer}).}
\label{tab:parallel-memory-impact}
\begin{tabular}{lcccc}
\toprule
\textbf{Strategy} & \textbf{Peak Activation} & \textbf{Weight Memory} & \textbf{Optimizer States} & \textbf{Comm (Per-Layer)} \\
\midrule
TP & $1/d$ (with SP) & $1/d$ & $1/d$ & High \\
EP & $\sim$1 (load-dependent) & $1/d$ (MoE only) & $1/d$ & Medium \\
PP & 1 ($>$1 with VPP) & $1/d$ & $1/d$ & Medium \\
CP & $1/d$ & 1 & $1/d^\dagger$ & Medium \\
DP & 1 & 1 & $1/d^\dagger$ & Low \\
\bottomrule
\end{tabular}
\end{table}

\begin{notebox}
\textbf{Quick feasibility testing.} Use \texttt{-{}-fake-init-process-group} to emulate distributed training on a single GPU, enabling rapid iteration on parallelism configurations without allocating a full cluster. See the Megatron-LM documentation\footnote{\url{https://github.com/NVIDIA/Megatron-LM/pull/2254}} for detailed usage.\\
\textbf{Interactive Memory Estimator}. We have developed an interactive memory simulator with a web GUI for quick parallelism tuning against memory footprints. See the blog\cite{nvidia_memory_estimator_blog} for details.

\end{notebox}

\noindent\textbf{Example.} For a 685B-parameter fine-grained MoE model, BF16 activations alone can exceed 130\,GB per GPU (see Table~\ref{tab:memory-breakdown}), immediately ruling out baseline configurations on 80\,GB devices even with parallelism, and motivating the memory optimizations in Phase~3.

\subsubsection{Phase 2: Select Optimal Parallelism Strategy}

Once memory-feasible configurations are identified, select the strategy that minimizes communication overhead while maintaining throughput. The optimal choice depends on model architecture, sequence length, and hardware topology.

\paragraph*{Guideline 1: Minimize Model Parallelism, Maximize Data Parallelism}
\begin{itemize}
    \item \textbf{Goal:} Keep TP/EP/PP/CP as small as possible while avoiding OOM.
    \item \textbf{Why:} Model parallelism introduces communication overhead that hurts performance.
    \item \textbf{How:} Use distributed optimizer (\texttt{-{}-use-distributed-optimizer}) to shard optimizer states across DP ranks, freeing memory for larger DP size.
\end{itemize}

\paragraph*{Guideline 2: Keep EP and TP Communication Within NVLink Domain}
\begin{itemize}
    \item \textbf{Goal:} Ensure EP$\times$TP fits within the NVLink Domain (typically 8 GPUs in a single node unless Multi-Node NVLink is used).
    \item \textbf{Why:} EP and TP are communication-intensive; NVLink provides much higher bandwidth than cross-node interconnects.
    \item \textbf{Scaling:} When scaling beyond the NVLink Domain, prefer PP over expanding TP/EP across nodes.
\end{itemize}

\noindent\textbf{Note:} For very large MoE models like DeepSeek-V3, EP communication volume may exceed what NVLink can hide. In this case, enable communication-computation overlap (see Section~\ref{sec:comm-wall}) to hide EP latency behind computation.

\paragraph*{Guideline 3: Use Pipeline Parallelism (PP) for Multi-Node Scaling}
\begin{itemize}
    \item \textbf{Goal:} Use PP to distribute layers across nodes while keeping EP$\times$TP within NVLink.
    \item \textbf{VPP:} Enable Virtual Pipeline Parallelism to reduce pipeline bubbles when PP $\geq 2$.
    \item \textbf{Config:} Set \texttt{-{}-pipeline-model-parallel-layout} to control Pipeline Parallelism settings (see Section~\ref{sec:flexible-vpp}). Larger VPP usually reduces pipeline bubbles but increases P2P communications; a middle value often provides the best balance. The workloads of each VPP rank should be balanced to maximize the throughput.
\end{itemize}

\paragraph*{Guideline 4: Prefer EP over TP for Expert Layers}
EP offers several advantages over TP for MoE layers:
\begin{itemize}
    \item \textbf{Better GEMM efficiency:} Larger local matrix sizes improve GPU utilization.
    \item \textbf{Lower communication:} EP has less communication overhead than TP for MoE layers.
    \item \textbf{Simpler computation graph:} Easier to overlap communication with computation.
    \item \textbf{Token permutation:} When EP $=$ \texttt{num\_experts}, local token permutation is eliminated.
\end{itemize}

\noindent\textbf{Example:} For Mixtral-8$\times$7B, EP8$\times$TP1 outperforms EP4$\times$TP2.

\paragraph*{Guideline 5: Enable Context Parallelism (CP) for Long Sequences}
\begin{itemize}
    \item \textbf{When to use:} Sequence length $\geq$ 8K tokens.
    \item \textbf{When to avoid:} For sequences $<$ 4K tokens, CP overhead typically exceeds benefits.
    \item \textbf{Key factor:} CP efficiency depends on overlapping communication with computation.
    \item \textbf{Config:} Set \texttt{-{}-context-parallel-size} to partition sequences across GPUs.
\end{itemize}

\noindent\textbf{Example.} Consider a 256-expert MoE model on an NVL72 system. Applying Guideline~4, Parallel Folding sets expert TP\,=\,1 so that each expert runs on a single GPU, maximizing GEMM efficiency. With Guideline~2, EP64 fits entirely within the NVLink domain. Guideline~1 then drives the remaining choices: the available memory budget determines TP and PP for attention layers, with DP filling the rest.

\subsubsection{Phase 3: Profile and Optimize Bottlenecks}

With a working parallelism configuration established, profile the training run to identify which wall dominates. Apply targeted optimizations based on the bottleneck type.

\paragraph*{Memory Bottleneck (Memory Wall)}
\textbf{Symptom:} Forced to use full recomputation or excessively large parallelism degrees to avoid OOM.

\noindent\textbf{Solutions:} Apply memory-saving techniques (Table~\ref{tab:memory-bottleneck}) to reduce memory consumption.

\begin{table}[ht]
\centering
\caption{Memory bottleneck solutions.}
\label{tab:memory-bottleneck}
\resizebox{\textwidth}{!}{%
\begin{tabular}{llll}
\toprule
\textbf{Optimization} & \textbf{Overhead} & \textbf{Config} & \textbf{Reference} \\
\midrule
FP8 Training & Low & \texttt{-{}-fp8-format -{}-fp8-recipe} & \S\ref{sec:fp8-activation} \\
Selective Recomputation & Low & \texttt{-{}-recompute-granularity -{}-recompute-modules} & \S\ref{sec:recomputation} \\
Precision-Aware Optimizer & Low & \texttt{-{}-use-precision-aware-optimizer} & \S\ref{sec:precision-opt} \\
Activation Offloading & Medium & \texttt{-{}-fine-grained-activation-offloading -{}-offload-modules} & \S\ref{sec:offloading} \\
Optimizer Offloading & Medium & \texttt{-{}-offload-optimizer-states} & \S\ref{sec:precision-opt} \\
\bottomrule
\end{tabular}%
}
\end{table}

\paragraph*{Communication Bottleneck (Communication Wall)}
\textbf{Symptom:} Profiling shows significant time spent in collective operations.

\noindent\textbf{Solutions:} Identify which communication is the bottleneck and apply the corresponding optimization. (Table~\ref{tab:comm-bottleneck})

\begin{table}[ht]
\centering
\caption{Communication bottleneck solutions.}
\label{tab:comm-bottleneck}
\resizebox{\textwidth}{!}{%
\begin{tabular}{lll}
\toprule
\textbf{Communication Type} & \textbf{Config} & \textbf{Reference} \\
\midrule
DP gradient reduce and param gather & \texttt{-{}-overlap-grad-reduce -{}-overlap-param-gather} & --- \\
TP communication & \texttt{-{}-tp-comm-overlap} & --- \\
EP dispatcher & \texttt{-{}-moe-token-dispatcher-type} & \S\ref{sec:deepep} \\
EP all-to-all hiding & \texttt{-{}-overlap-moe-expert-parallel-comm} & \S\ref{sec:comm-overlap} \\
PP send/recv & \texttt{-{}-pipeline-model-parallel-layout} & \S\ref{sec:flexible-vpp} \\
\bottomrule
\end{tabular}%
}
\end{table}

\paragraph*{CPU Overhead Bottleneck (Compute Efficiency Wall)}
The Compute Efficiency Wall manifests in two distinct forms: CPU overhead and computation inefficiency. CPU overhead occurs when the host cannot launch GPU kernels fast enough, creating gaps in GPU execution.

\noindent\textbf{Symptom:} Nsight Systems timeline shows gaps between GPU kernels where CPU cannot launch kernels fast enough.

\noindent\textbf{Solutions:} Reduce host-side overhead by minimizing kernel launches and enabling CUDA Graphs (Table~\ref{tab:cpu-bottleneck}).

\begin{table}[ht]
\centering
\caption{CPU overhead bottleneck solutions.}
\label{tab:cpu-bottleneck}
\begin{tabular}{lll}
\toprule
\textbf{Optimization} & \textbf{Config} & \textbf{Reference} \\
\midrule
Disable Python GC & \texttt{-{}-manual-gc -{}-manual-gc-interval 10} & --- \\
Reduce kernel launches & Decrease TP or increase MBS & --- \\
Enable CUDA Graphs & \texttt{-{}-cuda-graph-impl transformer\_engine} & \S\ref{sec:cuda-graphs} \\
\bottomrule
\end{tabular}
\end{table}

\paragraph*{Computation Bottleneck (Compute Efficiency Wall)}
Computation inefficiency occurs when GPU kernels themselves underutilize hardware resources, typically due to small GEMM sizes in fine-grained MoE architectures.

\noindent\textbf{Symptom:} GPU SM utilization is low despite no communication or CPU bottlenecks.

\noindent\textbf{Solutions:} Improve kernel efficiency through batching, fusion, and lower precision (Table~\ref{tab:compute-bottleneck}).

\begin{table}[ht]
\centering
\caption{Computation bottleneck solutions.}
\label{tab:compute-bottleneck}
\begin{tabular}{llp{4cm}}
\toprule
\textbf{Optimization} & \textbf{Config} & \textbf{Reference} \\
\midrule
Grouped GEMM & \texttt{-{}-moe-grouped-gemm} & \S\ref{sec:kernel-fusion} \\
Kernel fusions & \texttt{-{}-moe-router-fusion -{}-moe-permute-fusion} & \S\ref{sec:kernel-fusion} \\
FP8 precision & \texttt{-{}-fp8-format -{}-fp8-recipe} & \S\ref{sec:reduced-precision-training} \\
\bottomrule
\end{tabular}
\end{table}

\noindent\textbf{Example.} On an NVL8 system where EP spans across nodes, profiling may reveal all-to-all communication consuming 30--50\% of step time, pointing squarely at the Communication Wall. On an NVL72 system where the same EP degree stays within the NVLink domain, the dominant bottleneck after enabling FP8 often shifts to CPU overhead instead; the faster GPU computation exposes host-side launch latency. The same model on different hardware can require entirely different optimization strategies.

\subsubsection{Summary}

The ordering of the three phases matters: memory constraints are a hard barrier that must be resolved first (Phase~1), parallelism decisions determine the communication topology (Phase~2), and only then can profiling reveal which wall to attack (Phase~3). Crucially, this process is \textbf{iterative}. Memory optimizations may enable smaller parallelism degrees, returning to Phase~1. Some Phase~3 optimizations have their own memory cost (EP communication overlap requires extra buffers, CUDA Graphs consume additional memory), which may require revisiting earlier decisions. Continuous profiling after each round guides the next iteration.

\subsection{Case Study: Tuning DeepSeek-V3 on GB200 and H100}\label{sec:deepseek-case-study}

DeepSeek-V3 represents an extreme stress test for MoE training: 685B total parameters with Multi-Token Prediction (MTP) and Multi-Latent Attention (MLA). All three walls apply: memory pressure from 256 experts, high all-to-all volume from top-8 routing, and small per-expert GEMMs. The preceding workflow (Section~\ref{sec:best-practices-methodology}) showed how to approach each phase individually; this case study examines how those decisions interact to form a coherent optimization stack, and why the same model requires fundamentally different strategies on GB200 and H100.

\subsubsection{Final Optimized Configuration and Performance}

Table~\ref{tab:deepseekv3-config} summarizes the final optimized configurations and performance on GB200 and H100 platforms.

\begin{table}[ht]
\centering
\caption{DeepSeek-V3 final optimized configurations on GB200 and H100. $^\dagger$Parallel Folding is used; TP applies to the non-MoE modules only, expert TP is always 1.}
\label{tab:deepseekv3-config}
\begin{tabular}{lcc}
\toprule
\textbf{Configuration} & \textbf{GB200} & \textbf{H100} \\
\midrule
Hardware & 256$\times$GB200 & 1024$\times$H100 \\
Parallelism (TP/PP/EP)$^\dagger$ & 1/4/64 & 2/8/64 \\
VPP & 4 & 4 \\
GBS / MBS / SeqLen & 8192 / 1 / 4096 & 8192 / 1 / 4096 \\
Precision & MXFP8 & FP8-Blockwise \\
Dispatcher & HybridEP & DeepEP \\
Recompute & \texttt{mlp} & \texttt{mlp}, \texttt{mla\_up\_proj}, \texttt{moe\_act}, \texttt{layernorm} \\
CUDA Graphs & Enabled & --- \\
EP all-to-all Overlap & --- & Enabled \\
\midrule
\textbf{Performance (TFLOPS/GPU)} & \textbf{1048} & \textbf{368} \\
\bottomrule
\end{tabular}
\end{table}

The GB200 configuration uses HybridEP (optimized for NVL72) and CUDA Graphs to minimize CPU overhead, which becomes the dominant bottleneck in FP8 training on Blackwell. The H100 configuration uses FP8-blockwise precision with DeepEP and EP all-to-all overlap to hide communication latency.

Table~\ref{tab:deepseekv3-journey} summarizes the key optimizations applied on each platform.

\begin{table}[ht]
\centering
\caption{DeepSeek-V3 optimization summary by platform.}
\label{tab:deepseekv3-journey}
\begin{tabular}{lp{5cm}p{5cm}}
\toprule
\textbf{Category} & \textbf{GB200} & \textbf{H100} \\
\midrule
Parallelism & Parallel Folding \newline Flexible VPP & Parallel Folding \newline Flexible VPP \\
\midrule
Precision & MXFP8 & FP8-Blockwise \\
\midrule
Memory & Memory-efficient permutation \newline Fine-grained recomputation \newline FP8 primary weights \newline Low-precision optimizer states \newline Optimizer states offloading & Memory-efficient permutation \newline Fine-grained recomputation \newline FP8 primary weights \newline Low-precision optimizer states \\
\midrule
Communication & HybridEP & DeepEP \newline EP Communication overlap \\
\midrule
Compute Efficiency & Kernel fusions \newline CUDA Graphs \newline CPU-side optimizations & Kernel fusions \\
\bottomrule
\end{tabular}
\end{table}

\subsubsection{Anatomy of the Optimized Configuration}

\paragraph*{Why This Parallelism Layout?}
Both platforms use Parallel Folding (Guideline~4) to decouple attention and expert parallelism: TP applies only to attention layers for memory reduction, while experts use EP with TP\,=\,1 for better GEMM efficiency. EP64 is chosen so that each GPU holds exactly four of the 256 experts, eliminating local token permutation overhead.

The key platform difference is memory capacity. GB200's 192\,GB per GPU (vs.\ H100's 80\,GB) allows TP1/PP4 instead of TP2/PP8, halving the pipeline depth, which reduces pipeline bubbles and simplifies workload balancing (Guideline~1). Flexible VPP then fine-tunes the pipeline: on H100, the layout \texttt{Et*3|(tt|)*29m|L} groups embedding with 3 transformer layers in the first stage, places MTP in a standalone stage, and separates the loss; on GB200, \texttt{Et*4|(tttt|)*14tmL} distributes layers across 16 virtual stages for balanced computation across the shorter pipeline. These parallelism configurations establish the foundation for all subsequent optimizations.

\paragraph*{Memory Wall}
FP8 training is the first lever: it halves activation memory on both platforms, freeing GPU memory that would otherwise require aggressive recomputation. On H100, the freed memory budget is critical because it enables EP communication overlap, which requires extra buffers for pipelining dispatch and combine operations (see Communication Wall below). Fine-grained recomputation (\texttt{mlp}, \texttt{mla\_up\_proj}, \texttt{layernorm}, \texttt{moe\_act}) trims remaining memory pressure. Memory-efficient permutation eliminates redundant activation storage, and low-precision optimizer states (BF16) provide additional headroom.

On GB200, the memory chain resolves differently. NVL72 keeps EP local, so communication overlap is not critical and its memory budget is freed entirely. GB200's higher C2C bandwidth (NVLink-C2C) makes optimizer state offloading effective: the transfer overhead is low enough that offloading frees substantial GPU memory for activations. This in turn reduces the need for fine-grained recomputation to just \texttt{mlp}, far less aggressive than on H100.

\paragraph*{Communication Wall}
On H100/B200 (NVL8), EP64 spans 8 nodes, and cross-node all-to-all latency would consume nearly 50\% of step time with the standard all-to-all dispatcher. DeepEP reduces this overhead through fused permutation and optimized collective operations. Even so, EP communication overlap is essential to hide the remaining cross-node latency behind expert computation, and this overlap is only feasible because FP8 freed enough memory for the extra buffers (see Memory Wall above).

On GB200 (NVL72), EP64 stays entirely within the NVLink domain. HybridEP fully utilizes the 1.8\,TB/s bidirectional bandwidth without requiring communication overlap. The communication wall is effectively resolved by hardware topology alone, shifting the dominant bottleneck to compute efficiency.

\paragraph*{Compute Efficiency Wall}
FP8 accelerates GEMMs on both platforms (blockwise FP8 on H100, MXFP8 with native Blackwell Tensor Core support on GB200), but this speedup has a side effect: faster GPU computation exposes CPU overhead. On GB200, where NVL72 already eliminates the communication bottleneck, CPU overhead becomes the dominant constraint; the host cannot launch kernels fast enough to keep the GPU saturated.

Partial CUDA Graphs address this by capturing attention, router, and MoE preprocessing into static graphs, while leaving dynamic expert computation ungraphed. Kernel fusions (router fusion, permute fusion, MLA RoPE fusion) reduce kernel launch count. CPU/NUMA binding\footnote{The \texttt{bindpcie} script at \url{https://github.com/NVIDIA/mlperf-common/blob/main/client/bindpcie} automatically detects GPU/NUMA topology based on local rank and uses \texttt{numactl} to bind each process's CPU and memory to the NUMA node nearest its GPU.} reduces host-side memory access latency.

\medskip

The cross-cutting nature of these optimizations is evident throughout: FP8 simultaneously reduces memory (activations halved), and improves compute throughput, but introduces quantization kernels that amplify CPU overhead. The same model arrives at fundamentally different optimization stacks on different hardware: H100's stack centers on hiding communication latency (DeepEP + EP overlap), while GB200's centers on eliminating CPU overhead (CUDA Graphs + kernel fusions). This contrast validates the profile-driven iterative approach from Section~\ref{sec:best-practices-methodology}.

\subsubsection{Lessons Learned}

Four insights generalize beyond DeepSeek-V3:

\begin{enumerate}
    \item \textbf{Platform characteristics drive strategy}: GB200's larger memory (192GB vs 80GB) and higher C2C bandwidth enable more aggressive choices: smaller PP degree, less fine-grained recomputation, and effective optimizer offloading. H100's NVL8 requires EP communication overlap to hide cross-node latency, while GB200's NVL72 keeps EP64 within the NVLink domain.
    \item \textbf{Parallel Folding unlocks flexibility}: Decoupling attention TP from expert EP allows independent optimization of each layer type. Combined with flexible VPP, this enables fine-grained workload balancing across pipeline stages.
    \item \textbf{FP8 shifts bottlenecks}: FP8 training accelerates GEMMs and reduces memory, but amplifies CPU overhead as the dominant bottleneck. CUDA Graphs, kernel fusions, and CPU/NUMA binding become essential.
    \item \textbf{Iterative optimization}: The optimization process is inherently cyclical. Memory optimizations (recomputation, offloading) free GPU memory, enabling communication overlap. Communication optimizations expose compute efficiency bottlenecks. Some optimizations have cross-cutting effects: FP8 training reduces both memory and compute time but increases CPU overhead; CUDA Graphs reduce CPU overhead but consume additional memory. These interactions mean that applying one optimization may require revisiting earlier decisions. Continuous profiling after each change guides the next optimization target and helps identify when diminishing returns suggest moving to the next bottleneck.
\end{enumerate}

\section{Megatron-Core MoE in Reinforcement Learning}\label{sec:rl}
Reinforcement learning (RL) post-training has become a critical paradigm following OpenAI o1 and DeepSeek-R1~\cite{deepseekai2025deepseekr1}. Many leading RL-trained models are MoE architectures (DeepSeek-R1, Kimi-K2, etc.), and RL workloads introduce challenges that amplify the MoE-specific issues discussed throughout this report: highly variable sequence lengths stress the Memory Wall, interleaved inference and training phases demand rapid memory offloading, and routing discrepancies between inference and training engines threaten stability. To achieve high training throughput at scale, RL frameworks run Megatron-Core inside Ray workers as the distributed training backend, and use Megatron-Bridge for bidirectional Hugging Face to Megatron checkpoint conversion.

\subsection{Challenges for RL Post-Training}
RL post-training differs from pre-training in ways that impose new requirements on the training engine:

\begin{enumerate}
    \item \textbf{Variable-length sequences.} Pre-training typically operates on fixed-length sequences (e.g., 4K tokens). RL workloads, by contrast, produce highly variable sequence lengths: the maximum may reach 128K or even 1M tokens, while the mean within a mini-batch is often one-half to one-quarter of the maximum. This long-tailed distribution makes it difficult to balance compute efficiency against peak memory consumption.
    \item \textbf{Memory offloading.} RL frameworks typically co-locate a training engine and an inference engine on the same GPUs, offloading one engine's state while the other is active. This requires both engines to release and restore their full memory footprint quickly and completely---a non-trivial requirement when optimizer states, activations, and KV caches must all be managed.
    \item \textbf{Online weight export.} The inference engine must be updated with the latest parameters after each training step. This requires the training engine to export weights rapidly in a format the inference engine can load, across diverse model architectures.
    \item \textbf{Training stability.} Standard RL algorithms assume that sampled responses come from the current policy distribution. In practice, however, the inference and training engines use different optimized kernels, producing slightly different token probabilities even with identical parameters---effectively introducing off-policy bias. For MoE models, this problem is compounded: tokens from the same sequence may be routed to different experts in the inference and training engines, amplifying the discrepancy.
\end{enumerate}
\subsection{Megatron-Bridge}
A typical RL workflow starts from a pretrained Hugging Face model and fine-tunes it in an RL framework. At scale, these frameworks often run Megatron-Core inside Ray workers for distributed training, while rollout commonly uses an inference stack that expects Hugging Face-format checkpoints. This creates two integration needs: mapping model definitions to Megatron-Core modules, and converting checkpoints between Hugging Face and Megatron formats efficiently. Megatron-Bridge addresses the checkpoint interoperability layer, enabling fast HF-to-Megatron conversion for initialization, training, and export. This pattern is used across multiple RL frameworks, including veRL~\cite{sheng2024hybridflow}, Slime~\cite{slime_github}, and NeMo~RL.

\subsection{Megatron-Core Optimization for Reinforcement Learning}
\textbf{Packed Sequence Support.} As discussed in Section~\ref{subsec:packed_sequence_support}, we can pack sequences to remove the padding from a batch of variable sequence lengths. Building on packed sequence support, on the RL framework side, we recommend using a packing-aware dynamic batch size, which ensures that every batch has a similar total number of effective tokens. Moreover, we introduce an additional load–balancing strategy that explicitly accounts for the heterogeneous computational cost of the transformer blocks. Because the attention module scales quadratically with sequence length (\(O(L^{2})\)) whereas the feed-forward network (FFN) scales linearly (\(O(L)\)), a batch that is well balanced in terms of token count can still be highly unbalanced in wall-clock time. To mitigate this issue, we first compute, for every sample in the batch, the square of its sequence length and take the sum within the mini-batch. We then sort micro-batches according to this ``attention cost’’ metric and schedule them in a small-to-large-to-small pattern (i.e., increasing order followed by decreasing order).
This serpentine ordering delivers two benefits. (i) For data parallelism, pipeline parallelism, and expert parallelism, it reduces synchronization bubbles because consecutive micro-batches have comparable attention workloads. (ii) Within pipeline parallelism, the warm-up and cool-down phases, where stages are traditionally under-utilized, observe shorter idle periods, as lighter micro-batches arrive earlier and later in the schedule.

\textbf{Dynamic Context Parallelism.} As discussed in Section~\ref{subsec:dynamic-cp}, conventional training setups must pick a fixed context-parallel (CP) degree that guarantees the longest sequence in the workload will not trigger out-of-memory (OOM) failures. This conservative choice forces the majority of shorter sequences, whose memory footprints are far smaller, to run with an unnecessarily large CP degree, thereby wasting cross-device bandwidth and reducing overall throughput.
Dynamic CP removes this one-size-fits-all restriction by selecting the CP degree adaptively on a per-micro-batch basis. At runtime we bucket incoming sequences by length, choose the minimal CP degree that keeps each bucket within the device memory budget, and dispatch micro-batches accordingly. Long sequences receive a higher CP degree to stay memory-safe, whereas short sequences are executed with a lower CP degree that maximizes arithmetic intensity and reduces communication volume.

\textbf{CPU Optimizer Offloading.} To further relieve GPU memory pressure, we offload the optimizer states to host DRAM during the forward and backward passes, loading them back onto the GPU only for the parameter-update step. Because optimizer tensors can occupy multiple times the size of model parameters, temporarily evicting them frees a substantial amount of high-bandwidth GPU memory that can instead be used to cache activations or accommodate longer sequences.

\textbf{FP16 Training Support.} Although bfloat16 (BF16) is widely adopted for pre-training large language models, several studies have observed that, during reinforcement-learning training, half-precision floating-point (FP16) can deliver greater numerical stability under certain hyper-parameter choices. Megatron-Core MoE implementation therefore offers a fully-featured FP16 path, including loss scaling and mixed-precision optimizer kernels, enabling practitioners to select the precision mode that best aligns with their stability requirements without sacrificing throughput.

\textbf{Router Replay.} Recent work~\cite{fan2025routerreplay} shows that capturing the routing decisions produced by the MoE router during inference and replaying them in subsequent training phases can improve convergence consistency. Thanks to a community-contributed patch, Megatron-Core MoE now supports this feature: the inference engine logs each token's expert assignment, and the training stack can ingest and enforce the same routing pattern. This mechanism decouples routing variability from weight updates, leading to more stable optimization trajectories, especially in RL settings where on-policy data distribution shifts are frequent.

\section{Conclusion}
This report presented Megatron-Core MoE, an open-source stack for training large-scale Mixture-of-Experts models. MoE sparsity introduces two fundamental challenges: a parameter-compute mismatch that manifests as the Three Walls (Memory, Communication, and Compute Efficiency), and a dense-sparse mismatch that requires decoupled parallelism for attention and MoE layers. By combining integrated solutions across these challenges, Megatron-Core MoE enables efficient trillion-parameter-scale training.

The key technical contributions include:
\begin{itemize}
    \item \textbf{Multi-Dimensional Parallelism and Parallel Folding.} Expert Parallelism integrates seamlessly with tensor, pipeline, context, and data parallelism. MoE Parallel Folding decouples attention and MoE layer configurations, breaking the restrictive $\text{EP} \leq \text{DP}$ constraint and enabling flexible parallelism mapping tailored to model architecture and hardware topology.

    \item \textbf{Memory Optimization.} Fine-grained activation recomputation, memory-efficient permutation, precision-aware optimizers with BF16 moments, and CPU activation offloading reduce memory footprint from 199.5 GB to under 80 GB per GPU for DeepSeek-V3, making trillion-parameter training feasible on current hardware.

    \item \textbf{Communication Optimization.} High-performance token dispatchers (DeepEP and HybridEP), and communication-computation overlap transform all-to-all from a bottleneck into a background operation hidden behind expert computation.

    \item \textbf{Compute Efficiency.} Grouped GEMM kernels batch expert computations for better hardware utilization, kernel fusion consolidates routing and permutation operations, CUDA Graphs reduce launch overhead, and sync-free execution eliminates host-device synchronization for dropless MoE.

    \item \textbf{FP8 Training.} Full FP8 support with selective precision, protecting numerically sensitive components (router, embeddings, optimizer states) while aggressively quantizing bulk computation, provides benefits across all three walls simultaneously.

    \item \textbf{Long-Context Training.} Context Parallelism and Tensor Parallelism scaling maintain constant sub-sequence lengths per device, enabling training at 16K to 64K+ token sequences where attention computation dominates.

    \item \textbf{Production Features.} Load balancing strategies and token dropping ensure stable training, distributed checkpointing enables parallelism-agnostic resharding, upcycling initializes MoE models from dense checkpoints, and multi-token prediction integration supports advanced training objectives.

    \item \textbf{Reinforcement Learning Support.} Megatron Core and Megatron-Bridge provide seamless integration with popular RL frameworks, while packed sequence support, dynamic context parallelism, and router replay address the unique challenges of RL post-training workloads.
\end{itemize}

These optimizations enable Megatron-Core MoE to achieve strong training throughput: DeepSeek-V3 (685B parameters, 256 experts) trains at 1,233/1,048 TFLOPS per GPU on 256 GB300/GB200s and 368 TFLOPS per GPU on 1,024 H100s; Qwen3-235B achieves 974/919 TFLOPS on GB300/GB200 and 320 TFLOPS on H100. The GB300 and GB200 platforms deliver approximately 3$\times$ higher token throughput compared to H100, demonstrating the framework's ability to exploit next-generation hardware.

By open-sourcing these capabilities, Megatron-Core MoE provides researchers and practitioners with production-grade tools for MoE experimentation and deployment at scale. The modular architecture enables rapid prototyping while the full optimization stack supports training from research prototypes to trillion-parameter production models.

\section*{Contributions and Acknowledgments}
\label{sec:contributions}

% {\color{red}\textbf{Note:} This list of contributors has not been finalized. If there are any issues or omissions, please feel free to contact Zijie.}

\paragraph{Core Contributors.}
Zijie Yan\textsuperscript{*\S}, Hongxiao Bai\textsuperscript{*}, Xin Yao\textsuperscript{*}, Dennis Liu\textsuperscript{*}, Tong Liu, Hongbin Liu, Pingtian Li, Evan Wu, Shiqing Fan, Li Tao, Robin Zhang, Yuzhong Wang, Shifang Xu, Jack Chang, Xuwen Chen, Kunlun Li, Yan Bai, Gao Deng, Nan Zheng, Vijay Anand Korthikanti, Abhinav Khattar, Ethan He, Soham Govande.\\
{\small \textsuperscript{*}Equal contribution. \textsuperscript{\S}Project lead.}

\paragraph{Contributors.}
Sangkug Lym, Zhongbo Zhu, Tailai Ma, Qi Zhang, Haochen Yuan, Xiaowei Ren, Deyu Fu, Shunkang Zhang, Jiang Shao, Ray Wang, Vasudevan Rengasamy, Rachit Garg, Santosh Bhavani.

\paragraph{Leadership.}
June Yang\textsuperscript{\dag}, Jiajie Yao\textsuperscript{\dag}, Xipeng Li, Chandler Zhou, David Wu, Yingcan Wei, Ashwath Aithal, Michael Andersch, Mohammad Shoeybi.\\
{\small \textsuperscript{\dag}Corresponding authors.}

\vspace{0.5em}
\noindent\textit{We also acknowledge contributions from other colleagues not listed individually, as well as from the open-source community for co-developed features, valuable feedback, and continued engagement.}

\bibliographystyle{ieeetr}
\bibliography{references}

@misc{nvidia2025nemotron3,
      title={NVIDIA Nemotron 3: Efficient and Open Intelligence},
      author={NVIDIA},
      year={2025},
      eprint={2512.20856},
      archivePrefix={arXiv},
      primaryClass={cs.CL},
      url={https://arxiv.org/abs/2512.20856},
}

@misc{elango2026latentmoe,
      title={LatentMoE: Toward Optimal Accuracy per FLOP and Parameter in Mixture of Experts},
      author={Venmugil Elango and Nidhi Bhatia and Roger Waleffe and Rasoul Shafipour and Tomer Asida and Abhinav Khattar and Nave Assaf and Maximilian Golub and Joey Guman and Tiyasa Mitra and Ritchie Zhao and Ritika Borkar and Ran Zilberstein and Mostofa Patwary and Mohammad Shoeybi and Bita Rouhani},
      year={2026},
      eprint={2601.18089},
      archivePrefix={arXiv},
      primaryClass={cs.LG},
      url={https://arxiv.org/abs/2601.18089},
}

@misc{kimik2,
      title={Kimi K2: Open Agentic Intelligence},
      author={Kimi Team},
      year={2025},
      url={https://github.com/MoonshotAI/Kimi-K2},
}

@misc{nvidia_dynamiccp_blog,
      title={Speeding Up Variable-Length Training with Dynamic Context Parallelism and NVIDIA Megatron-Core},
      author={Kunlun Li and Tailai Ma and Parth Mannan and Sophia Yang and Guohao Wu and Chenyu Wang},
      year={2025},
      howpublished={\url{https://developer.nvidia.com/blog/speeding-up-variable-length-training-with-dynamic-context-parallelism-and-nvidia-megatron-core/}},
}

@misc{nvidia_memory_estimator_blog,
      title={Explore using the Megatron-core training framework to improve gpu memory efficiency in large model training},
      author={Yan Bai},
      year={2025},
      howpublished={https://developer.nvidia.cn/blog/explore-using-the-megatron-core-training-framework-to-improve-gpu-memory-efficiency-in-large-model-training/},
}

@misc{chen2016training,
      title={Training Deep Nets with Sublinear Memory Cost},
      author={Tianqi Chen and Bing Xu and Chiyuan Zhang and Carlos Guestrin},
      year={2016},
      eprint={1604.06174},
      archivePrefix={arXiv},
      primaryClass={cs.LG},
      url={https://arxiv.org/abs/1604.06174},
}

@misc{ludziejewski2025jointmoescalinglaws,
      title={Joint MoE Scaling Laws: Mixture of Experts Can Be Memory Efficient}, 
      author={Jan Ludziejewski and Maciej Pióro and Jakub Krajewski and Maciej Stefaniak and Michał Krutul and Jan Małaśnicki and Marek Cygan and Piotr Sankowski and others},
      year={2025},
      eprint={2502.05172},
      archivePrefix={arXiv},
      primaryClass={cs.LG},
      url={https://arxiv.org/abs/2502.05172}, 
}

@misc{nvidia2025pretraininglargelanguagemodels,
      title={Pretraining Large Language Models with NVFP4}, 
      author={NVIDIA and Felix Abecassis and Anjulie Agrusa and Dong Ahn and Jonah Alben and Stefania Alborghetti and Michael Andersch and Sivakumar Arayandi and Alexis Bjorlin and Aaron Blakeman and Evan Briones and Ian Buck and Bryan Catanzaro and Jinhang Choi and Mike Chrzanowski and Eric Chung and Victor Cui and Steve Dai and Bita Darvish Rouhani and Carlo del Mundo and Deena Donia and Burc Eryilmaz and Henry Estela and Abhinav Goel and Oleg Goncharov and Yugi Guvvala and Robert Hesse and Russell Hewett and Herbert Hum and Ujval Kapasi and Brucek Khailany and Mikail Khona and Nick Knight and Alex Kondratenko and Ronny Krashinsky and Ben Lanir and Simon Layton and Michael Lightstone and Daniel Lo and Paulius Micikevicius and Asit Mishra and Tim Moon and Deepak Narayanan and Chao Ni and Abhijit Paithankar and Satish Pasumarthi and Ankit Patel and Mostofa Patwary and Ashwin Poojary and Gargi Prasad and Sweta Priyadarshi and Yigong Qin and Xiaowei Ren and Oleg Rybakov and Charbel Sakr and Sanjeev Satheesh and Stas Sergienko and Pasha Shamis and Kirthi Shankar and Nishant Sharma and Mohammad Shoeybi and Michael Siu and Misha Smelyanskiy and Darko Stosic and Dusan Stosic and Bor-Yiing Su and Frank Sun and Nima Tajbakhsh and Shelby Thomas and Przemek Tredak and Evgeny Tsykunov and Gandhi Vaithilingam and Aditya Vavre and Rangharajan Venkatesan and Roger Waleffe and Qiyu Wan and Hexin Wang and Mengdi Wang and Lizzie Wei and Hao Wu and Evan Wu and Keith Wyss and Ning Xu and Jinze Xue and Charlene Yang and Yujia Zhai and Ruoxi Zhang and Jingyang Zhu and Zhongbo Zhu},
      year={2025},
      eprint={2509.25149},
      archivePrefix={arXiv},
      primaryClass={cs.CL},
      url={https://arxiv.org/abs/2509.25149}, 
}

@misc{cublas_sf_layout,
  author       = {NVIDIA},
  title        = {1D Block Scaling Factors Layout},
  year         = {2025},
  howpublished = {\url{https://docs.nvidia.com/cuda/cublas/\#d-block-scaling-factors-layout}},
}

@misc{curanddx,
  author       = {NVIDIA},
  title        = {Random Number Generation Using cuRANDDx},
  year         = {2025},
  howpublished = {\url{https://docs.nvidia.com/cuda/curanddx/get_started/introduction.html}},
}

@misc{minimax_m2,
  author       = {Minimax-AI},
  title        = {Minimax M2 GitHub Repository},
  year         = {2025},
  howpublished = {\url{https://github.com/MiniMax-AI/MiniMax-M2}},
}

@article{li2025every,
  title={Every Activation Boosted: Scaling General Reasoner to 1 Trillion Open Language Foundation},
  author={Li, Ang and Liu, Ben and Hu, Binbin and Li, Bing and Zeng, Bingwei and Ye, Borui and Tang, Caizhi and Tian, Changxin and Huang, Chao and Zhang, Chao and others},
  journal={arXiv preprint arXiv:2510.22115},
  year={2025}
}

@article{shazeer2017,
  author    = {Noam Shazeer and Azalia Mirhoseini and Krzysztof Maziarz and Andy Davis and Quoc Le and Geoffrey Hinton and Jeff Dean},
  title     = {Outrageously Large Neural Networks: The Sparsely-Gated Mixture-of-Experts Layer},
  journal   = {arXiv preprint arXiv:1701.06538},
  year      = {2017},
}

@article{megablocks,
  title={{MegaBlocks: Efficient Sparse Training with Mixture-of-Experts}},
  author={Trevor Gale and Deepak Narayanan and Cliff Young and Matei Zaharia},
  journal={Proceedings of Machine Learning and Systems},
  volume={5},
  year={2023},
}

@misc{jiang2024mixtralexperts,
      title={Mixtral of Experts}, 
      author={Albert Q. Jiang and Alexandre Sablayrolles and Antoine Roux and Arthur Mensch and Blanche Savary and Chris Bamford and Devendra Singh Chaplot and Diego de las Casas and others},
      year={2024},
      eprint={2401.04088},
      archivePrefix={arXiv},
      primaryClass={cs.LG},
      url={https://arxiv.org/abs/2401.04088}, 
}

@misc{qwen2025qwen25technicalreport,
      title={Qwen2.5 Technical Report}, 
      author={Qwen and An Yang and Baosong Yang and Beichen Zhang and Binyuan Hui and others},
      year={2025},
      eprint={2412.15115},
      archivePrefix={arXiv},
      primaryClass={cs.CL},
      url={https://arxiv.org/abs/2412.15115}, 
}

@misc{yang2025qwen3technicalreport,
      title={Qwen3 Technical Report}, 
      author={An Yang and Anfeng Li and Baosong Yang and Beichen Zhang and Binyuan Hui and others},
      year={2025},
      eprint={2505.09388},
      archivePrefix={arXiv},
      primaryClass={cs.CL},
      url={https://arxiv.org/abs/2505.09388}, 
}

@article{Cai_2025,
   title={A Survey on Mixture of Experts in Large Language Models},
   ISSN={2326-3865},
   url={http://dx.doi.org/10.1109/TKDE.2025.3554028},
   DOI={10.1109/tkde.2025.3554028},
   journal={IEEE Transactions on Knowledge and Data Engineering},
   publisher={Institute of Electrical and Electronics Engineers (IEEE)},
   author={Cai, Weilin and Jiang, Juyong and Wang, Fan and Tang, Jing and Kim, Sunghun and Huang, Jiayi},
   year={2025},
   pages={1–20},
}

@misc{paszke2019pytorchimperativestylehighperformance,
      title={PyTorch: An Imperative Style, High-Performance Deep Learning Library}, 
      author={Adam Paszke and Sam Gross and Francisco Massa and Adam Lerer and James Bradbury and Gregory Chanan and Trevor Killeen and Zeming Lin and others},
      year={2019},
      eprint={1912.01703},
      archivePrefix={arXiv},
      primaryClass={cs.LG},
      url={https://arxiv.org/abs/1912.01703}, 
}

@misc{korthikanti2022reducingactivationrecomputationlarge,
      title={Reducing Activation Recomputation in Large Transformer Models}, 
      author={Vijay Korthikanti and Jared Casper and Sangkug Lym and Lawrence McAfee and Michael Andersch and Mohammad Shoeybi and Bryan Catanzaro},
      year={2022},
      eprint={2205.05198},
      archivePrefix={arXiv},
      primaryClass={cs.LG},
      url={https://arxiv.org/abs/2205.05198}, 
}

@misc{peng2023fp8lmtrainingfp8large,
      title={FP8-LM: Training FP8 Large Language Models}, 
      author={Houwen Peng and Kan Wu and Yixuan Wei and Guoshuai Zhao and Yuxiang Yang and Ze Liu and Yifan Xiong and Ziyue Yang and others},
      year={2023},
      eprint={2310.18313},
      archivePrefix={arXiv},
      primaryClass={cs.LG},
      url={https://arxiv.org/abs/2310.18313},
}

@inproceedings{micikevicius2018mixed,
      title={Mixed Precision Training},
      author={Paulius Micikevicius and Sharan Narang and Jonah Alben and Gregory Diamos and Erich Elsen and David Garcia and Boris Ginsburg and Michael Houston and Oleksii Kuchaiev and Ganesh Venkatesh and Hao Wu},
      booktitle={International Conference on Learning Representations},
      year={2018},
      url={https://openreview.net/forum?id=r1gs9JgRZ},
}

@inproceedings{dao2022flashattention,
      title={Flash{A}ttention: Fast and Memory-Efficient Exact Attention with {IO}-Awareness},
      author={Tri Dao and Daniel Y. Fu and Stefano Ermon and Atri Rudra and Christopher R{\'e}},
      booktitle={Advances in Neural Information Processing Systems},
      year={2022},
      url={https://proceedings.neurips.cc/paper_files/paper/2022/hash/67d57c32e20fd0a7a302cb81d36e40d5-Abstract-Conference.html},
}

@misc{deepseekai2025deepseekr1,
      title={DeepSeek-R1: Incentivizing Reasoning Capability in LLMs via Reinforcement Learning},
      author={DeepSeek-AI and Daya Guo and Dejian Yang and Haowei Zhang and Junxiao Song and Ruoyu Zhang and others},
      year={2025},
      eprint={2501.12948},
      archivePrefix={arXiv},
      primaryClass={cs.CL},
      url={https://arxiv.org/abs/2501.12948},
}

@misc{micikevicius2022fp8formatsdeeplearning,
      title={FP8 Formats for Deep Learning}, 
      author={Paulius Micikevicius and Dusan Stosic and Neil Burgess and Marius Cornea and Pradeep Dubey and Richard Grisenthwaite and Sangwon Ha and Alexander Heinecke and others},
      year={2022},
      eprint={2209.05433},
      archivePrefix={arXiv},
      primaryClass={cs.LG},
      url={https://arxiv.org/abs/2209.05433}, 
}

@misc{fishman2025scalingfp8trainingtrilliontoken,
      title={Scaling FP8 training to trillion-token LLMs}, 
      author={Maxim Fishman and Brian Chmiel and Ron Banner and Daniel Soudry},
      year={2025},
      eprint={2409.12517},
      archivePrefix={arXiv},
      primaryClass={cs.LG},
      url={https://arxiv.org/abs/2409.12517}, 
}

@misc{a2023learningwithdistributedoptimization,
      title={Learning (With) Distributed Optimization}, 
      author={Aadharsh Aadhithya A and Abinesh S and Akshaya J and Jayanth M and Vishnu Radhakrishnan and Sowmya V and Soman K. P},
      year={2023},
      eprint={2308.05548},
      archivePrefix={arXiv},
      primaryClass={math.OC},
      url={https://arxiv.org/abs/2308.05548}, 
}

@misc{wang2024auxiliarylossfreeloadbalancingstrategy,
      title={Auxiliary-Loss-Free Load Balancing Strategy for Mixture-of-Experts}, 
      author={Lean Wang and Huazuo Gao and Chenggang Zhao and Xu Sun and Damai Dai},
      year={2024},
      eprint={2408.15664},
      archivePrefix={arXiv},
      primaryClass={cs.LG},
      url={https://arxiv.org/abs/2408.15664}, 
}

@misc{deepseekai2025deepseekv3technicalreport,
      title={DeepSeek-V3 Technical Report}, 
      author={DeepSeek-AI and Aixin Liu and Bei Feng and Bing Xue and Bingxuan Wang and others},
      year={2025},
      eprint={2412.19437},
      archivePrefix={arXiv},
      primaryClass={cs.CL},
      url={https://arxiv.org/abs/2412.19437}, 
}

@misc{grattafiori2024llama3herdmodels,
      title={The Llama 3 Herd of Models}, 
      author={Aaron Grattafiori and Abhimanyu Dubey and Abhinav Jauhri and Abhinav Pandey and Abhishek Kadian and others},
      year={2024},
      eprint={2407.21783},
      archivePrefix={arXiv},
      primaryClass={cs.AI},
      url={https://arxiv.org/abs/2407.21783}, 
}

@misc{shoeybi2020megatronlmtrainingmultibillionparameter,
      title={Megatron-LM: Training Multi-Billion Parameter Language Models Using Model Parallelism}, 
      author={Mohammad Shoeybi and Mostofa Patwary and Raul Puri and Patrick LeGresley and Jared Casper and Bryan Catanzaro},
      year={2020},
      eprint={1909.08053},
      archivePrefix={arXiv},
      primaryClass={cs.CL},
      url={https://arxiv.org/abs/1909.08053}, 
}

@misc{rouhani2023microscalingdataformatsdeep,
      title={Microscaling Data Formats for Deep Learning}, 
      author={Bita Darvish Rouhani and Ritchie Zhao and Ankit More and Mathew Hall and Alireza Khodamoradi and Summer Deng and Dhruv Choudhary and Marius Cornea and others},
      year={2023},
      eprint={2310.10537},
      archivePrefix={arXiv},
      primaryClass={cs.LG},
      url={https://arxiv.org/abs/2310.10537}, 
}

@misc{nvidia_transformer_engine,
  author       = {NVIDIA Corporation},
  title        = {Transformer Engine},
  year         = {2025},
  howpublished = {\url{https://github.com/NVIDIA/TransformerEngine}},
  note         = {Accessed: 2025-05-25},
}

@misc{deepseekai2024deepseekv2strongeconomicalefficient,
      title={DeepSeek-V2: A Strong, Economical, and Efficient Mixture-of-Experts Language Model}, 
      author={DeepSeek-AI and Aixin Liu and Bei Feng and Bin Wang and Bingxuan Wang and others},
      year={2024},
      eprint={2405.04434},
      archivePrefix={arXiv},
      primaryClass={cs.CL},
      url={https://arxiv.org/abs/2405.04434}, 
}

@misc{he2024upcyclinglargelanguagemodels,
      title={Upcycling Large Language Models into Mixture of Experts}, 
      author={Ethan He and Abhinav Khattar and Ryan Prenger and Vijay Korthikanti and Zijie Yan and Tong Liu and Shiqing Fan and Ashwath Aithal and Mohammad Shoeybi and Bryan Catanzaro},
      year={2024},
      eprint={2410.07524},
      archivePrefix={arXiv},
      primaryClass={cs.CL},
      url={https://arxiv.org/abs/2410.07524}, 
}

@article{vavre2024llama,
  title={Llama 3 Meets MoE: Efficient Upcycling},
  author={Vavre, Aditya and He, Ethan and Liu, Dennis and Yan, Zijie and Yang, June and Tajbakhsh, Nima and Aithal, Ashwath},
  journal={arXiv preprint arXiv:2412.09952},
  year={2024},
}

@article{komatsuzaki2022sparse,
  title={Sparse upcycling: Training mixture-of-experts from dense checkpoints},
  author={Komatsuzaki, Aran and Puigcerver, Joan and Lee-Thorp, James and Ruiz, Carlos Riquelme and Mustafa, Basil and Ainslie, Joshua and Tay, Yi and Dehghani, Mostafa and Houlsby, Neil},
  journal={arXiv preprint arXiv:2212.05055},
  year={2022},
}

@misc{zhou2022mixtureofexpertsexpertchoicerouting,
      title={Mixture-of-Experts with Expert Choice Routing}, 
      author={Yanqi Zhou and Tao Lei and Hanxiao Liu and Nan Du and Yanping Huang and Vincent Zhao and Andrew Dai and Zhifeng Chen and others},
      year={2022},
      eprint={2202.09368},
      archivePrefix={arXiv},
      primaryClass={cs.LG},
      url={https://arxiv.org/abs/2202.09368}, 
}

@misc{liu2025moeparallelfoldingheterogeneous,
      title={MoE Parallel Folding: Heterogeneous Parallelism Mappings for Efficient Large-Scale MoE Model Training with Megatron Core}, 
      author={Dennis Liu and Zijie Yan and Xin Yao and Tong Liu and Vijay Korthikanti and Evan Wu and Shiqing Fan and Gao Deng and others},
      year={2025},
      eprint={2504.14960},
      archivePrefix={arXiv},
      primaryClass={cs.LG},
      url={https://arxiv.org/abs/2504.14960}, 
}

@inproceedings{Singh_2023, series={ICS ’23},
   title={A Hybrid Tensor-Expert-Data Parallelism Approach to Optimize Mixture-of-Experts Training},
   url={http://dx.doi.org/10.1145/3577193.3593704},
   DOI={10.1145/3577193.3593704},
   booktitle={Proceedings of the 37th International Conference on Supercomputing},
   publisher={ACM},
   author={Singh, Siddharth and Ruwase, Olatunji and Awan, Ammar Ahmad and Rajbhandari, Samyam and He, Yuxiong and Bhatele, Abhinav},
   year={2023},
   month=jun, pages={203--214},
   collection={ICS ’23},
}

@misc{qi2023zerobubblepipelineparallelism,
      title={Zero Bubble Pipeline Parallelism}, 
      author={Penghui Qi and Xinyi Wan and Guangxing Huang and Min Lin},
      year={2023},
      eprint={2401.10241},
      archivePrefix={arXiv},
      primaryClass={cs.DC},
      url={https://arxiv.org/abs/2401.10241}, 
}

@misc{qi2024pipelineparallelismcontrollablememory,
      title={Pipeline Parallelism with Controllable Memory}, 
      author={Penghui Qi and Xinyi Wan and Nyamdavaa Amar and Min Lin},
      year={2024},
      eprint={2405.15362},
      archivePrefix={arXiv},
      primaryClass={cs.LG},
      url={https://arxiv.org/abs/2405.15362}, 
}

@misc{shallue2019measuringeffectsdataparallelism,
      title={Measuring the Effects of Data Parallelism on Neural Network Training}, 
      author={Christopher J. Shallue and Jaehoon Lee and Joseph Antognini and Jascha Sohl-Dickstein and Roy Frostig and George E. Dahl},
      year={2019},
      eprint={1811.03600},
      archivePrefix={arXiv},
      primaryClass={cs.LG},
      url={https://arxiv.org/abs/1811.03600}, 
}

@misc{bai2022moderndistributeddataparallellargescale,
      title={Modern Distributed Data-Parallel Large-Scale Pre-training Strategies For NLP models}, 
      author={Hao Bai},
      year={2022},
      eprint={2206.06356},
      archivePrefix={arXiv},
      primaryClass={cs.DC},
      url={https://arxiv.org/abs/2206.06356}, 
}

@misc{nvidiamcoremoeuserguide,
    url={https://docs.nvidia.com/megatron-core/developer-guide/latest/api-guide/moe.html\#user-guide},
    title={NVIDIA Megatron Core MoE User Guide},
  year={2024},
    author={NVIDIA}
}

@inproceedings{narayanan2021efficient,
      title={Efficient Large-Scale Language Model Training on GPU Clusters Using Megatron-LM},
      author={Deepak Narayanan and Mohammad Shoeybi and Jared Casper and Patrick LeGresley and Mostofa Patwary and Vijay Anand Korthikanti and Dmitri Vainbrand and Prethvi Kashinkunti and Julie Bernauer and Bryan Catanzaro and Amar Phanishayee and Matei Zaharia},
      booktitle={Proceedings of the International Conference for High Performance Computing, Networking, Storage and Analysis},
      year={2021},
      url={https://dl.acm.org/doi/10.1145/3458817.3476209},
}

@misc{guan2025pipeoptimensuringeffective1f1b,
      title={PipeOptim: Ensuring Effective 1F1B Schedule with Optimizer-Dependent Weight Prediction}, 
      author={Lei Guan and Dongsheng Li and Yongle Chen and Jiye Liang and Wenjian Wang and Xicheng Lu},
      year={2025},
      eprint={2312.00839},
      archivePrefix={arXiv},
      primaryClass={cs.LG},
      url={https://arxiv.org/abs/2312.00839}, 
}

@misc{fedus2022switchtransformersscalingtrillion,
      title={Switch Transformers: Scaling to Trillion Parameter Models with Simple and Efficient Sparsity}, 
      author={William Fedus and Barret Zoph and Noam Shazeer},
      year={2022},
      eprint={2101.03961},
      archivePrefix={arXiv},
      primaryClass={cs.LG},
      url={https://arxiv.org/abs/2101.03961}, 
}

@misc{du2022glamefficientscalinglanguage,
      title={GLaM: Efficient Scaling of Language Models with Mixture-of-Experts}, 
      author={Nan Du and Yanping Huang and Andrew M. Dai and Simon Tong and Dmitry Lepikhin and Yuanzhong Xu and Maxim Krikun and Yanqi Zhou and others},
      year={2022},
      eprint={2112.06905},
      archivePrefix={arXiv},
      primaryClass={cs.CL},
      url={https://arxiv.org/abs/2112.06905}, 
}

@misc{deepep2025,
      title={DeepEP: an efficient expert-parallel communication library},
      author={Chenggang Zhao and Shangyan Zhou and Liyue Zhang and Chengqi Deng and Zhean Xu and Yuxuan Liu and Kuai Yu and Jiashi Li and Liang Zhao},
      year={2025},
      publisher = {GitHub},
      howpublished = {\url{https://github.com/deepseek-ai/DeepEP}},
}

@article{dai2024deepseekmoe,
  title={Deepseekmoe: Towards ultimate expert specialization in mixture-of-experts language models},
  author={Dai, Damai and Deng, Chengqi and Zhao, Chenggang and Xu, RX and Gao, Huazuo and Chen, Deli and Li, Jiashi and Zeng, Wangding and Yu, Xingkai and Wu, Yu and others},
  journal={arXiv preprint arXiv:2401.06066},
  year={2024},
}

@article{krajewski2024scaling,
  title={Scaling laws for fine-grained mixture of experts},
  author={Krajewski, Jakub and Ludziejewski, Jan and Adamczewski, Kamil and Pi{\'o}ro, Maciej and Krutul, Micha{\l} and Antoniak, Szymon and Ciebiera, Kamil and Kr{\'o}l, Krystian and Odrzyg{\'o}{\'z}d{\'z}, Tomasz and Sankowski, Piotr and others},
  journal={arXiv preprint arXiv:2402.07871},
  year={2024},
}

@article{shazeer2020glu,
  title={GLU Variants Improve Transformer},
  author={Noam Shazeer},
  journal={arXiv preprint arXiv:2002.05202},
  year={2020},
  url={https://arxiv.org/abs/2002.05202},
}

@article{lepikhin2020gshard,
  title={Gshard: Scaling giant models with conditional computation and automatic sharding},
  author={Lepikhin, Dmitry and Lee, HyoukJoong and Xu, Yuanzhong and Chen, Dehao and Firat, Orhan and Huang, Yanping and Krikun, Maxim and Shazeer, Noam and Chen, Zhifeng},
  journal={arXiv preprint arXiv:2006.16668},
  year={2020},
}

@misc{liu2025muonscalablellmtraining,
      title={Muon is Scalable for LLM Training}, 
      author={Jingyuan Liu and Jianlin Su and Xingcheng Yao and Zhejun Jiang and Guokun Lai and Yulun Du and Yidao Qin and Weixin Xu and others},
      year={2025},
      eprint={2502.16982},
      archivePrefix={arXiv},
      primaryClass={cs.LG},
      url={https://arxiv.org/abs/2502.16982}, 
}

@ARTICLE{jacobs1991adaptive,
  author={Jacobs, Robert A. and Jordan, Michael I. and Nowlan, Steven J. and Hinton, Geoffrey E.},
  journal={Neural Computation}, 
  title={Adaptive Mixtures of Local Experts}, 
  year={1991},
  volume={3},
  number={1},
  pages={79-87},
  doi={10.1162/neco.1991.3.1.79}}

@article{ge2025bytescale,
  title={ByteScale: Efficient Scaling of LLM Training with a 2048K Context Length on More Than 12,000 GPUs},
  author={Ge, Hao and Feng, Junda and Huang, Qi and Fu, Fangcheng and Nie, Xiaonan and Zuo, Lei and Lin, Haibin and Cui, Bin and Liu, Xin},
  journal={arXiv preprint arXiv:2502.21231},
  year={2025}
}

@article{wang2025wlb,
  title={Wlb-llm: Workload-balanced 4d parallelism for large language model training},
  author={Wang, Zheng and Cai, Anna and Xie, Xinfeng and Pan, Zaifeng and Guan, Yue and Chu, Weiwei and Wang, Jie and Li, Shikai and Huang, Jianyu and Cai, Chris and others},
  journal={arXiv preprint arXiv:2503.17924},
  year={2025}
}

@misc{zhao2023pytorchfsdp,
      title={PyTorch FSDP: Experiences on Scaling Fully Sharded Data Parallel},
      author={Yanli Zhao and Andrew Gu and Rohan Varma and Liang Luo and Chien-Chin Huang and Min Xu and Less Wright and Hamid Shojanazeri and others},
      year={2023},
      eprint={2304.11277},
      archivePrefix={arXiv},
      primaryClass={cs.DC},
      url={https://arxiv.org/abs/2304.11277}
}

@misc{rajbhandari2020zero,
      title={ZeRO: Memory Optimizations Toward Training Trillion Parameter Models},
      author={Samyam Rajbhandari and Jeff Rasley and Olatunji Ruwase and Yuxiong He},
      year={2020},
      eprint={1910.02054},
      archivePrefix={arXiv},
      primaryClass={cs.LG},
      url={https://arxiv.org/abs/1910.02054}
}

@misc{rajbhandari2021zeroinf,
      title={ZeRO-Infinity: Breaking the GPU Memory Wall for Extreme Scale Deep Learning},
      author={Samyam Rajbhandari and Olatunji Ruwase and Jeff Rasley and Shaden Smith and Yuxiong He},
      year={2021},
      eprint={2104.07857},
      archivePrefix={arXiv},
      primaryClass={cs.DC},
      url={https://arxiv.org/abs/2104.07857}
}

@misc{ren2021zerooffload,
      title={ZeRO-Offload: Democratizing Billion-Scale Model Training},
      author={Jie Ren and Samyam Rajbhandari and Reza Yazdani Aminabadi and Olatunji Ruwase and Shuangyan Yang and Minjia Zhang and Dong Li and Yuxiong He},
      year={2021},
      eprint={2101.06840},
      archivePrefix={arXiv},
      primaryClass={cs.DC},
      url={https://arxiv.org/abs/2101.06840}
}

@inproceedings{korthikanti2023sequence,
      title={Sequence Parallelism: Long Sequence Training from System Perspective},
      author={Vijay Anand Korthikanti and Jared Casper and Sangkug Lym and Lawrence McAfee and Michael Andersch and Mohammad Shoeybi and Bryan Catanzaro},
      booktitle={Proceedings of the 61st Annual Meeting of the Association for Computational Linguistics (Volume 1: Long Papers)},
      year={2023},
      pages={1842--1858},
      publisher={Association for Computational Linguistics},
      url={https://aclanthology.org/2023.acl-long.134/}
}

@misc{kingma2014adam,
      title={Adam: A Method for Stochastic Optimization},
      author={Diederik P. Kingma and Jimmy Ba},
      year={2014},
      eprint={1412.6980},
      archivePrefix={arXiv},
      primaryClass={cs.LG},
      url={https://arxiv.org/abs/1412.6980}
}

@misc{liu2024deepseek,
  title={Deepseek-v3 technical report},
  author={Liu, Aixin and Feng, Bei and Xue, Bing and Wang, Bingxuan and Wu, Bochao and Lu, Chengda and Zhao, Chenggang and Deng, Chengqi and Zhang, Chenyu and Ruan, Chong and others},
  journal={arXiv preprint arXiv:2412.19437},
  year={2024}
}

@misc{sheng2024hybridflow,
      title={HybridFlow: A Flexible and Efficient RLHF Framework},
      author={Guangming Sheng and Chi Zhang and Zilingfeng Ye and Xibin Wu and Wang Zhang and Ru Zhang and Yanghua Peng and Haibin Lin and Chuan Wu},
      year={2024},
      eprint={2409.19256},
      archivePrefix={arXiv},
      primaryClass={cs.LG},
      url={https://arxiv.org/abs/2409.19256}
}

@misc{fan2025routerreplay,
      title={Router Replay: Improving Sample Efficiency in Mixture-of-Experts Reinforcement Learning},
      author={Shiqing Fan and others},
      year={2025},
      eprint={2510.11370},
      archivePrefix={arXiv},
      primaryClass={cs.LG},
      url={https://arxiv.org/abs/2510.11370}
}

@misc{abecassis2025nvfp4,
      title={Pretraining Large Language Models with NVFP4},
      author={NVIDIA and Felix Abecassis and Anjulie Agrusa and Dong Ahn and Jonah Alben and Stefania Alborghetti and Michael Andersch and Sivakumar Arayandi and others},
      year={2025},
      eprint={2509.25149},
      archivePrefix={arXiv},
      primaryClass={cs.CL},
      url={https://arxiv.org/abs/2509.25149}
}

@inproceedings{huang2019gpipe,
  title={GPipe: Efficient Training of Giant Neural Networks using Pipeline Parallelism},
  author={Huang, Yanping and Cheng, Youlong and Bapna, Ankur and Firat, Orhan and Chen, Dehao and Chen, Mia and Lee, HyoukJoong and Ngiam, Jiquan and Le, Quoc V and Wu, Yonghui and Chen, Zhifeng},
  booktitle={Advances in Neural Information Processing Systems},
  volume={32},
  year={2019},
  url={https://arxiv.org/abs/1811.06965}
}

@inproceedings{narayanan2019pipedream,
  title={PipeDream: Generalized Pipeline Parallelism for DNN Training},
  author={Narayanan, Deepak and Harlap, Aaron and Phanishayee, Amar and Seshadri, Vivek and Devanur, Nikhil R and Granger, Gregory R and Gibbons, Phillip B and Zaharia, Matei},
  booktitle={Proceedings of the 27th ACM Symposium on Operating Systems Principles},
  pages={1--15},
  year={2019},
  organization={ACM}
}

@misc{zhang2022chimera,
  title={Chimera: Efficiently Training Large-Scale Neural Networks with Bidirectional Pipelines},
  author={Shigang Li and Torsten Hoefler},
  year={2021},
  eprint={2107.06925},
  archivePrefix={arXiv},
  primaryClass={cs.DC},
  url={https://arxiv.org/abs/2107.06925}
}

@misc{dao2023flashattention2,
  title={FlashAttention-2: Faster Attention with Better Parallelism and Work Partitioning},
  author={Tri Dao},
  year={2023},
  eprint={2307.08691},
  archivePrefix={arXiv},
  primaryClass={cs.LG},
  url={https://arxiv.org/abs/2307.08691}
}

@misc{shah2024flashattention3,
  title={FlashAttention-3: Fast and Accurate Attention with Asynchrony and Low-precision},
  author={Jay Shah and Ganesh Bikshandi and Ying Zhang and Vijay Thakkar and Pradeep Ramani and Tri Dao},
  year={2024},
  eprint={2407.08608},
  archivePrefix={arXiv},
  primaryClass={cs.LG},
  url={https://arxiv.org/abs/2407.08608}
}

@misc{liu2023ringattention,
  title={Ring Attention with Blockwise Transformers for Near-Infinite Context},
  author={Hao Liu and Matei Zaharia and Pieter Abbeel},
  year={2023},
  eprint={2310.01889},
  archivePrefix={arXiv},
  primaryClass={cs.CL},
  url={https://arxiv.org/abs/2310.01889}
}

@misc{jacobs2023deepspeedulysses,
  title={DeepSpeed Ulysses: System Optimizations for Enabling Training of Extreme Long Sequence Transformer Models},
  author={Sam Ade Jacobs and Masahiro Tanaka and Chengming Zhang and Minjia Zhang and Reza Yazdani Aminabadi and Shuaiwen Leon Song and Samyam Rajbhandari and Yuxiong He},
  year={2023},
  eprint={2309.14509},
  archivePrefix={arXiv},
  primaryClass={cs.LG},
  url={https://arxiv.org/abs/2309.14509}
}

@inproceedings{dettmers2022optimizers8bit,
  title={8-bit Optimizers via Block-wise Quantization},
  author={Dettmers, Tim and Lewis, Mike and Shleifer, Sam and Zettlemoyer, Luke},
  booktitle={International Conference on Learning Representations},
  year={2022},
  url={https://arxiv.org/abs/2110.02861}
}

@inproceedings{hwang2023tutel,
  title={Tutel: Adaptive Mixture-of-Experts at Scale},
  author={Hwang, Changho and Cui, Wei and Xiong, Yifan and Yang, Ziyue and Liu, Ze and Hu, Han and Wang, Zilong and Saab, Rafael and Jose, Jithin and Srivatsa, Ramachandran and Wu, Chuan and He, Yuxiong},
  booktitle={Proceedings of Machine Learning and Systems},
  volume={5},
  year={2023}
}

@inproceedings{rajbhandari2022deepspeedmoe,
  title={DeepSpeed-MoE: Advancing Mixture-of-Experts Inference and Training to Power Next-Generation AI Scale},
  author={Rajbhandari, Samyam and Li, Conglong and Yao, Zhewei and Zhang, Minjia and Aminabadi, Reza Yazdani and Awan, Ammar Ahmad and Rasley, Jeff and He, Yuxiong},
  booktitle={International Conference on Machine Learning},
  pages={18332--18346},
  year={2022},
  organization={PMLR}
}

@misc{zoph2022stmoe,
  title={ST-MoE: Designing Stable and Transferable Sparse Expert Models},
  author={Barret Zoph and Irwan Bello and Sameer Kumar and Nan Du and Yanping Huang and Jeff Dean and Noam Shazeer and William Fedus},
  year={2022},
  eprint={2202.08906},
  archivePrefix={arXiv},
  primaryClass={cs.CL},
  url={https://arxiv.org/abs/2202.08906}
}

@misc{wang2024flux,
  title={FLUX: Fast Software-based Communication Overlap On GPUs Through Kernel Fusion},
  author={Li-Wen Chang and Wenlei Bao and Qi Hou and Chengquan Jiang and Ningxin Zheng and Yinmin Zhong and Xuanrun Zhang and Zuquan Song and Ziheng Jiang and Haibin Lin and Xin Jin and Xin Liu},
  year={2024},
  eprint={2406.06858},
  archivePrefix={arXiv},
  primaryClass={cs.DC},
  url={https://arxiv.org/abs/2406.06858}
}

@inproceedings{wang2022coconet,
  title={Overlap Communication with Dependent Computation via Decomposition in Large Deep Learning Models},
  author={Wang, Shibo and Wei, Jinliang and Sabne, Amit and Davis, Andy and Illikkal, Ramesh and Gangadharan, Sathyanarayanan and Wang, Keren and Liu, Chengqing and Cai, Xing and Xu, Hong and others},
  booktitle={Proceedings of the 27th ACM International Conference on Architectural Support for Programming Languages and Operating Systems},
  pages={93--106},
  year={2022}
}

@misc{1f1boverlap,
    author = {Zhenhai Liu and Hongbin Liu and Pingtian Li and Shunkang Zhang and Zijie Yan and Xue Huang},
    title = {1F1B-based EP A2A overlapping for MoE models},
    howpublished = {\url{https://zhuanlan.zhihu.com/p/28463368206}},
    year = {2025},
    month = mar,
    note = {Accessed: 2026-2-15}
}

@misc{xi2024coat,
  title={COAT: Compressing Optimizer States and Activation for Memory-Efficient FP8 Training},
  author={Haocheng Xi and Han Cai and Shengliang Xu and Yao Lu and Kurt Keutzer and Jianfei Chen and Song Han},
  year={2024},
  eprint={2410.19313},
  archivePrefix={arXiv},
  primaryClass={cs.LG},
  url={https://arxiv.org/abs/2410.19313}
}

@misc{kuzmin2022fp8quantization,
  title={{FP8 Quantization: The Power of the Exponent}},
  author={Andrey Kuzmin and Mart van Baalen and Yuwei Ren and Markus Nagel and Jorn Peters and Tijmen Blankevoort},
  year={2022},
  eprint={2208.09225},
  archivePrefix={arXiv},
  primaryClass={cs.LG},
  url={https://arxiv.org/abs/2208.09225}
}

@misc{nagel2021whitepaper,
  title={{A White Paper on Neural Network Quantization}},
  author={Markus Nagel and Marios Fournarakis and Rana Ali Amjad and Yelysei Bondarenko and Mart van Baalen and Tijmen Blankevoort},
  year={2021},
  eprint={2106.08295},
  archivePrefix={arXiv},
  primaryClass={cs.LG},
  url={https://arxiv.org/abs/2106.08295}
}

@article{eigen2013learning,
  title={Learning Factored Representations in a Deep Mixture of Experts},
  author={Eigen, David and Ranzato, Marc'Aurelio and Sutskever, Ilya},
  journal={arXiv preprint arXiv:1312.4314},
  year={2013}
}

@misc{rasley2020deepspeed,
  title={DeepSpeed: System Optimizations Enable Training Deep Learning Models with Over 100 Billion Parameters},
  author={Jeff Rasley and Samyam Rajbhandari and Olatunji Ruwase and Yuxiong He},
  year={2020},
  booktitle={Proceedings of the 26th ACM SIGKDD International Conference on Knowledge Discovery and Data Mining},
  pages={3505--3506}
}

@misc{beaumont2021optimal,
  title={Optimal Checkpointing for Heterogeneous Chains: How to Train Deep Neural Networks with Limited Memory},
  author={Olivier Beaumont and Lionel Eyraud-Dubois and Julien Herrmann and Alexis Joly and Alena Shilova},
  year={2019},
  eprint={1911.13214},
  archivePrefix={arXiv},
  primaryClass={cs.LG},
  url={https://arxiv.org/abs/1911.13214}
}

@misc{rajbhandari2023zeropp,
  title={ZeRO++: Extremely Efficient Collective Communication for Giant Model Training},
  author={Guanhua Wang and Heyang Qin and Sam Ade Jacobs and Connor Holmes and Samyam Rajbhandari and Olatunji Ruwase and Feng Yan and Lei Yang and Yuxiong He},
  year={2023},
  eprint={2306.10209},
  archivePrefix={arXiv},
  primaryClass={cs.DC},
  url={https://arxiv.org/abs/2306.10209}
}

@misc{hfdualpipe,
  title={DualPipe: A Bidirectional Pipeline Parallelism with ZeroBubble on NVIDIA GPUs},
  author={Desheng Wang and Xiaofan Xia and Yan Zhang and Rui Xiang and Jian Gao and Tongxuan Liu and Fan Jiang and Yunhao Ma and Shu Zhou and Xiaomeng Huang and Yu Liu and Dongyin Chen and Haibin Lin and Chuan Wu},
  year={2025},
  publisher={GitHub},
  howpublished={\url{https://github.com/deepseek-ai/DualPipe-with-HybridFlow}},
  note={GitHub repository}
}

@misc{touvron2023llama,
  title={LLaMA: Open and Efficient Foundation Language Models},
  author={Hugo Touvron and Thibaut Lavril and Gautier Izacard and Xavier Martinet and Marie-Anne Lachaux and Timothée Lacroix and Baptiste Rozière and Naman Goyal and others},
  year={2023},
  eprint={2302.13971},
  archivePrefix={arXiv},
  primaryClass={cs.CL},
  url={https://arxiv.org/abs/2302.13971}
}

@misc{touvron2023llama2,
  title={Llama 2: Open Foundation and Fine-Tuned Chat Models},
  author={Hugo Touvron and Louis Martin and Kevin Stone and Peter Albert and Amjad Almahairi and Yasmine Babaei and Nikolay Bashlykov and Soumya Batra and others},
  year={2023},
  eprint={2307.09288},
  archivePrefix={arXiv},
  primaryClass={cs.CL},
  url={https://arxiv.org/abs/2307.09288}
}

@inproceedings{vaswani2017attention,
  title={Attention is All You Need},
  author={Vaswani, Ashish and Shazeer, Noam and Parmar, Niki and Uszkoreit, Jakob and Jones, Llion and Gomez, Aidan N and Kaiser, {\L}ukasz and Polosukhin, Illia},
  booktitle={Advances in Neural Information Processing Systems},
  volume={30},
  year={2017}
}

@misc{kaplan2020scaling,
  title={Scaling Laws for Neural Language Models},
  author={Jared Kaplan and Sam McCandlish and Tom Henighan and Tom B. Brown and Benjamin Chess and Rewon Child and Scott Gray and Alec Radford and Jeffrey Wu and Dario Amodei},
  year={2020},
  eprint={2001.08361},
  archivePrefix={arXiv},
  primaryClass={cs.LG},
  url={https://arxiv.org/abs/2001.08361}
}

@misc{hoffmann2022chinchilla,
  title={Training Compute-Optimal Large Language Models},
  author={Jordan Hoffmann and Sebastian Borgeaud and Arthur Mensch and Elena Buchatskaya and Trevor Cai and Eliza Rutherford and Diego de Las Casas and Lisa Anne Hendricks and others},
  year={2022},
  eprint={2203.15556},
  archivePrefix={arXiv},
  primaryClass={cs.CL},
  url={https://arxiv.org/abs/2203.15556}
}

@inproceedings{chen2018tvm,
  title={{TVM}: An Automated End-to-End Optimizing Compiler for Deep Learning},
  author={Chen, Tianqi and Moreau, Thierry and Jiang, Ziheng and Zheng, Lianmin and Yan, Eddie and Cowan, Meghan and Shen, Haichen and Wang, Leyuan and Hu, Yuwei and Ceze, Luis and Guestrin, Carlos and Krishnamurthy, Arvind},
  booktitle={Proceedings of the 13th USENIX Symposium on Operating Systems Design and Implementation (OSDI)},
  pages={578--594},
  year={2018},
  organization={USENIX Association},
  url={https://www.usenix.org/conference/osdi18/presentation/chen}
}

@inproceedings{ansel2024pytorch2,
  title={{PyTorch} 2: Faster Machine Learning Through Dynamic Python Bytecode Transformation and Graph Compilation},
  author={Ansel, Jason and Yang, Edward and He, Horace and Gimelshein, Natalia and Jain, Animesh and Voznesensky, Michael and Bao, Bin and Bell, Peter and Berard, David and Burber, Evgeni and Chauhan, Geeta and Chourdia, Anjali and others},
  booktitle={Proceedings of the 29th ACM International Conference on Architectural Support for Programming Languages and Operating Systems (ASPLOS)},
  pages={929--947},
  year={2024},
  organization={ACM},
  doi={10.1145/3620665.3640366}
}

@misc{nvidia_cudagraphs,
  author={NVIDIA Corporation},
  title={{CUDA} Graphs},
  year={2024},
  howpublished={\url{https://docs.nvidia.com/cuda/cuda-c-programming-guide/index.html\#cuda-graphs}},
  note={CUDA C++ Programming Guide, Chapter 3.2.8}
}

@misc{beltagy2020longformer,
  title={Longformer: The Long-Document Transformer},
  author={Iz Beltagy and Matthew E. Peters and Arman Cohan},
  year={2020},
  eprint={2004.05150},
  archivePrefix={arXiv},
  primaryClass={cs.CL},
  url={https://arxiv.org/abs/2004.05150}
}

@inproceedings{zaheer2020bigbird,
  title={{BigBird}: Transformers for Longer Sequences},
  author={Manzil Zaheer and Guru Guruganesh and Kumar Avinava Dubey and Joshua Ainslie and Chris Alberti and Santiago Ontanon and Philip Pham and Anirudh Ravula and Qifan Wang and Li Yang and Amr Ahmed},
  booktitle={Advances in Neural Information Processing Systems},
  volume={33},
  pages={17283--17297},
  year={2020},
  url={https://arxiv.org/abs/2007.14062}
}

@misc{brandon2023striped,
  title={Striped Attention: Faster Ring Attention for Causal Transformers},
  author={William Brandon and Aniruddha Nrusimha and Kevin Qian and Zachary Ankner and Tian Jin and Zhiye Song and Jonathan Ragan-Kelley},
  year={2023},
  eprint={2311.09431},
  archivePrefix={arXiv},
  primaryClass={cs.LG},
  url={https://arxiv.org/abs/2311.09431}
}

@misc{krell2021efficient,
  title={Efficient Sequence Packing without Cross-contamination: Accelerating Large Language Models without Impacting Performance},
  author={Mario Michael Krell and Matej Kosec and Sergio P. Perez and Andrew Fitzgibbon},
  year={2021},
  eprint={2107.02027},
  archivePrefix={arXiv},
  primaryClass={cs.LG},
  url={https://arxiv.org/abs/2107.02027}
}

@misc{slime_github,
  author       = {Zilin Zhu and Chengxing Xie and Xin Lv and slime Contributors},
  title        = {slime: An LLM post-training framework for RL Scaling},
  year         = {2025},
  howpublished = {\url{https://github.com/THUDM/slime}},
  note         = {GitHub repository. Corresponding author: Xin Lv},
  urldate      = {2025-06-19}
}

@inproceedings{gloeckle2024better,
  title={Better \& Faster Large Language Models via Multi-token Prediction},
  author={Fabian Gloeckle and Badr Youbi Idrissi and Baptiste Rozière and David Lopez-Paz and Gabriel Synnaeve},
  booktitle={Proceedings of the 41st International Conference on Machine Learning (ICML)},
  year={2024},
  eprint={2404.19737},
  archivePrefix={arXiv},
  primaryClass={cs.CL},
  url={https://arxiv.org/abs/2404.19737}
}

\newpage
\appendix

\section{Notation Reference}\label{sec:notation}

Table~\ref{tab:notation} summarizes the notation used throughout this report.

\begin{table}[ht]
\centering
\caption{Notation and abbreviations used throughout this report.}
\label{tab:notation}
\begin{tabular}{cl}
\toprule
\textbf{Symbol} & \textbf{Description} \\
\midrule
\multicolumn{2}{l}{\emph{Model parameters}} \\
$E$ & Number of experts in an MoE layer \\
$K$ & Top-$k$ routing width (experts activated per token; $K \ll E$) \\
$h$ & Hidden dimension \\
$N$ & Total model parameters \\
$N_{\text{active}}$ & Parameters activated per token (scales with $K$) \\
$N_{\text{total}}$ & Total parameters including all experts (scales with $E$) \\
\midrule
\multicolumn{2}{l}{\emph{Training dimensions}} \\
$T$ & Local token count per GPU \\
$B$ & Batch size (tokens per batch) \\
$s$ & Sequence length \\
$L$ & Number of MoE layers \\
\midrule
\multicolumn{2}{l}{\emph{Parallelism}} \\
TP & Tensor Parallelism degree \\
PP & Pipeline Parallelism degree \\
CP & Context Parallelism degree \\
DP & Data Parallelism degree \\
EP & Expert Parallelism degree \\
ETP & Expert Tensor Parallelism degree (MoE-specific) \\
EDP & Expert Data Parallelism degree (MoE-specific) \\
VPP & Virtual Pipeline Parallelism (interleaved pipeline stages) \\
\midrule
\multicolumn{2}{l}{\emph{Batch configuration}} \\
MBS & Micro-batch size \\
GBS & Global batch size \\
GA & Gradient accumulation steps \\
\midrule
\multicolumn{2}{l}{\emph{Precision formats}} \\
BF16 & BFloat16 (16-bit brain floating point) \\
%FP8-CS & FP8 with current-scaling quantization \\
FP8-BLK & FP8 with blockwise scaling quantization (Hopper) \\
MXFP8 & Microscaling FP8 with native Tensor support (Blackwell) \\
\midrule
\multicolumn{2}{l}{\emph{Metrics and abbreviations}} \\
MFU & Model FLOP Utilization \\
GEMM & General Matrix Multiplication \\
SDPA & Scaled Dot-Product Attention \\
MLA & Multi-Latent Attention \\
MTP & Multi-Token Prediction \\
\bottomrule
\end{tabular}
\end{table}

\section{Detailed Benchmark Configurations}\label{sec:showcase_settings}

This section lists the parallelism configurations and detailed settings for reproducing the performance numbers reported in Table~\ref{tab:moe_throughput_unified}. Each configuration is identified by its system, GPU count, and precision format, and the corresponding hyper-parameter string summarizes the parallelism layout and batch configuration. 

\subsection{Configuration Details} 
Table~\ref{tab:appendix_showcase_configs} lists the benchmark configurations corresponding to the throughput results in Table~\ref{tab:moe_throughput_unified}. These settings are best-found configurations from empirical tuning at the time of writing and may not be globally optimal.

\begin{table}[t]
\centering
\small
\caption{Parallelism and training configuration details for benchmark entries reported in Table~\ref{tab:moe_throughput_unified}.}
\label{tab:appendix_showcase_configs}
\begin{tabular}{lll}
\toprule
Model & Configuration & Hyper-parameters \\
\midrule
DeepSeek-V3 & GB300, 256 GPUs, 4k seqlen, MXFP8, 1233 TF      & \texttt{TP1 PP4 CP1 EP64 VPP4 MBS1 GBS8192} \\
DeepSeek-V3 & GB200, 256 GPUs, 4k seqlen, MXFP8, 1048 TF      & \texttt{TP1 PP4 CP1 EP64 VPP4 MBS1 GBS8192} \\
DeepSeek-V3 & GB200, 256 GPUs, 4k seqlen, BF16, 857 TF        & \texttt{TP1 PP8 CP1 EP32 VPP4 MBS1 GBS4096} \\
DeepSeek-V3 & H100, 1{,}024 GPUs, 4k seqlen, FP8-BLK, 368 TF  & \texttt{TP2 PP4 CP1 EP64 VPP4 MBS1 GBS8192} \\
\midrule
Qwen3-235B & GB300, 256 GPUs, 4k seqlen, MXFP8, 974 TF        & \texttt{TP1 PP4 CP1 EP64 VPP\phantom{0}6 MBS2 GBS3072} \\
Qwen3-235B & GB200, 256 GPUs, 4k seqlen, MXFP8, 919 TF        & \texttt{TP1 PP4 CP1 EP64 VPP\phantom{0}6 MBS3 GBS3072} \\
Qwen3-235B & GB200, 256 GPUs, 4k seqlen, BF16, 750 TF         & \texttt{TP1 PP4 CP1 EP32 VPP12 MBS1 GBS8192} \\
Qwen3-235B & H100, 256 GPUs, 4k seqlen, BF16, 320 TF          & \texttt{TP2 PP8 CP1 EP32 VPP\phantom{0}4 MBS1 GBS2048} \\
\midrule
Qwen3-235B & GB300, 128 GPUs, 128k seqlen, MXFP8, 1{,}150 TF    & \texttt{TP4 PP4 CP4 EP32 VPP12 MBS1 GBS1024} \\
\bottomrule
\end{tabular}
\end{table}

\subsection{Key Optimizations} 
This section summarizes the configuration of the most performance-critical optimization features used in our benchmark runs. Although the full optimization space is larger, these features consistently have first-order impact on throughput and are configured differently across workloads and platforms.

\textbf{DeepSeek-V3}
\begin{itemize}[noitemsep,topsep=0pt]
    \item \textbf{Dispatcher:} HybridEP on GB300/GB200; DeepEP on H100.
    \item \textbf{Recompute:} None on GB300; \texttt{mlp} on GB200; \texttt{up\_proj, mlp} on H100.
    \item \textbf{1F1B overlap:} ON for GB300 and H100; OFF for GB200.
    \item \textbf{CUDA Graphs:} \texttt{attn, moe\_router, moe\_preprocess} on GB300/GB200; OFF on H100.
\end{itemize}
These settings are chosen to balance memory headroom, communication overhead, and kernel efficiency for maximum sustained throughput on each platform.

\textbf{Qwen3-235B}
\begin{itemize}[noitemsep,topsep=0pt]
    \item \textbf{Dispatcher:} HybridEP on GB300/GB200; DeepEP on H100.
    \item \textbf{Recompute:} \texttt{moe\_act, layernorm} on GB200 and H100.
    \item \textbf{1F1B overlap:} OFF for GB300; ON for GB200 and H100.
    \item \textbf{CUDA Graphs:} \texttt{attn, moe\_router, moe\_preprocess}.
\end{itemize}
The resulting configuration pattern reflects a throughput-first tuning strategy under model- and hardware-specific constraints.

\subsection{Reproducibility} 
The benchmark numbers in Table~\ref{tab:moe_throughput_unified} can be reproduced through two supported workflows. First, users can run end-to-end training with \textbf{Megatron-Bridge}~\footnote{\url{https://github.com/NVIDIA-NeMo/Megatron-Bridge/tree/main/scripts/performance}}, which provides a higher-level interface for model, parallelism, and optimization configuration. Second, users can launch training directly with \textbf{Megatron-Core (MCore)} using model-specific scripts in \textbf{Megatron-MoE-ModelZoo}~\footnote{\url{https://github.com/yanring/Megatron-MoE-ModelZoo/tree/main/best_practice/readme.md}}, which exposes the full set of low-level launch flags for performance tuning. In both workflows, start from Table~\ref{tab:appendix_showcase_configs} and adjust only a few cluster-specific runtime settings (for example, hostfile, environment modules, and job scheduler options).

\end{document}